\documentclass{aa}
\usepackage{graphicx}
\usepackage{lscape}
\usepackage{natbib}
\usepackage{xtab}
\usepackage{txfonts}
\begin{document}
   \title{Internal kinematics of spiral galaxies in distant clusters
   }

   \subtitle{IV. Gas kinematics of spiral galaxies in intermediate redshift clusters and in the field\thanks{Based on observations collected at the European
   Southern Observatory (ESO), Cerro Paranal, Chile (ESO Nos. 74.B--0592 \& 75.B-0187) and observations of the Hubble Space Telescope (HST No
   10635).}}

   \author{E. Kutdemir\inst{1,2}
	  \and
	  B. L. Ziegler\inst{2}
	  \and
	  R. F. Peletier\inst{1}
	  \and
	  C. Da Rocha\inst{2,3}	 
	  \and
	  A. B\"ohm\inst{4}
	  \and
	  M. Verdugo\inst{5}
	  }

   \institute{Kapteyn Astronomical Institute, PO BOX 800, 9700 AV Groningen, 
         The Netherlands \\
              \email{kutdemir@astro.rug.nl}
	 \and 
	 European-Southern Observatory, Karl-Schwarzschild Str. 2, 85748 Garching, Germany 
         \and
	 N\'ucleo de Astrof\'{\i}sica Te\'orica, Universidade Cruzeiro do Sul, R. Galv\~ao Bueno 868, 01506-000, S\~ao Paulo, SP, Brazil
	 \and
	 Institute of Astro- and Particle Physics, Technikerstrasse 25/8, 6020 Innsbruck, Austria
	 \and
	 Max-Planck-Institut f\"ur extraterrestrische Physik, Giessenbachstra\ss{}e 85748 Garching be M\"unchen, Germany
             \vspace*{3.48mm}}

   \date{Received ;  }

 
  \abstract
  {}
   {We trace the interaction processes of galaxies at intermediate redshift by measuring the irregularity of their ionized gas kinematics,
and investigate these irregularities as a function of the environment (cluster versus field) and of morphological type (spiral versus irregular).
} 
   {We obtain the gas velocity fields by placing three parallel and adjacent VLT/FORS2 slits on each galaxy.  To quantify irregularities in the gas kinematics, we use three indicators: the standard deviation of the
kinematic position angle ($\sigma_{\rm PA}$), the mean deviation of the line of sight velocity profile from the
cosine form which is measured using high order Fourier terms ($k_{3,5}/k_{1}$) and the average misalignment
between the kinematical and photometric major axes ($\Delta \phi$).  These indicators are then examined together
with some photometric and structural parameters {(measured from HST and FORS2 images in the optical)} such as the disk scale length, rest-frame colors, asymmetry,
concentration, Gini coefficient and $M_{20}$.  Our sample consists of 92 distant galaxies.  16 cluster ($z \sim0.3$ and
$z \sim0.5$) and 29 field galaxies ($0.10 \le z \le 0.91$, mean z=0.44) of these have velocity fields
with sufficient signal to be analyzed.  To compare our sample with the local universe, we also analyze a
sample from the SINGS survey.}
 {We find that the fraction of galaxies that have irregular gas kinematics is remarkably similar in galaxy clusters and in the field at intermediate redshifts (according to $\sigma_{\rm PA}
\approx 10\%, k_{3,5}/k_{1} \approx 30\%, \Delta \phi \approx 70\% $).  The distribution of the field and cluster galaxies in (ir)regularity parameters space is also similar.  {On the
other hand galaxies with small central concentration of light, that we see in the field sample, are absent in the cluster sample.  We find that field galaxies at intermediate redshifts have
more irregular velocity fields as well as more clumpy and less centrally concentrated light distributions than their local counterparts.  Comparison with a SINS sample of 11 $z \sim 2$
galaxies shows that these distant galaxies have more irregular gas kinematics than our intermediate redshift cluster and field sample.}  We do not find a dependence of the irregularities in
gas kinematics on morphological type.  We find that two different indicators of star formation correlate with irregularity in the gas kinematics.
}
{More irregular gas kinematics, also more clumpy and less centrally concentrated light distributions of spiral field galaxies at intermediate redshifts in comparison to their local counterparts
indicate that these galaxies are probably still in the process of building their disks via mechanisms such as accretion and mergers.  On the other hand, they have less irregular gas kinematics
compared to galaxies at $z \sim 2$. }

   \maketitle
%

\section{Introduction}

Galaxy clusters are important laboratories for understanding the origin of different morphological types of galaxies. 
The main reason for that is the relation between local galaxy density and morphological type \citep{Dress80}.  For
nearby rich clusters, the spiral galaxy fraction decreases from $80\%$ in the field to $60\%$ in the cluster outskirts
and to virtually zero in the core region.  This relation is redshift dependent and while the fraction of elliptical
galaxies \citep[$\sim15\%$,][]{VHGH04} does not change with redshift, the S0 fraction increases with decreasing
redshift and the spiral fraction, on the contrary, decreases \citep{CBSES98}.  The fraction of spirals with no current
star formation activity is significantly larger in clusters than in the field
\citep{Bergh76,PSDCB99,CBBSG01,GOSBB03,VZG08,Sikkema09}.  Also, distant clusters have a larger fraction of star
forming galaxies compared to nearby clusters \citep{BO78b,BO84a,ELYC01,KB01}.  It is well established with these
observational studies that spiral galaxies have been transformed into S0s mostly in denser regions of the universe. 
Then the question is what are the physical processes that are responsible for this morphological transformation. 
Several mechanisms have been proposed such as gas stripping mechanisms: ram pressure stripping
\citep{GG72,QMB00,KKUSZ08,KKFRS08,KSSFZ09}, viscous stripping \citep{N82} and thermal evaporation \citep{CS77}; tidal
forces due to the cumulative effect of many weak encounters: harassment \citep{R76,MLK98}; removal of the outer
gaseous halos by the hydrodynamic interaction with the intracluster medium (ICM) plus the global tidal field of the
cluster: starvation \citep{LTC80}; mergers and strong galaxy-galaxy interactions which are efficient when relative
velocities are low, and therefore, they mostly occur in galaxy groups or in the outskirts of clusters \citep{MH97}.  It
is known that cluster galaxies loose gas because of their interactions with the ICM and as a result, their star
formation gets switched off. Local studies show that cluster galaxies are deficient in neutral hydrogen compared to
their field counterparts and that becomes significant within the Abell radius \citep{DL73,GH85,G87,G89,SMGGG01}.  The
HI distribution of these galaxies frequently shows asymmetries and displacement from the optical disk as well
\citep{G89,CKBG94,BCBG00,VHGH04}.  It was proposed that passive spiral galaxies in clusters might be the intermediate
phase before becoming an S0 \citep{Bergh76}.  Later on, it was argued that S0's can not be formed by removing gas from
disks of spirals via mechanisms such as ram pressure stripping, since S0's have systematically larger bulge sizes and
bulge to disk ratios (B/D) compared to spiral galaxies in all density regimes (\citealt{B79,Dress80,G80} but see also
\citealt{AS08}).  Tidal interactions (e.g. harassment) on the other hand are expected to trigger gas accretion into
the circumnuclear regions \citep{MLKDO96} and therefore increase the bulge size and B/D ratio.  Therefore, S0's might
have formed via minor mergers, harassment or a combination of the two \citep{DS83,NMRT99,ABP01,HRC03}.  

All these discussions are pointing out that studying stellar populations and morphologies of cluster galaxies is
crucial in understanding the interaction processes.  What about their kinematics?  In the Virgo cluster, for
example, half of 89 spiral galaxies that were observed by \citet{RWK99} turned out to have disturbed gas
kinematics.  Mergers and tidal processes such as harassment are capable of causing disturbances also in stellar
velocity fields \citep{MLQS99,M04}.  ICM-related processes on the other hand, even ram-pressure stripping are
insufficient to be able to affect stellar kinematics of spiral galaxies \citep{QMB00}.  Some attempts have been
made to evaluate the effectiveness of the interaction processes as a function of location in the galaxy cluster. 
\citet{DGHHC01} measured Tully-Fisher Relation (TFR) residuals for 510 cluster spirals and concluded that they do not show a dependence
on distance from the cluster center. \citet{MMTES07} constructed the TFR in both $K_S$ and $V$ bands for 40
cluster, 37 field spirals at intermediate redshift and found that the cluster TFR exhibits significantly
larger scatter than the field relation in both bands and the residuals do not show a clear trend with
$R/R_{vir}$.  They found that the TFR residuals do not correlate with the star formation rate and dust content. 
They also checked whether central surface mass density of galaxies, which can be used to probe the action of
harassment, shows a trend as a function of radius.  They found that it shows a break at approximately $1R_{vir}$,
outside of which spirals exhibit  nearly uniformly low central density values.  They argue that a combination of
merging in the cluster outskirts with harassment in the intermediate and inner cluster regions might explain both
the TFR scatter and the radial trend in density which persist up to $2R_{vir}$.

\begin{table*}
\caption{Basic galaxy cluster information.}
\label{cltable}
\setlength{\tabcolsep}{6.5pt}
\small\centerline {
\begin{tabular}{lccccccccccc}
\hline
name &      $z$ & ref   &  $R_{vir} $ & ref & $\sigma$ & ref & $L_{X} $ & ref &  dynamical state	& ref & N \\
 &	& & $(h^{-1}~Mpc)$ & & $km s^{-1}$ & &	$(10^{44} erg s^{-1})$ &  & &	&\\
\hline
MS~0451.6-0305     &      0.540&(1)	   &	  1.17&(5)  	&	1371&(7)  &    10.19&(10)	& relatively relaxed	&(12)&  4 \\	    
MS~2137.3-2353     &      0.313&(2)	   &	  1.95&(6) 	&	960&(9)   &    7.97&(10)  	& relaxed		&(13)&  5 \\
MS~1008.1-1224     &      0.3061&(3)	   &	  1.18&(5) 	&	1042&(8)  &    2.29&(10)  	& non-relaxed 		&(14)&  5 \\
Cl~0412-65         &      0.507&(4)	   &	  0.62&(5) 	&	700&(4)   &    0.16&(11)  	& 			&    &	2 \\
\hline
\end{tabular}}
\medskip   
References:   (1): \citep{D96}; (2): \citep{SMGMSWFH91}; (3): \citep{YEMAC98}; (4): \citep{DSPBC99}; (5): \citep{GM01}; (6): \citep{ASFE03};
(7):\citep{CYEAGMP96}; (8):\citep{BGCYE99}; (9):\citep{KMPM95}; (10): \citep{LGHLA99}; (11): \citep{SEDCOSB97}; (12): \citep{Don96}; (13): \citep{JCBB05}; (14): \citep{AMWP02}.\\
The last column (N) gives the number of galaxies in each cluster that are used in the kinematical analysis.

\end{table*}

As discussed above, galaxy evolution in clusters is rather complex, since there are several interaction mechanisms
involved.  To understand the nature of these mechanisms, it is important to examine together  morphological and kinematical
properties of cluster galaxies.  In this series of papers we make use of both gas velocity fields and high resolution
images of galaxies in four intermediate-redshift clusters and their field to do that.  Most studies in the literature rely
on long-slit data for identifying kinematical disturbances.  Using a velocity map enables us to have a more accurate
measure of the kinematical (ir)regularity.  A velocity field can be decomposed into velocity, position angle and
inclination of circular orbits at each radius \citep[see][]{KCZC06}.  The deviation of the kinematic major axis (KMA)
around its mean value and the misalignment between KMA and the photometric major axes (PMA) both indicate kinematical
disturbance.  We also make a simple rotating model that has the mean position angle and inclination of the observed
velocity map. The residual of the observed and the simple rotating map is fitted with high order Fourier terms and the
squared sum of these terms is used as another indicator.  We measure these irregularity indicators for both field galaxies
and cluster members and compare them with each other to search for the environmental imprint on gas kinematics.   We then
combine this information with the morphological and photometric properties of these galaxies and investigate whether
certain characteristics make galaxies more sensitive to environmental effects.  We also use the relations between intrinsic
galaxy properties and efficiency of interaction processes, that are known from theory, to investigate which mechanisms are
at work on the cluster galaxies in our sample.

We also investigate the evolution of field galaxies by studying their gas kinematics as a function of redshift.  We measure the irregularities in their gas kinematics both at
intermediate redshifts and in the local universe and compare them with each other.  These results are then compared with the studies of spatially resolved gas kinematics at similar
or higher redshifts:  \citet[][hereafter S08]{SGSTB08} analyzed gas velocity fields and velocity dispersion maps of 11 galaxies at $z\sim2$, observed with SINFONI, and
classified these systems into two categories: merging and non-merging.  They found that more than $50\%$ of these galaxies are consistent with a single rotating disk
interpretation.  With FLAMES at the VLT, \citet[][hereafter Y08]{YFHN08} studied 63 intermediate-mass field galaxies at $0.4<z<0.75$.  Using spatially resolved gas kinematics of these objects,
($2\arcsec\times3\arcsec$ field of view) they find that both velocity fields and velocity dispersion maps of $26\%$ of these galaxies are incompatible with disk rotation.

Spatially resolved velocity fields are essential for quantifying irregularities in the kinematics.  Because the inclination and the position angle of the orbits at each radii can be
assessed and the velocity profile along each orbit can be analyzed and compared with a simple rotating case.  For Tully-Fisher studies the main problem with the long slit data is the fact
that it can be misleading in case the kinematic and photometric axes are misaligned.  The conventional way of obtaining the spatially resolved spectra is using integral field (IFU) spectroscopy
\citep[e.g. S08, Y08,][]{PFHY08}.  We use another approach and place three parallel, adjacent VLT/FORS2 slits on each galaxy.  This novel method has the advantage that we can explore
the velocity fields up to large radii ($\approx 3 \arcsec$, which corresponds to 16 kpc at z$=0.4$) and it is more efficient than IFUs in terms of observing time.  The spatial sampling
along the minor and major axes is $1\arcsec$ and $0\farcs25$ respectively.  We described our method and presented the analysis of our MS~0451 sample in \citet[][hereafter Paper
III]{KZPR08}.  Here, we include in our analysis galaxies in three more intermediate redshift clusters and their field.  A plan of the paper follows. In Sect.~\ref{obs}, we describe our
sample and the improvements in our data reduction technique in comparison with Paper III.  In Sect.~\ref{anal}, we explain the analysis of the data. In Sect.~\ref{discuss}, we
discuss our results and compare them with the literature. Sect.~\ref{conclu} summarizes the results and our conclusions.

Throughout this paper, we assume that the Hubble constant, the matter density and the cosmological constant
are $H_0 = 70$~km\,s$^{-1}$~Mpc$^{-1}$, $\Omega_{\rm m}=0.3$ and $\Omega_{\lambda}=0.7$ respectively
\citep{TSBCC03}.

   \begin{figure}
   \centering
   \includegraphics[width=8cm,clip]{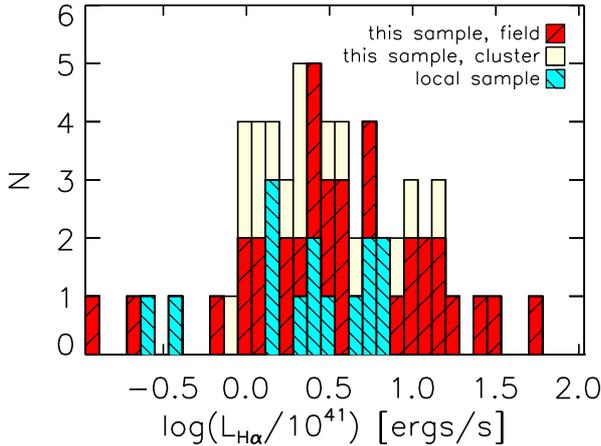}
 \caption{Distribution of $H_\alpha$ luminosities of both our galaxies and a local sample from SINGS.}
         \label{lum_ha}
   \end{figure}
   
\section{Sample and data properties \label{obs}}
\subsection{Sample \label{samp}}

Our sample includes four galaxy clusters which have different properties (Table~\ref{cltable}).  To be able to
compare the galaxies that experience similar environmental conditions, we scale cluster-centric distance by each
cluster's virial radius.  It is known that even galaxy clusters at the same redshift can be very different from
one another. Two well-studied rich clusters in the local universe, Coma and Virgo are a good example for that.
While Coma is dynamically relaxed and spiral poor, Virgo on the contrary is unrelaxed and spiral rich
\citep{Pog06}. 

Here we give some information about each cluster in our sample: MS~0451 is a massive cluster with very high X-ray
luminosity. $28\%$ of its spiral population are passive \citep{METSRS07}; MS~1008 is a very regular and rich cluster
\citep{LEMC99,LGHLA99}; MS~2137 is a rich and dynamically relaxed cluster \citep{JCBB05}; Cl~0412 (F1557.19TC) is a
poor cluster that is not well-studied.  Our cluster selection depended on the availability of their HST/WFPC2 imaging
when the project was initiated in 1999.  \citealp[See][hereafter Paper I]{ZBJHM03} and \citealp[][hereafter Paper II]{JZBHM04} for more detailed
information about the sample selection.

During the target selection, we gave priority to galaxies from our previous studies, both in field and cluster environments, that have detectable emission lines for extracting velocities.  Further objects were drawn from a catalog provided by the CNOC survey \citep{EYAMC98}
with either redshift information or measured (g-r) color that matches expectations for spiral templates at $z\approx0.5$.  If there was an
unused slitlet in the MXU setup, and no suitable candidate was available, a galaxy was picked at random.  For our analysis, we also use a sample
from SINGS as a local reference for comparison (see Paper III).  SINGS galaxies are a diverse set of local normal galaxies \citep{KABCDDE03}.
\citet{DCAHCBK06}'s subsample that we use in our analysis consists of galaxies that have star forming regions, so that their $H_{\alpha}$
kinematics could be extracted.  We excluded from our analysis the galaxies in this sample that have luminosities that are very different from
the luminosities of our intermediate redshift sample, so that the two samples have comparable stellar masses.  In Fig.~\ref{lum_ha}, we compare
the $H_\alpha$ luminosities of galaxies in our sample and in the local sample.  $H_\alpha$ luminosities of the galaxies in our sample were calculated using the available emission lines in their spectra as explained in Sect.~\ref{sfr}.  Since the galaxies in our sample were selected mainly based on
the strength of their emission, several of them have larger $H_\alpha$ luminosities, and therefore, higher star formation rates than the galaxies
of the local sample.  This raises the question whether our sample is biased towards disturbed galaxies, since perturbations are expected to
trigger star formation and consequently increase the strength of emission lines.  The Kolmogorov-Smirnov (K-S) test of these distributions does
not indicate a significant difference between the $H_{\alpha}$ luminosities of the two samples (Table~\ref{tab_kshalpha}).  However, we check
our results by repeating the analysis for a subsample that is in the same $H_{\alpha}$ luminosity interval as the local sample.

We presented the analysis of our MS~0451 sample in Paper III.  We give the basic information about the rest of the sample in
Table~\ref{centdist}.  The first character of a galaxy name indicates the sample
(1:MS~0451~;~2:MS~1008~;~3:MS~2137~;~4:Cl~0412).  The second character is ``C'' for cluster members and ``F'' for field
galaxies.  The last part of the name assigns a number to each galaxy.  We identified galaxies with redshifts between
$3\sigma$ below and above the cluster redshifts as cluster members.  Only for MS~0451, which was analyzed in Paper III, we
used the redshift interval defined by \citet{MMTES07} using the redshift distribution of over 500 objects. That gives an
interval which is $\Delta z=0.006$ larger on both sides than what the $3\sigma$ definition gives.  For Cl~0412,
\citet{DSPBC99} determined cluster membership using the redshift distribution of 22 galaxies.  The redshift interval they
define selects the same galaxies as the $3\sigma$ criterion to be cluster members.

\begin{table}
\caption{K-S statistics of $H_\alpha$ luminosities of our sample and the local sample.}\label{tab_kshalpha}
\small\centerline {
\begin{tabular}{lcccc}
\hline
&D & P 	\\
\hline
log($H_\alpha$ luminosity) & 0.23 & 0.580 \\
\hline
\end{tabular}}
\medskip
D: K-S statistics specifying the maximum deviation between the cumulative distribution of the $H_\alpha$ luminosity for the local and intermediate redshift samples; P: significance level of the K-S statistics.
\end{table}

\begin{table*}
\renewcommand{\arraystretch}{0.68}
\caption{Basic galaxy information.}
\label{centdist}
\setlength{\tabcolsep}{13.5pt}
\small\centerline {
\begin{tabular}{lcccccccc}
\hline
ID &      $z$   &  $d$ & NED name & Type & Type-Ref. &  Type  & $z$ & z-Ref. 	\\
$(1)$ &	$(2)$ &	$(3)$ & $(4)$ &	$(5)$ &	$(6)$ &	$(7)$  & $(8)$ & $(9)$\\
\hline
2C1	&	0.2958	     &       0.7     &       PPP 001575       &       Sb-Sc    &  1   &     Irr/Pec    &       0.2968    	&     1         \\
2C2	&	0.3115	     &       1.0     &       PPP 001149       &       Sb-Sc    &  1   &     S/Pec     &       0.2963     	&     1         \\
2C3	&	0.3024	     &       0.9     &       PPP 000726       &       Sb-Sc    &  1   &     S	  &	 0.3026	   	&     1  	    \\  	
2C4	&	0.2975	     &       0.6     &       --  	      &       --	 &     &     Irr/Pec	  &	    --     	&           	 \\
2C5	&	0.2981	     &       0.2     &       PPP 000847       &       Sc-Irr   &  1   &     Irr/Pec	  &	  0.2935      &     1	       \\
2C6	&	0.3121	     &       0.1     &       [SED2002] 049    &       --  	 &     &     Irr/Pec	  &	    --        & 	       \\
2C7	&	0.3136	     &       0.9     &       PPP 000596       &       Sc-Irr   &   1  &     Irr/Pec	  &	  0.3120      &     1	       \\
2C8	&	0.3164	     &       0.9     &       PPP 001521       &       Sb-Sc	 &   1  &     Irr/Pec	  &	  0.3176      &     1	       \\
2C9	&	0.3082	     &       0.7     &       PPP 001560       &       E        &   1  &     E	  &	 0.3076 	   	&     1  	    \\
2C10	&	0.3093	     &       0.9     &       PPP 001378       &       E        &   1  &     S0/E     &       0.3077	   	&     1         \\
2C11	&	0.3049	     &       0.8     &       PPP 001673       &       E        &   1  &     S	   &	  0.3049	      &     1		  \\
2C12	&	--	     &       0.7     &       FPG 0100 NED02   &       E        &   1  &     S	   &	  0.3134	      &     1		  \\
2F1	&	0.6792	     &       --      &       -- 	       &       --	  &     &    Irr/Pec	&     --	   	&         \\
2F2	&	0.2082	     &       --      &       -- 	       &       --	  &     &    S/S0	&     --	   	&         \\
2F3	&	0.6809	     &       --      &       -- 	       &       --	  &     &    Irr/Pec	 &    --	   	&         \\
2F4	&	0.6857	     &       --      &       -- 	       &       --	  &     &    S       &        --	   	&         \\
2F5	&	0.4642	     &       --      &       PPP 000566       &       --	  &     &    S0       &       0.4644     	&     4    \\
2F6	&	0.1669	     &       --      &       PPP 001815       &       Sc-Irr   &   1  &     S      &	    0.1675	   	&     1    \\
2F7	&	0.4021	     &       --      &       -- 	       &       --	  &     &    Irr/Pec	 &    --	   	&         \\
2F8	&	0.6781	     &       --      &       -- 	       &       --	  &     &    Irr/Pec	 &    --	   	&         \\
2F9	&	0.4352	     &       --      &       -- 	       &       --	  &     &    Irr/Pec	   &  --	   	&         \\
2F10	&	0.3632	     &       --      &       FPG 0100 NED01   &       Sc-Irr   &   1  &     Irr/Pec	  &	  0.3645 	&     1      	  \\
2F11	&	0.3618	     &       --      &       PPP 001627       &       --	  &     &    S       &       0.3623	   	&     4    \\
2F12	&	0.3220	     &       --      &       PPP 000772       &       Sc-Irr    &  1   &    S       &       0.3216	   	&     1    \\
2F13	&	--   	     &       --      &       -- 	       &       --	  &     &    Irr/Pec	   &  --	   	&         \\
2F14	&	--   	     &       --      &       --  	      &       --  	 &     &     Irr/Pec	 &	   --	   	&        \\
2F15	&	--   	     &       --      &       --  	      &       --  	 &     &     --      &         --	   	&        \\
2F16	&	--   	     &       --      &       --  	      &       --  	 &     &     --      &         --	   	&        \\
2F17	&	--   	     &       --      &       --  	      &       --  	 &     &     Irr/Pec	  &	    --     	&           	 \\
2F18	&	0.1413	     &       --      &       -- 	       &       --	  &     &    E       &        --	   	&         \\
2F19	&	0.0052	     &       --      &       -- 	       &       --	  &     &    S0       &       --	   	&         \\
2F20	&	0.4247	     &       --      &       PPP 001823       &       --	  &     &    S       &       0.4256	   	&     4    \\
2F21	&	0.2381	     &       --      &       -- 	       &       --	  &     &    S0       &       --	   	&         \\
3C1	&	0.3095	     &       0.7     &       --	 	      &       --  	 &     &     S      &	      --	   	&        \\
3C2	&	0.3152	     &       0.7     &       [SED2002] 072    &       --  	 &     &     Irr/Pec	  &	    --     	&           	 \\
3C3	&	0.3095       &       0.7     &       [SED2002] 065    &       --  	 &     &     S      &	      --	   	&        \\
3C4	&	0.3164	     &       0.2     &       [SED2002] 009    &       --  	 &     &     S      &	      --	   	&        \\
3C5	&	0.3172	     &       1.0     &       --   	      &       --  	 &     &     S      &	      --	   	&        \\
3C6	&	0.3155	     &       0.7     &       --   	      &       --  	 &     &     Irr/Pec	  &	    --     	&           	 \\
3C7	&	0.3230	     &       0.9     &       --   	      &       --  	 &     &     S      &	      --	   	&        \\
3C8	&	0.3137	     &       0.9     &       -- 	      &       --  	 &     &     S0      &         --	   	&        \\
3C9	&	0.3141	     &       0.9     &       --   	      &       --  	 &     &     S      &	      --	   	&        \\
3F1	&	0.4528	     &       --      &       -- 	       &       --	  &     &    S/Pec	 &    --	   	&         \\
3F2	&	0.1501	     &       --      &        --	       &       --	  &     &    Irr/Pec	   &  --	   	&         \\
3F3	&	0.1951	     &       --      &       --        &       --	       	   &     &   Irr/Pec	    & --	   	&         \\
3F4	&	0.5675	     &       --      &       [SED2002] 069    &       --	  &     &    Irr/Pec	   &  --	   	&         \\
3F5	&	0.2859	     &       --      &        --       &       --	       	   &     &   Irr/Pec	    & --	   	&         \\
3F6	&	0.2822	     &       --      &        --	       &       --	  &     &    Irr/Pec	   &  --	   	&         \\
3F7	&	0.1876	     &       --      &        --	       &       --	  &     &    S       &        --	   	&         \\
3F8	&	0.5037	     &       --      &       [SED2002] 053    &       Irr	  &   2  &    S       &        --	   	&         \\
3F9	&	0.1880	     &       --      &       [SED2002] 141    &       --	  &     &    Irr/Pec	  &   --	   	&         \\
3F10	&	0.7498	     &       --      &        --	       &       --	  &     &    E       &        --	   	&         \\
3F11	&	0.8872	     &       --      &        --	       &       --	  &     &    S0       &       --	   	&         \\
3F12	&	--   	     &       --      &       [SED2002] 121    &       --  	 &     &     E      &	      --	   	&        \\
3F13	&	0.4421	     &       --      &       [SED2002] 104    &       E/S0	  &   2  &    S       &        --	   	&         \\
4C1	&	0.5027	     &       0.1     &        --  	      &       --  	 &     &     S        &       --	   	&         \\
4C2	&	0.5085	     &       1.3     &        --	      &       --  	 &     &     Irr/Pec	  &	    --     	&           	 \\
4C3	&	0.5099	     &       0.6     &        --  	      &       --  	 &     &     Irr/Pec	  &	    --     	&           	 \\
4F1	&	0.2918	     &       --      &        --	       &       --	  &     &    Irr/Pec	   &  --	   	&         \\
4F2	&	0.8478	     &       --      &        --       &       --	       	   &     &   S/Pec	  &   --	   	&         \\
4F3	&	0.8916	     &       --      &        --	       &       --	  &     &    Irr/Pec	   &  --	   	&         \\
4F4	&	0.3599	     &       --      &       [DSP99] 024      &       --	  &     &    Irr/Pec	   &	   0.3600	&     3    \\
4F5	&	0.6073	     &       --      &        --       &       --	       	   &     &   S        &       --	   	&         \\
4F6	&	0.6071	     &       --      &       [DSP99] 017      &       Sc	  &  3   &    S       &       0.6060	   	&     3    \\
4F7	&	0.6083	     &       --      &       [DSP99] 023      &       --	  &     &    Irr/Pec	   &	   0.6080	&     3    \\
4F8	&	0.4335	     &       --      &       [DSP99] 022      &       --	  &     &    S       &       0.4331	   	&     3    \\
4F9	&	0.4737	     &       --      &       [DSP99] 021      &       --	  &     &    Irr/Pec	   &	   0.4738	&     3    \\
4F10	&	0.5478	     &       --      &       --        &       --	       	   &     &   Irr/Pec	    & --	   	&         \\
4F11	&	--	     &       --      &       -- 	      &       --  	 &     &     Irr/Pec	  &	    --     	&           	 \\
4F12	&	0.5481	     &       --      &       --        &       --	       	   &     &   S        &       --	   	&         \\
4F13	&	0.4993       &       --      &       --        &       --	       	   &     &   Irr/Pec	    & --	   	&         \\
4F14	&	0.5646	     &       --      &       --        &       --	       	   &     &   S        &       --	   	&         \\
\hline
\end{tabular}}
\medskip
Column (1): object ID; Col.~(2): redshift; Col.~(3): projected distance from the cluster center in Mpc; Col.~(4): name of the galaxy in Nasa
Extragalactic Database (NED); Col.~(5): morphological type of the galaxy; Col.~(6) the reference for the morphological type; Col.~(7): eye-ball morphological
classification (this paper); Col.~(8): redshift of the galaxy; Col.~(9) the reference for the redshift.    
\\ References:   (1): \citep{YEMAC98};(2): \citep{SEDHP02};(3): \citep{DSPBC99}; (4): \citep{JZBHM04}.  
NED names given in column (4) begin with ``MS 1008.1-1224:'', ``MS 2137.3-2353:'' and ``F1557.19TC:'' for object IDs in the first column
that begin with ``2'', ``3'' and ``4'' respectively.  
For galaxies 2F13, 2F14, 2F17, 2F15, 2F16, 3F12 and 4F11 the redshift could not be determined.  Different
possibilities for identification of the emission line visible in their spectra rules out that these galaxies are cluster members.
\end{table*}

\begin{table*}
\caption{Eye-ball morphological classification of the MS~0451 sample.}
\label{eyeball}
\setlength{\tabcolsep}{8.5pt}
\small\centerline {
\begin{tabular}{lcccccccccccc}
\hline
ID & \vline & 1C1 & 1C2 & 1C3 & 1C4 & 1C5 & 1C6     & 1C7 & 1C8     & 1C9     & 1C10    & 1C11   \\
\hline
Type & \vline & S & S   & S/S0   & S   & S0  & Irr/Pec & S   & Irr/Pec & Irr/Pec & Irr/Pec & E    \\
\hline
\hline
ID & \vline & 1F1     & 1F2 & 1F3     & 1F4     & 1F5 & 1F6 & 1F7     & 1F8 & 1F9 & 1F10    & 1F11    \\
\hline
Type & \vline &  Irr/Pec & S   & Irr/Pec & Irr/Pec & S   & S   & Irr/Pec & S   & S0  & Irr/Pec & Irr/Pec \\
\hline
\end{tabular}}
\medskip
\end{table*}

In Sect.~\ref{app_indiv} in the Appendix, we give some information about each galaxy.  In case the galaxy has emission lines, we present: 
\begin{description}[g--]
\item[\textit{a--}] the HST-ACS image of the galaxy in the $F606W$ (broad V band filter); 

\item[\textit{b--}] rotation curves of different emission lines (and for some cases based on the absorption lines) extracted along the central slit without correction for inclination and seeing; 

\item[\textit{c--}] position angles of kinematic and photometric axes as a function of radius; 

\item[\textit{d--}] rotation curves extracted along the central slit and the kinematic major axis; 

\item[\textit{e--}] velocity field obtained using the strongest line in the \mbox{spectrum}; 

\item[\textit{f--}] normalized flux map of the line used for constructing the velocity field; 

\item[\textit{g--}] velocity map reconstructed using 6~harmonic terms;  

\item[\textit{h--}] residual of the velocity map and the reconstructed map;   

\item[\textit{i--}] simple rotation map constructed for position angle and ellipticity fixed to their global values;  

\item[\textit{j--}] residual of the velocity map and the simple rotation model;  

\item[\textit{k--}] position angle and flattening as a function of radius;  

\item[\textit{l--}] $k_{3}/k_{1}$ and $k_{5}/k_{1}$ (from the analysis
where position angle and ellipticity are fixed to their global values) as a
function of radius.
\end{description}

\subsection{Spectroscopic data}
Our observations were spread across 5 nights in October and November 2004 for Cl~0412 (seeing $0\farcs57$ ({\it FWHM})); 7 nights in
December 2004 and February 2005 for MS~1008 (seeing $0\farcs73$); 5 nights between May and July 2005 for MS~2137 (seeing $0\farcs76$). 
Each sample was observed using three masks and the integration time of each mask was split into three exposures.  Even in cases where all
three exposures were taken during the same night, the frames were not perfectly aligned, therefore, we completed the reduction of each
frame before combining them.

The spectral data reduction was done in the same way as explained in Paper III, apart from using a different sky subtraction method, which improved the results considerably.  In Paper III, the
sky is modelled in spectra that are interpolated along the X axis for wavelength calibration.  Here we use the algorithm described in \citet{K03} which is based on modelling the sky in the
original data frame as a function of the rectified coordinates.  Modelling the sky before applying any rectification/rebinning to the data reduces the amount of noise that is introduced to the
data during the sky-subtraction process \citep[see also][]{JNHPJ08}.  To quantify the difference, we reduced one of our spectra using both the old and the new methods.  We averaged 15
spatial rows, that are far from the galaxy spectrum, and therefore include sky-line residuals only.  A third order polynomial was fitted to and then subtracted from this distribution across the
wavelength axis and the root mean square of the counts was calculated for both spectra. A comparison of the two shows that the noise is $30\%$ less in case we use the new sky subtraction
method.

To be able to compare the $H_\alpha$ emission line fluxes of our sample with the local sample, we applied a rough flux calibration to our data. 
We used the spectrum of a star that we observed together with our MS~0451 sample for the calibration of all our data, since they were observed
with the same instrument.  The star that we used is $U0825\_01208341$ in the PMM USNO-A2.0 catalogue of \citet{M98}.  We transformed the B and R
magnitudes of the star given in the catalog onto the standard Johnson-Cousins system using the conversions provided by \citet{Kidger03}.\\

\subsection{Photometric data}
Environmental effects on how a galaxy evolves depend on its intrinsic properties.  For example it is known that harassment is more efficient on
low central surface mass density galaxies \citep{MLQS99}.  In this context, it is important to test whether the abnormalities that we see in gas
kinematics of a galaxy correlate with its photometrical properties.  To investigate this issue, we use both VLT/FORS2 and HST/ACS images.  We
obtained imaging of the MS~1008, MS~2137 and Cl~0412 samples in the ACS/$F606W$ filter while we exploited existing imaging of MS~0451 in the
ACS/$F814W$ filter from the ST$-$ECF HST~archive.  Ground based images were taken in the $B$, $V$, $R$ and $I$~filters for the whole sample. 
The FORS2 filters B, V and I are close approximations to the Johnson-Cousins \citep{bes90} photometric system while the R filter is a special filter for
FORS2 that is similar to the Cousins R \footnote{The FORS2 filter curves are given at http://www.eso.org/instruments/fors/inst/Filters/curves.html}.\\

   \begin{figure*}
   \centering
   \includegraphics[width=18cm,clip]{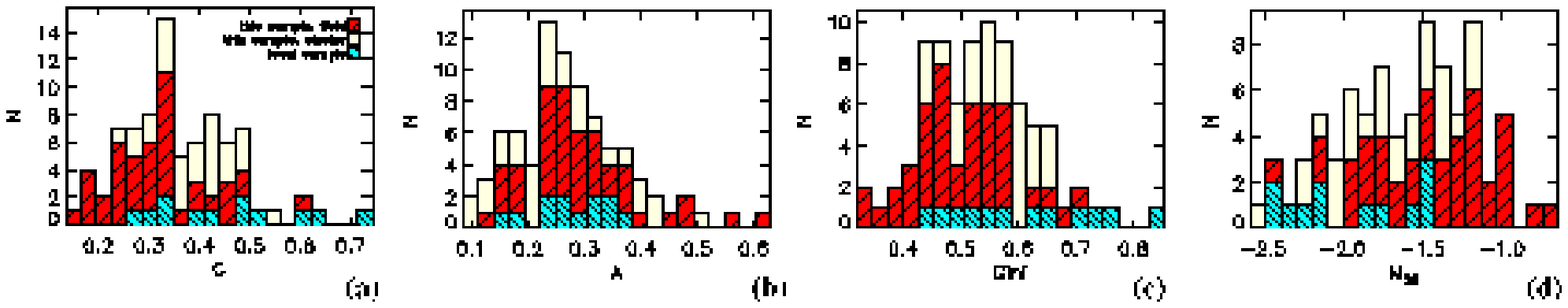}
 \caption{ Distribution of morphological parameters for our sample and the local sample.  a: Concentration parameter.  b: Asymmetry parameter.  c: Gini coefficient.  d:
$M_{20}$ index.}
         \label{histmorph}
   \end{figure*}

{\section{Analysis} \label{anal}}

{\subsection{Photometry} \label{anphot}}
Surface photometry analysis, magnitude measurements, extinction and k-correction were done in the same manner
as explained in Paper III.  We have not applied an internal dust (inclination) correction.  Galactic
extinction and k-corrected $M_B$~magnitudes, rest-frame $B-V$, $V-R$ and $R-I$ colors of the galaxies in our
sample are given in Table~\ref{tabrun2} in the Appendix.  $k$-correction was done using the kcorrect algorithm by \citet{BR07}.

\citet{AVYB94} defined concentration and asymmetry parameters to be able to do the morphological classification of galaxies in a quantitative and automated way.  The first parameter quantifies how concentrated the light
distribution of an object is, and it is larger for earlier type galaxies.  The second parameter measures how asymmetric the light distribution of a galaxy is and becomes larger for later type galaxies.  We use
slightly different definitions for asymmetry and concentration parameters than in Paper III.  Here we give the new definitions that are based on \citet{ABGES96} and \citet{CBJ00}.  The concentration is the ratio of
the flux within $G_{1}$, the area inside the $1\sigma$ isophote of the sky level and $G_{2}$, the region which has the same axis-ratio as $G_{1}$, but has a major-axis size that is 0.3 times the major-axis size of
$G_{1}$:

\begin{equation}
C=\frac{\sum_{i,j\in G_{2}}I_{ij}}{\sum_{k,l\in G_{1}}I_{kl}}\cdot
\label{conc}
\end{equation}

The asymmetry parameter A is the normalized residual of a galaxy image and its 180 degrees
rotated counterpart.  It is calculated within the $1\sigma$ isophote of the galaxy
(Eq.~\ref{asym}).  The central pixel for the asymmetry measurement is determined by shifting
the galaxy on a $50\times50$ grid and finding the minimum A. The asymmetry of a blank area was
measured in the same way in the vicinity of the object to correct for the contribution of the
background noise. 

\begin{equation}
A=\left(\frac{\sum_{i,j}|I(i,j)-I_{180}(i,j)|}{\sum_{i,j}|I(i,j)|}\right)-\left(\frac{\sum_{k,l}|B(k,l)-B_{180}(k,l)|}{\sum_{i,j}|I(i,j)|}\right)\cdot
\label{asym}
\end{equation}

We measure two additional parameters that we did not use in Paper III: the Gini coefficient and the $M_{20}$ index \citep{ABN03,LPM04}.  The Gini coefficient quantifies the non-uniformity in the
light distribution and strongly correlates with the concentration index for local galaxies.  Since the Gini coefficient has no dependence on the definition of the center of an object
(Eq.~\ref{gini}), it is often used as an alternative to the concentration parameter in studies of high-redshift galaxies, a large fraction of which are peculiar.

\begin{equation}
G=\left(\frac{1}{|\overline f|n(n-1)}\right)\sum_{i=1}^n(2i-n-1)|f_i|,
\label{gini}
\end{equation}

\noindent where $|f_i|$ are the absolute flux values of a galaxy's constitutent pixels sorted in increasing order, 
$|\overline f|$ is their mean value and n is the number of pixels.

The $M_{20}$ index is based on the total second-order moment $M_{TOT}$, which is the flux in each pixel $f_i$
multiplied by its squared distance to the galaxy center, summed over all pixels of the galaxy
(Eq.~\ref{mtot}).  

   \begin{figure}
   \centering
   \includegraphics[width=8cm,clip]{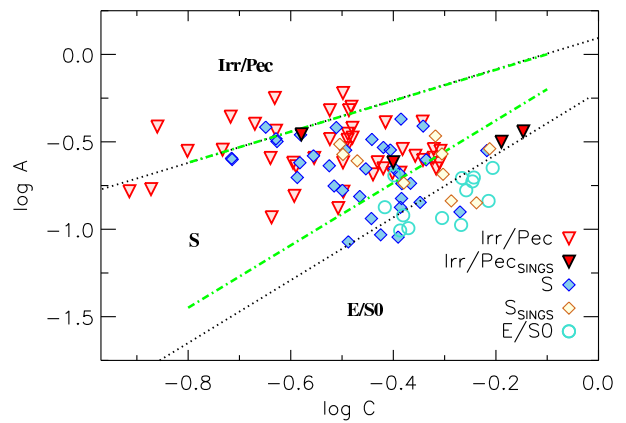}
 \caption{Distribution of galaxies in our sample and a local sample from SINGS on the $A-C$ plane.  Their morphological types, that are determined
by our eye-ball classification (Table~\ref{centdist}, Col.7.), are indicated with different symbols as shown in the legend.  The
dash-dotted green lines are the selection limits separating different morphological types determined by \citet{MFMBMBI06}.  The
borders that are adjusted by minimizing the amount of contamination from different types in each region are shown with black dotted lines.}
         \label{asymconc}
   \end{figure}

\begin{equation}
M_{TOT}=\sum_{i=1}^nM_i=\sum_{i}^nf_i((x_i-x_c)^2+(y_i-y_c)^2)\cdot
\label{mtot}
\end{equation}

$x_c$ and $y_c$ are the coordinates of the galaxy center which are determined in a way to minimize $M_{TOT}$.
$M_{20}$ is the normalized second order moment of the brightest $20\%$ of the galaxy's flux
(Eq.~\ref{m20}).  To compute $M_{20}$, the pixels are ordered such that i increases with decreasing flux,
and $M_i$ is summed over the brightest pixels until the integrated value reaches $20\%$ of the total galaxy
flux: 

\begin{equation}
M_{20}=log(\frac{\sum_{i}M_{i}}{M_{tot}})  while \sum_{i}f_i<0.2f_{tot}\cdot 
\label{m20}
\end{equation}

$M_{20}$ correlates with the square of the distance of the brightest regions of a galaxy from its center, which makes it sensitive to merger
signatures.  $M_{20}$ is smaller for centrally concentrated objects (early types) and increases in case of off-center light concentrations,
spiral arms, bright nuclei, bars, etc.

We present the asymmetry, concentration, Gini and $M_{20}$ parameters of our galaxies in Table~\ref{tabrun2} in the Appendix.  The same parameters for the SINGS galaxies are given in Table~\ref{singsphot}.  We applied a
K-S test to the asymmetry, concentration, Gini and $M_{20}$ distributions of the local versus distant field samples as well as the distant cluster versus field samples.  Only the galaxies for which we have spectroscopic
redshifts (see Table~\ref{centdist} here and Table~1 in Paper III) and that were classified as late types (spiral or irregular) according to our eye-ball classification (Table~\ref{centdist} and Table~\ref{eyeball}) were
used in this analysis.  Galaxy 2F19 was also excluded since it is not distant ($z=0.0052$).  The results are given in Table~\ref{tab_ksmorph} and the distributions are shown in Fig.\ref{histmorph}.  The distant cluster and
field samples have significantly different distributions for the concentration, Gini and $M_{20}$ parameters.  For the asymmetry, the difference between the two samples is considerable, but not very significant.  The
distributions of the local and distant field samples are significantly different for the $M_{20}$ and concentration parameters.  The difference is large for the Gini coefficient while the significance level of the statistic
is not very high.  The asymmetry distributions are similar for the two samples.

The rest-frame wavelength of the ACS images of our sample corresponds
to the B band, therefore we used blue KPNO, CTIO, Palomar and Isaac Newton images of the SINGS galaxies for the measurements.  These images
were convolved with a point spread function and then rebinned to have the same seeing and pixel size in kpc as the HST images of our sample at
the mean redshift of our clusters ($z=0.4$).  The asymmetry and concentration parameters can be used for morphological classification.  We
present our galaxies and local galaxies from SINGS on the $A-C$~plane together with our eye-ball classification of their morphologies in
Fig.~\ref{asymconc}.  Selection limits which separate different morphological types on the plane, as determined by \citet{MFMBMBI06}, are shown
on top of this plot. The borders are then adjusted by minimizing the amount of contamination from different types in each region.

\begin{table}
\caption{Photometric parameters for the SINGS sample.}
\label{singsphot}
\small\centerline {
\begin{tabular}{ccccc}
\hline
ID  &    $A$  &    $C$  &     $G$  & $M_{20}$ \\
$(1)$ & $(2)$ &$(3)$ & $(4)$ &$(5)$ \\
\hline
NGC0628	 &     0.31  &     0.31  &     0.45  &      -1.54    \\	  
NGC3031  &     0.18  &     0.42  &     0.57  &      -2.33    \\
NGC3049  &     0.27  &     0.50  &     0.76  &      -2.29    \\
NGC3184  &     0.27  &     0.32  &     0.53  &      -1.45    \\
NGC3521  &     0.29  &     0.61  &     0.70  &      -2.41    \\
NGC3938  &     0.34  &     0.48  &     0.66  &      -1.83    \\
NGC4536  &     0.25  &     0.34  &     0.50  &      -1.50    \\
NGC4569  &     0.21  &     0.50  &     0.65  &      -1.88    \\
NGC4579  &     0.14  &     0.58  &     0.70  &      -2.48    \\
NGC4625  &     0.36  &     0.71  &     0.74  &      -1.81    \\
NGC4725  &     0.24  &     0.40  &     0.54  &      -2.41    \\
NGC5055  &     0.15  &     0.52  &     0.63  &      -2.13    \\
NGC5194  &     0.35  &     0.26  &     0.47  &      -1.42    \\
NGC5713  &     0.32  &     0.65  &     0.83  &      -2.13    \\
\hline
\end{tabular}}
\medskip
Column~(1): object ID;
Col.~(2): asymmetry index;
Col.~(3): concentration index;
Col.~(4): Gini coefficient;
Col.~(5): $M_{20}$ index. \\
We could not obtain reliable measurements of the photometric parameters of NGC3621, NGC4236, NGC2976 and NGC7331 because of the
large number of stars
and artifacts on the images, therefore they are not used in our analysis.
\end{table}

\begin{table}
\caption{K-S statistics of the morphological parameters of our sample and the local sample.}\label{tab_ksmorph}
\small\centerline {
\begin{tabular}{lcccc}
\hline
&D & P 	\\
\hline
&distant galaxies: cluster versus field & \\
\hline
\hline
$M_{20}$ & 0.36 & 0.013 \\
\hline
$Gini$ & 0.46 & 0.000 \\
\hline
$A$ & 0.31 & 0.048 \\
\hline
$C$ & 0.50 & 0.000 \\
\hline
&field galaxies: distant versus local & \\
\hline
\hline
$M_{20}$ & 0.50 & 0.011 \\
\hline
$Gini$ & 0.43 & 0.048 \\
\hline
$A$ & 0.18 & 0.898 \\
\hline
$C$ & 0.48 & 0.017 \\
\hline
\end{tabular}}
\medskip
D: K-S statistics specifying the maximum deviation between the cumulative distribution of the morphological parameters for distant
galaxies: cluster versus field and for field galaxies: distant versus local; P: significance level of the K-S statistics.  In this analysis, only the
galaxies for which spectroscopic redshifts are available (see Table~\ref{centdist} here and Table~1 in Paper III) and that are late morphological types
(spiral/irregular) according to our eye-ball classification (see Table~\ref{centdist} and Table~\ref{eyeball}) were used.  Galaxy 2F19 was also excluded since it is not distant
($z=0.0052$).
\end{table}

{\subsection{Kinematics}}
We analyze the gas kinematics of our whole sample in the way that was described
in Paper III, using a sample from SINGS as a local
reference for comparison (see Paper III).  As stated in the introduction section,
we look for indications of disturbance in velocity fields to be able to examine
environmental effects.  There are three parameters that we use for quantifying
these abnormalities: 
\begin{description}[g--]
\item[\textit{a--}] the standard deviation of the kinematic position angle
($\sigma_{\rm PA}$); 
\item[\textit{b--}] the average misalignment between the photometric and
kinematic axes ($\Delta \phi$);
\item[\textit{c--}] the mean deviation of the velocity field from a
simple rotating disk ($ k_{3,5}/k_{1}$).  
\end{description}
The line of sight (LOS) velocity profile of a simple rotating disk is a cosine
function \citep{KCZC06}.  Our last (ir)regularity parameter measures the deviation
from the cosine form that can be represented with the third and the fifth order terms
of the Fourier series.  The exact definition of each of the (ir)regularity parameters
is given in Paper III.  Their values for each galaxy are listed in Table~\ref{tab2}. 
Examination of the data shows that in some special cases these measurements have to be
excluded from the analysis (explained in Sect.~\ref{special}).

\begin{table}
\caption{Parameters quantifying the (ir)regularity of the gas kinematics measured
for our sample.}
\label{tab2}
\small\centerline {
\begin{tabular}{lccccc}
\hline
ID &  $R_{\rm max}$ &  $\sigma_{\rm PA}$    &   $\Delta \phi$    &  $ k_{3,5}/k_{1}$ & $V_{asym}$ \\
$(1)$ & $(2)$ & $(3)$  &  $(4)$  & $(5)$ & $(6)$   \\
\hline
 1C7 &  13.8	     &  23 &  68 $\pm$ 26   & 0.32 $\pm$ 0.20  &  0.25  $\pm$  0.08  \\
 1C8 &  20.9	     &   2 &   9 $\pm$  6   & 0.06 $\pm$ 0.05  &  0.08  $\pm$  0.02  \\
 1C9 &	9.2	     &  19 &  66 $\pm$  20  & 0.10 $\pm$ 0.03  &  0.13  $\pm$  0.02   \\ 
 1C10 &  11.7	     &   9 &  14 $\pm$  9   & 0.26 $\pm$ 0.08  &  0.19  $\pm$  0.08  \\
 1F2 &  10.5	     &   2 &  35 $\pm$ 37   & 0.08 $\pm$ 0.02  &  0.08  $\pm$  0.02  \\
 1F3 &	10.0	     &   7 &  39 $\pm$  9   & 0.07 $\pm$ 0.05  &  0.05  $\pm$  0.02  \\
 1F4 &	5.5	     &  21 &  18 $\pm$  20  & 0.30 $\pm$ 0.19  &  0.21  $\pm$  0.10  \\
 1F5 &  11.3	     &   5 &  46 $\pm$  9   & 0.08 $\pm$ 0.03  &  0.09  $\pm$  0.03  \\ 
 1F6 &	4.4	     &  29 &  57 $\pm$ 44   & 0.27 $\pm$ 0.18  &  0.20  $\pm$  0.21  \\ 
 1F7 &  14.1  	     &   3 &   0 $\pm$ 11   & 0.05 $\pm$ 0.02  &  0.06  $\pm$  0.01  \\
 1F10 &	8.4	     &   8 &   1 $\pm$  7   & 0.25 $\pm$ 0.05  &  0.23  $\pm$  0.10  \\
 1F11 &  1\farcs7    &   5 &  18 $\pm$  5   & 0.05 $\pm$ 0.02  &  0.20  $\pm$  0.16  \\
 2C3	    &        9.7	  &  21	  &	47  $\pm$     60	  &   0.11	   $\pm$      0.04	&   0.10    $\pm$	0.04    \\
 2C5	    &        7.5	  &   5	  &	38  $\pm$     37	  &   0.06	   $\pm$      0.06	&   0.02    $\pm$	0.02    \\
 2C6	    &        8.8	  &  15	  &	84  $\pm$     14	  &   0.10	   $\pm$      0.03	&   0.12    $\pm$	0.04    \\
 2C7	    &        8.3	  &   6	  &	61  $\pm$     18	  &   0.09	   $\pm$      0.09	&   0.03    $\pm$	0.02    \\
 2C8	    &        7.6	  &   4	  &	77  $\pm$     17	  &   0.16	   $\pm$      0.08	&   0.11    $\pm$	0.05    \\
 2F1	    &        14.2	  &   6	  &	73  $\pm$      5	  &   0.06	   $\pm$      0.03	&   0.08    $\pm$	0.04    \\
 2F2	    &        4.2	  &   6	  &	72  $\pm$     34	  &   0.09	   $\pm$      0.08	&   0.12    $\pm$	0.14    \\
 2F3	    &        14.4	  &  17	  &	 1  $\pm$     34	  &   0.12	   $\pm$      0.05	&   0.17    $\pm$	0.08    \\
 2F4	    &        8.0	  &   7	  &	24  $\pm$     35	  &   0.06	   $\pm$      0.01	&   0.07    $\pm$	0.03    \\
 2F5	    &        11.3	  &  37	  &	 4  $\pm$     36	  &   0.06	   $\pm$      0.01	&   0.10    $\pm$	0.07    \\
 2F6	    &        4.6	  &   9	  &	62  $\pm$     13	  &   0.13	   $\pm$      0.08	&   0.11    $\pm$	0.06    \\
 2F9	    &        7.4	  &   8	  &	22  $\pm$      8	  &   0.18	   $\pm$      0.13	&   0.17    $\pm$	0.12    \\
 2F10	    &        11.6	  &   7	  &	37  $\pm$     10	  &   0.09	   $\pm$      0.04	&   0.11    $\pm$	0.03    \\
 2F11	    &        17.0	  &  43	  &	29  $\pm$     49	  &   0.83	   $\pm$      0.83	&   1.05    $\pm$	1.86    \\
 2F12	    &        11.4	  &   3	  &	30  $\pm$     36	  &   0.07	   $\pm$      0.02	&   0.05    $\pm$	0.02    \\
 2F15\&16    &        1\farcs9 	  &  19   &    --   		 	  &   0.77	   $\pm$      0.50	&   2.38    $\pm$	6.15    \\
 3C3	    &        12.2	  &   4	  &	 3  $\pm$      4	  &   0.17	   $\pm$      0.11	&   0.13    $\pm$	0.05    \\
 3C4	    &        12.7	  &   8	  &	 1  $\pm$      9	  &   0.15	   $\pm$      0.05	&   0.08    $\pm$	0.03    \\
 3C5	    &        7.5	  &   7	  &	84  $\pm$      7	  &   0.07	   $\pm$      0.03	&   0.11    $\pm$	0.10    \\
 3C6	    &        6.5	  &  11   &	52  $\pm$      8	  &   0.10	   $\pm$      0.07	&   0.18    $\pm$	0.11    \\
 3C7	    &        8.8	  &   3	  &	46  $\pm$      2	  &   0.09	   $\pm$      0.05	&   0.20    $\pm$	0.13    \\
 3F3	    &        3.7	  &   5	  &	48  $\pm$      5	  &   0.11	   $\pm$      0.02	&   0.10    $\pm$	0.07    \\
 3F6	    &        6.7	  &   3	  &	55  $\pm$      3	  &   0.16	   $\pm$      0.18	&   0.15    $\pm$	0.13    \\
 3F7	    &        4.8	  &   9	  &	97  $\pm$     25	  &   0.06	   $\pm$      0.04	&   0.06    $\pm$	0.03    \\
 3F8	    &        13.2	  &   4	  &	17  $\pm$     37	  &   0.20	   $\pm$      0.05	&   0.30    $\pm$	0.43    \\
 3F9	    &        7.7	  &   9	  &	46  $\pm$      9	  &   0.13	   $\pm$      0.02	&   0.13    $\pm$	0.11    \\
 4C2	    &        7.4	  &   3	  &	65  $\pm$      2	  &   0.08	   $\pm$      0.01	&   0.07    $\pm$	0.04    \\
 4C3	    &        8.9	  &  13	  &	63  $\pm$     12	  &   0.22	   $\pm$      0.14	&   0.14    $\pm$	0.03    \\
 4F3	    &        15.0	  &   9	  &	40  $\pm$     20	  &   0.07	   $\pm$      0.03	&   0.14    $\pm$	0.03    \\
 4F4	    &        10.8	  &   8	  &	60  $\pm$      9	  &   0.82	   $\pm$      1.21	&   0.57    $\pm$	0.64    \\
 4F5	    &        16.5	  &   9	  &	17  $\pm$     32	  &   0.07	   $\pm$      0.04	&   0.04    $\pm$	0.02    \\
 4F6	    &        10.7	  &   6	  &	28  $\pm$     31	  &   0.12	   $\pm$      0.05	&   0.09    $\pm$	0.04    \\
 4F7	    &        27.5	  &  29	  &	 7  $\pm$     53	  &   0.20	   $\pm$      0.06	&   0.17    $\pm$	0.05    \\
 4F8	    &        14.5	  &  18	  &	39  $\pm$     31	  &   0.18	   $\pm$      0.09	&   0.10    $\pm$	0.03    \\
 4F9	    &        8.7	  &   3	  &	55  $\pm$      3	  &   0.07	   $\pm$      0.05	&   0.05    $\pm$	0.02    \\
 4F12	    &        6.4  	  &  12	  &	24  $\pm$     18	  &   0.29	   $\pm$      0.16	&   0.17    $\pm$	0.07    \\
 4F13	    &        9.1	  &   2	  &	53  $\pm$      6	  &   0.10	   $\pm$      0.11	&   0.04    $\pm$	0.03    \\
\hline
\end{tabular}}
\medskip
Column (1): object ID; Col.~(2): maximum radius for which the kinematic parameters could be calculated.  The conversion from arcsecond into kpc was done as
explained in \citet{W06}; Col.~(3): standard deviation of the kinematic position angle~($\sigma_{\rm PA}$); Col.~(4): mean misalignment between the
kinematic and photometric position angles~($\Delta \phi$); Col.~(5): mean $k_{3,5}/k_{1}$ of the analysis done while fixing the position angle and the
ellipticity to their global values; Col.~(6): $V_{asym}$ parameter of S08 measured as described in Sect.~\ref{shap}. The error is the standard deviation of
the parameter in the range of observations.  \\
For 1F11 and 2F15\&16, the spectroscopic redshifts are not available, therefore we give $R_{\rm max}$ of these objects in
arcseconds.  $k_{3,5}/k_{1}$ of galaxy~1F5, $\Delta \phi$ of the galaxies that
have $\epsilon \leq 0.1$ (galaxies~1F2,~2F4,~2F12~and~4C3) and all parameters for galaxies~1F10,~2F5,~2F11~and~2F15\&16 are
rather meaningless as explained in Paper III for the MS~0451 sample and here, in Sect.~\ref{special} for the rest, therefore they are
excluded from the analysis.
\end{table}

\subsection{Star Formation Rates \label{sfr}}

Here we analyze the star formation properties of the galaxies in our sample.  We use the total fluxes of available emission lines in the spectra to measure
the star formation rates (SFR).  The luminosities are calculated using these fluxes and corrected for extinction following \citet{TF85}.  The inclinations
are measured from our HST/ACS imaging (Table~\ref{tab4}).  For a comparison between different extinction corrections, we have applied the definitions of
\citet{GHSWCF94} and \citet{TPHSVW98}, which makes a factor of $10\%$ difference at most in SFRs.  \citet{TF85} better matches the extinction law
\citep{CCM89} for reasonable values of E(B-V).  Star formation rates that rely on [OII]3727 or $H_\alpha$ line were calculated applying \citet{K92} and for
$H_\beta$, case B recombination was assumed, which implies a factor 2.86 difference in comparison with $H_\alpha$.  Note that we have not corrected the
$H_\beta$ luminosities for underlying stellar absorption.  For the calculations based on [OIII]5007 we have followed \citet{MHVRM08} and \citet{TMSMBF00}.

\begin{figure}
   \centering
   \includegraphics[width=8cm,clip]{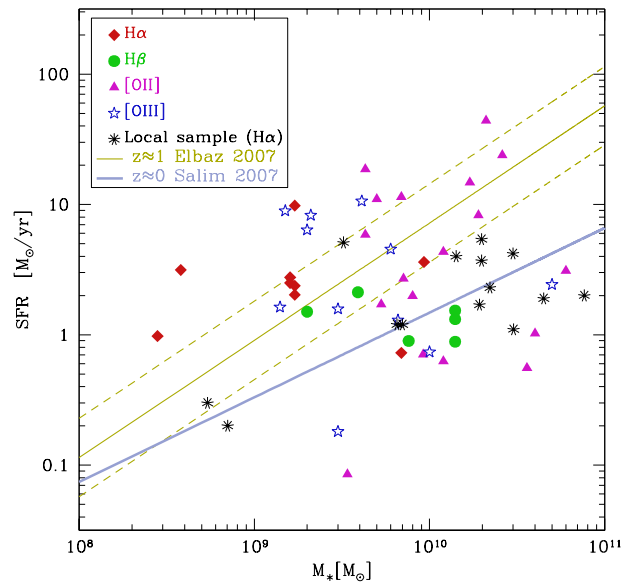}
 \caption{SFR versus stellar mass for our sample and the local sample from SINGS.  The SFR-stellar mass relation based on $z\sim1$ galaxies from GOODS
 \citet{EDLDA07} and the local relation from \citet{SRCBJ07} are overplotted for comparison.  The stellar masses of the SINGS galaxies are calculated using their $M_B$ and
 B-V as explained in \citet{BPWLC05}.}
\label{sfrmass}
\end{figure}

\begin{figure}
   \centering
   \includegraphics[width=8cm,clip]{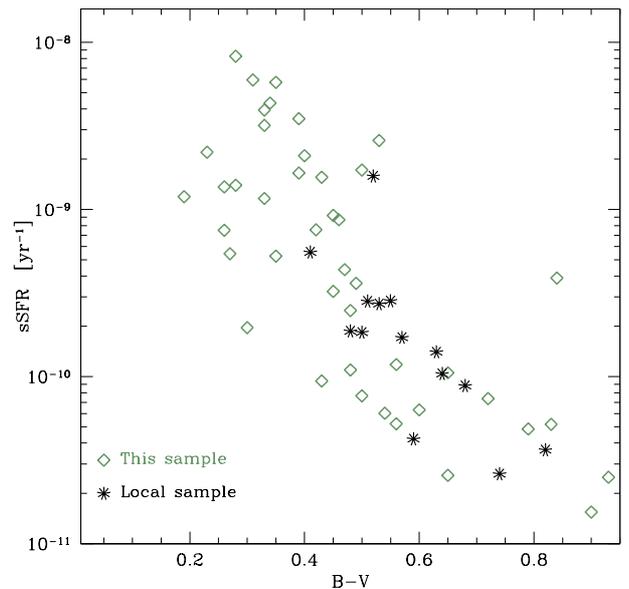}
 \caption{Specific star formation rate versus rest-frame B-V color for our sample and the local sample from SINGS.}
\label{ssfr_color}
\end{figure}

In Fig.\ref{sfrmass}, we plot the SFR versus stellar mass for the galaxies in our sample.  A comparison with the relations for $z\sim1$ and $z\sim0$
galaxies and the SINGS local sample shows that SF properties of the galaxies in our sample cover a wide range and some galaxies have higher star formation
rates than the $z\sim1$ relation.  It should be noted that this relation has a large intrinsic scatter at all redshifts.  In Fig.\ref{ssfr_color} we show
specific SFR versus rest-frame B-V which follows the expected trend.

\begin{figure*}
   \centering
   \includegraphics[width=17cm,clip]{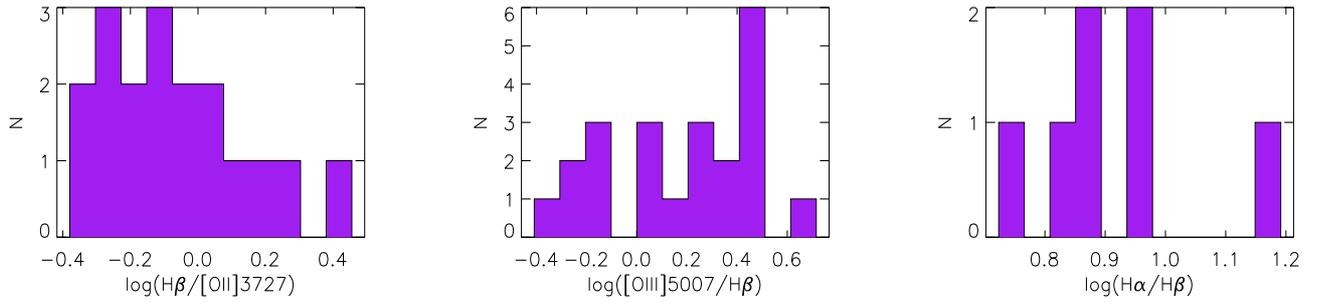}
 \caption{Histograms of the emission line flux ratios}
         \label{fl_rat}
   \end{figure*}

Since we calculated the star formation rates using the integrated flux from three adjacent slits that cover a galaxy, aperture effects are negligible.  On the other hand, we are
forced to use different emission lines for calculating SFRs due to the different rest-frame wavelength coverage of the spectra from galaxies at different redshifts.  The conversions
that are used for this purpose are likely to cause some systematic errors \citep{M06}.  To check for our sample, how successful it would be to use a constant factor for conversion
from one emission line flux to the other, we plot the frequency distribution of emission line flux ratios in Fig.~\ref{fl_rat}.  This exercise shows that the uncertainty in
$H_\alpha$ luminosities (Table~\ref{gasamount}) that is caused by these differences (sigma of the distribution) is about a factor 2.

   \begin{figure*}
   \centering
   \includegraphics[width=17cm,clip]{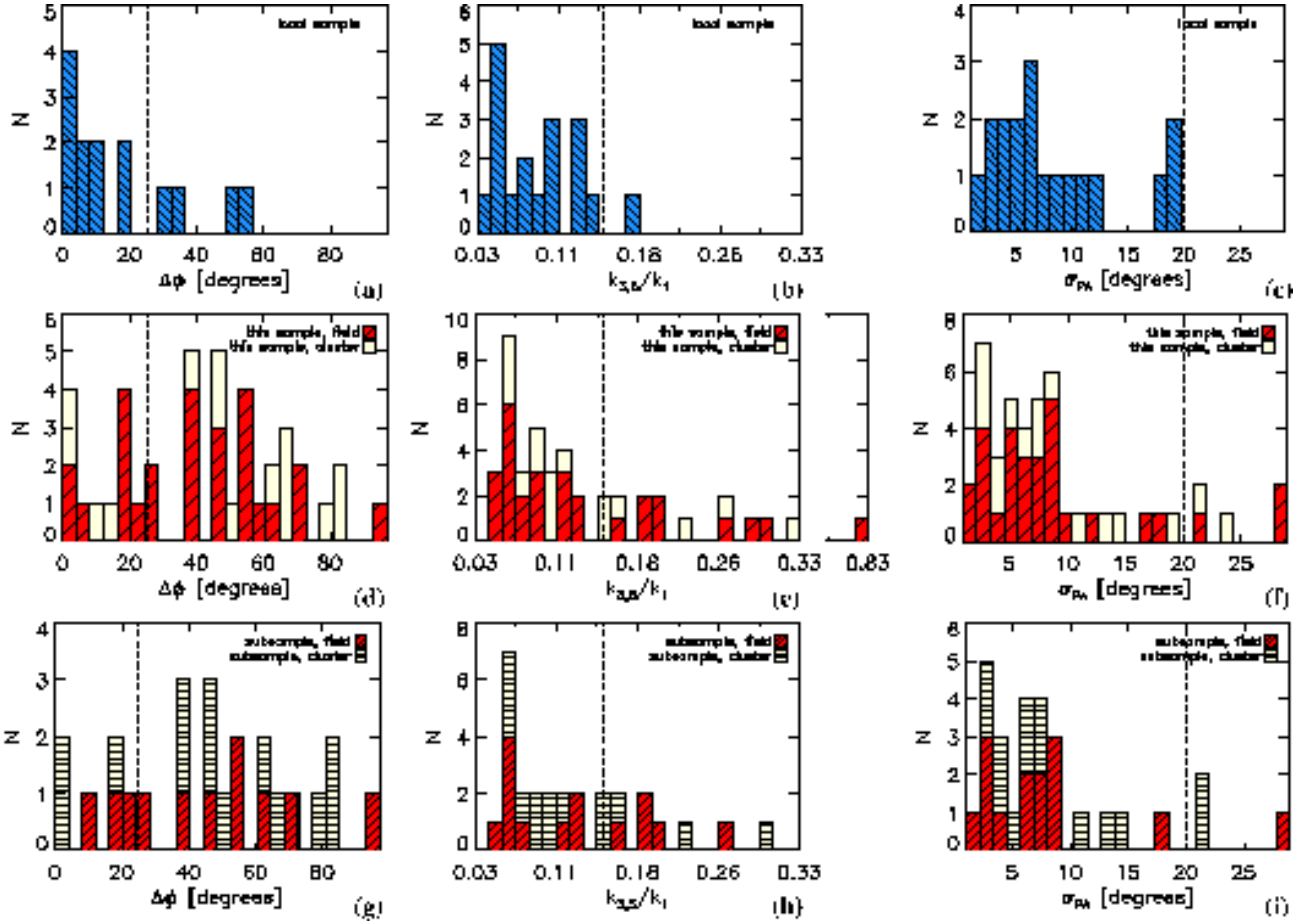}
 \caption{Histograms of the mean misalignment between kinematic and photometric major axes~($\Delta \phi$), mean~$k_{3,5}/k_{1}$ and standard deviation of kinematic position
angle~($\sigma_{\rm PA}$).  The local sample is given in panel.~(a), (b) and (c). Our sample is given in panel~(d), (e) and (f).  In panel~(g), (h) and (i) we show the galaxies in our
sample that have $H_{\alpha}$ luminosities in the same interval as the local sample.  The regularity borders, that are indicated on each plot, were determined in Paper III using the
local sample except for the peculiar galaxies and Virgo members.  $\Delta \phi$ of the galaxies that have $\epsilon \leq 0.1$ (galaxies~1F2, 2F4, 2F12 and 4C3 in ``this sample'',
NGC~628, NGC~3184, NGC~3938 and NGC~5713 in the local sample), $k_{3,5}$ of galaxy~1F5 (this sample) and all parameters for galaxies~1F10, 2F5, 2F11 and 2F15\&16 (this sample) are
doubtful as explained in Paper III for the MS~0451 sample and here, in Sect.~\ref{special} for the rest, therefore, they are excluded from the histograms.}
         \label{cop1}
   \end{figure*}

    \begin{figure*}
   \centering
   \includegraphics[width=17cm,clip]{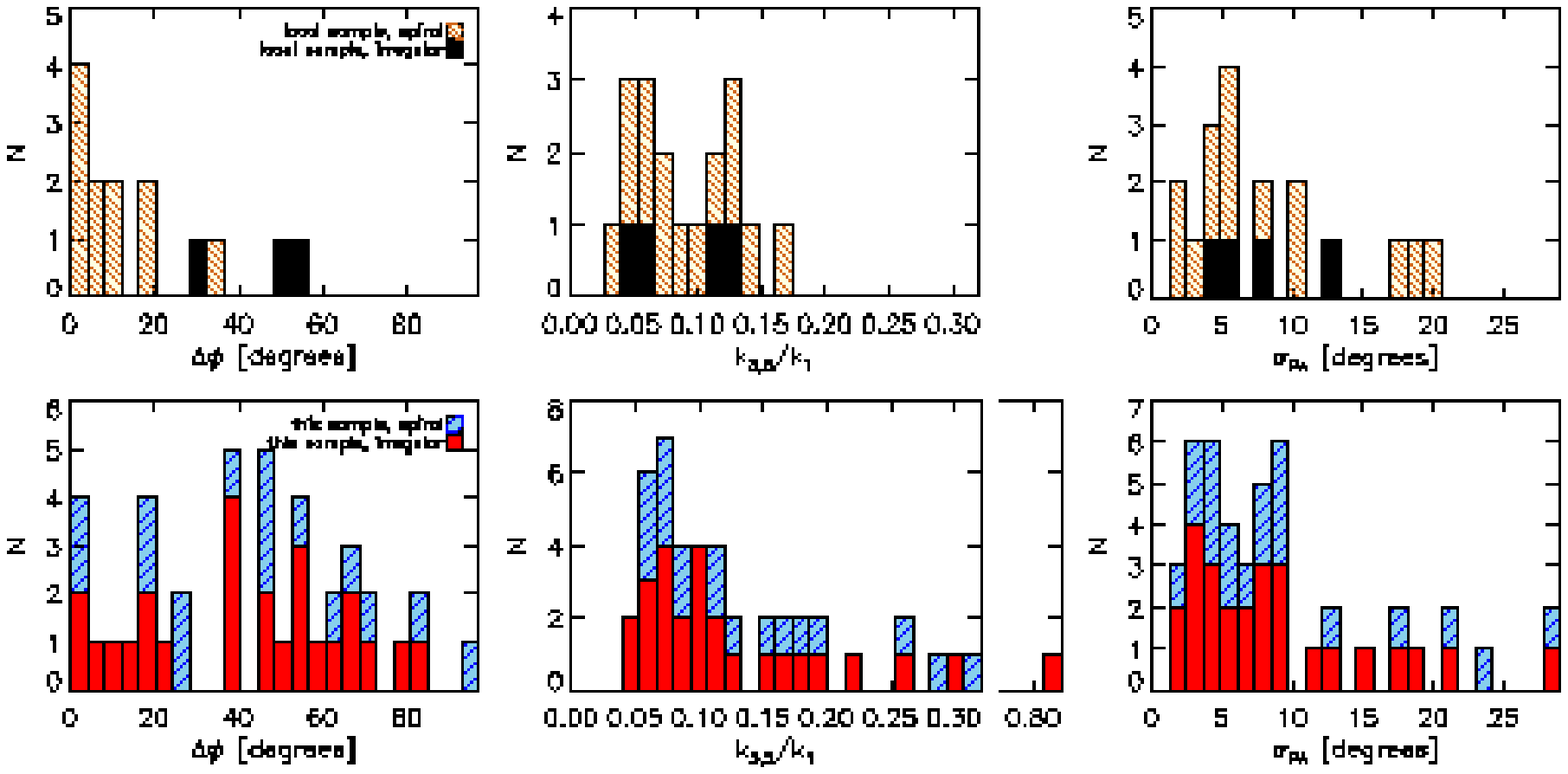}
 \caption{Histograms of the mean misalignment between the kinematic and photometric major axes~($\Delta \phi$), the mean~$k_{3,5}/k_{1}$ and the standard deviation of the kinematic
position angle~($\sigma_{\rm PA}$) for spiral versus irregular/peculiar galaxies in the local sample (top) and in our sample (bottom).  The morphological types are determined by
our eye-ball classification (see Table~\ref{centdist} and Table~\ref{eyeball}) $\Delta \phi$ of the galaxies that have $\epsilon \leq 0.1$ (galaxies~1F2, 2F4, 2F12 and 4C3 in ``this sample'',
NGC~628, NGC~3184, NGC~3938 and NGC~5713 in the local sample), $k_{3,5}$ of galaxy~1F5 (this sample) and all parameters for galaxies~1F10, 2F5, 2F11 and 2F15\&16 (this sample) are
doubtful as explained in Paper III for the MS~0451 sample and here, in Sect.~\ref{special} for the rest; therefore, they are excluded from the histograms.}
         \label{cop1_morph}
   \end{figure*}  
   
\subsection{Frequency Distribution of the Kinematic Irregularities \label{frq}}

In Fig.~\ref{cop1} d, e and f, we show the frequency distribution of each (ir)regularity parameter for the field and cluster galaxies in our sample.  The same information for local galaxies
is given above each plot for comparison (Fig.~\ref{cop1} a, b and c).  Cluster and field galaxies are distributed in a similar manner in the (ir)regularity parameters space.  Both cluster
and field galaxies populate regions inside and outside the area where regular velocity fields of local galaxies are located.  The Kolmogorov-Smirnov (K-S) test of the distributions also confirms that field
and cluster populations are not significantly different from one another (see Table~\ref{tab_ks1}).  Here we discuss the origin of the largest parameter values:  The two galaxies that have
the largest $\sigma_{\rm PA}$ values are 1F6 and 4F7.  1F6 has a kinematically decoupled core, therefore, it is probably a merger remnant (Paper III, Fig.B.13).  4F7 seems to be a merger
too.  The residual of its velocity field and reconstructed velocity map reveals the existence of a counter-rotating component in the outer part (see Fig.\ref{gal4F7}.g and j).  There are
tidal structures visible on its HST image as well (Fig.\ref{gal4F7}.a).  The largest $\Delta \phi$ belongs to 3F7 which has a strong bar (Fig.\ref{gal3F7}).  Although the kinematic and the
photometric position angles match quite well in the disk region, the extent of the observed velocity field does not go far outside the bar (see Fig.\ref{gal3F7}.a,c,e), therefore, this
galaxy has a very large $\Delta \phi$ value.  $\Delta \phi$ clearly has an important contribution of a bar in case of two other galaxies in our sample: 1F6 and 2C3.  So a large $\Delta
\phi$ either indicates a misalignment between the stellar disk and the kinematic axis of the gas, or the presence of a bar.  

In Sect.~\ref{anphot}, we determined the morphological type of the galaxies by our eye-ball classification (Table~\ref{centdist} \& Table~\ref{eyeball}).  Here we check
how the (ir)regularity parameter values of different morphological types are distributed (Fig.~\ref{cop1_morph}) and find that
irregularities in gas kinematics of spiral and irregular /peculiar galaxies are very similar (see Table~\ref{tabks} for the K-S test results).

\begin{table}
\renewcommand{\arraystretch}{0.9}
\caption{$H_\alpha$ luminosities.}\label{gasamount}
\small\centerline {
\begin{tabular}{lccc}
\hline
Name &  $L_{\rm H\alpha} / 10^{40} [ergs/s]$  & Line	   \\
$(1)$ & $(2)$ &   $(3)$    	   \\
\hline
1C7	&	100	&	O[II]3727 	\\
1C8 	&	20	&	H$\beta$ 	\\
1C9 	&	100	&	O[II]3727 	\\
1C10 	&	100	&	H$\beta$	\\
1F2 	&	20	&	H$\beta$	\\
1F3 	&	20	&	H$\beta$	\\
1F4 	&	40     &       H$\alpha$       \\
1F5 	&	100     &       H$\alpha$       \\
1F6 	&	10	&	H$\alpha$	\\
1F7 	&	600	&	O[II]3727	\\
1F10 	&	20	&	O[III]5007	\\
2C3   &       10    &       H$\beta$        \\
2C5   &       10    &       H$\beta$        \\
2C6   &       30    &       O[II]3727       \\
2C7   &       20    &       O[III]5007      \\
2C8   &       9    &       H$\alpha$       \\
2F1   &       30   &       H$\beta$        \\
2F2   &       1    &       O[II]3727       \\
2F3   &       100   &       O[II]3727       \\
2F4   &       10    &       O[II]3727       \\
2F5   &       7    &       O[II]3727       \\
2F6   &       30    &	    H$\alpha$	    \\
2F9   &       9    &	    O[III]5007      \\
2F10  &       100   &       O[III]5007      \\
2F11  &       9    &       H$\alpha$       \\
2F12  &       30    &       O[II]3727       \\
2F15\&16  &       --      &       O[II]3727       \\
3C3   &       50    &	    H$\alpha$	    \\
3C4   &       30    &	    O[III]5007      \\
3C5   &       20    &	    O[III]5007      \\
3C6   &       9    &	    O[II]3727	    \\
3C7   &       15    &	    O[III]5007      \\
3F3   &       2    &       O[III]5007      \\
3F6   &       30    &	    H$\alpha$	    \\
3F7   &       30    &	    H$\alpha$	    \\
3F8   &       40    &	    O[II]3727	    \\
3F9   &       30    &	    H$\alpha$	    \\
4C2   &       20    &	    O[II]3727	    \\
4C3   &       8    &	    O[II]3727	    \\
4F3   &       300   &	    O[II]3727	    \\ 
4F4   &       100   &	    O[III]5007      \\ 
4F5   &       100   &	    O[II]3727	    \\ 
4F6   &       50   &	    O[II]3727	    \\ 
4F7   &       200   &	    O[II]3727	    \\  
4F8   &       60   &	    O[III]5007      \\ 
4F9   &       80   &	    O[III]5007      \\ 
4F12  &       200   &	    O[II]3727	    \\ 
4F13  &       100   &	    O[III]5007      \\  
NGC0628	&	50     &		       \\
NGC2976 &	3     &		       \\
NGC3031 &	14     &		       \\
NGC3049 &	  --	&			\\
NGC3184 &	15     &		       \\
NGC3521 &	20     &		       \\
NGC3621 &	60     &		       \\
NGC3938 &	15     &		       \\
NGC4236 &	4     &		       \\
NGC4536 &	50     &		       \\
NGC4569 &	20     &		       \\
NGC4579 &	30     &		       \\
NGC4625 &	  --	&			\\
NGC4725 &	  --	&			\\
NGC5055 &	30    & 		      \\
NGC5194 &	70    & 		      \\
NGC5713	&	  --	&			\\
NGC7331	&	50	&			\\
\hline
\end{tabular}}
\medskip
Column (1): galaxy ID;  Col.~(2): $H_\alpha$~luminosity; Col.~(3): emission line that is used for calculating the $H_\alpha$~luminosity.\\
The errors in $H_\alpha$~luminosities are estimated to be $\sim50\%$ based on the width of the histograms in Fig.~\ref{fl_rat}.
\end{table}

\begin{table}
\caption{K-S statistics of (ir)regularity parameters of the cluster and field galaxies in our sample.}\label{tab_ks1}
\small\centerline {
\begin{tabular}{lcccc}
\hline
&D & P 	\\
\hline
${\sigma_{\rm PA}}$ & 0.18 & 0.874 \\
${k_{3,5}/k_{1}}$ & 0.18 & 0.856  \\
${\Delta \phi}$ & 0.32 & 0.209 \\
\hline
\end{tabular}}
\medskip
D: K-S statistics specifying the maximum deviation between the cumulative distribution of the given parameter for cluster and field galaxies in our sample; P: significance level of the
K-S statistics. \\ Note: $k_{3,5}/k_{1}$ of galaxy~1F5, $\Delta \phi$ of the galaxies that have $\epsilon \leq 0.1$ (galaxies~1F2,~2F4,~2F12~and~4C3) and all parameters for
galaxies~1F10,~2F5,~2F11 and 2F15\&16 are excluded from the calculations as explained in Paper III for the MS~0451 sample and here, in Sect.~\ref{special} for the rest.
\end{table}

\begin{table}
\caption{K-S statistics comparing the kinematic (ir)regularity parameters of the galaxies in our sample classified as spiral or irregular/ peculiar using their photometry.}\label{tabks}
\small\centerline {
\begin{tabular}{lcccc}
\hline
&D & P 	\\
\hline
${\sigma_{\rm PA}}$ & 0.15 & 0.995 \\
${k_{3,5}/k_{1}}$ & 0.10 & 1.000  \\
${\Delta \phi}$ & 0.25 & 0.786 \\
\hline
\end{tabular}}
\medskip
D: K-S statistics specifying the maximum deviation between the cumulative distribution of the given parameter for spiral and
irregular galaxies in our sample; P: significance level of the K-S statistics. \\ Note: $k_{3,5}/k_{1}$ of galaxy~1F5, $\Delta \phi$ of the
galaxies that have $\epsilon \leq 0.1$ (1F2, 2F4, 2F12 and 4C3) and all parameters for 1F10, 2F5,
2F11 and 2F15\&16 have been excluded from the calculations as explained in Paper III for the MS~0451 sample and here, in Sect.~\ref{special} for the rest.
\end{table}

\begin{figure}
   \centering
   \includegraphics[width=8cm,clip]{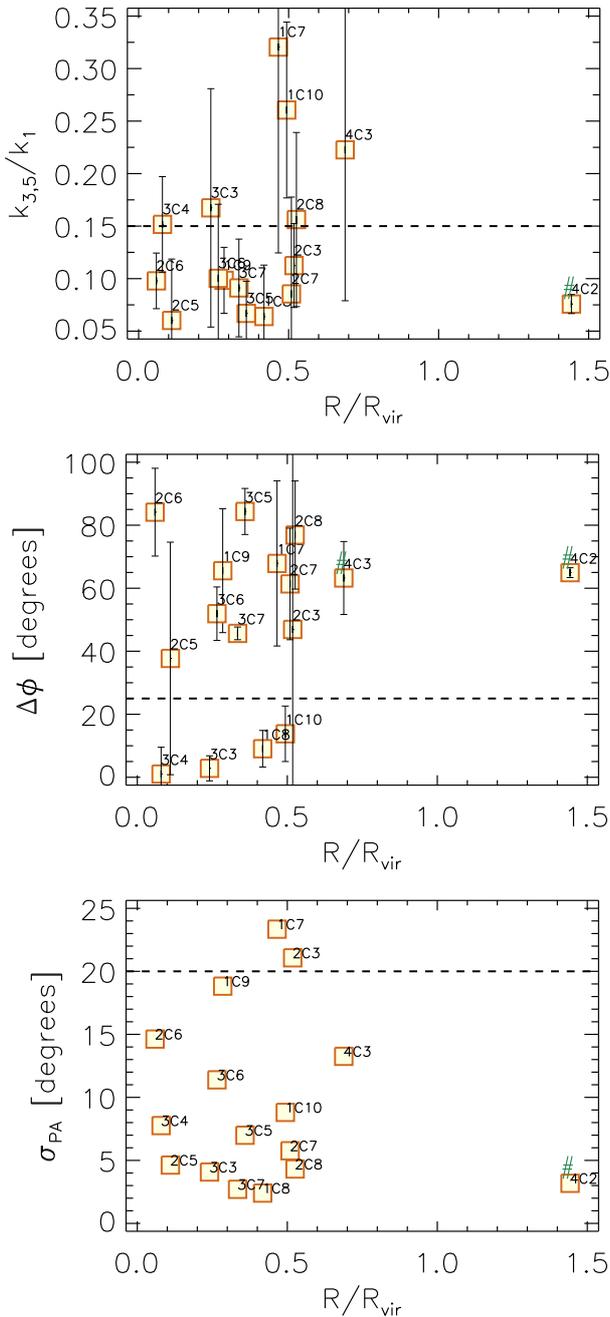}
 \caption{(Ir)regularity parameters versus $R/R_{vir}$.  The regularity threshold of each parameter is shown with a dashed
line.  \#~symbol indicates the galaxies that are not used in correlation measurements.  4C2 is excluded because the
correlations are measured inside 1 $R_{vir}$.  4C3 is excluded from the correlation measurement of $\Delta \phi$ since
it has $\epsilon \leq 0.1$ (see Sect.~\ref{special}).  \textit{Top}: mean~$k_{3,5}/k_{1}$ versus $R/R_{vir}$. \textit{Middle}: mean
misalignment between kinematic and  photometric axes~($\Delta \phi$) versus $R/R_{vir}$. \textit{Bottom}: standard
deviation of kinematic position angle~($\sigma_{\rm PA}$) versus $R/R_{vir}$.}
\label{cl_dist}
\end{figure}

\subsection{Dependence On the Clustercentric Distance}

All cluster members in our sample, except for 4C2, are well inside the virial radius, where both tidal processes and ICM-related
mechanisms are effective.  In Fig.~\ref{cl_dist}, we show the distance of each galaxy from the cluster center in projection (in virial
radii) and plot that against (ir)regularity parameters.  There are quite a few galaxies within half a virial radius from the center,
that are below the irregularity threshold of $k_{3,5}/k_{1}$ and $\sigma_{\rm PA}$.  $69\pm11\%$ of the cluster galaxies are regular
according to both of these parameters while most of them have large $\Delta \phi$ values.  The fraction of galaxies within 1 $R_{vir}$
that have regular gas kinematics according to all the three criteria is $13\pm8\%$.

\subsection{Correlations}

Here we measure the correlations of the irregularity parameters with the $H_\alpha$ luminosity, with each other and with some photometric parameters using an outlier
resistant linear regression fitting technique (Table~\ref{tab3}).

\subsubsection{Correlations With $H_\alpha$ Luminosity}

$H_{\alpha}$ luminosities are listed in Table~\ref{gasamount}.  In case the $H_{\alpha}$ line was outside the observed wavelength interval, we converted the fluxes of
available emission lines to $H_{\alpha}$ flux as explained in Sect.~\ref{sfr}.  

For the cluster members, we find significant correlations between the $log(L_{\rm H_\alpha})$ and two indicators of kinematical irregularities: $\sigma_{\rm PA}$ and
$k_{3,5}/k_{1}$  (Fig.\ref{halphalum}, a and b).  H$\alpha$ emission mainly stems from HII regions and it indicates star formation \citep[e.g.][]{KTC94}.  B-V color and
H$\alpha$ luminosity are expected to anti-correlate with each other for a given morphological type since galaxies with bluer B-V colors have a larger ratio of blue to red
stars, and therefore, a better capability of ionizing the gas to form HII regions \citep{Cohen76}.  However, dust extinction can weaken this correlation.  For our data, we
find a weak anti-correlation between these two quantities only for cluster members (Fig.\ref{halphalum}, c).  The irregularity parameters become larger for bluer galaxies.
However the trends are very weak (see Table~\ref{tab3}).

\subsubsection{Correlations Between (Ir)regularity Parameters}

To be able to use the (ir)regularity parameters as a tool to distinguish disturbed velocity fields from regular ones, it is
necessary to determine a threshold value for each of them.  We did that by measuring each parameter for local, regular velocity
fields from SINGS (see Paper III, Sect.~4).  In Fig.~\ref{cop2} in the Appendix, we show how the galaxies in our sample and in the local sample are
distributed in the plane of one parameter versus another.  Regularity borders that are defined using the local galaxies are
indicated on each plot. We find a weak correlation between $k_{3,5}/k_{1}$ and $\sigma_{\rm PA}$ (see Table~\ref{tab3}) which agrees
with what we found in Paper III using only the MS~0451 sample.  If we look at the galaxies that have large $k_{3,5}/k_{1}$ and
$\sigma_{\rm PA}$ parameters, most of them show signs of an additional kinematic component in the residual of the simple rotating
model and the original velocity field (residual maps are presented in Sect.~\ref{app_indiv}, part (j) of each figure).  These galaxies are
1F4, 1F6, 4F7 and 4F8.  In all cases, the existence of a secondary component is clear in the residual map.  $k_{5}$ is sensitive to
extra kinematic components and $\sigma_{\rm PA}$ is sensitive to their misalignment with the main component.  Therefore, the
outliers of the $k_{3,5}/k_{1}$ versus $\sigma_{\rm PA}$ plot mostly consist of velocity fields that have multiple kinematic
components.  This explains the weak correlation between these two parameters.

\begin{table}
\renewcommand{\arraystretch}{0.78}
\caption{Linear Pearson correlation coefficients.}\label{tab3}
\setlength{\tabcolsep}{2.5pt}
\small\centerline {
\begin{tabular*}{\columnwidth}{lccc}
\hline
&$\sigma_{\rm PA}-k_{3,5}/k_{1}$ & $\Delta \phi-\sigma_{\rm PA}$ & $k_{3,5}/k_{1}-\Delta \phi$   \\
c+f &   	\bf{0.5} (Fig.~\ref{cop2},a)		   & 		0.0       & 		0.0	         \\
\hline
&$\sigma_{\rm PA}-R/R_{vir}$ & $\Delta \phi-R/R_{vir}$ & $k_{3,5}/k_{1}-R/R_{vir}$   \\
c &   		0.1	   & 		0.2       & 	0.4		         \\
\hline
& $R_{\rm d}-k_{3,5}/k_{1}$ & $R_{\rm d}-\sigma_{\rm PA}$ & $R_{\rm d}-\Delta \phi$		 \\
c+f &  	0.2	 &  	0.3       & 		--0.4		     \\
c &  	0.1	 &  	 \bf{0.5} (Fig.~\ref{cor_clus},b)    & 	        --0.1		     \\
\hline
&   $M_{B}-k_{3,5}/k_{1}$	 &  $M_{B}-\sigma_{\rm PA}$   &  $M_{B}-\Delta \phi$		 \\
c+f &  	0.0	     &	      	--0.1	  &	   	0.4	    \\
c &  	--0.3	 &  	\bf{--0.5} (Fig.~\ref{cor_clus},c)      & 		0.1		     \\
\hline
& $A-k_{3,5}/k_{1}$ &  $A-\sigma_{\rm PA}$ & $A-\Delta \phi$		 \\
c+f &  	 --0.2    & 	0.1	 &   	--0.1	     \\
c &  	--0.1	 &  	 0.2      & 	0.0			     \\
\hline
& $C-k_{3,5}/k_{1}$ & $C-\sigma_{\rm PA}$  & 	$C-\Delta \phi$	 \\
c+f & 	0.0  &  	--0.1    & 	   0.2 		      \\
c &  	0.1	 &  	 0.0      & 	   --0.2			\\
\hline
 & $log(L_{\rm H_\alpha})-k_{3,5}/k_{1}$ & $log(L_{\rm H_\alpha})-\sigma_{\rm PA}$ & 	$log(L_{\rm H_\alpha})-\Delta \phi$\\
c+f &    0.1		 &	0.1		& 		--0.1		     \\
c &  	\bf{0.8} (Fig.~\ref{halphalum},b)	 &  	 \bf{0.8} (Fig.~\ref{halphalum},a)     & 		0.0		     \\
\hline
 & $z-k_{3,5}/k_{1}$ & $z-\sigma_{\rm PA}$ & 	$z-\Delta \phi$\\
 f & 0.1 & 0.0 & 0.1 \\
\hline
 & $(B-V)-k_{3,5}/k_{1}$ & $(B-V)-\sigma_{\rm PA}$ & 	$(B-V)-\Delta \phi$\\
c+f & --0.1 & --0.2 & --0.2\\
c &  --0.2	 &  	 --0.1      & 		--0.2		     \\
\hline
 & $(V-I)-k_{3,5}/k_{1}$ & $(V-I)-\sigma_{\rm PA}$ & 	$(V-I)-\Delta \phi$\\
c+f & 0.0 & 0.2 & 0.0 \\
c &  	0.4	 &  	 0.2      & 		0.1		     \\
\hline
 & $(R-I)-k_{3,5}/k_{1}$ & $(R-I)-\sigma_{\rm PA}$ & 	$(R-I)-\Delta \phi$\\
c+f & --0.1 & --0.2 & 0.0 \\
c &  --0.3	 &  	--0.3       & 		--0.2		     \\
\hline
 & $Gini-k_{3,5}/k_{1}$ & $Gini-\sigma_{\rm PA}$ & 	$Gini-\Delta \phi$\\
c+f & --0.1 & 0.0 & 0.3 \\
c &  	0.2	 &  	 0.1      & 	0.1			     \\
\hline
 & $M_{20}-k_{3,5}/k_{1}$ & $M_{20}-\sigma_{\rm PA}$ & 	$M_{20}-\Delta \phi$\\
c+f & 0.1 & 0.2  & 0.0 \\
c &  	0.1	 &  	 0.3      & 	\bf{0.5} (Fig.~\ref{cor_clus},a)			     \\
\hline
 & $log(M_{*}[M_{\sun}])-k_{3,5}/k_{1}$ & $log(M_{*}[M_{\sun}])-\sigma_{\rm PA}$ & 	$log(M_{*}[M_{\sun}])-\Delta \phi$\\
c+f & --0.3 	 & 	 --0.2  	&	      --0.4				\\
c &   --0.2	 &  	   0.0      	& 	  --0.3			     	\\
\hline
\end{tabular*}}
\medskip
The figures where $\Delta \phi$ of the galaxies that have $\epsilon \leq 0.1$ (galaxies~1F2,~2F4,~2F12~and~4C3 in
``this sample'', NGC~628, NGC~3184, NGC~3938 and NGC~5713 in the local sample),
$k_{3,5}/k_{1}$ of galaxy~1F5 (this sample) and all parameters for galaxies~1F10,~2F5,~2F11~and~2F15\&16
(this sample) are doubtful as explained in Paper III for the MS~0451 sample and here, in Sect.~\ref{special} for the rest.  Therefore, they are
excluded while calculating the correlation coefficients.  For the calculation
of the correlations with the redshift, only field galaxies were used, so the
results do not have the bias of the environment.
\end{table}

\subsubsection{Correlations With Photometric Parameters}

Apart from the Gini coefficient, $M_{20}$, photometric asymmetry and concentration parameters that are defined in Sect.~\ref{anphot}, the methods we use for
measuring the morphological/photometric parameters are explained in Paper III.  Photometric and morphological parameters of the galaxies in our sample are given in Table~\ref{tabrun2} and Table~\ref{tab4}
respectively (see Paper III, Table~C.2 for the morphological parameters of the MS~0451 sample.).  The (ir)regularity parameters of the local
sample galaxies are given in Paper III, Table~3.  Their photometric parameters are given here, in Table~\ref{singsphot}.

To focus on the effects of the interactions that take place only in clusters, we now restrict ourselves to our cluster sample, where most
galaxies are within half a virial radius from the cluster center.  In this region, mergers are rare, while harassment and ICM related
mechanisms such as ram pressure stripping are expected to be effective \citep{9MGGLQS7}.  We give the correlation measurements of the
cluster members that are located within $1 R_{vir}$ from the cluster center in Table~\ref{tab3}.  The parameters that correlate with
each other are plotted in Fig.~\ref{cor_clus} and discussed in Sect.~\ref{photcordis}.

  \begin{figure*}
   \centering
   \includegraphics[width=17cm,clip]{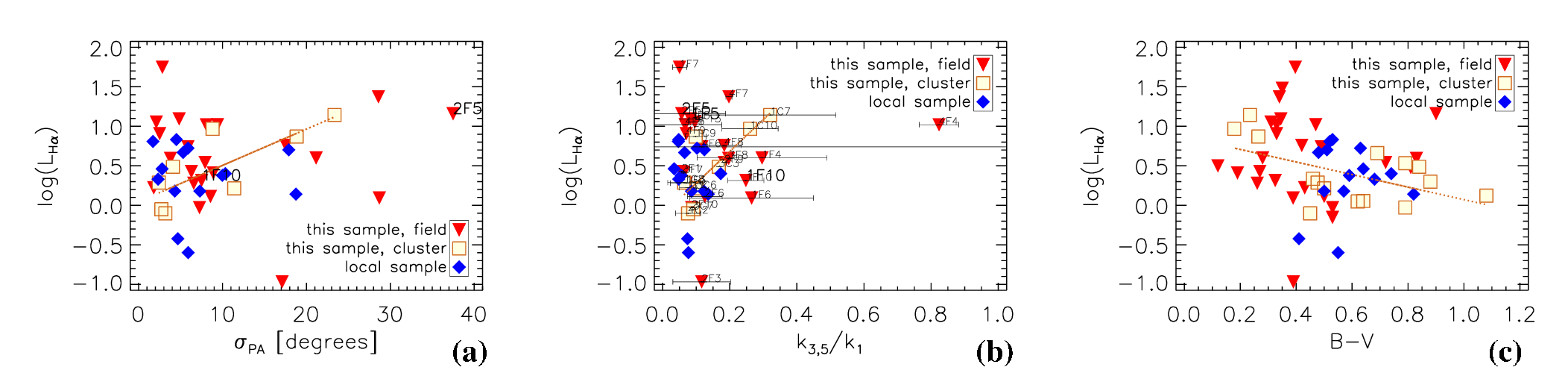}
 \caption{{\bf a)}~H$\alpha$ luminosity versus standard deviation of kinematic position angle~($\sigma_{\rm PA}$).  Correlation coefficient for cluster members is $0.8$.  {\bf b)}~H$\alpha$ luminosity
 versus mean~$k_{3,5}/k_{1}$.  Correlation coefficient for cluster members is $0.8$.  {\bf c)}~H$\alpha$ luminosity versus B-V color for our sample and the local sample.  Correlation coefficient for
 cluster members is $-0.5$.  Correlations are shown on top of each plot.}
	 \label{halphalum}
   \end{figure*}

\subsection{The Fraction of Irregular Gas Kinematics}

We quantified irregularities in gas kinematics using three different parameters, and for each of them, we compared the number distribution of field and
cluster galaxies.  Now we will look at the fraction of galaxies that have irregular gas kinematics.  Fractions that are measured for each irregularity type
separately and also without distinguishing between the three types are given in Table~\ref{tab5}.  We obtain very similar fractions of irregular gas
kinematics for cluster and field environments.  Each irregularity parameter gives a very different fraction compared to the others, which will be discussed
in Sect.~\ref{irregfracdis}.  

\begin{table}
\caption{Irregularity fraction.}\label{tab5}
\setlength{\tabcolsep}{1.5pt}
\small\centerline {
\begin{tabular*}{\columnwidth}{lccccc}
\hline
& ${\rm frac}_{\sigma_{\rm PA}}$ & ${\rm frac}_{\Delta \phi}$ & ${\rm frac}_{k_{3,5}/k_{1}}$ & ${\rm frac}_{any}$& ${\rm frac}_{all}$	\\
& $(1)$ & $(2)$ & $(3)$ & $(4)$& $(5)$	\\
\hline
field \& cluster & 11 $\pm$ 5 $\%$ & 68 $\pm$ 7 $\%$ & 32 $\pm$ 7 $\%$ & 80 $\pm$ 6 $\%$ & 4 $\pm$ 3 $\%$\\
only field & 10 $\pm$ 6 $\%$ & 65 $\pm$ 9 $\%$ & 32 $\pm$ 9 $\%$  & 76 $\pm$ 8 $\%$ & 3 $\pm$ 3 $\%$ \\
only cluster & 13 $\pm$ 8 $\%$ & 73 $\pm$ 11 $\%$ & 31 $\pm$ 12 $\%$  & 88 $\pm$ 8 $\%$ & 6 $\pm$ 6 $\%$ \\
\hline
\end{tabular*}}
\medskip
Column (1): fraction of irregular velocity fields according to $\sigma_{\rm PA}$ criterion; Col.~(2): fraction of irregular velocity fields according to $\Delta \phi$~criterion; Col.~(3): fraction of irregular velocity fields according to $k_{3,5}/k_{1}$ criterion; Col.~(4): fraction of irregular velocity fields according to at least one of the three criteria; Col.~(5): fraction of irregular velocity fields according to all the three criteria together. \\
Poisson errors are given for each fraction. \\ Note: $k_{3,5}/k_{1}$ of galaxy~1F5, $\Delta \phi$ of the
galaxies that have $\epsilon \leq 0.1$ (1F2, 2F4, 2F12 and 4C3) and all parameters for 1F10, 2F5,
2F11 and 2F15\&16 have been excluded from the calculations as explained in Paper III for the MS~0451 sample and here, in Sect.~\ref{special} for the rest.

\end{table}

\subsection{Special Cases \label{special}}

Here we explain the cases that we exclude from our analysis.  For the same information on the MS~0451 sample, see Paper III, Sect.~4.1.  For
face-on galaxies, photometric position angle measurements are very uncertain. Since LOS velocities are very small in such cases, the effect of
noise becomes more pronounced in velocity fields.  This causes $\Delta \phi$ to be unreliable.  Therefore, we excluded such cases from our
analysis (2F4, 2F11, 2F12, 4C3).  Among those, 2F11 is an extreme case which is completely excluded from the analysis (see
Fig.\ref{gal2F14}.e).  The other galaxies that we did not use in our analysis are 2F5, 2F15\&16.  2F5 does not have any signal in the
upper slit, which affects the measurements.  Looking at the iso-velocity lines on the receding side (Fig.\ref{gal2F7}.e), it looks as if
the highest positive velocities are located in the top right corner, which is missing on the map.  2F15\&16 (see Fig.\ref{gal2F20}) are
at the same redshift, however it is not clear what kind of objects they are and the emission line they have could not be identified.  The
velocity field includes information from both, but most of it comes from 2F15.  The [OIII]5007 velocity field of 4F4 (see Fig.\ref{gal4F4}.e)
looks quite disturbed, although the flux map of the same emission line looks rather regular.  Emission from this galaxy is very strong and
therefore, we can rely on its (ir)regularity parameters.

\section{Discussion \label{discuss}}

\subsection{Frequency Distribution of the Kinematic Irregularities}

We analyze together gas kinematics and stellar photometry of spiral galaxies in clusters and in the field.  We find that the fraction of galaxies that have irregular gas
kinematics is very similar in our cluster and field samples.  These two samples also give a very similar frequency distribution of each (ir)regularity parameter.  When
interpreting the results we have to consider that our sample selection is based on the emission line flux of galaxies.  A comparison of our sample with a local sample from
SINGS shows that some galaxies in both our cluster and field samples have higher $H_{\alpha}$ luminosities, and therefore, higher star formation rates (Fig.\ref{lum_ha}). 
In some of these cases, high star formation activity might be the result of some type of interaction.  It has to be considered, however, that $H\alpha$ luminosity of a
galaxy can also increase due to facts that are unrelated to interactions such as regular starbursts \citep{K98}.

In Fig.~\ref{cop1} (g), (h) and (i) we show how the subsample of galaxies that have $H_{\alpha}$ luminosities within the same interval as the SINGS sample
is distributed in the (ir)regularity parameters space.  The field galaxies that are populating the high irregularity end of the plots are not necessarily
the ones with high $H_{\alpha}$ luminosities.  So, independent from whether the high star formation galaxies are included or not, the distribution of cluster
and field galaxies in irregularity space is very similar.  The majority of the field galaxies in our sample are more irregular than local field galaxies according to at
least one of the three (ir)regularity criteria.  This is the case even if we take into account only the ones that have $H_{\alpha}$ luminosities within the
same interval as the SINGS sample, which are mostly in the interval $0.1\le z \le 0.5$.  This could be the result of a higher occurence of disk building
processes such as mergers and accretion events at these redshifts.  Using N-body simulations, \citet{GKK01} investigate the relative major merger rate of
the population of cluster, group and isolated halos as a function of redshift.  They find that for cluster galaxies, the relative merger rate increases
with redshift while it decreases for isolated galaxies.  At $z\sim0.5$, they find the major merger rate in the field to be two times as high as that in
clusters \citep[see][Fig.~9]{GKK01}.

In the local universe, evidence has been accumulating, mainly from HI studies, on the importance of cold gas accretion:  A large number of galaxies are accompanied by
gas-rich dwarfs or are surrounded by HI cloud complexes, tails and filaments \citep{SFOH08}.  Most of the high-velocity clouds around the Milky Way are now widely accepted to
belong to its halo and direct evidence for infall of intergalactic gas \citep{WYHBW07,WYWBR08}.  Recently, accretion of satellites has also been revealed by studies of the
distribution and kinematics of stars in the halos of the Milky Way and of M31.  The discovery of the Sgr Dwarf galaxy \citep{ibata94} is regarded as proof that accretion is
still taking place.  It is also possible that the warped outer layers, lopsidedness and the extra-planar gas, which are very common features in galaxies, are related to the
accretion process.  Observational results suggest cold gas accretion to be a likely formation mechanism for the polar disks \citep{BABBWK04,SPAVG09}. Simulations support
this picture \citep{MMS06}.

The radial velocity difference and angular separation of some galaxies in our sample suggest that they may be gravitationally bound to each other.  Since we do not 
have the spectra of the objects surrounding the galaxies in our sample, we can not make a definite statement of whether they are group members or not.   The number
statistics in \citet{HG82} show that velocity dispersions up to 400 $km s^{-1}$ and sizes up to 2 Mpc are likely (with median values of 155 $km s^{-1}$ and 0.7 Mpc) for
galaxy groups.  For galaxy pairs, it is expected that at least 35 percent of the ones with projected separation of less than 20 $h^{-1}$ kpc and velocity difference of less
than 500 $km s^{-1}$ are physically bound \citep{PCLDPGA07}.  On the other hand there are several interacting pairs with a projected separation of around 50 $h^{-1}$ kpc
\citep{BBG99,PCMPCP00}.  \citet{LTAC03} find that star formation in galaxy pairs is significantly enhanced over that of isolated galaxies with similar redshifts in the field
for projected separations less than 25 $h^{-1}$ kpc and velocity differences of less than 100 $km s^{-1}$.

Based on this information, the galaxies in our sample that might be gravitationally interacting with each other are listed in Table~\ref{grinfo}.  The average values of each irregularity
parameter for these galaxies (excluding the unreliable values that are mentioned in Paper III for the MS~0451 sample and here, in Sect.~\ref{special} for the rest) are $k_{3,5}/k_{1}=0.15$,
$\sigma_{\rm PA}=13$ and $\Delta \phi=23$.  Among these galaxies, 4F7 has a very large $\sigma_{\rm PA}$ and 4F12 has a very high $k_{3,5}/k_{1}$.  Excluding the galaxies in Table~\ref{grinfo}
from the comparison between the irregularity distributions of the cluster and field galaxies (see Sect.~\ref{frq}) does not change the results (see Table~\ref{tab_ks2}).

\begin{table}
\caption{Possible dynamical pairs.}\label{grinfo}
\setlength{\tabcolsep}{15pt}
\small\centerline {
\begin{tabular*}{\columnwidth}{lcc}
\hline
Pair & Projected distance [kpc] & $\Delta$ V [km/s] 	\\
\hline
\hline
2F10 \& 2F11 & 262 & 420 \\
\hline
4F10 \& 4F12 & 737 & 90 \\
\hline
4F5 \& 4F6 & 842 & 60 \\
\hline
4F6 \& 4F7 & 775 & 360 \\
\hline
4F5 \&4 F7 & 1170 & 230 \\
\hline
\end{tabular*}}
\medskip
Column (1): Names of the galaxies; Col.~(2): the projected distance between them; Col.~(3): the difference between their radial velocities.
\end{table}

\begin{table}
\caption{K-S statistics of (ir)regularity parameters of the cluster and field galaxies in our sample, excluding possible dynamical pairs.}\label{tab_ks2}
\small\centerline {
\begin{tabular}{lcccc}
\hline
&D & P 	\\
\hline
${\sigma_{\rm PA}}$ & 0.21 & 0.740 \\
${k_{3,5}/k_{1}}$ & 0.20 & 0.791  \\
${\Delta \phi}$ & 0.30 & 0.317 \\
\hline
\end{tabular}}
\medskip
D: K-S statistics specifying the maximum deviation between the cumulative distribution of the given parameter for cluster and field galaxies in our sample excluding the galaxies in
Table~\ref{grinfo}; P: significance level of the K-S statistics. \\ Note: $k_{3,5}/k_{1}$ of galaxy~1F5, $\Delta \phi$ of the galaxies that have $\epsilon \leq 0.1$
(galaxies~1F2,~2F4,~2F12~and~4C3) and all parameters for galaxies~1F10,~2F5,~2F11~and~2F15\&16 are excluded from the calculations as explained in Paper III for the MS~0451
sample and here, in Sect.~\ref{special} for the rest.
\end{table}

\subsection{Frequency Distribution of the Morphological Parameters}

In Sect.~\ref{anphot} we compare the distributions of some morphological parameters: asymmetry, concentration, $M_{20}$ and Gini coefficient for distant cluster versus field samples as well
as for local versus distant field samples (see Fig.\ref{histmorph} and Table~\ref{tab_ksmorph}).  We find a significant difference between the distribution of the concentration, the Gini
coefficient and the $M_{20}$ parameter for the cluster versus field galaxies at intermediate redshifts, in the sense that the cluster sample lacks galaxies with low concentration index.  This
might be due to the activity of interaction processes such as harassment which causes matter to migrate towards the center.

The local and distant field samples are also different: the concentration, the Gini coefficient and the $M_{20}$ index all suggest that local galaxies have a more centrally-concentrated and less
clumpy light distribution with respect to the distant galaxies.  This is consistent with what we find studying gas kinematics: it looks as if field galaxies at intermediate redshifts are
still in the process of building up their disks.

\subsection{The Fraction of Irregular Gas Kinematics \label{irregfracdis}}

In Y08, the [OII] doublet velocity fields of 63 field galaxies, that are at $z=0.4-0.75$ and that have $M_{stellar} \ge 1.5\times10^{10}M_{\sun}$, are analyzed.  They classify galaxy
kinematics based on an eye-inspection of the gas velocity map, gas velocity dispersion map and high resolution image together.  Their study of velocity fields however is limited to
$2\arcsec\times3\arcsec$ field of view while it can be as large as $\sim 3\arcsec\times6\arcsec$ in our case.  They call a galaxy ``rotating disk'' if its velocity field has an ordered gradient, the
photometric and the kinematic major axes are aligned and the velocity dispersion map has a single peak close to the kinematic center.  If the velocity dispersion map has no peak or has a peak that
is offset from the center while the other criteria are satisfied, they classify the case as ``perturbed rotation''.  If both the velocity dispersion map and the velocity map deviate from the
regular case,  they classify it as ``complex kinematics''.  Deviation from the regular case for a velocity field corresponds here to an irregular velocity gradient and/or a misalignment between
the photometric and kinematic axes.  Therefore, when the three indications we use are at the level of being detectable by eye within the central part of a galaxy, its kinematics can be
classified as complex according to this scheme.  Even then they find that $26\%$ of their sample have velocity fields and velocity dispersion maps that are both incompatible with disk
rotation.  When we calculate the irregularity fractions of the field galaxies in our sample that are within the same redshift interval as their sample, we find $10 \pm 9\%$ (according to
$\sigma_{\rm PA}$), $60 \pm 15\%$ ($\Delta \phi$), $30 \pm 14\%$ ($k_{3,5}/k_{1}$).  Their result is very close to what we find using the $k_{3,5}/k_{1}$ criterion.  However it should be
noted that most of the galaxies in our sample are less massive (see Table~\ref{tabrun2}, Column 8).

  \begin{figure*}
   \centering
   \includegraphics[width=17cm,clip]{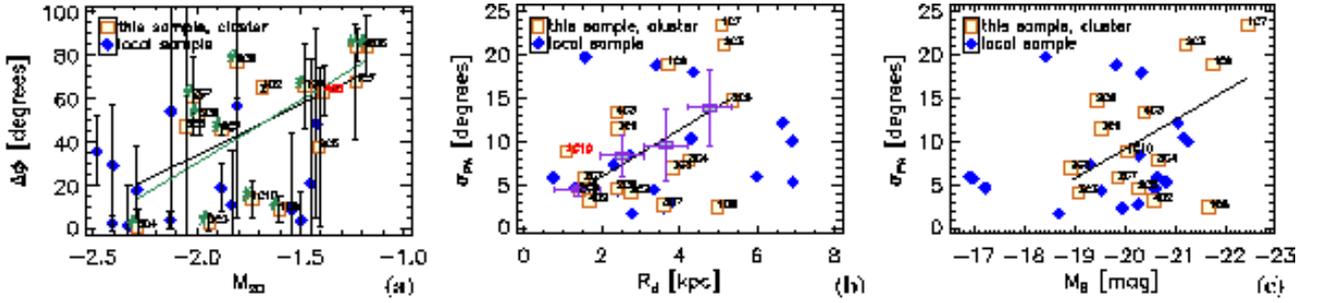}
 \caption{(Ir)regularity parameters versus some photometric quantities.  For each plot, the line shows the correlation between the given parameters.  The local galaxies are not used in correlation
measurements, but shown as a reference in each plot.  The names that are written in red and italic belong to the galaxies that are excluded from the correlation as explained in Paper III for the
MS~0451 sample and here, in Sect.~\ref{special} for the rest.  \textit{a}: $M_{20}$ index versus the mean misalignment between the photometric and kinematic major axes~($\Delta \phi$) (correlation coefficient=0.5).  The thick green line shows the correlation obtained excluding the cases where the $\Delta \phi$ value mainly indicates a clumpy light distribution or a bar instead
of an abnormality in the position of the kinematic axis.  The galaxies for which this is the case are indicated with \#~symbol next to them.  Measuring the correlation after excluding these galaxies
gives the same result.  \textit{b}: The disk scale length~($R_{d}$) versus the standard deviation of the kinematic position angle~($\sigma_{\rm PA}$) (correlation coefficient=0.5).  The
standard deviation of $\sigma_{\rm PA}$ is measured within equal $R_{d}$ intervals that are indicated with horizontal error bars and the mean $\sigma_{\rm PA}$ in each bin is given with a
rectangular green symbol while the deviation from the mean is given with vertical error bars.  \textit{c}: Absolute magnitude in the $B$~band ($M_{B}$) versus the standard deviation of the kinematic
position angle~($\sigma_{\rm PA}$) (correlation coefficient$=-0.5$).}
         \label{cor_clus}
   \end{figure*}

It is known from the local Universe that most galaxies in the central parts of galaxy clusters lack gas.  To be able to study velocity fields
of galaxies, priority was given to emission line galaxies in our sample selection.  Therefore, most cluster galaxies in our sample are perhaps
just infalling and have not been severely affected by the cluster environment yet.  This would explain the similarity between the gas
kinematics of cluster and field galaxies in our sample.

We use $k_{3,5}/k_{1}$, $\sigma_{\rm PA}$ and $\Delta \phi$ to trace the effects of the interaction processes on gas kinematics.  We find that the irregularity fractions measured using each of
these parameters are very different from one another: $\sigma_{\rm PA}$ gives a value around $10\%$, $k_{3,5}/k_{1}$ $\sim30\%$ and $\Delta \phi$ $\sim70\%$ for both cluster and field galaxies. 
This may have a number of different reasons.  One is the effect of lower spatial resolution for intermediate redshift galaxies.  Our simulations in Paper III, Appendix A, indicate that small
scale irregularities may be smeared out as a result of the resolution effects.  A misalignment between the stellar disk and the rotation plane of the gas on the other hand is unlikely to be
affected much by low resolution.  One should also realize that not all galaxies with high $\Delta \phi$ are irregular.  For example, galaxies with bars can have larger $\Delta \phi$ values. 
Also, some galaxies in the local universe are found showing regular kinematics with an HI polar disk (perpendicular to the stellar disk) \citep{VGS97,SPAVG09}.  \citet{PMCKD02} observed 18
blue compact dwarf galaxies and for 8 of these they found strong misalignments between the photometric and kinematic position angles although the isovelocity contours do not indicate strong
irregularities in gas motions.  There are even merger remnants in the local universe that have very regular gas velocity fields such as NGC 3921 \citep{HVG96}.

\subsection{Correlations With the $H_\alpha$ Luminosity}

Larger irregularities ($k_{3,5}/k_{1}$ and $\sigma_{\rm PA}$) we find for higher $log(L_{\rm H_\alpha})$ probably show that galaxies which have more irregular gas kinematics have higher star formation rates.  These
correlations are valid only for cluster members (see Fig.\ref{halphalum}).  According to models, most interaction processes in clusters increase star formation activity at the
beginning, before they eventually suppress it.  Gravitational interactions are expected first to trigger nuclear gas infall.  Models by \citet{Fujit98} show that increased star
formation activity is expected in case of harassment, since high-speed encounters between galaxies cause gas to accumulate to centers of galaxies.  Ram-pressure stripping, which is
the hydrodynamic interaction between the hot ICM and the cold ISM, leads to an increase of the external pressure, shock formation, thermal instabilities and turbulent motions within
the disk.  \citet{Evra91} and \citet{BeCo03}, for example, show that all these events increase cloud-cloud collisions and cloud collapse, and therefore, enhance star formation
activity.  However, in case of ram-pressure stripping, there are not many observations supporting this picture.  Some examples in A1367 that experience ram-pressure stripping are
CGCG 97-023, where enhanced star formation activity per unit mass, compared to galaxies of similar type and luminosity is confirmed \citep{GCCBK95}, CGCG 97-073 and CGCG 97-079
\citep{bosgav06}.  Models of \citet{Fujit98} and \citet{FN99} that quantify the variations of the star formation activity, show that on short timescales ($\sim10^8$ yr) in
high-density, rich clusters, the star formation activity can increase by up to a factor of 2 at most.  But on longer timescales, removal of the HI gas leads to a decrease of the fuel
feeding the star formation, and galaxies become quiescent \citep{Fujit98,FN99,ON01}.

\subsection{Correlations With Photometric Parameters \label{photcordis}}

We find a weak correlation between $M_{20}$ and $\Delta \phi$ for cluster members (Fig.~\ref{cor_clus}a).  $M_{20}$ becomes very large in case the galaxy light has a clumpy distribution. 
Since clumpiness is mainly caused by star forming regions, it means that galaxies that have irregular gas kinematics have more star formation.  $M_{20}$ also increases towards later types.  Since
galaxies that have high mass concentration (earlier types) are more resistant to tidal mechanisms, the correlation we find is expected as a result of this fact as well.  The concentration parameter
on the other hand, which is another indicator of galaxy type, does not give any correlation with the irregularity in gas kinematics.  Therefore, the substructures must be the main cause of the
correlation that we find.

We need to note that $\Delta \phi$ cannot be considered as a pure indicator of interactions since it is sensitive to
bars that are misaligned with the disk.  Even though the formation of a bar can be triggered by environmentally induced gravitational instabilities, such as tidal
interactions between galaxies and the cluster potential well, it can also just be the result of a misalignment between the disk and the triaxial halo of the galaxy
itself \citep{KS01,BF02}.  Since a bar is not necessarily formed by an interaction process, we remeasured the correlation of $\Delta \phi$ with the other parameters
excluding the cases where we see that a bar has an important contribution to the $\Delta \phi$ value (3F7, 1F6 and 2C3).  The results remained the same.  

We find a correlation between the disk scale length and $\sigma_{\rm PA}$ (Fig.~\ref{cor_clus}b).  However, what we see on the plot is an increasing deviation of
$\sigma_{\rm PA}$ values with increasing $R_{d}$ rather than a correlation.  While small galaxies are all regular, there are both regular and irregular cases among
larger galaxies.  It is known that some interaction processes are more effective on larger galaxies such as interactions between galaxies and the cluster potential
well, viscous stripping and thermal evaporation.  Tidal interactions between galaxies, on the other hand, are more efficient on smaller galaxies \citep{BV90}.  What
we see in our data might be an indication of the activity of some of the first group processes in the central $1R_{vir}$ of the clusters in our sample.  

We find that $M_B$ and $\sigma_{\rm PA}$ anti-correlate with each other (Fig.~\ref{cor_clus}c) showing that more massive galaxies have more irregular gas
kinematics.  Since larger galaxies are also more massive \citep{TRR04}, the interpretation of this correlation is the same as the correlation that we find for the
disk scale length.  One or a combination of the following mechanisms might be effective on the cluster members in our sample: viscous stripping, thermal
evaporation and tidal interactions between galaxies and the cluster potential well.  

We investigated whether morphological peculiarities correlate with irregularities in the gas kinematics.  Among the galaxies that have irregular/peculiar morphology according to our eye-ball
classification (see Table~\ref{centdist} and Table~\ref{eyeball}), there is no trend towards larger kinematical irregularities (see Fig.~\ref{cop1_morph}).  We find very similar distributions of the irregularity
parameters for spirals and irregular/peculiar galaxies (Table~\ref{tabks}).  We also do not find a correlation between the photometric asymmetry and irregularities in the gas kinematics. 
\citet{NHPF08} on the other hand find a good agreement between their morphological and kinematical classifications.

For the correlations that we find for cluster galaxies, where the $\sigma_{\rm PA}$ parameter is involved, it should be considered that there are only two
galaxies that are slightly above the irregularity threshold of $\sigma_{\rm PA}$ (Fig.\ref{cor_clus}b\&c).  Therefore, although we discuss the possibility of
these correlations being a result of cluster specific interaction mechanisms, this is not a strong result.

\subsection{Comparison With High Redshift Studies \label{shap}}

\begin{figure}
   \centering
   \includegraphics[width=8cm,clip]{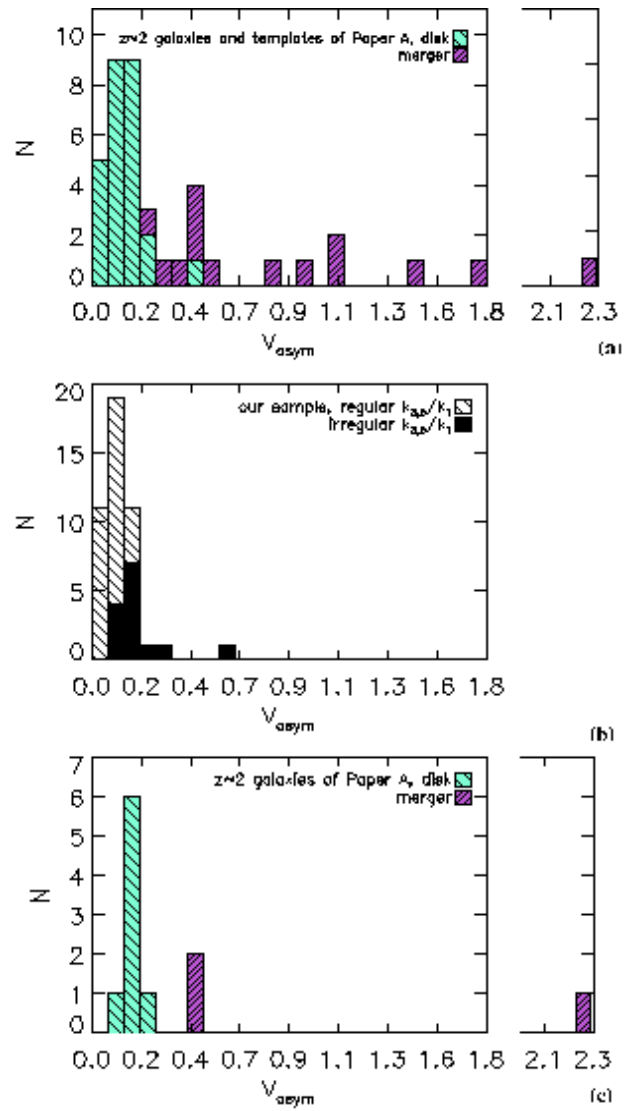}
 \caption{Number distribution of the $V_{asym}$ parameter that is described in S08. \textit{a}: Galaxies and the templates in S08 that are classified as disk and
 merger. \textit{b}: The galaxies in our sample that are classified as regular or irregular depending on their $k_{3,5}/k_{1}$ values.  The galaxies that have doubtful
 $k_{3,5}/k_{1}$ values as explained in Paper III for the MS~0451 sample and here, in Sect.~\ref{special} for the rest are excluded from this histogram. \textit{c}: $z
 \sim 2$ SINS galaxies in S08 that are classified as disk and merger.}
\label{shapiro_asym}
\end{figure}

S08 uses a method that is based on kinemetry to distinguish merging and non-merging systems.  They quantify the
asymmetries in both the velocity field and the velocity dispersion map of ionized gas.  Using the measurements of these two
parameters for template galaxies, they determine where merging and non-merging systems are located on the plane of these
parameters versus each other and define a criterion to distinguish them from one another:
$K_{asym}=\sqrt{V^{2}_{asym}+\sigma^{2}_{asym}}=0.5$.  We measured the $V_{asym}$ parameter of our velocity fields
(Table~\ref{tab2}) and compared these with the $V_{asym}$ of the merging and non-merging galaxies and models in S08 (see
Fig.~\ref{shapiro_asym}).  The quality of our sigma-maps is not satisfactory for an analysis.  However, the possibility that
$\sigma_{asym}$ and $V_{asym}$ give contradictory results is very low (see Fig.~5 in S08).  Therefore, we use $V_{asym}$
alone with the purpose of making a comparison between the methods.  To measure a global ellipticity and a global
inclination, which are used while measuring the $V_{asym}$ parameter, we calculate the median of their values outside half the
full width at half maximum of the seeing.

   \begin{figure}
   \centering
   \includegraphics[width=8cm,clip]{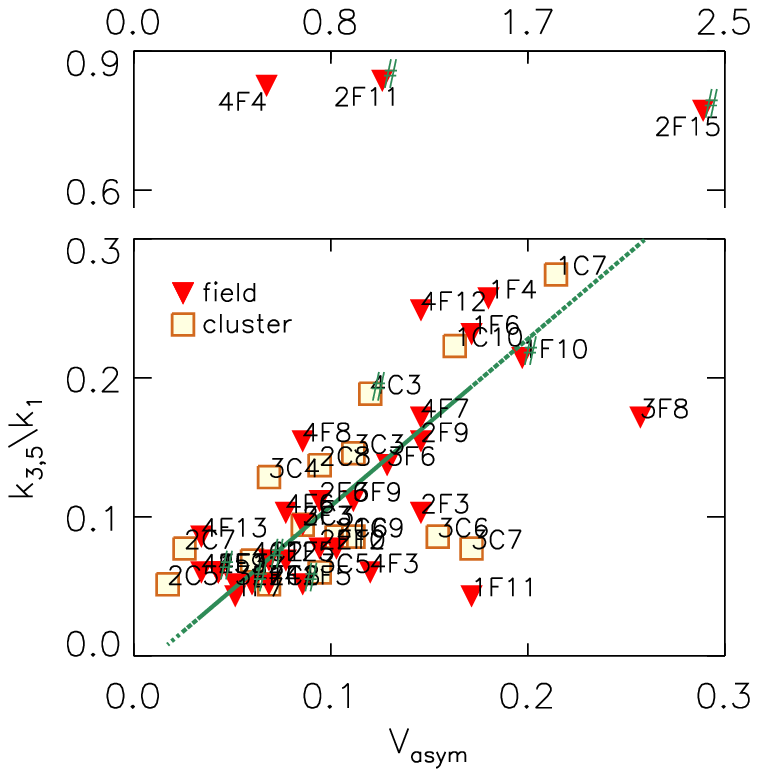}
 \caption{$V_{asym}$ parameter that is described in S08 versus $k_{3,5}/k_{1}$.  Data points that are far out of the general
 distribution are shown separately, on top of the main plot.  \# sign indicates the galaxies for which, both parameters are unreliable (see
 Sect.~\ref{special}).  These galaxies are not used in the correlation measurement.}  
\label{shapiro_asym_k35}
\end{figure}

If we look at the $V_{asym}$ distribution of our galaxies, we see that most of them are located in the region of the non-merging galaxies, as defined by
S08 (Fig.~\ref{shapiro_asym}a, b).  $V_{asym}$ is similar to $k_{3,5}/k_{1}$ and a comparison between these parameters for our galaxies shows that they
correlate with each other (Fig.~\ref{shapiro_asym_k35}, correlation coefficient=0.86).  On the other hand, the classification thresholds are quite different for
the two criteria.  Following S08, most of our objects would be regular while many more galaxies are classified as irregular with our method.  This is
visible in Fig.~\ref{shapiro_asym} where the $V_{asym}$ distribution of our galaxies is given together with non-mergers and mergers in S08. 
Classification of these galaxies according to the $k_{3,5}/k_{1}$ criterion is indicated on the same plot.  While the method of S08 aims at separating
mergers from non-mergers, it would not fulfill our requirement of tracing the imprints of environmental processes.  4F7 (in the non-merger region with
$V_{asym}$=0.17) is a good example to explain that since the deviation of the position angle, the residual map of the velocity field and the simple rotating disk
model and also the galaxy image itself provide signs of a merger.

We find that the $z \sim 2$ galaxies in S08 are more irregular than our complete sample that includes both cluster and field galaxies in
$0.10 \le z \le 0.91$ (see Fig.~\ref{shapiro_asym}b and c).  The K-S test results show that the maximum deviation between these two distributions is
$59\%$ and the probability that the samples are similar is 0.002.

\section{Summary and conclusions \label{conclu}}

Using gas velocity fields, we trace the activity of interaction processes both in galaxy clusters and in the field.  To measure the irregularities in
velocity fields, we use three different indicators: the standard deviation of the kinematic position angle ($\sigma_{\rm PA}$), the mean deviation of
the LOS velocity profile from a cosine function which is measured using high order Fourier terms ($k_{3,5}/k_{1}$) and the average misalignment
between the kinematical and photometric major axes ($\Delta \phi$).  A regularity threshold for each of these parameters is defined using local field
galaxies from SINGS.  16 cluster members (z $\sim 0.3$ and z $\sim 0.5$) and 29 field galaxies (at $0.10\leq$z$\leq0.91$) are then analyzed and
studied with respect to the local galaxies in the field and compared with each other to evaluate the effect of interaction processes on gas
kinematics.  

Our analysis shows that distant field galaxies have more irregular gas kinematics than their local counterparts.  This suggests a higher frequency of disk-building
processes such as accretion events and mergers in the distant universe.  Morphological properties of these galaxies lead us to the same conclusion. The
concentration, the Gini coefficient and the $M_{20}$ index measurements indicate that distant field galaxies are more clumpy and less centrally concentrated than local
field galaxies.

We make a comparison between gas kinematics of our intermediate redshift sample and the z $\sim 2$ sample of S08 using the $V_{asym}$ parameter, which is defined
in S08 to distinguish merging and non merging systems.  We find that our sample, that includes both field and cluster galaxies within $0.10 \le z \le 0.91$, have more
regular gas kinematics than the z $\sim 2$ galaxies.

Y08 shows that a large fraction of spiral galaxies with $M_{stellar} \ge 1.5\times10^{10}M_{\sun}$ at $z=0.4-0.75$ have irregular gas kinematics. 
When they exclude these galaxies with irregular kinematics, they find no evolution in the scatter and the slope of the K-band TFR \citep{PFHY08}.  Our analysis shows that a
large fraction of less massive distant spirals at a median redshift of $z=0.36$ also have irregular gas kinematics.

We find that cluster and field galaxies are distributed in a similar manner in the (ir)regularity parameters space.  We also measure the fraction of irregular
velocity fields.  For each parameter, we find remarkably similar fractions for cluster members and field galaxies.  This shows that the cluster galaxies in our sample
are not severely affected by the cluster environment.  Galaxies in the central parts of clusters are expected to have an imporant fraction of their gas stripped via
interaction processes and it is difficult to obtain velocity fields of these galaxies, especially at high redshifts.  Therefore, it is probable that some cluster
galaxies in our sample, for which the velocity fields could be analyzed, are just infalling.  If this is the case, that explains the similarity we find between the
gas kinematics of cluster and field galaxies.  On the other hand, a comparison between the morphological properties of the cluster and field galaxies in our
sample reveals a clear difference between them: the cluster sample lacks galaxies with low concentration index.  This might be due to the
activity of interaction processes such as harassment which causes matter to migrate towards the center.  We compare the gas
kinematics of spiral and irregular galaxies as well and find no significant difference between these morphological classes.

We find that galaxies with higher $H_{\alpha}$ luminosities have larger $k_{3,5}/k_{1}$ and $\sigma_{\rm PA}$ values.  In addition to that, galaxies with
more substructures have larger average misalignment between their kinematic and photometric axes ($\Delta \phi$).  Since substructures mostly are star forming
regions, all these correlations mean that galaxies which have more irregular gas kinematics have high star formation rates.  This is consistent with the theory since
most interaction mechanisms in clusters increase star formation activity at the beginning, before they eventually supress it.

\vspace*{3.5mm}

\begin{acknowledgements} We would like to thank Daniel Kelson for making his sky subtraction algorithm public, and Scott Trager
for his help in using it. We thank Davor Krajnovi{\'c} for the Kinemetry software.  We are grateful to the authors of
\citet{DCAHCBK06} for kindly providing us with the H$\alpha$~velocity fields of the galaxies in the SINGS local sample.  We
thank Jacqueline van Gorkom for fruitful discussions.  We thank the referee for the valuable comments.  We appreciate the efficient support of ESO and the Paranal staff. This
work has been financially supported by the \emph{Volkswagen Foundation} (I/76\,520), the \emph{Deut\-sche
For\-schungs\-ge\-mein\-schaft, DFG\/}  (project number~ZI\,663/6) within the Priority Program~1177,  the Kapteyn Astronomical
Institute of the University of Groningen and the German Space Agency \emph{DLR\/} (project number 50\,OR\,0602).  EK thanks
the Netherlands organization for international cooperation in higher education for the Huygens grant.  This
research has made use of the NASA/IPAC Extragalactic Database (NED) which is operated by the Jet
Propulsion Laboratory, California Institute of Technology, under contract with the National
Aeronautics and Space Administration.
\end{acknowledgements}

\bibliographystyle{aa}
\bibliography{abb,all}

\begin{thebibliography}{129}
\expandafter\ifx\csname natexlab\endcsname\relax\def\natexlab#1{#1}\fi

\bibitem[{{Abraham} {et~al.}(1994){Abraham}, {Valdes}, {Yee}, \& {van den
  Bergh}}]{AVYB94}
{Abraham}, R.~G., {Valdes}, F., {Yee}, H. K.~C., \& {van den Bergh}, S. 1994,
  ApJ, 432, 75

\bibitem[{{Abraham} {et~al.}(1996){Abraham}, {van den Bergh}, {Glazebrook},
  {Ellis}, {Santiago}, {Surma}, \& {Griffiths}}]{ABGES96}
{Abraham}, R.~G., {van den Bergh}, S., {Glazebrook}, K., {et~al.} 1996, ApJS,
  107, 1

\bibitem[{{Abraham} {et~al.}(2003){Abraham}, {van den Bergh}, \&
  {Nair}}]{ABN03}
{Abraham}, R.~G., {van den Bergh}, S., \& {Nair}, P. 2003, \apj, 588, 218

\bibitem[{{Aguerri} {et~al.}(2001){Aguerri}, {Balcells}, \& {Peletier}}]{ABP01}
{Aguerri}, J.~A.~L., {Balcells}, M., \& {Peletier}, R.~F. 2001, \aap, 367, 428

\bibitem[{{Allen} {et~al.}(2003){Allen}, {Schmidt}, {Fabian}, \&
  {Ebeling}}]{ASFE03}
{Allen}, S.~W., {Schmidt}, R.~W., {Fabian}, A.~C., \& {Ebeling}, H. 2003,
  \mnras, 342, 287

\bibitem[{{Arag{\'o}n-Salamanca}(2008)}]{AS08}
{Arag{\'o}n-Salamanca}, A. 2008, in IAU Symposium, Vol. 245, IAU Symposium, ed.
  M.~{Bureau}, E.~{Athanassoula}, \& B.~{Barbuy}, 285--288

\bibitem[{{Athreya} {et~al.}(2002){Athreya}, {Mellier}, {van Waerbeke},
  {Pell{\'o}}, {Fort}, \& {Dantel-Fort}}]{AMWP02}
{Athreya}, R.~M., {Mellier}, Y., {van Waerbeke}, L., {et~al.} 2002, \aap, 384,
  743

\bibitem[{{Barton} {et~al.}(1999){Barton}, {Bromley}, \& {Geller}}]{BBG99}
{Barton}, E.~J., {Bromley}, B.~C., \& {Geller}, M.~J. 1999, \apjl, 511, L25

\bibitem[{{Bekki} \& {Couch}(2003)}]{BeCo03}
{Bekki}, K. \& {Couch}, W.~J. 2003, \apjl, 596, L13

\bibitem[{{Bekki} \& {Freeman}(2002)}]{BF02}
{Bekki}, K. \& {Freeman}, K.~C. 2002, \apjl, 574, L21

\bibitem[{{Bell} {et~al.}(2005){Bell}, {Papovich}, {Wolf}, {Le Floc'h},
  {Caldwell}, {Barden}, {Egami}, {McIntosh}, {Meisenheimer},
  {P{\'e}rez-Gonz{\'a}lez}, {Rieke}, {Rieke}, {Rigby}, \& {Rix}}]{BPWLC05}
{Bell}, E.~F., {Papovich}, C., {Wolf}, C., {et~al.} 2005, \apj, 625, 23

\bibitem[{{Bessell}(1990)}]{bes90}
{Bessell}, M.~S. 1990, \pasp, 102, 1181

\bibitem[{{Blanton} \& {Roweis}(2007)}]{BR07}
{Blanton}, M.~R. \& {Roweis}, S. 2007, AJ, 133, 734

\bibitem[{{Borgani} {et~al.}(1999){Borgani}, {Girardi}, {Carlberg}, {Yee}, \&
  {Ellingson}}]{BGCYE99}
{Borgani}, S., {Girardi}, M., {Carlberg}, R.~G., {Yee}, H.~K.~C., \&
  {Ellingson}, E. 1999, \apj, 527, 561

\bibitem[{{Boselli} \& {Gavazzi}(2006)}]{bosgav06}
{Boselli}, A. \& {Gavazzi}, G. 2006, \pasp, 118, 517

\bibitem[{{Bravo-Alfaro} {et~al.}(2004){Bravo-Alfaro}, {Brinks}, {Baker},
  {Walter}, \& {Kunth}}]{BABBWK04}
{Bravo-Alfaro}, H., {Brinks}, E., {Baker}, A.~J., {Walter}, F., \& {Kunth}, D.
  2004, \aj, 127, 264

\bibitem[{{Bravo-Alfaro} {et~al.}(2000){Bravo-Alfaro}, {Cayatte}, {van Gorkom},
  \& {Balkowski}}]{BCBG00}
{Bravo-Alfaro}, H., {Cayatte}, V., {van Gorkom}, J.~H., \& {Balkowski}, C.
  2000, \aj, 119, 580

\bibitem[{{Burstein}(1979)}]{B79}
{Burstein}, D. 1979, \apj, 234, 435

\bibitem[{{Butcher} \& {Oemler}(1978)}]{BO78b}
{Butcher}, H. \& {Oemler}, Jr., A. 1978, \apj, 226, 559

\bibitem[{{Butcher} \& {Oemler}(1984)}]{BO84a}
{Butcher}, H. \& {Oemler}, Jr., A. 1984, \apj, 285, 426

\bibitem[{{Byrd} \& {Valtonen}(1990)}]{BV90}
{Byrd}, G. \& {Valtonen}, M. 1990, \apj, 350, 89

\bibitem[{{Cardelli} {et~al.}(1989){Cardelli}, {Clayton}, \& {Mathis}}]{CCM89}
{Cardelli}, J.~A., {Clayton}, G.~C., \& {Mathis}, J.~S. 1989, \apj, 345, 245

\bibitem[{{Carlberg} {et~al.}(1996){Carlberg}, {Yee}, {Ellingson}, {Abraham},
  {Gravel}, {Morris}, \& {Pritchet}}]{CYEAGMP96}
{Carlberg}, R.~G., {Yee}, H.~K.~C., {Ellingson}, E., {et~al.} 1996, \apj, 462,
  32

\bibitem[{{Cayatte} {et~al.}(1994){Cayatte}, {Kotanyi}, {Balkowski}, \& {van
  Gorkom}}]{CKBG94}
{Cayatte}, V., {Kotanyi}, C., {Balkowski}, C., \& {van Gorkom}, J.~H. 1994,
  \aj, 107, 1003

\bibitem[{{Cohen}(1976)}]{Cohen76}
{Cohen}, J.~G. 1976, \apj, 203, 587

\bibitem[{{Conselice} {et~al.}(2000){Conselice}, {Bershady}, \&
  {Jangren}}]{CBJ00}
{Conselice}, C.~J., {Bershady}, M.~A., \& {Jangren}, A. 2000, ApJ, 529, 886

\bibitem[{{Couch} {et~al.}(2001){Couch}, {Balogh}, {Bower}, {Smail},
  {Glazebrook}, \& {Taylor}}]{CBBSG01}
{Couch}, W.~J., {Balogh}, M.~L., {Bower}, R.~G., {et~al.} 2001, \apj, 549, 820

\bibitem[{{Couch} {et~al.}(1998){Couch}, {Barger}, {Smail}, {Ellis}, \&
  {Sharples}}]{CBSES98}
{Couch}, W.~J., {Barger}, A.~J., {Smail}, I., {Ellis}, R.~S., \& {Sharples},
  R.~M. 1998, \apj, 497, 188

\bibitem[{{Cowie} \& {Songaila}(1977)}]{CS77}
{Cowie}, L.~L. \& {Songaila}, A. 1977, \nat, 266, 501

\bibitem[{{Daigle} {et~al.}(2006){Daigle}, {Carignan}, {Amram}, {Hernandez},
  {Chemin}, {Balkowski}, \& {Kennicutt}}]{DCAHCBK06}
{Daigle}, O., {Carignan}, C., {Amram}, P., {et~al.} 2006, \mnras, 367, 469

\bibitem[{{Dale} {et~al.}(2001){Dale}, {Giovanelli}, {Haynes}, {Hardy}, \&
  {Campusano}}]{DGHHC01}
{Dale}, D.~A., {Giovanelli}, R., {Haynes}, M.~P., {Hardy}, E., \& {Campusano},
  L.~E. 2001, \aj, 121, 1886

\bibitem[{{Davies} \& {Lewis}(1973)}]{DL73}
{Davies}, R.~D. \& {Lewis}, B.~M. 1973, \mnras, 165, 231

\bibitem[{{De Propris} {et~al.}(2007){De Propris}, {Conselice}, {Liske},
  {Driver}, {Patton}, {Graham}, \& {Allen}}]{PCLDPGA07}
{De Propris}, R., {Conselice}, C.~J., {Liske}, J., {et~al.} 2007, \apj, 666,
  212

\bibitem[{{Donahue}(1996{\natexlab{a}})}]{Don96}
{Donahue}, M. 1996{\natexlab{a}}, in Astronomical Society of the Pacific
  Conference Series, Vol.~88, Clusters, Lensing, and the Future of the
  Universe, ed. {V.~Trimble \& A.~Reisenegger}, 76--+

\bibitem[{{Donahue}(1996{\natexlab{b}})}]{D96}
{Donahue}, M. 1996{\natexlab{b}}, \apj, 468, 79

\bibitem[{{Dressler}(1980)}]{Dress80}
{Dressler}, A. 1980, \apj, 236, 351

\bibitem[{{Dressler} \& {Sandage}(1983)}]{DS83}
{Dressler}, A. \& {Sandage}, A. 1983, \apj, 265, 664

\bibitem[{{Dressler} {et~al.}(1999){Dressler}, {Smail}, {Poggianti}, {Butcher},
  {Couch}, {Ellis}, \& {Oemler}}]{DSPBC99}
{Dressler}, A., {Smail}, I., {Poggianti}, B.~M., {et~al.} 1999, \apjs, 122, 51

\bibitem[{{Elbaz} {et~al.}(2007){Elbaz}, {Daddi}, {Le Borgne}, {Dickinson},
  {Alexander}, {Chary}, {Starck}, {Brandt}, {Kitzbichler}, {MacDonald},
  {Nonino}, {Popesso}, {Stern}, \& {Vanzella}}]{EDLDA07}
{Elbaz}, D., {Daddi}, E., {Le Borgne}, D., {et~al.} 2007, \aap, 468, 33

\bibitem[{{Ellingson} {et~al.}(2001){Ellingson}, {Lin}, {Yee}, \&
  {Carlberg}}]{ELYC01}
{Ellingson}, E., {Lin}, H., {Yee}, H.~K.~C., \& {Carlberg}, R.~G. 2001, \apj,
  547, 609

\bibitem[{{Ellingson} {et~al.}(1998){Ellingson}, {Yee}, {Abraham}, {Morris}, \&
  {Carlberg}}]{EYAMC98}
{Ellingson}, E., {Yee}, H.~K.~C., {Abraham}, R.~G., {Morris}, S.~L., \&
  {Carlberg}, R.~G. 1998, \apjs, 116, 247

\bibitem[{{Evrard}(1991)}]{Evra91}
{Evrard}, A.~E. 1991, \mnras, 248, 8P

\bibitem[{{Fujita}(1998)}]{Fujit98}
{Fujita}, Y. 1998, \apj, 509, 587

\bibitem[{{Fujita} \& {Nagashima}(1999)}]{FN99}
{Fujita}, Y. \& {Nagashima}, M. 1999, \apj, 516, 619

\bibitem[{{Gavazzi}(1987)}]{G87}
{Gavazzi}, G. 1987, \apj, 320, 96

\bibitem[{{Gavazzi}(1989)}]{G89}
{Gavazzi}, G. 1989, \apj, 346, 59

\bibitem[{{Gavazzi} {et~al.}(1995){Gavazzi}, {Contursi}, {Carrasco}, {Boselli},
  {Kennicutt}, {Scodeggio}, \& {Jaffe}}]{GCCBK95}
{Gavazzi}, G., {Contursi}, A., {Carrasco}, L., {et~al.} 1995, \aap, 304, 325

\bibitem[{{Giovanelli} \& {Haynes}(1985)}]{GH85}
{Giovanelli}, R. \& {Haynes}, M.~P. 1985, \apj, 292, 404

\bibitem[{{Giovanelli} {et~al.}(1994){Giovanelli}, {Haynes}, {Salzer},
  {Wegner}, {da Costa}, \& {Freudling}}]{GHSWCF94}
{Giovanelli}, R., {Haynes}, M.~P., {Salzer}, J.~J., {et~al.} 1994, \aj, 107,
  2036

\bibitem[{{Girardi} \& {Mezzetti}(2001)}]{GM01}
{Girardi}, M. \& {Mezzetti}, M. 2001, \apj, 548, 79

\bibitem[{{Gisler}(1980)}]{G80}
{Gisler}, G.~R. 1980, \aj, 85, 623

\bibitem[{{Goto} {et~al.}(2003){Goto}, {Okamura}, {Sekiguchi}, {Bernardi},
  {Brinkmann}, {G{\'o}mez}, {Harvanek}, {Kleinman}, {Krzesinski}, {Long},
  {Loveday}, {Miller}, {Neilsen}, {Newman}, {Nitta}, {Sheth}, {Snedden}, \&
  {Yamauchi}}]{GOSBB03}
{Goto}, T., {Okamura}, S., {Sekiguchi}, M., {et~al.} 2003, \pasj, 55, 757

\bibitem[{{Gottl{\"o}ber} {et~al.}(2001){Gottl{\"o}ber}, {Klypin}, \&
  {Kravtsov}}]{GKK01}
{Gottl{\"o}ber}, S., {Klypin}, A., \& {Kravtsov}, A.~V. 2001, \apj, 546, 223

\bibitem[{{Gunn} \& {Gott}(1972)}]{GG72}
{Gunn}, J.~E. \& {Gott}, J.~R.~I. 1972, \apj, 176, 1

\bibitem[{{Hibbard} \& {van Gorkom}(1996)}]{HVG96}
{Hibbard}, J.~E. \& {van Gorkom}, J.~H. 1996, \aj, 111, 655

\bibitem[{{Hinz} {et~al.}(2003){Hinz}, {Rieke}, \& {Caldwell}}]{HRC03}
{Hinz}, J.~L., {Rieke}, G.~H., \& {Caldwell}, N. 2003, \aj, 126, 2622

\bibitem[{{Huchra} \& {Geller}(1982)}]{HG82}
{Huchra}, J.~P. \& {Geller}, M.~J. 1982, \apj, 257, 423

\bibitem[{{Ibata} {et~al.}(1994){Ibata}, {Gilmore}, \& {Irwin}}]{ibata94}
{Ibata}, R.~A., {Gilmore}, G., \& {Irwin}, M.~J. 1994, Nature, 370, 194

\bibitem[{{J{\"a}ger} {et~al.}(2004){J{\"a}ger}, {Ziegler}, {B{\"o}hm},
  {Heidt}, {M{\"o}llenhoff}, {Hopp}, {Mendez}, \& {Wagner}}]{JZBHM04}
{J{\"a}ger}, K., {Ziegler}, B.~L., {B{\"o}hm}, A., {et~al.} 2004, \aap, 422,
  907 (Paper II)

\bibitem[{{Jeltema} {et~al.}(2005){Jeltema}, {Canizares}, {Bautz}, \&
  {Buote}}]{JCBB05}
{Jeltema}, T.~E., {Canizares}, C.~R., {Bautz}, M.~W., \& {Buote}, D.~A. 2005,
  \apj, 624, 606

\bibitem[{{Kapferer} {et~al.}(2008){Kapferer}, {Kronberger}, {Ferrari},
  {Riser}, \& {Schindler}}]{KKFRS08}
{Kapferer}, W., {Kronberger}, T., {Ferrari}, C., {Riser}, T., \& {Schindler},
  S. 2008, \mnras, 389, 1405

\bibitem[{{Kapferer} {et~al.}(2009){Kapferer}, {Sluka}, {Schindler}, {Ferrari},
  \& {Ziegler}}]{KSSFZ09}
{Kapferer}, W., {Sluka}, C., {Schindler}, S., {Ferrari}, C., \& {Ziegler}, B.
  2009, \aap, 499, 87

\bibitem[{{Kelson}(2003)}]{K03}
{Kelson}, D.~D. 2003, \pasp, 115, 688

\bibitem[{{Kennicutt}(1992)}]{K92}
{Kennicutt}, Jr., R.~C. 1992, \apj, 388, 310

\bibitem[{{Kennicutt}(1998)}]{K98}
{Kennicutt}, Jr., R.~C. 1998, \araa, 36, 189

\bibitem[{{Kennicutt} {et~al.}(2003){Kennicutt}, {Armus}, {Bendo}, {Calzetti},
  {Dale}, {Draine}, {Engelbracht}, {Gordon}, {Grauer}, {Helou}, {Hollenbach},
  {Jarrett}, {Kewley}, {Leitherer}, {Li}, {Malhotra}, {Regan}, {Rieke},
  {Rieke}, {Roussel}, {Smith}, {Thornley}, \& {Walter}}]{KABCDDE03}
{Kennicutt}, Jr., R.~C., {Armus}, L., {Bendo}, G., {et~al.} 2003, \pasp, 115,
  928

\bibitem[{{Kennicutt} {et~al.}(1994){Kennicutt}, {Tamblyn}, \&
  {Congdon}}]{KTC94}
{Kennicutt}, Jr., R.~C., {Tamblyn}, P., \& {Congdon}, C.~E. 1994, \apj, 435, 22

\bibitem[{{Kidger}(2003)}]{Kidger03}
{Kidger}, M.~R. 2003, http://www.britastro.org/asteroids/USNO photometry.htm

\bibitem[{{Kneib} {et~al.}(1995){Kneib}, {Mellier}, {Pello}, {Miralda-Escude},
  {Le Borgne}, {Boehringer}, \& {Picat}}]{KMPM95}
{Kneib}, J.~P., {Mellier}, Y., {Pello}, R., {et~al.} 1995, \aap, 303, 27

\bibitem[{{Kodama} \& {Bower}(2001)}]{KB01}
{Kodama}, T. \& {Bower}, R.~G. 2001, \mnras, 321, 18

\bibitem[{{Kodama} \& {Smail}(2001)}]{KS01}
{Kodama}, T. \& {Smail}, I. 2001, \mnras, 326, 637

\bibitem[{{Krajnovi{\'c}} {et~al.}(2006){Krajnovi{\'c}}, {Cappellari}, {de
  Zeeuw}, \& {Copin}}]{KCZC06}
{Krajnovi{\'c}}, D., {Cappellari}, M., {de Zeeuw}, P.~T., \& {Copin}, Y. 2006,
  MNRAS, 366, 787

\bibitem[{{Kronberger} {et~al.}(2008){Kronberger}, {Kapferer},
  {Unterguggenberger}, {Schindler}, \& {Ziegler}}]{KKUSZ08}
{Kronberger}, T., {Kapferer}, W., {Unterguggenberger}, S., {Schindler}, S., \&
  {Ziegler}, B.~L. 2008, \aap, 483, 783

\bibitem[{{Kutdemir} {et~al.}(2008){Kutdemir}, {Ziegler}, {Peletier}, {Da
  Rocha}, {Kronberger}, {Kapferer}, {Schindler}, {B{\"o}hm}, {J{\"a}ger},
  {Kuntschner}, \& {Verdugo}}]{KZPR08}
{Kutdemir}, E., {Ziegler}, B.~L., {Peletier}, R.~F., {et~al.} 2008, \aap, 488,
  117 (Paper III)

\bibitem[{{Lambas} {et~al.}(2003){Lambas}, {Tissera}, {Alonso}, \&
  {Coldwell}}]{LTAC03}
{Lambas}, D.~G., {Tissera}, P.~B., {Alonso}, M.~S., \& {Coldwell}, G. 2003,
  \mnras, 346, 1189

\bibitem[{{Larson} {et~al.}(1980){Larson}, {Tinsley}, \& {Caldwell}}]{LTC80}
{Larson}, R.~B., {Tinsley}, B.~M., \& {Caldwell}, C.~N. 1980, ApJ, 237, 692

\bibitem[{{Lewis} {et~al.}(1999){Lewis}, {Ellingson}, {Morris}, \&
  {Carlberg}}]{LEMC99}
{Lewis}, A.~D., {Ellingson}, E., {Morris}, S.~L., \& {Carlberg}, R.~G. 1999,
  \apj, 517, 587

\bibitem[{{Lotz} {et~al.}(2004){Lotz}, {Primack}, \& {Madau}}]{LPM04}
{Lotz}, J.~M., {Primack}, J., \& {Madau}, P. 2004, \aj, 128, 163

\bibitem[{{Luppino} {et~al.}(1999){Luppino}, {Gioia}, {Hammer}, {Le F{\`e}vre},
  \& {Annis}}]{LGHLA99}
{Luppino}, G.~A., {Gioia}, I.~M., {Hammer}, F., {Le F{\`e}vre}, O., \& {Annis},
  J.~A. 1999, \aaps, 136, 117

\bibitem[{{Macci{\`o}} {et~al.}(2006){Macci{\`o}}, {Moore}, \&
  {Stadel}}]{MMS06}
{Macci{\`o}}, A.~V., {Moore}, B., \& {Stadel}, J. 2006, \apjl, 636, L25

\bibitem[{{Makino} \& {Hut}(1997)}]{MH97}
{Makino}, J. \& {Hut}, P. 1997, \apj, 481, 83

\bibitem[{{Maschietto} {et~al.}(2008){Maschietto}, {Hatch}, {Venemans},
  {R{\"o}ttgering}, {Miley}, {Overzier}, {Dopita}, {Eisenhardt}, {Kurk},
  {Meurer}, {Pentericci}, {Rosati}, {Stanford}, {van Breugel}, \&
  {Zirm}}]{MHVRM08}
{Maschietto}, F., {Hatch}, N.~A., {Venemans}, B.~P., {et~al.} 2008, \mnras,
  389, 1223

\bibitem[{{Menanteau} {et~al.}(2006){Menanteau}, {Ford}, {Motta},
  {Ben{\'{\i}}tez}, {Martel}, {Blakeslee}, \& {Infante}}]{MFMBMBI06}
{Menanteau}, F., {Ford}, H.~C., {Motta}, V., {et~al.} 2006, \aj, 131, 208

\bibitem[{{Mihos}(2004)}]{M04}
{Mihos}, J.~C. 2004, in Clusters of Galaxies: Probes of Cosmological Structure
  and Galaxy Evolution, ed. J.~S. {Mulchaey}, A.~{Dressler}, \& A.~{Oemler},
  277--+

\bibitem[{{Milvang-Jensen} {et~al.}(2008){Milvang-Jensen}, {Noll}, {Halliday},
  {Poggianti}, {Jablonka}, {Arag{\'o}n-Salamanca}, {Saglia}, {Nowak}, {von der
  Linden}, {De Lucia}, {Pell{\'o}}, {Moustakas}, {Poirier}, {Bamford}, {Clowe},
  {Dalcanton}, {Rudnick}, {Simard}, {White}, \& {Zaritsky}}]{JNHPJ08}
{Milvang-Jensen}, B., {Noll}, S., {Halliday}, C., {et~al.} 2008, \aap, 482, 419

\bibitem[{{Monet et al.}(1998)}]{M98}
{Monet et al.}, D. 1998, VizieR Online Data Catalog, 1252, 0

\bibitem[{{Moore} {et~al.}(1997){Moore}, {Ghigna}, {Governato}, {Lake},
  {Quinn}, \& {Stadel}}]{9MGGLQS7}
{Moore}, B., {Ghigna}, S., {Governato}, F., {et~al.} 1997, ArXiv Astrophysics
  e-prints

\bibitem[{{Moore} {et~al.}(1996){Moore}, {Katz}, {Lake}, {Dressler}, \&
  {Oemler}}]{MLKDO96}
{Moore}, B., {Katz}, N., {Lake}, G., {Dressler}, A., \& {Oemler}, A. 1996,
  \nat, 379, 613

\bibitem[{{Moore} {et~al.}(1998){Moore}, {Lake}, \& {Katz}}]{MLK98}
{Moore}, B., {Lake}, G., \& {Katz}, N. 1998, \apj, 495, 139

\bibitem[{{Moore} {et~al.}(1999){Moore}, {Lake}, {Quinn}, \& {Stadel}}]{MLQS99}
{Moore}, B., {Lake}, G., {Quinn}, T., \& {Stadel}, J. 1999, \mnras, 304, 465

\bibitem[{{Moran} {et~al.}(2007{\natexlab{a}}){Moran}, {Ellis}, {Treu},
  {Smith}, {Rich}, \& {Smail}}]{METSRS07}
{Moran}, S.~M., {Ellis}, R.~S., {Treu}, T., {et~al.} 2007{\natexlab{a}}, \apj,
  671, 1503

\bibitem[{{Moran} {et~al.}(2007{\natexlab{b}}){Moran}, {Miller}, {Treu},
  {Ellis}, \& {Smith}}]{MMTES07}
{Moran}, S.~M., {Miller}, N., {Treu}, T., {Ellis}, R.~S., \& {Smith}, G.~P.
  2007{\natexlab{b}}, \apj, 659, 1138

\bibitem[{{Moustakas} {et~al.}(2006){Moustakas}, {Kennicutt}, \&
  {Tremonti}}]{M06}
{Moustakas}, J., {Kennicutt}, Jr., R.~C., \& {Tremonti}, C.~A. 2006, \apj, 642,
  775

\bibitem[{{Neichel} {et~al.}(2008){Neichel}, {Hammer}, {Puech}, {Flores},
  {Lehnert}, {Rawat}, {Yang}, {Delgado}, {Amram}, {Balkowski}, {Cesarsky},
  {Dannerbauer}, {Fuentes-Carrera}, {Guiderdoni}, {Kembhavi}, {Liang},
  {Nesvadba}, {{\"O}stlin}, {Pozzetti}, {Ravikumar}, {di Serego Alighieri},
  {Vergani}, {Vernet}, \& {Wozniak}}]{NHPF08}
{Neichel}, B., {Hammer}, F., {Puech}, M., {et~al.} 2008, \aap, 484, 159

\bibitem[{{Neistein} {et~al.}(1999){Neistein}, {Maoz}, {Rix}, \&
  {Tonry}}]{NMRT99}
{Neistein}, E., {Maoz}, D., {Rix}, H.-W., \& {Tonry}, J.~L. 1999, \aj, 117,
  2666

\bibitem[{{Nulsen}(1982)}]{N82}
{Nulsen}, P.~E.~J. 1982, \mnras, 198, 1007

\bibitem[{{Okamoto} \& {Nagashima}(2001)}]{ON01}
{Okamoto}, T. \& {Nagashima}, M. 2001, \apj, 547, 109

\bibitem[{{Patton} {et~al.}(2000){Patton}, {Carlberg}, {Marzke}, {Pritchet},
  {da Costa}, \& {Pellegrini}}]{PCMPCP00}
{Patton}, D.~R., {Carlberg}, R.~G., {Marzke}, R.~O., {et~al.} 2000, \apj, 536,
  153

\bibitem[{{Petrosian} {et~al.}(2002){Petrosian}, {Movsessian}, {Comte},
  {Kunth}, \& {Dodonov}}]{PMCKD02}
{Petrosian}, A.~R., {Movsessian}, T., {Comte}, G., {Kunth}, D., \& {Dodonov},
  S. 2002, \aap, 391, 487

\bibitem[{{Poggianti}(2006)}]{Pog06}
{Poggianti}, B.~M. 2006, in The Many Scales in the Universe: JENAM 2004
  Astrophysics Reviews, ed. J.~C. {Del Toro Iniesta}, E.~J. {Alfaro}, J.~G.
  {Gorgas}, E.~{Salvador-Sole}, \& H.~{Butcher}, 71--+

\bibitem[{{Poggianti} {et~al.}(1999){Poggianti}, {Smail}, {Dressler}, {Couch},
  {Barger}, {Butcher}, {Ellis}, \& {Oemler}}]{PSDCB99}
{Poggianti}, B.~M., {Smail}, I., {Dressler}, A., {et~al.} 1999, \apj, 518, 576

\bibitem[{{Puech} {et~al.}(2008){Puech}, {Flores}, {Hammer}, {Yang}, {Neichel},
  {Lehnert}, {Chemin}, {Nesvadba}, {Epinat}, {Amram}, {Balkowski}, {Cesarsky},
  {Dannerbauer}, {di Serego Alighieri}, {Fuentes-Carrera}, {Guiderdoni},
  {Kembhavi}, {Liang}, {{\"O}stlin}, {Pozzetti}, {Ravikumar}, {Rawat},
  {Vergani}, {Vernet}, \& {Wozniak}}]{PFHY08}
{Puech}, M., {Flores}, H., {Hammer}, F., {et~al.} 2008, \aap, 484, 173

\bibitem[{{Quilis} {et~al.}(2000){Quilis}, {Moore}, \& {Bower}}]{QMB00}
{Quilis}, V., {Moore}, B., \& {Bower}, R. 2000, Science, 288, 1617

\bibitem[{{Richstone}(1976)}]{R76}
{Richstone}, D.~O. 1976, \apj, 204, 642

\bibitem[{{Rubin} {et~al.}(1999){Rubin}, {Waterman}, \& {Kenney}}]{RWK99}
{Rubin}, V.~C., {Waterman}, A.~H., \& {Kenney}, J.~D.~P. 1999, \aj, 118, 236

\bibitem[{{Salim} {et~al.}(2007){Salim}, {Rich}, {Charlot}, {Brinchmann},
  {Johnson}, {Schiminovich}, {Seibert}, {Mallery}, {Heckman}, {Forster},
  {Friedman}, {Martin}, {Morrissey}, {Neff}, {Small}, {Wyder}, {Bianchi},
  {Donas}, {Lee}, {Madore}, {Milliard}, {Szalay}, {Welsh}, \& {Yi}}]{SRCBJ07}
{Salim}, S., {Rich}, R.~M., {Charlot}, S., {et~al.} 2007, \apjs, 173, 267

\bibitem[{{Sancisi} {et~al.}(2008){Sancisi}, {Fraternali}, {Oosterloo}, \& {van
  der Hulst}}]{SFOH08}
{Sancisi}, R., {Fraternali}, F., {Oosterloo}, T., \& {van der Hulst}, T. 2008,
  ARAA, 15, 189

\bibitem[{{Shapiro} {et~al.}(2008){Shapiro}, {Genzel}, {F{\"o}rster Schreiber},
  {Tacconi}, {Bouch{\'e}}, {Cresci}, {Davies}, {Eisenhauer}, {Johansson},
  {Krajnovi{\'c}}, {Lutz}, {Naab}, {Arimoto}, {Arribas}, {Cimatti}, {Colina},
  {Daddi}, {Daigle}, {Erb}, {Hernandez}, {Kong}, {Mignoli}, {Onodera},
  {Renzini}, {Shapley}, \& {Steidel}}]{SGSTB08}
{Shapiro}, K.~L., {Genzel}, R., {F{\"o}rster Schreiber}, N.~M., {et~al.} 2008,
  \apj, 682, 231 (S08)

\bibitem[{Sikkema(2009)}]{Sikkema09}
Sikkema, G. 2009, Phd thesis, Kapteyn Astronomical Institute, Groningen

\bibitem[{{Smail} {et~al.}(1997){Smail}, {Ellis}, {Dressler}, {Couch},
  {Oemler}, {Sharples}, \& {Butcher}}]{SEDCOSB97}
{Smail}, I., {Ellis}, R.~S., {Dressler}, A., {et~al.} 1997, \apj, 479, 70

\bibitem[{{Solanes} {et~al.}(2001){Solanes}, {Manrique},
  {Garc{\'{\i}}a-G{\'o}mez}, {Gonz{\'a}lez-Casado}, {Giovanelli}, \&
  {Haynes}}]{SMGGG01}
{Solanes}, J.~M., {Manrique}, A., {Garc{\'{\i}}a-G{\'o}mez}, C., {et~al.} 2001,
  \apj, 548, 97

\bibitem[{{Stanford} {et~al.}(2002){Stanford}, {Eisenhardt}, {Dickinson},
  {Holden}, \& {De Propris}}]{SEDHP02}
{Stanford}, S.~A., {Eisenhardt}, P.~R., {Dickinson}, M., {Holden}, B.~P., \&
  {De Propris}, R. 2002, \apjs, 142, 153

\bibitem[{{Stanonik} {et~al.}(2009){Stanonik}, {Platen}, {Arag{\'o}n-Calvo},
  {van Gorkom}, {van de Weygaert}, {van der Hulst}, \& {Peebles}}]{SPAVG09}
{Stanonik}, K., {Platen}, E., {Arag{\'o}n-Calvo}, M.~A., {et~al.} 2009, \apjl,
  696, L6

\bibitem[{{Stocke} {et~al.}(1991){Stocke}, {Morris}, {Gioia}, {Maccacaro},
  {Schild}, {Wolter}, {Fleming}, \& {Henry}}]{SMGMSWFH91}
{Stocke}, J.~T., {Morris}, S.~L., {Gioia}, I.~M., {et~al.} 1991, \apjs, 76, 813

\bibitem[{{Teplitz} {et~al.}(2000){Teplitz}, {Malkan}, {Steidel}, {McLean},
  {Becklin}, {Figer}, {Gilbert}, {Graham}, {Larkin}, {Levenson}, \&
  {Wilcox}}]{TMSMBF00}
{Teplitz}, H.~I., {Malkan}, M.~A., {Steidel}, C.~C., {et~al.} 2000, \apj, 542,
  18

\bibitem[{{Tonry} {et~al.}(2003){Tonry}, {Schmidt}, {Barris}, {Candia},
  {Challis}, {Clocchiatti}, {Coil}, {Filippenko}, {Garnavich}, {Hogan},
  {Holland}, {Jha}, {Kirshner}, {Krisciunas}, {Leibundgut}, {Li}, {Matheson},
  {Phillips}, {Riess}, {Schommer}, {Smith}, {Sollerman}, {Spyromilio},
  {Stubbs}, \& {Suntzeff}}]{TSBCC03}
{Tonry}, J.~L., {Schmidt}, B.~P., {Barris}, B., {et~al.} 2003, ApJ, 594, 1

\bibitem[{{Trujillo} {et~al.}(2004){Trujillo}, {Rudnick}, {Rix}, {Labb{\'e}},
  {Franx}, {Daddi}, {van Dokkum}, {F{\"o}rster Schreiber}, {Kuijken},
  {Moorwood}, {R{\"o}ttgering}, {van de Wel}, {van der Werf}, \& {van
  Starkenburg}}]{TRR04}
{Trujillo}, I., {Rudnick}, G., {Rix}, H.-W., {et~al.} 2004, \apj, 604, 521

\bibitem[{{Tully} \& {Fouqu\'e}(1985)}]{TF85}
{Tully}, R.~B. \& {Fouqu\'e}, P. 1985, \apjs, 58, 67

\bibitem[{{Tully} {et~al.}(1998){Tully}, {Pierce}, {Huang}, {Saunders},
  {Verheijen}, \& {Witchalls}}]{TPHSVW98}
{Tully}, R.~B., {Pierce}, M.~J., {Huang}, J.-S., {et~al.} 1998, \aj, 115, 2264

\bibitem[{{van den Bergh}(1976)}]{Bergh76}
{van den Bergh}, S. 1976, \apj, 206, 883

\bibitem[{{van Gorkom} \& {Schiminovich}(1997)}]{VGS97}
{van Gorkom}, J. \& {Schiminovich}, D. 1997, in Astronomical Society of the
  Pacific Conference Series, Vol. 116, The Nature of Elliptical Galaxies; 2nd
  Stromlo Symposium, ed. M.~{Arnaboldi}, G.~S. {Da Costa}, \& P.~{Saha}, 310--+

\bibitem[{{Verdugo} {et~al.}(2008){Verdugo}, {Ziegler}, \& {Gerken}}]{VZG08}
{Verdugo}, M., {Ziegler}, B.~L., \& {Gerken}, B. 2008, \aap, 486, 9

\bibitem[{{Vogt} {et~al.}(2004){Vogt}, {Haynes}, {Giovanelli}, \&
  {Herter}}]{VHGH04}
{Vogt}, N.~P., {Haynes}, M.~P., {Giovanelli}, R., \& {Herter}, T. 2004, \aj,
  127, 3300

\bibitem[{{Wakker} {et~al.}(2007){Wakker}, {York}, {Howk}, {Barentine},
  {Wilhelm}, {Peletier}, {van Woerden}, {Beers}, {Ivezi{\'c}}, {Richter}, \&
  {Schwarz}}]{WYHBW07}
{Wakker}, B.~P., {York}, D.~G., {Howk}, J.~C., {et~al.} 2007, ApJL, 670, L113

\bibitem[{{Wakker} {et~al.}(2008){Wakker}, {York}, {Wilhelm}, {Barentine},
  {Richter}, {Beers}, {Ivezi{\'c}}, \& {Howk}}]{WYWBR08}
{Wakker}, B.~P., {York}, D.~G., {Wilhelm}, R., {et~al.} 2008, ApJ, 672, 298

\bibitem[{{Wright}(2006)}]{W06}
{Wright}, E.~L. 2006, \pasp, 118, 1711

\bibitem[{{Yang} {et~al.}(2008){Yang}, {Flores}, {Hammer}, {Neichel}, {Puech},
  {Nesvadba}, {Rawat}, {Cesarsky}, {Lehnert}, {Pozzetti}, {Fuentes-Carrera},
  {Amram}, {Balkowski}, {Dannerbauer}, {di Serego Alighieri}, {Guiderdoni},
  {Kembhavi}, {Liang}, {{\"O}stlin}, {Ravikumar}, {Vergani}, {Vernet}, \&
  {Wozniak}}]{YFHN08}
{Yang}, Y., {Flores}, H., {Hammer}, F., {et~al.} 2008, \aap, 477, 789 (Y08)

\bibitem[{{Yee} {et~al.}(1998){Yee}, {Ellingson}, {Morris}, {Abraham}, \&
  {Carlberg}}]{YEMAC98}
{Yee}, H.~K.~C., {Ellingson}, E., {Morris}, S.~L., {Abraham}, R.~G., \&
  {Carlberg}, R.~G. 1998, \apjs, 116, 211

\bibitem[{Ziegler {et~al.}(2003)Ziegler, B{\"o}hm, {J{\" a}ger}, {Heidt}, \&
  M{\"o}llenhoff}]{ZBJHM03}
Ziegler, B.~L., B{\"o}hm, A., {J{\" a}ger}, K., {Heidt}, J., \& M{\"o}llenhoff,
  C. 2003, ApJL, 598, L87 (Paper I)

\end{thebibliography}

\Online

\appendix
\section{Relations between kinematic (ir)regularity parameters \label{app_rel}}

   \begin{figure*}
   \centering
   \includegraphics[width=16cm,clip]{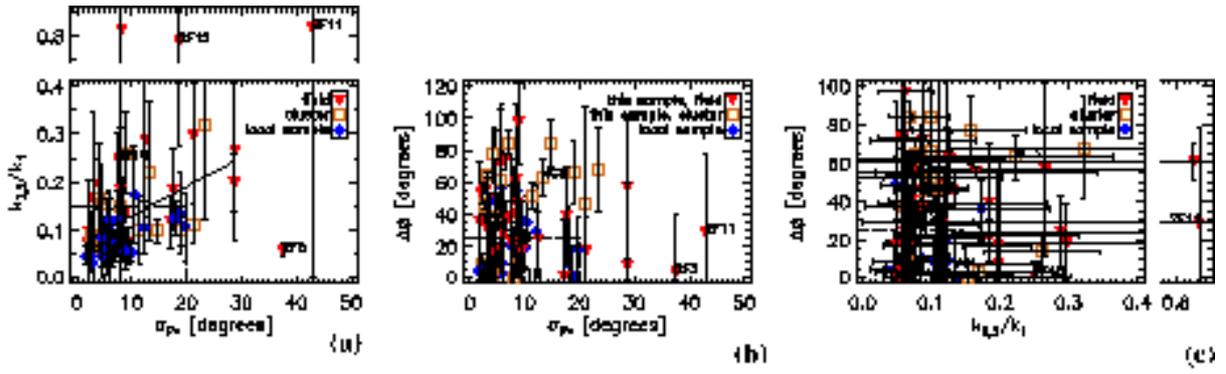}
 \caption{(Ir)regularity parameters versus each other.  The regularity threshold of each parameter is shown with a dashed line.  The galaxies in
the local sample are used only for determination of the regularity thresholds as explained in Paper III.  They are not used in correlation
measurements.  The galaxies that are excluded from the correlation as explained in Sect.~\ref{special} are indicated with their names on the plot. 
\textit{a}: standard deviation of the kinematic position angle~($\sigma_{\rm PA}$) versus mean~$k_{3,5}/k_{1}$.  The correlation we find
between the two for our cluster$+$field sample (correlation coefficient=0.5) is indicated on the plot.  Data points that are far out of the general distribution are shown
separately, on top of the main plot. \textit{b}: standard deviation of kinematic position angle~($\sigma_{\rm PA}$) versus mean misalignment
between kinematic and  photometric axes~($\Delta \phi$). \textit{c}: mean $k_{3,5}/k_{1}$ versus mean misalignment between kinematic and
photometric axes~($\Delta \phi$).  Data points that are far out of the general distribution are shown separately, on the right side of the main
plot.}
\label{cop2}
\end{figure*}

\clearpage
\section{Individual galaxies \label{app_indiv}}\vspace*{-2mm}

Here we present some figures showing the data and its analysis for each object in our sample.  The HST~image and the velocity field of each
galaxy have the same orientation (the slit position is parallel to the \mbox{$x$-axis}).  The velocities and the positions are given with
respect to the continuum center of the galaxies.

\subsection{Cluster galaxies}

   \begin{figure*}
   \centering
   \includegraphics[angle=0,width=13cm,clip]{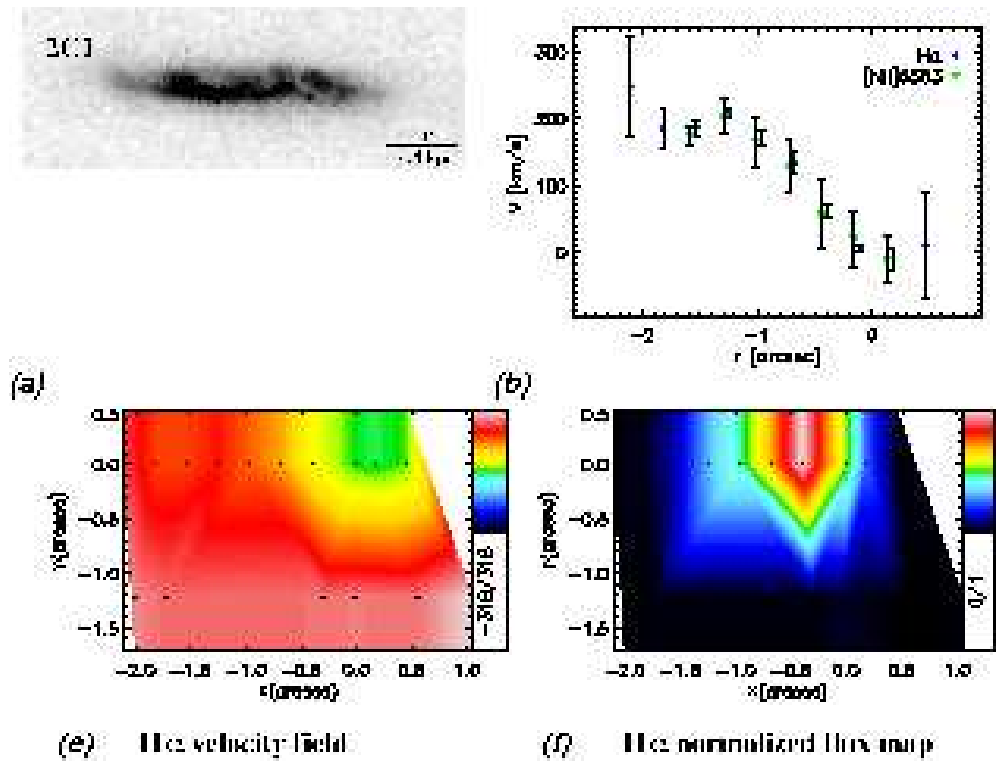}
 \caption{{\bf a)} HST-ACS image of the galaxy in the $V$~band. {\bf b)}~Rotation curves of different emission lines extracted along the central slit.
{\bf e)}~H$\alpha$~velocity field. 
{\bf f)}~Normalized H$\alpha$~flux map.}
         \label{gal2C3}
   \end{figure*}

   \begin{figure*}
   \centering
   \includegraphics[angle=0,width=13cm,clip]{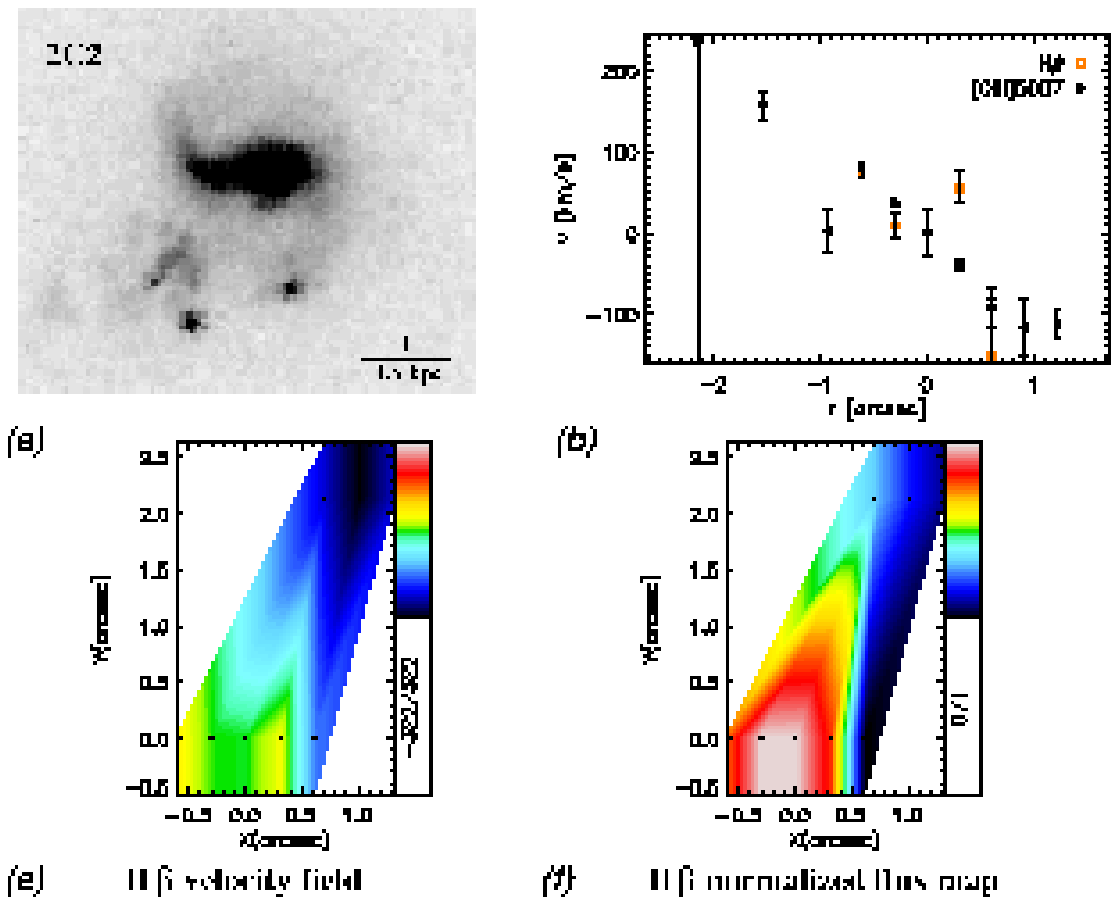}
 \caption{{\bf a)} HST-ACS image of the galaxy in the $V$~band. 
 {\bf b)}~Rotation curves of different emission lines extracted along the central slit.
{\bf e)}~H$\beta$~velocity field.
{\bf f)}~Normalized H$\beta$~flux map.}
         \label{gal2C5}
   \end{figure*}

\clearpage

   \begin{figure*}
   \centering
   \includegraphics[angle=0,width=14.5cm,clip]{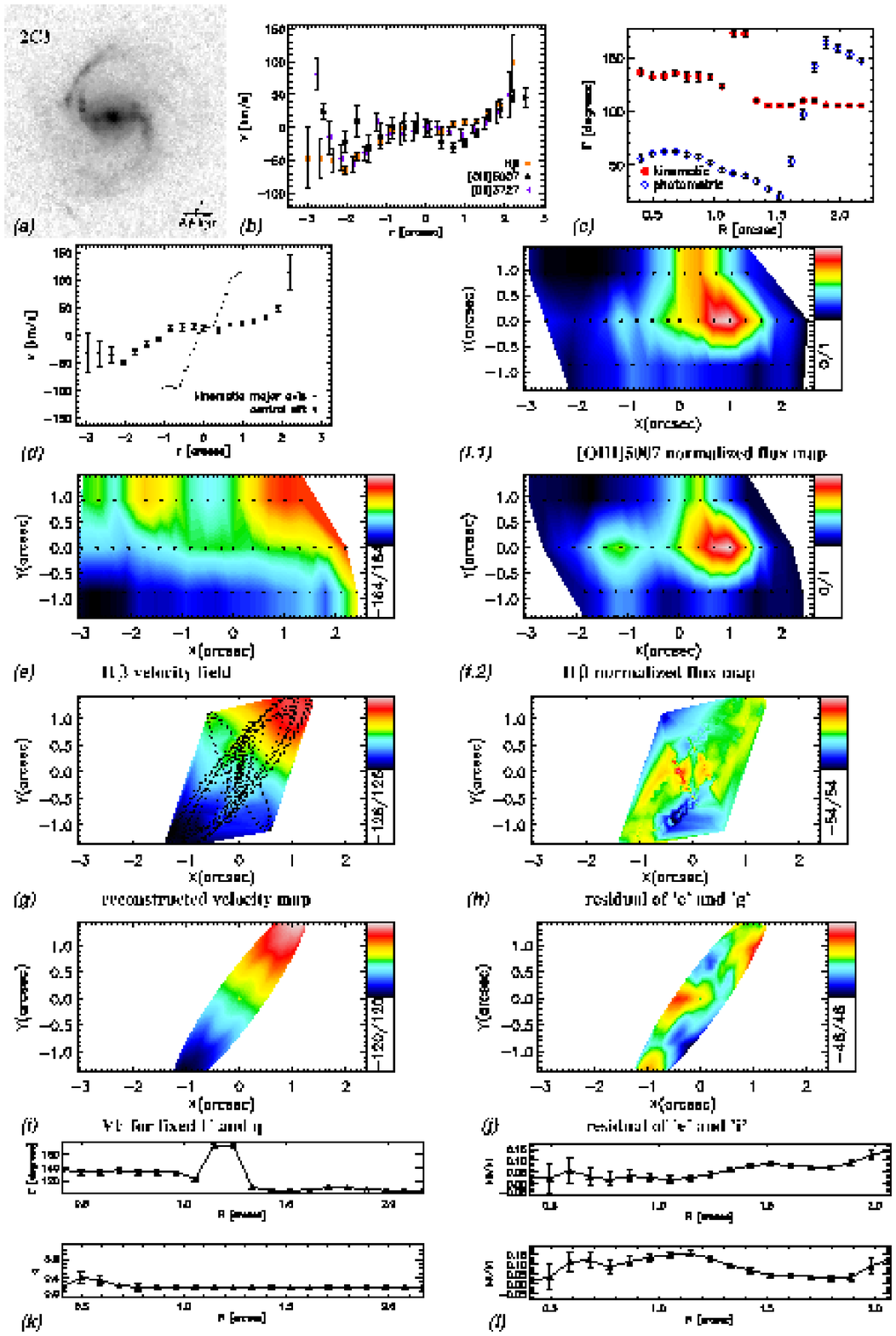}
 \caption{{\bf a)}~HST-ACS image of the galaxy in the $V$~band. {\bf b)}~Rotation curves of different emission lines extracted along the central slit.
{\bf c)}~Position angles of kinematic and photometric axes as a function of radius.
{\bf d)}~Rotation curves extracted along the central slit and the kinematic major axis.
{\bf e)}~H$\beta$~velocity field. 
{\bf f.1)}~Normalized [OIII]5007~flux map. 
{\bf f.2)}~Normalized H$\beta$~flux map. 
{\bf g)}~Velocity map reconstructed using 6~harmonic terms.
{\bf h)}~Residual of the velocity map and the reconstructed map. 
{\bf i)}~Simple rotation map constructed for position angle and ellipticity fixed to their global values.
{\bf j)}~Residual of the velocity map and the simple rotation map.
{\bf k)}~Position angle and flattening as a function of radius. 
{\bf l)}~$k_{3}/k_{1}$ and $k_{5}/k_{1}$ (from the analysis where position angle and ellipticity are fixed to their global values) as a function of radius.}
         \label{gal2C1}
         \end{figure*}

   \begin{figure*}
   \centering
   \includegraphics[angle=0,width=13cm,clip]{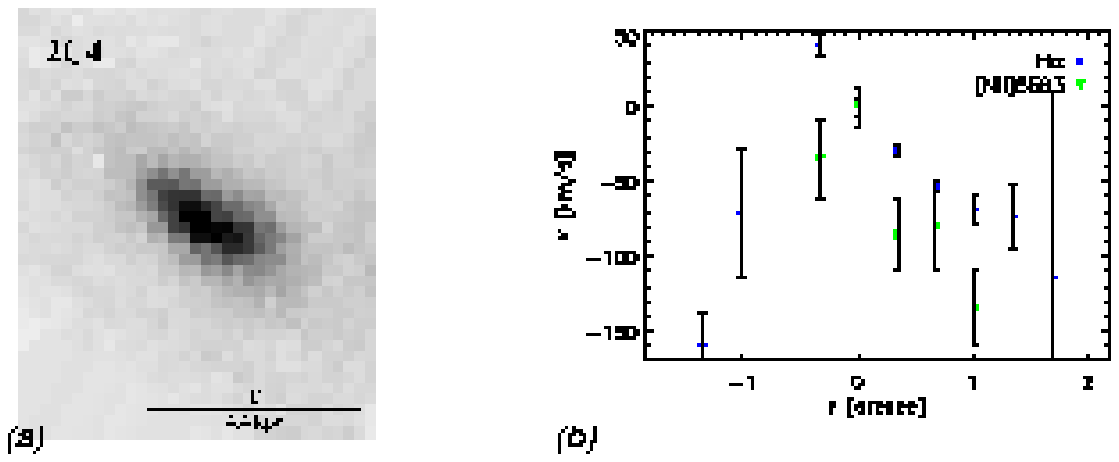}
 \caption{{\bf a)} HST-ACS image of the galaxy in the $V$~band. 
{\bf b)}~Rotation curves of different emission lines extracted along the centralslit.}
         \label{gal2C8}
   \end{figure*}

   \begin{figure*}
   \centering
   \includegraphics[angle=0,width=13.5cm,clip]{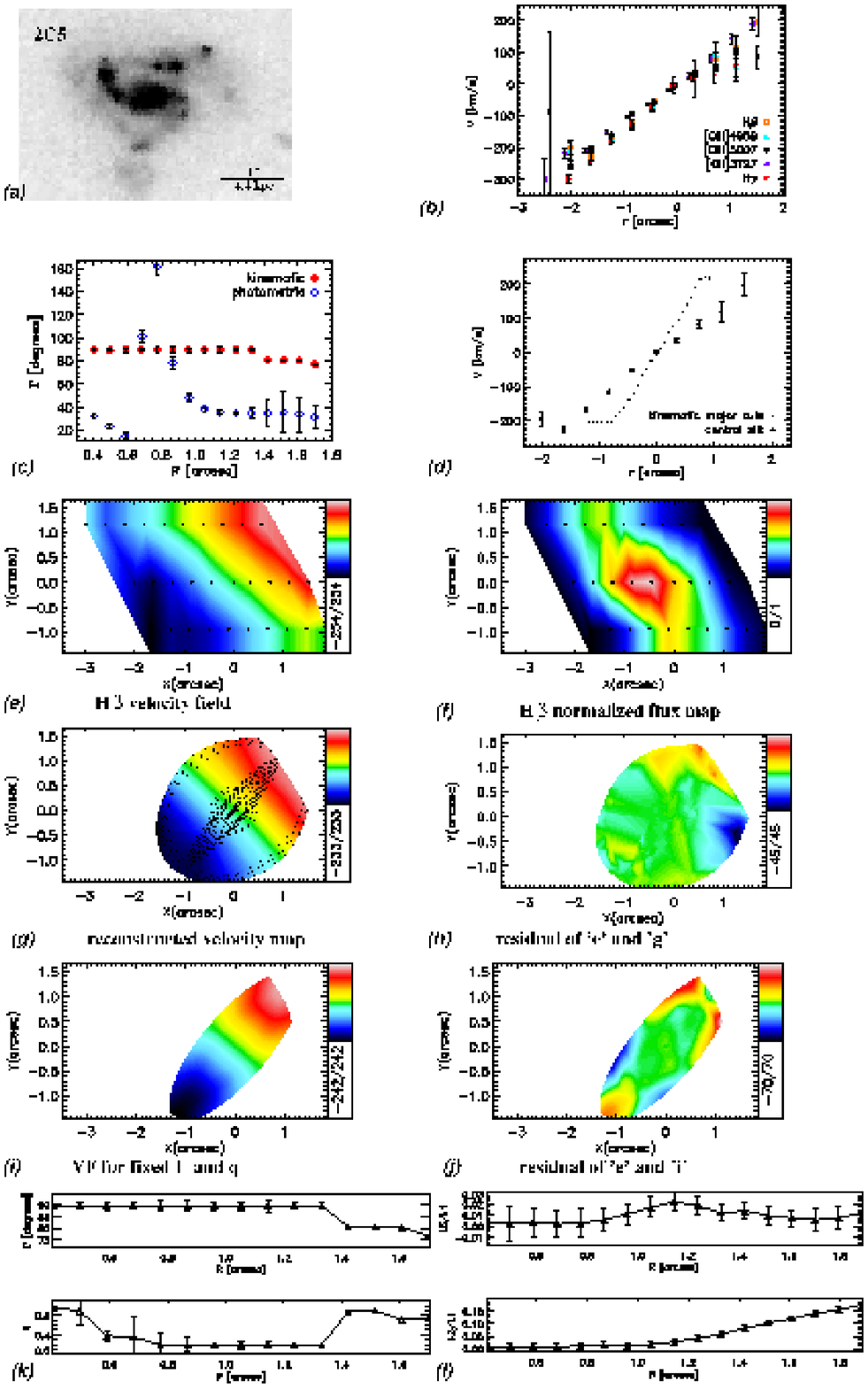}
 \caption{{\bf a)} HST-ACS image of the galaxy in the $V$~band. 
 {\bf b)}~Rotation curves of different emission lines extracted along the central slit.
{\bf c)}~Position angles of kinematic and photometric axes as a function of radius.
{\bf d)}~Rotation curves extracted along the central slit and the kinematic major axis.
{\bf e)}~H$\beta$~velocity field.
{\bf f)}~Normalized H$\beta$~flux map. 
{\bf g)}~Velocity map reconstructed using 6~harmonic terms.
{\bf h)}~Residual of the velocity map and the reconstructed map. 
{\bf i)}~Simple rotation map constructed for position angle and ellipticity fixed to their global values.
{\bf j)}~Residual of the velocity map and the simple rotation map.
{\bf k)}~Position angle and flattening as a function of radius. 
{\bf l)}~$k_{3}/k_{1}$ and $k_{5}/k_{1}$ (from the analysis where position angle and ellipticity are fixed to their global values) as a function of radius.}
         \label{gal2C2}
   \end{figure*}

   \begin{figure*}
   \centering
   \includegraphics[angle=0,width=14.5cm,clip]{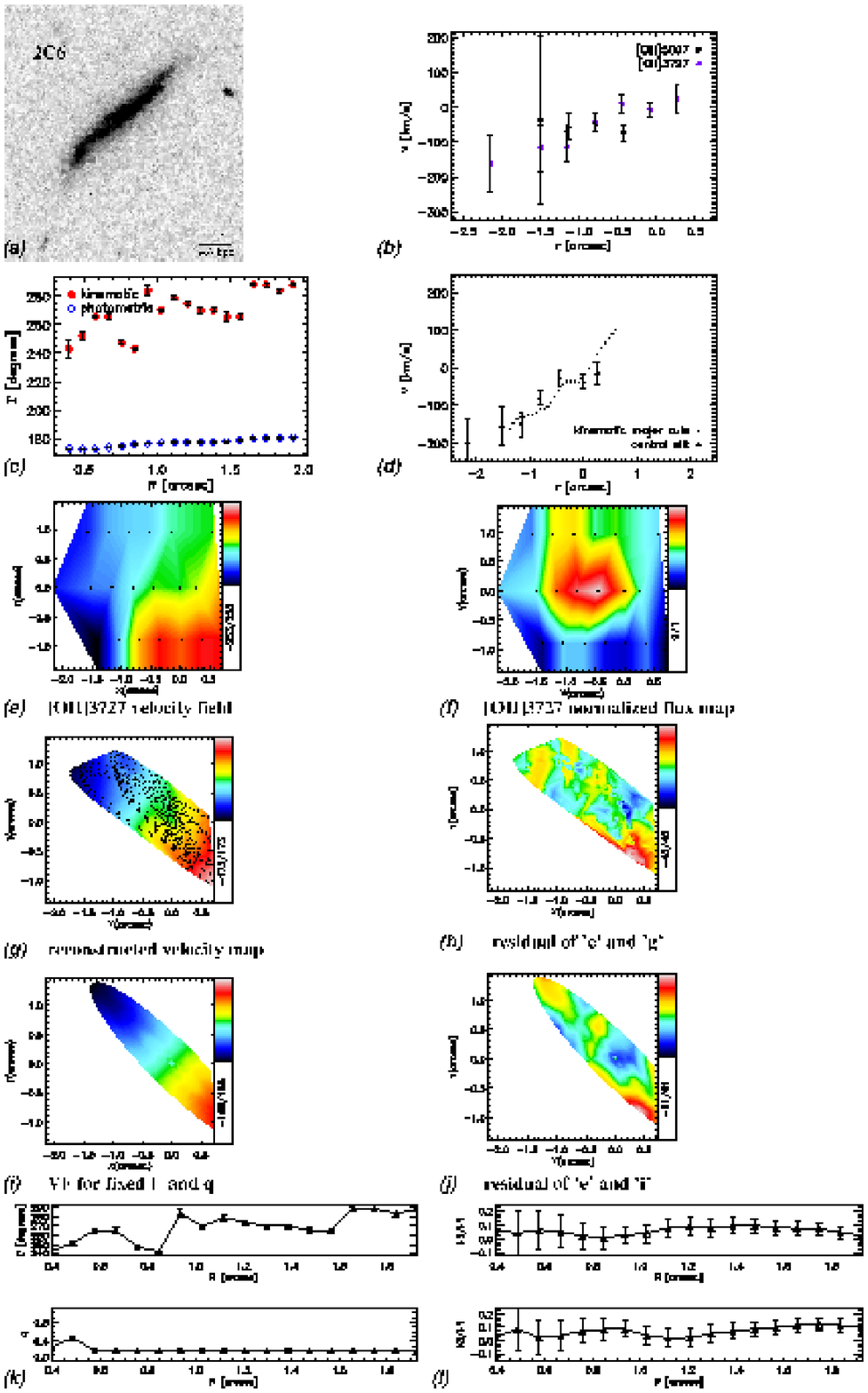}
\caption{{\bf a)} HST-ACS image of the galaxy in the $V$~band. 
{\bf b)}~Rotation curves of different emission lines extracted along the central slit.
{\bf c)}~Position angles of kinematic and photometric axes as a function of radius.
{\bf d)}~Rotation curves extracted along the central slit and the kinematic major axis.
{\bf e)}~[OII]3727~velocity field. 
{\bf f)}~Normalized [OII]3727~flux map. 
{\bf g)}~Velocity map reconstructed using 6~harmonic terms.
{\bf h)}~Residual of the velocity map and the reconstructed map. 
{\bf i)}~Simple rotation map constructed for position angle and ellipticity fixed to their global values.
{\bf j)}~Residual of the velocity map and the simple rotation map.
{\bf k)}~Position angle and flattening as a function of radius. 
{\bf l)}~$k_{3}/k_{1}$ and $k_{5}/k_{1}$ (from the analysis where position angle and ellipticity are fixed to their global values) as a function of radius.}
         \label{gal2C9}
   \end{figure*}

\clearpage

       \begin{figure*}
   \centering
   \includegraphics[angle=0,width=15.5cm,clip]{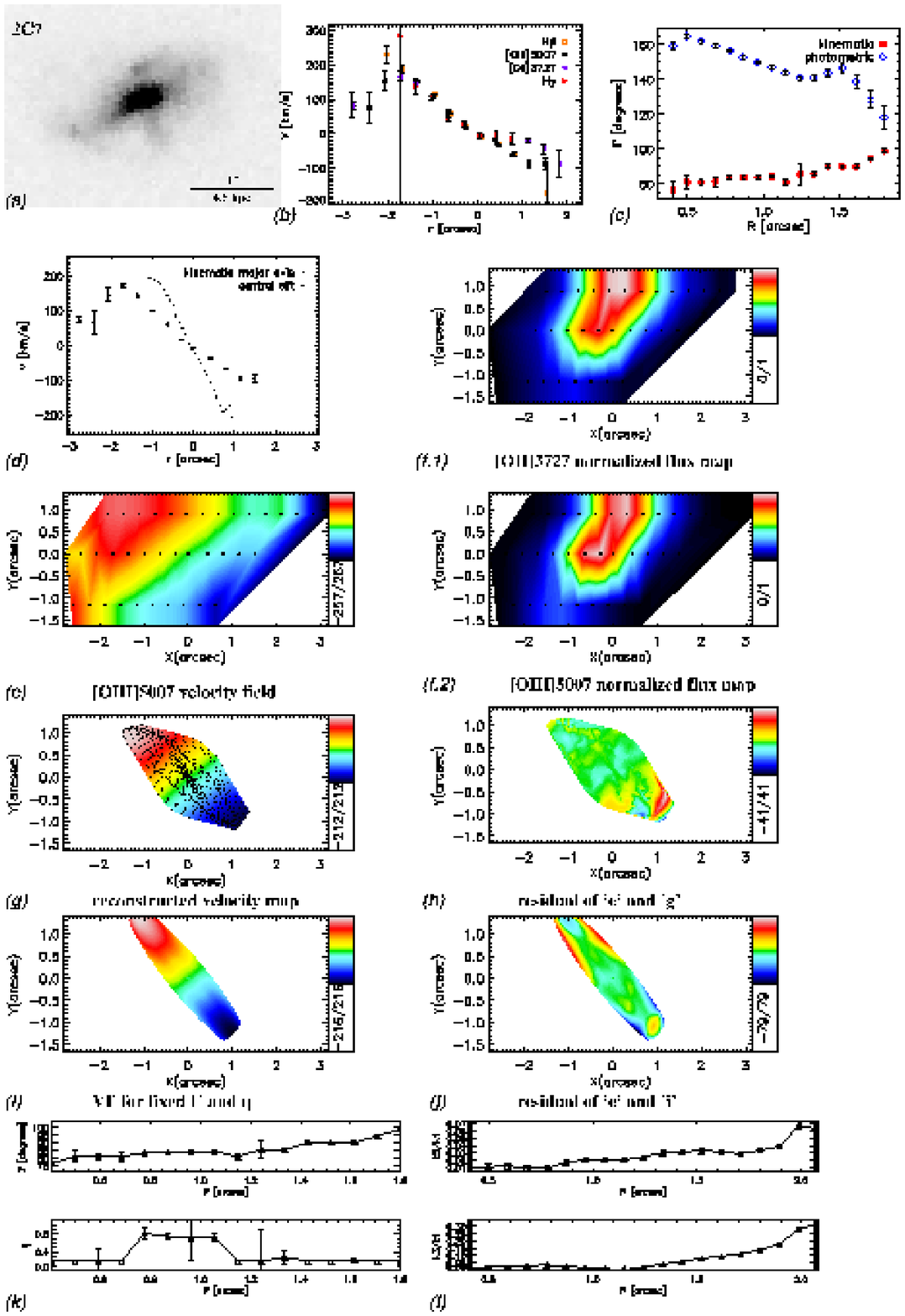}
\caption{{\bf a)} HST-ACS image of the galaxy in the $V$~band. 
{\bf b)}~Rotation curves of different emission lines extracted along the central slit.
{\bf c)}~Position angles of kinematic and photometric axes as a function of radius.
{\bf d)}~Rotation curves extracted along the central slit and the kinematic major axis.
{\bf e)}~[OIII]5007~velocity field. 
{\bf f.1)}~Normalized [OII]3727~flux map.
{\bf f.2)}~Normalized [OIII]5007~flux map.
{\bf g)}~Velocity map reconstructed using 6~harmonic terms.
{\bf h)}~Residual of the velocity map and the reconstructed map. 
{\bf i)}~Simple rotation map constructed for position angle and ellipticity fixed to their global values.
{\bf j)}~Residual of the velocity map and the simple rotation map.
{\bf k)}~Position angle and flattening as a function of radius. {\bf l)}~$k_{3}/k_{1}$ and $k_{5}/k_{1}$ (from the analysis where position angle and ellipticity are fixed to their global values) as a function of radius.}
         \label{gal2C11}
   \end{figure*}

       \begin{figure*}
   \centering
   \includegraphics[angle=0,width=13.5cm,clip]{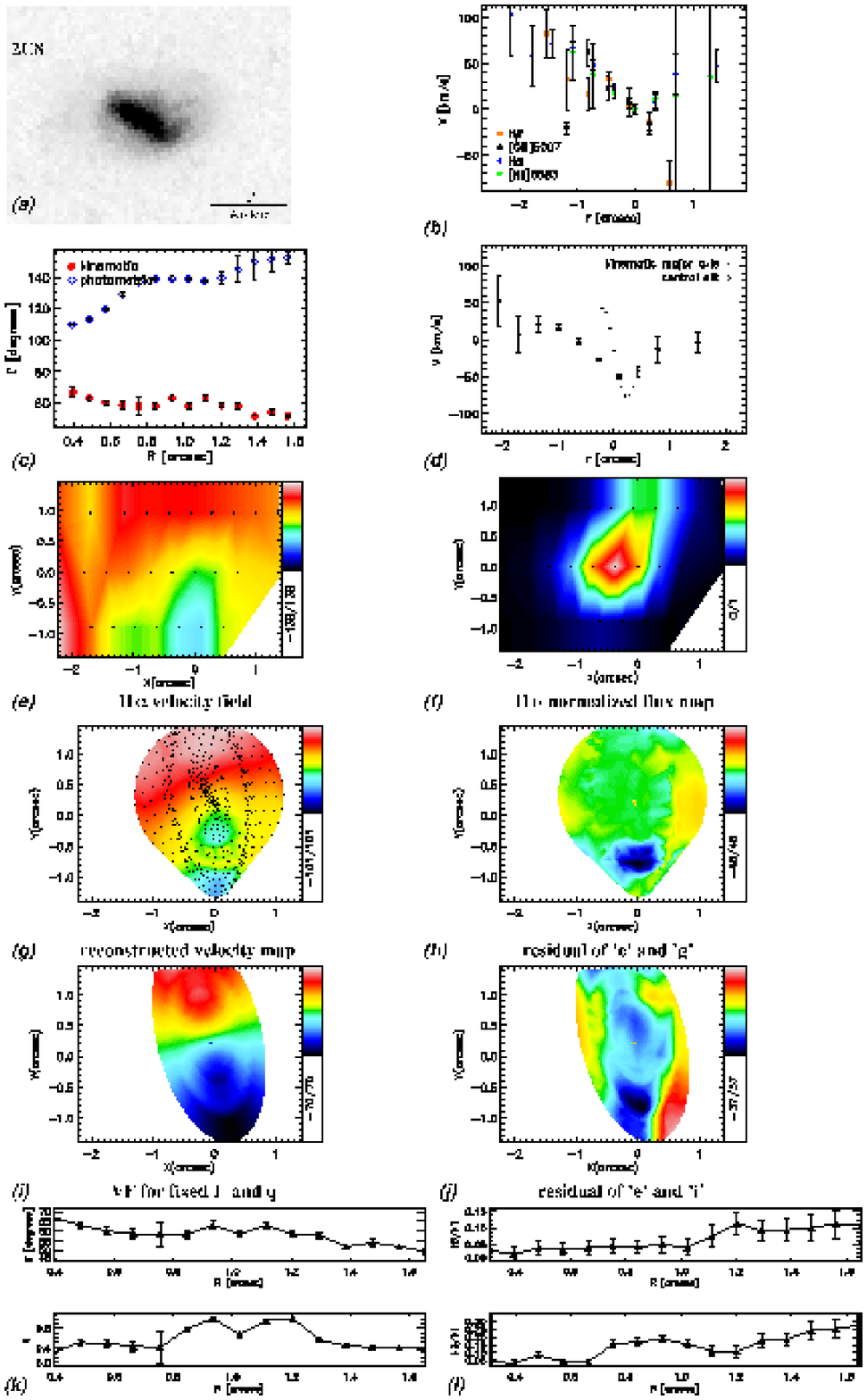}
\caption{{\bf a)} HST-ACS image of the galaxy in the $V$~band. 
{\bf b)}~Rotation curves of different emission lines extracted along the central slit.
{\bf c)}~Position angles of kinematic and photometric axes as a function of radius.
{\bf d)}~Rotation curves extracted along the central slit and the kinematic major axis.
{\bf e)}~H$\alpha$~velocity field. 
{\bf f)}~Normalized H$\alpha$~flux map. 
{\bf g)}~Velocity map reconstructed using 6~harmonic terms.
{\bf h)}~Residual of the velocity map and the reconstructed map. 
{\bf i)}~Simple rotation map constructed for position angle and ellipticity fixed to their global values.
{\bf j)}~Residual of the velocity map and the simple rotation map.
{\bf k)}~Position angle and flattening as a function of radius. 
{\bf l)}~$k_{3}/k_{1}$ and $k_{5}/k_{1}$ (from the analysis where position angle and ellipticity are fixed to their global values) as a function of radius.}
         \label{gal2C12}
   \end{figure*}

   \begin{figure*}
   \centering
\includegraphics[angle=0,width=13cm,clip]{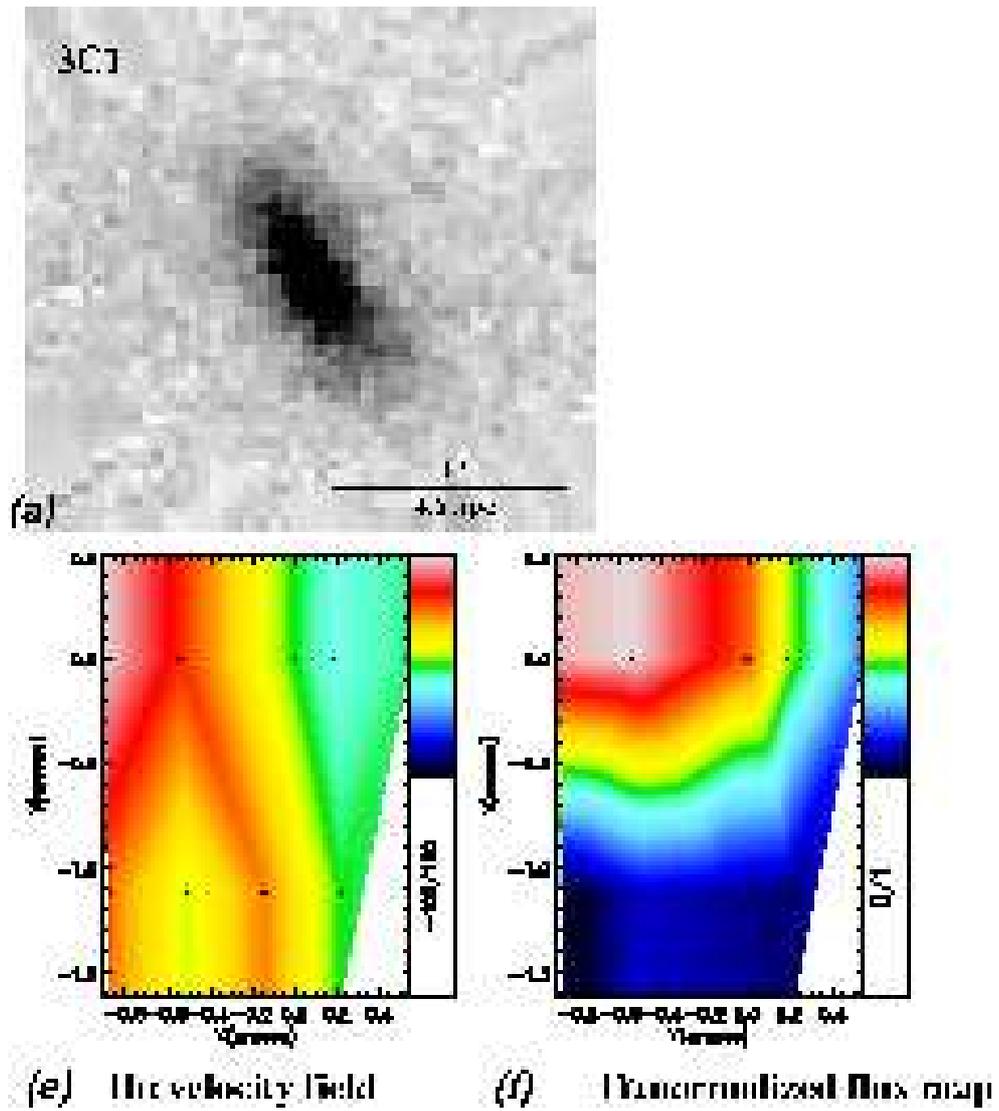}
 \caption{{\bf a)} HST-ACS image of the galaxy in the $V$~band.
{\bf e)}~H$\alpha$~velocity field.
{\bf f)}~Normalized H$\alpha$~flux map.}
         \label{gal3C2}
   \end{figure*}

   \begin{figure*}
   \centering
\includegraphics[angle=0,width=16cm,clip]{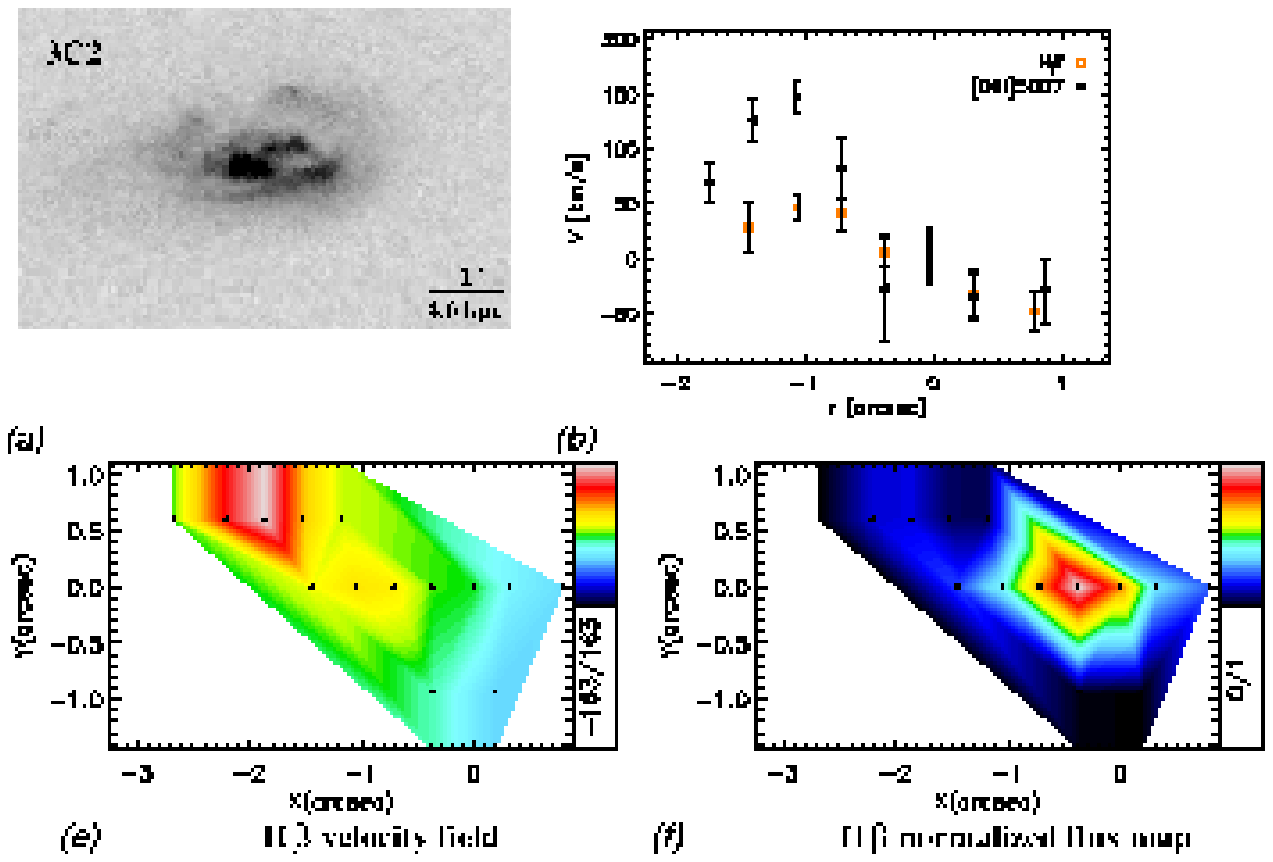}
 \caption{{\bf a)} HST-ACS image of the galaxy in the $V$~band. {\bf b)}~Rotation curves of different emission lines extracted along the central slit.
{\bf e)}~H$\beta$~velocity field. {\bf f)}~Normalized H$\beta$~flux map.}
         \label{gal3C4}
   \end{figure*}
    
\clearpage
     
   \begin{figure*}
   \centering
   \includegraphics[angle=0,width=14cm,clip]{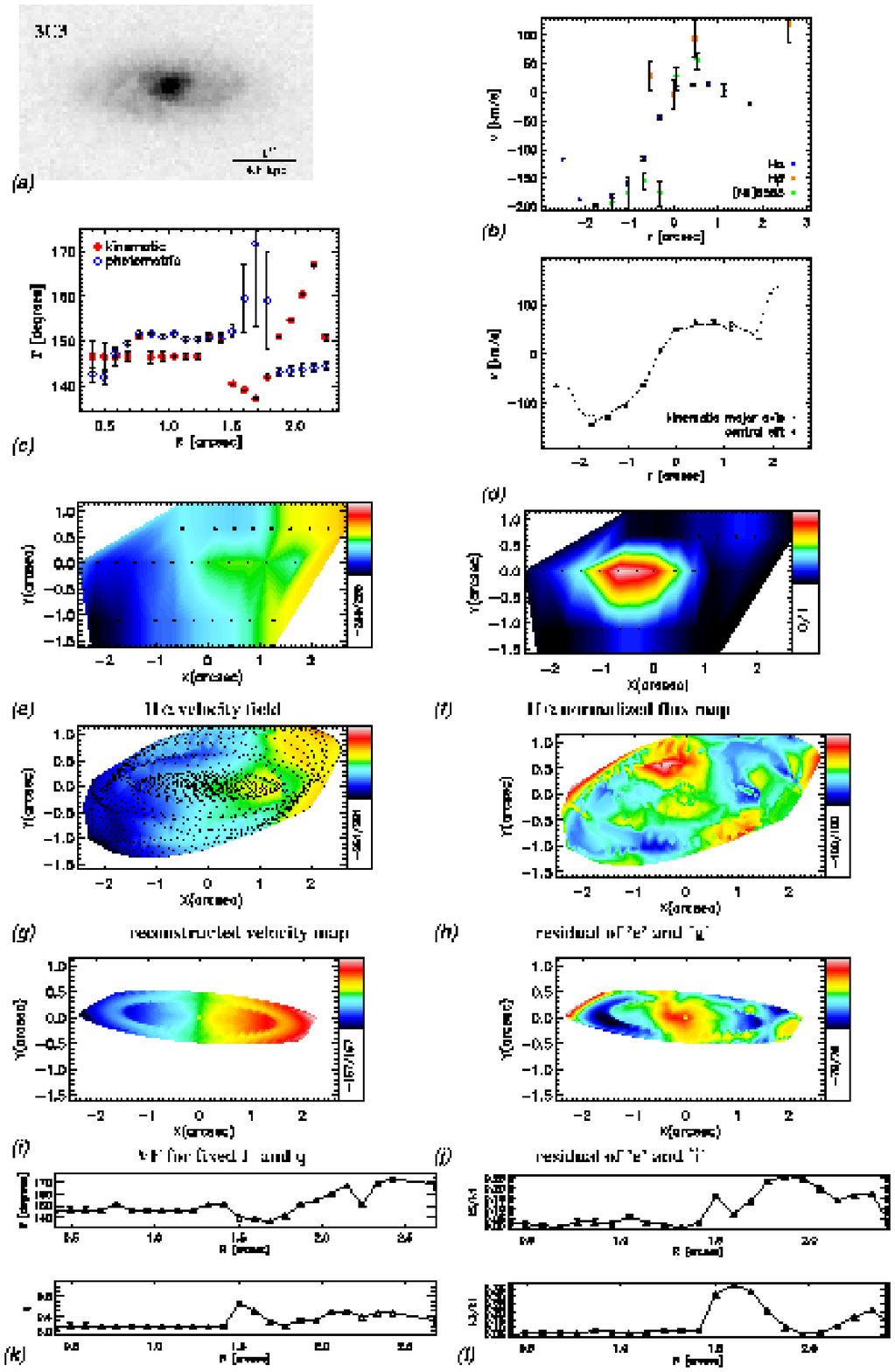}
 \caption{{\bf a)} HST-ACS image of the galaxy in the $V$~band. 
{\bf b)}~Rotation curves of different emission lines extracted along the central slit.
{\bf c)}~Position angles of kinematic and photometric axes as a function of radius.
{\bf d)}~Rotation curves extracted along the central slit and the kinematic major axis.
{\bf e)}~H$\alpha$~velocity field. 
{\bf f)}~Normalized H$\alpha$~flux map. 
{\bf g)}~Velocity map reconstructed using 6~harmonic terms.
{\bf h)}~Residual of the velocity map and the reconstructed map. 
{\bf i)}~Simple rotation map constructed for position angle and ellipticity fixed to their global values.
{\bf j)}~Residual of the velocity map and the simple rotation map.
{\bf k)}~Position angle and flattening as a function of radius. {\bf l)}~$k_{3}/k_{1}$ and $k_{5}/k_{1}$ (from the analysis where position angle and ellipticity are fixed to their global values) as a function of radius.}
         \label{gal3C1}
   \end{figure*}

   \begin{figure*}
   \centering
   \includegraphics[angle=0,width=16cm,clip]{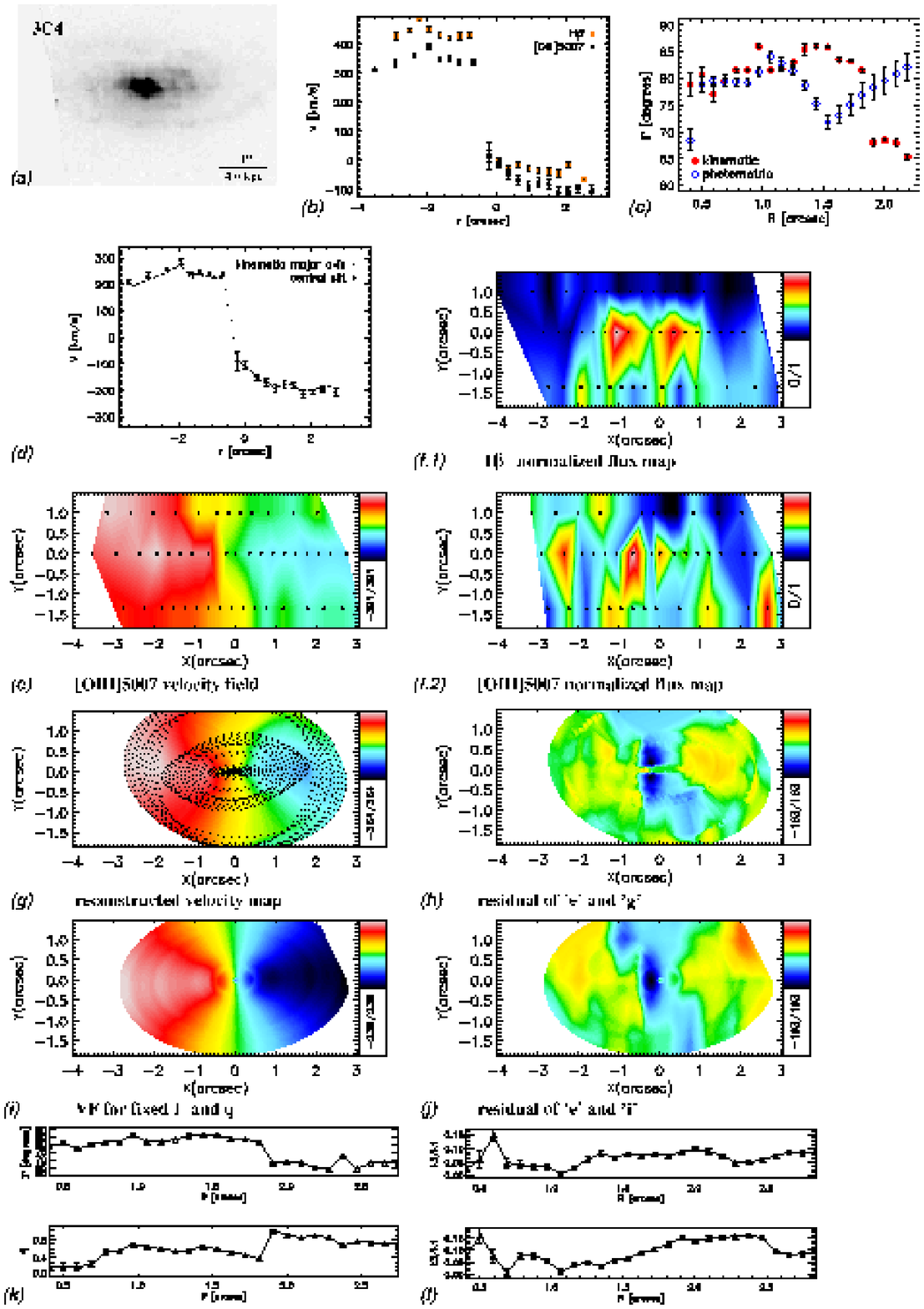}
 \caption{{\bf a)} HST-ACS image of the galaxy in the $V$~band. {\bf b)}~Rotation curves of different emission lines extracted along the central slit.
{\bf c)}~Position angles of kinematic and photometric axes as a function of radius.
{\bf d)}~Rotation curves extracted along the central slit and the kinematic major axis.
{\bf e)}~[OIII]5007~velocity field. 
{\bf f.1)}~Normalized H$\beta$~flux map. 
{\bf f.2)}~Normalized [OIII]5007~flux map. 
{\bf g)}~Velocity map reconstructed using 6~harmonic terms.
{\bf h)}~Residual of the velocity map and the reconstructed map. 
{\bf i)}~Simple rotation map constructed for position angle and ellipticity fixed to their global values.
{\bf j)}~Residual of the velocity map and the simple rotation map.
{\bf k)}~Position angle and flattening as a function of radius. {\bf l)}~$k_{3}/k_{1}$ and $k_{5}/k_{1}$ (from the analysis where position angle and ellipticity are fixed to their global values) as a function of radius.}
         \label{gal3C3}
   \end{figure*}

   \begin{figure*}
   \centering
   \includegraphics[angle=0,width=16.5cm,clip]{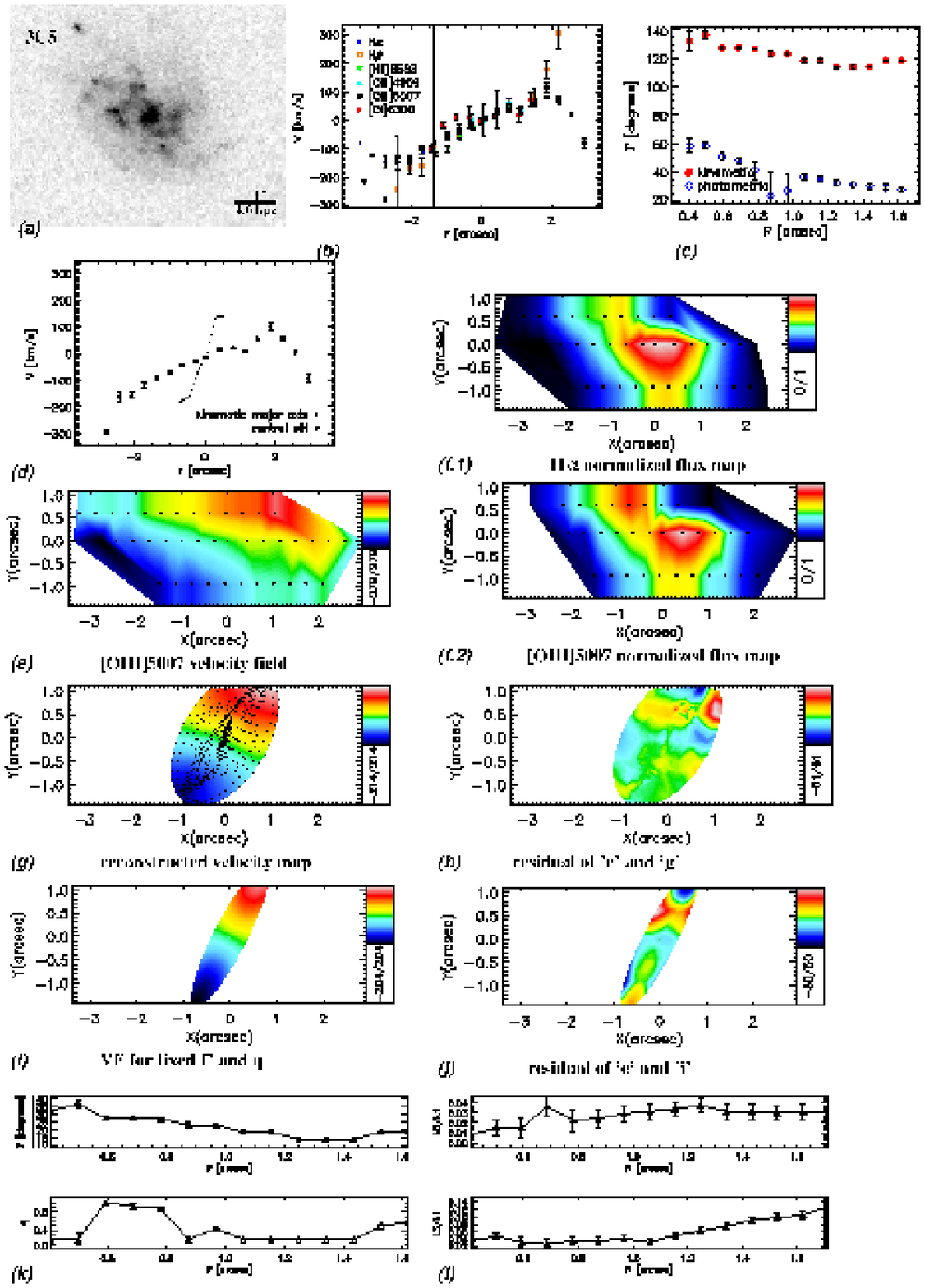}
 \caption{{\bf a)} HST-ACS image of the galaxy in the $V$~band. 
 {\bf b)}~Rotation curves of different emission lines extracted along the central slit.
{\bf c)}~Position angles of kinematic and photometric axes as a function of radius.
{\bf d)}~Rotation curves extracted along the central slit and the kinematic major axis.
{\bf e)}~[OIII]5007~velocity field. 
{\bf f.1)}~Normalized H$\alpha$~flux map.
{\bf f.2)}~Normalized [OIII]5007~flux map.
{\bf g)}~Velocity map reconstructed using 6~harmonic terms.
{\bf h)}~Residual of the velocity map and the reconstructed map. 
{\bf i)}~Simple rotation map constructed for position angle and ellipticity fixed to their global values.
{\bf j)}~Residual of the velocity map and the simple rotation map.
{\bf k)}~Position angle and flattening as a function of radius. 
{\bf l)}~$k_{3}/k_{1}$ and $k_{5}/k_{1}$ (from the analysis where position angle and ellipticity are fixed to their global values) as a function of radius.}
         \label{gal3C6}
   \end{figure*}

   \begin{figure*}
   \centering
   \includegraphics[angle=0,width=14.5cm,clip]{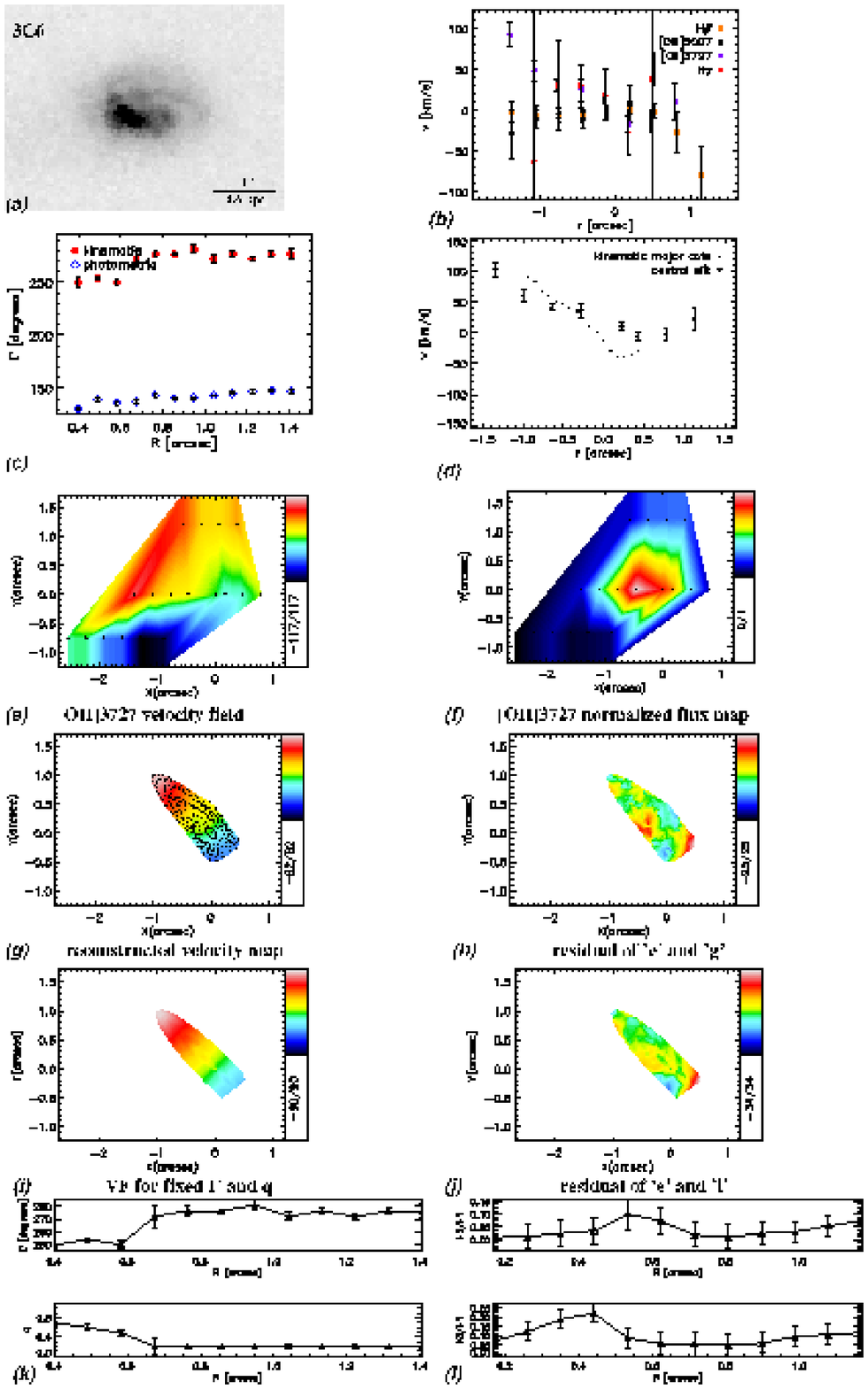}
 \caption{{\bf a)} HST-ACS image of the galaxy in the $V$~band. 
{\bf b)}~Rotation curves of different emission lines extracted along the central slit.
{\bf c)}~Position angles of kinematic and photometric axes as a function of radius.
{\bf d)}~Rotation curves extracted along the central slit and the kinematic major axis.
{\bf e)}~[OII]3727~velocity field. 
{\bf f)}~Normalized [OII]3727~flux map. 
{\bf g)}~Velocity map reconstructed using 6~harmonic terms.
{\bf h)}~Residual of the velocity map and the reconstructed map. 
{\bf i)}~Simple rotation map constructed for position angle and ellipticity fixed to their global values.
{\bf j)}~Residual of the velocity map and the simple rotation map.
{\bf k)}~Position angle and flattening as a function of radius. {\bf l)}~$k_{3}/k_{1}$ and $k_{5}/k_{1}$ (from the analysis where position angle and ellipticity are fixed to their global values) as a function of radius.}
         \label{gal3C7}
   \end{figure*}

   \begin{figure*}
   \centering
   \includegraphics[angle=0,width=15.2cm,clip]{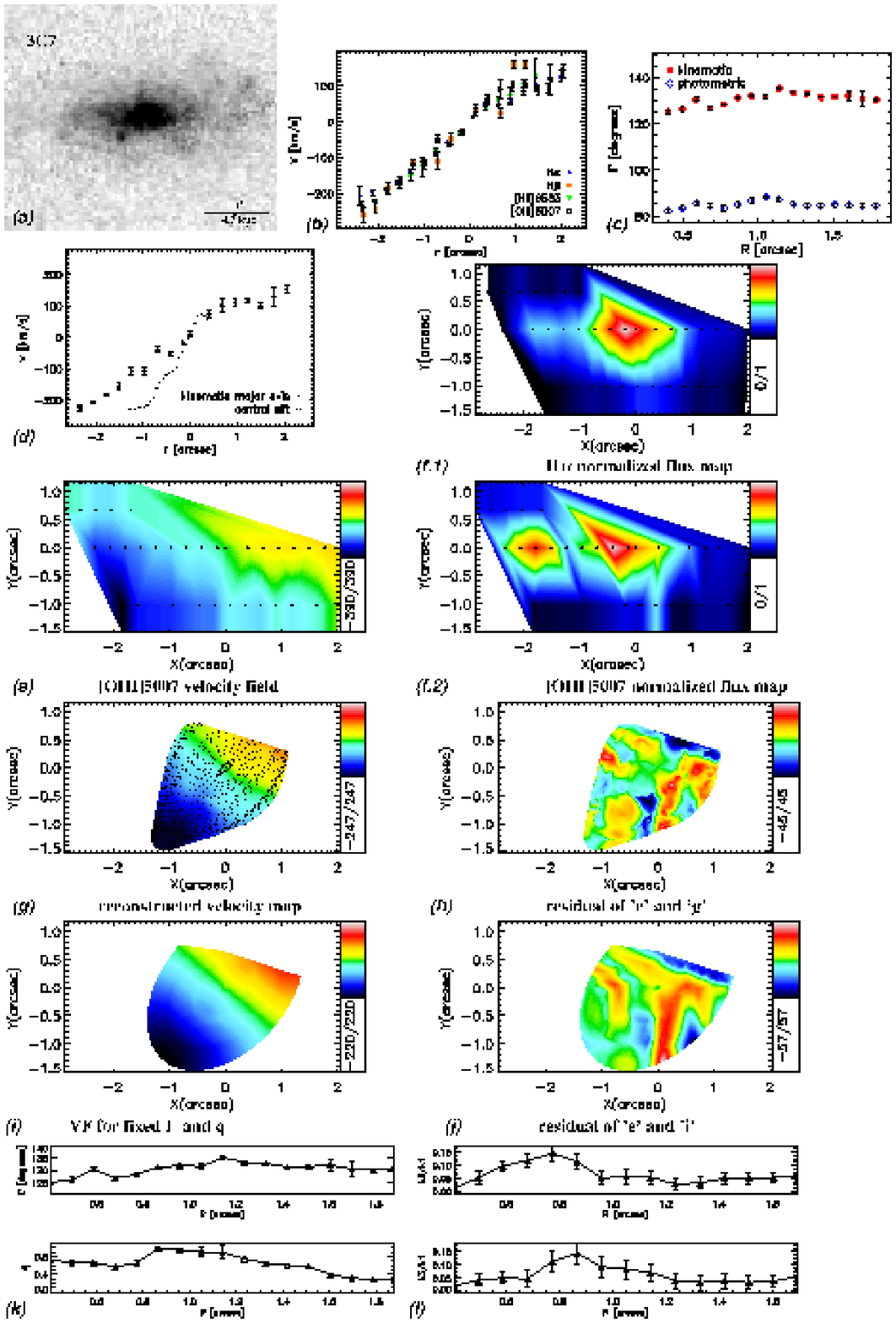}
 \caption{{\bf a)} HST-ACS image of the galaxy in the $V$~band. 
{\bf b)}~Rotation curves of different emission lines extracted along the central slit.
{\bf c)}~Position angles of kinematic and photometric axes as a function of radius.
{\bf d)}~Rotation curves extracted along the central slit and the kinematic major axis.
{\bf e)}~[OIII]5007~velocity field. 
{\bf f.1)}~Normalized H$\alpha$~flux map.
{\bf f.2)}~Normalized [OIII]5007~flux map.
{\bf g)}~Velocity map reconstructed using 6~harmonic terms.
{\bf h)}~Residual of the velocity map and the reconstructed map.  
{\bf i)}~Simple rotation map constructed for position angle and ellipticity fixed to their global values.
{\bf j)}~Residual of the velocity map and the simple rotation map.
{\bf k)}~Position angle and flattening as a function of radius. 
{\bf l)}~$k_{3}/k_{1}$ and $k_{5}/k_{1}$ (from the analysis where position angle and ellipticity are fixed to their global values) as a function of radius.}
         \label{gal3C9}
   \end{figure*}

   \begin{figure*}
   \centering
   \includegraphics[angle=0,width=15cm,clip]{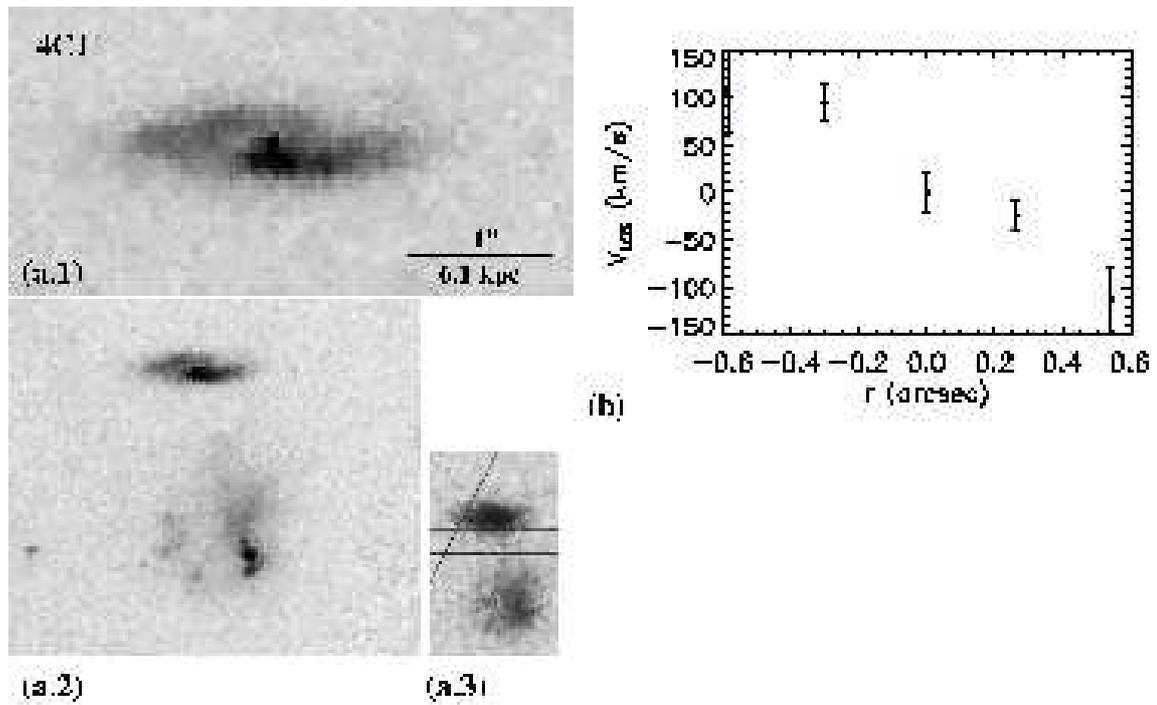}
 \caption{{\bf a.1)} HST-ACS image of the galaxy in the $V$~band. 
{\bf a.2)}~This cluster galaxy does not have emission lines.  The spectra include an emission line from the galaxy below the target that is
seen in the HST image.  The redshift of this galaxy could not be determined from the composite spectrum. 
{\bf a.3)}~FORS2 image of the galaxies. Parallel lines represent the bottom slit which gives a composite spectrum of both galaxies together.
{\bf b)}~Rotation curve of the galaxy that is below the target.  It is obtained from the slit positioned inbetween the two galaxies (see
a.3).  Velocities were measured with respect to the velocity at the continuum center of the bottom slit.  Signal level in the other two slits
was not high enough to obtain a rotation curve.}
         \label{gal4C3}
   \end{figure*}

\clearpage
   \begin{figure*}
   \centering
   \includegraphics[angle=0,width=15.2cm,clip]{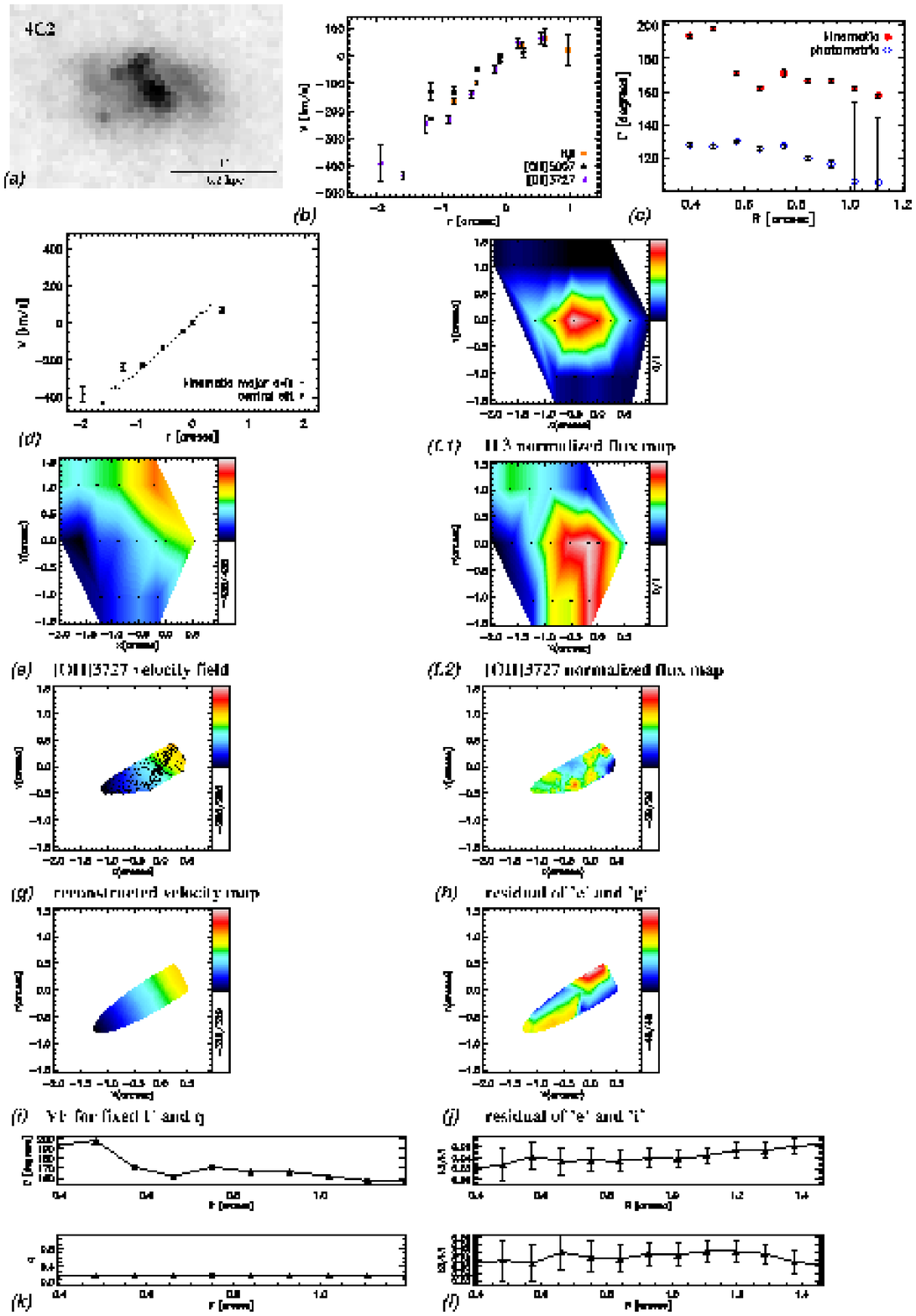}
 \caption{{\bf a)} HST-ACS image of the galaxy in the $V$~band. 
{\bf b)}~Rotation curves of different emission lines extracted along the central slit.
{\bf c)}~Position angles of kinematic and photometric axes as a function of radius.
{\bf d)}~Rotation curves extracted along the central slit and the kinematic major axis.
{\bf e)}~[OII]3727~velocity field. 
{\bf f.1)}~Normalized H$\beta$~flux map. 
{\bf f.2)}~Normalized [OII]3727~flux map. 
{\bf g)}~Velocity map reconstructed using 6~harmonic terms.
{\bf h)}~Residual of the velocity map and the reconstructed map. 
{\bf i)}~Simple rotation map constructed for position angle and ellipticity fixed to their global values.
{\bf j)}~Residual of the velocity map and the simple rotation map.
{\bf k)}~Position angle and flattening as a function of radius. {\bf l)}~$k_{3}/k_{1}$ and $k_{5}/k_{1}$ (from the analysis where position angle and ellipticity are fixed to their global values) as a function of radius.}
         \label{gal4C1}
   \end{figure*}

   \begin{figure*}
   \centering
   \includegraphics[angle=0,width=13.5cm,clip]{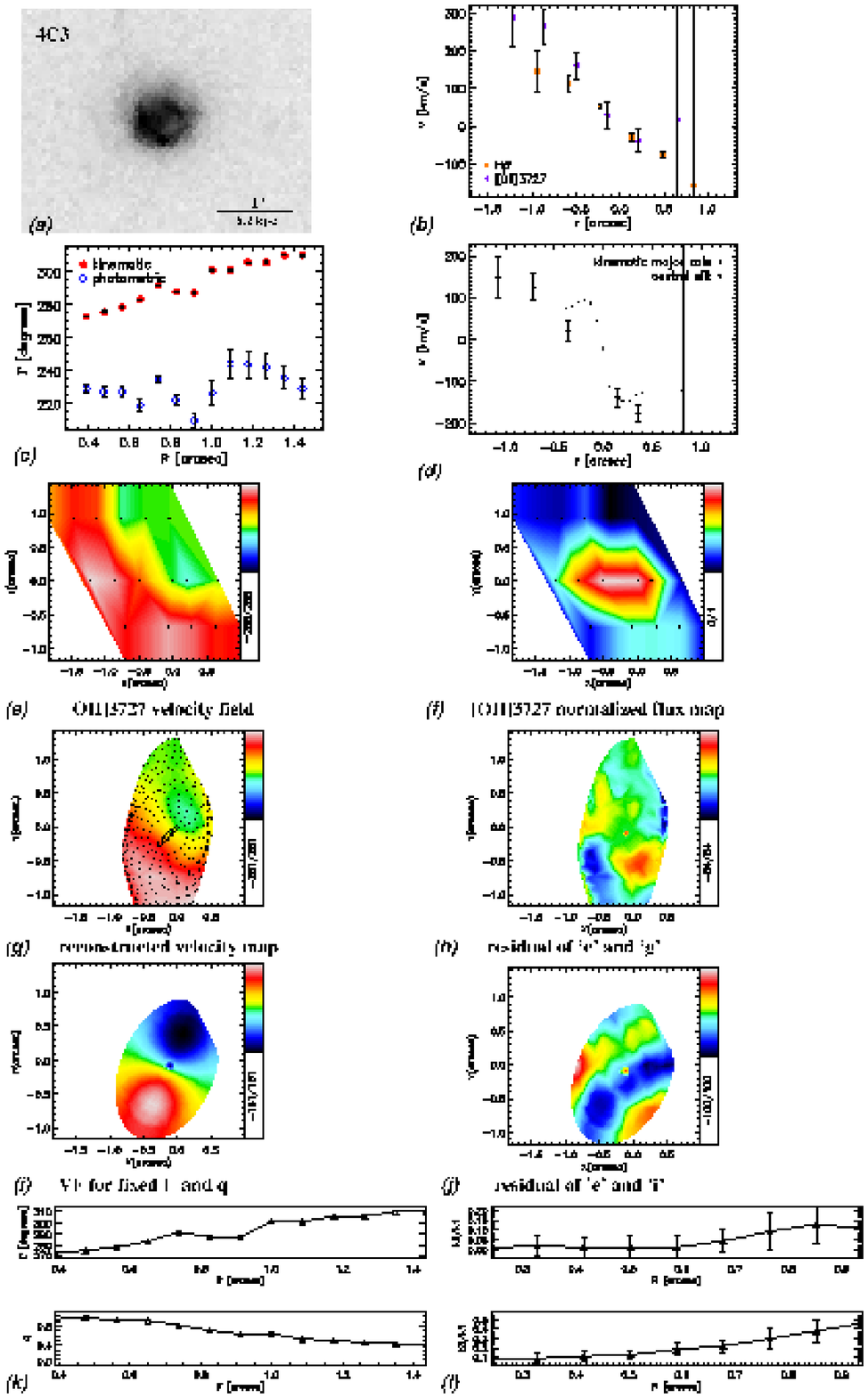}
 \caption{{\bf a)} HST-ACS image of the galaxy in the $V$~band. 
{\bf b)}~Rotation curves of different emission lines extracted along the central slit.
{\bf c)}~Position angles of kinematic and photometric axes as a function of radius.
{\bf d)}~Rotation curves extracted along the central slit and the kinematic major axis.
{\bf e)}~[OII]3727~velocity field. 
{\bf f)}~Normalized [OII]3727~flux map. 
{\bf g)}~Velocity map reconstructed using 6~harmonic terms.
{\bf h)}~Residual of the velocity map and the reconstructed map. 
{\bf i)}~Simple rotation map constructed for position angle and ellipticity fixed to their global values.
{\bf j)}~Residual of the velocity map and the simple rotation map.
{\bf k)}~Position angle and flattening as a function of radius. 
{\bf l)}~$k_{3}/k_{1}$ and $k_{5}/k_{1}$ (from the analysis where position angle and ellipticity are fixed to their global values) as a function of radius.}
         \label{gal4C2}
   \end{figure*}

\clearpage

\subsection{Field galaxies}

   \begin{figure*}
   \centering
   \includegraphics[angle=0,width=14.2cm,clip]{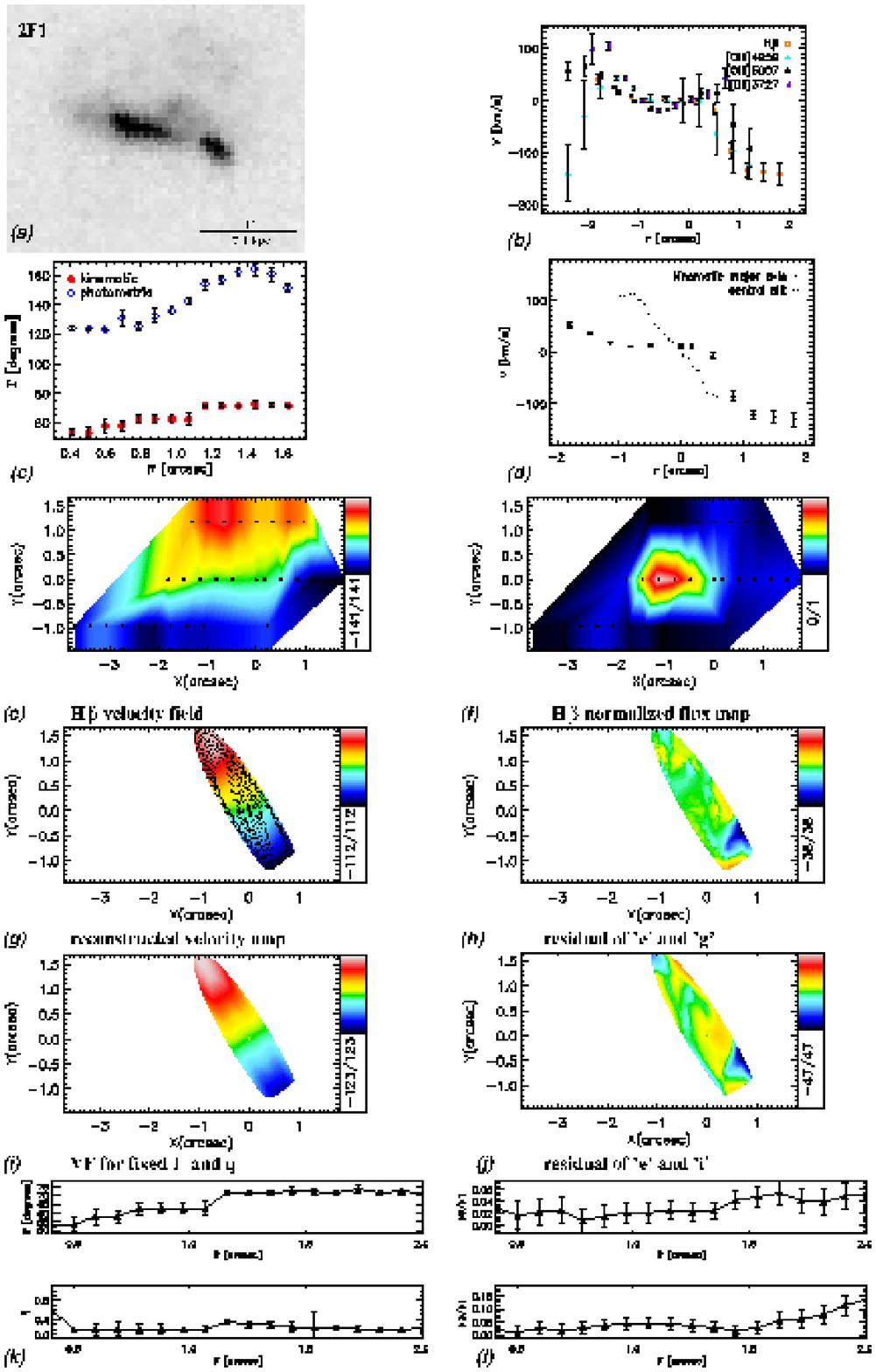}
 \caption{{\bf a)} HST-ACS image of the galaxy in the $V$~band. 
{\bf b)}~Rotation curves of different emission lines extracted along the central slit.
{\bf c)}~Position angles of kinematic and photometric axes as a function of radius.
{\bf d)}~Rotation curves extracted along the central slit and the kinematic major axis.
{\bf e)}~H$\beta$~velocity field.
{\bf f)}~Normalized H$\beta$~flux map.  
{\bf g)}~Velocity map reconstructed using 6~harmonic terms.
{\bf h)}~Residual of the velocity map and the reconstructed map. 
{\bf i)}~Simple rotation map constructed for position angle and ellipticity fixed to their global values.
{\bf j)}~Residual of the velocity map and the simple rotation map.
{\bf k)}~Position angle and flattening as a function of radius. 
{\bf l)}~$k_{3}/k_{1}$ and $k_{5}/k_{1}$ (from the analysis where position angle and ellipticity are fixed to their global values) as a function of radius.}
         \label{gal2F1}
   \end{figure*}

   \begin{figure*}
   \centering
   \includegraphics[angle=0,width=14cm,clip]{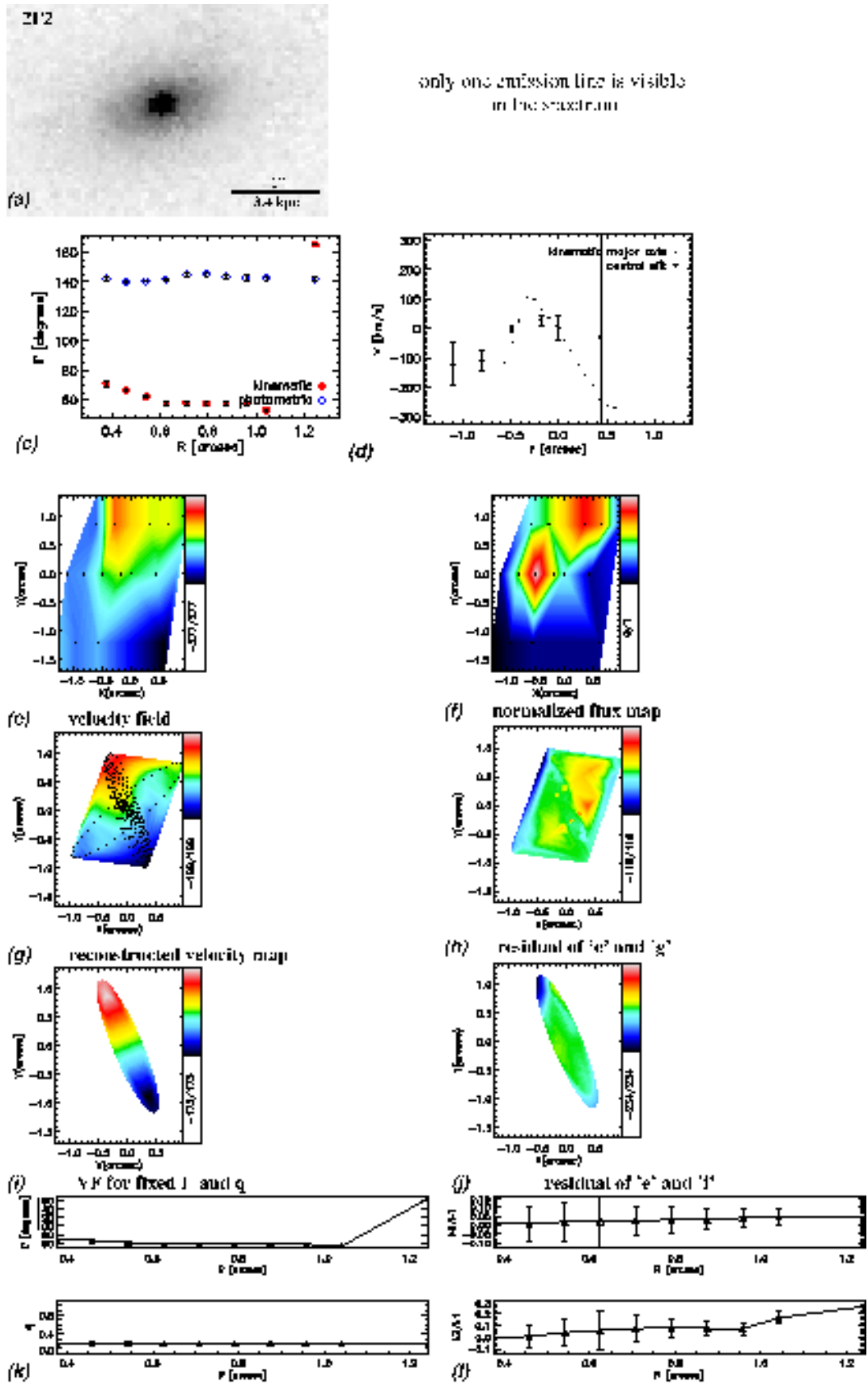}
 \caption{{\bf a)} HST-ACS image of the galaxy in the $V$~band. 
{\bf c)}~Position angles of kinematic and photometric axes as a function of radius.
{\bf d)}~Rotation curves extracted along the central slit and the kinematic major axis.
{\bf e)}~Velocity field constructed using the emission line which could not be identified. 
{\bf f)}~Normalized flux map of the emission line which could not be identified. 
{\bf g)}~Velocity map reconstructed using 6~harmonic terms.
{\bf h)}~Residual of the velocity map and the reconstructed map. 
{\bf i)}~Simple rotation map constructed for position angle and ellipticity fixed to their global values.
{\bf j)}~Residual of the velocity map and the simple rotation map.
{\bf k)}~Position angle and flattening as a function of radius. 
{\bf l)}~$k_{3}/k_{1}$ and $k_{5}/k_{1}$ (from the analysis where position angle and ellipticity are fixed to their global values) as a function of radius.}
          \label{gal2F3}
   \end{figure*}

   \begin{figure*}
   \centering
   \includegraphics[angle=0,width=14cm,clip]{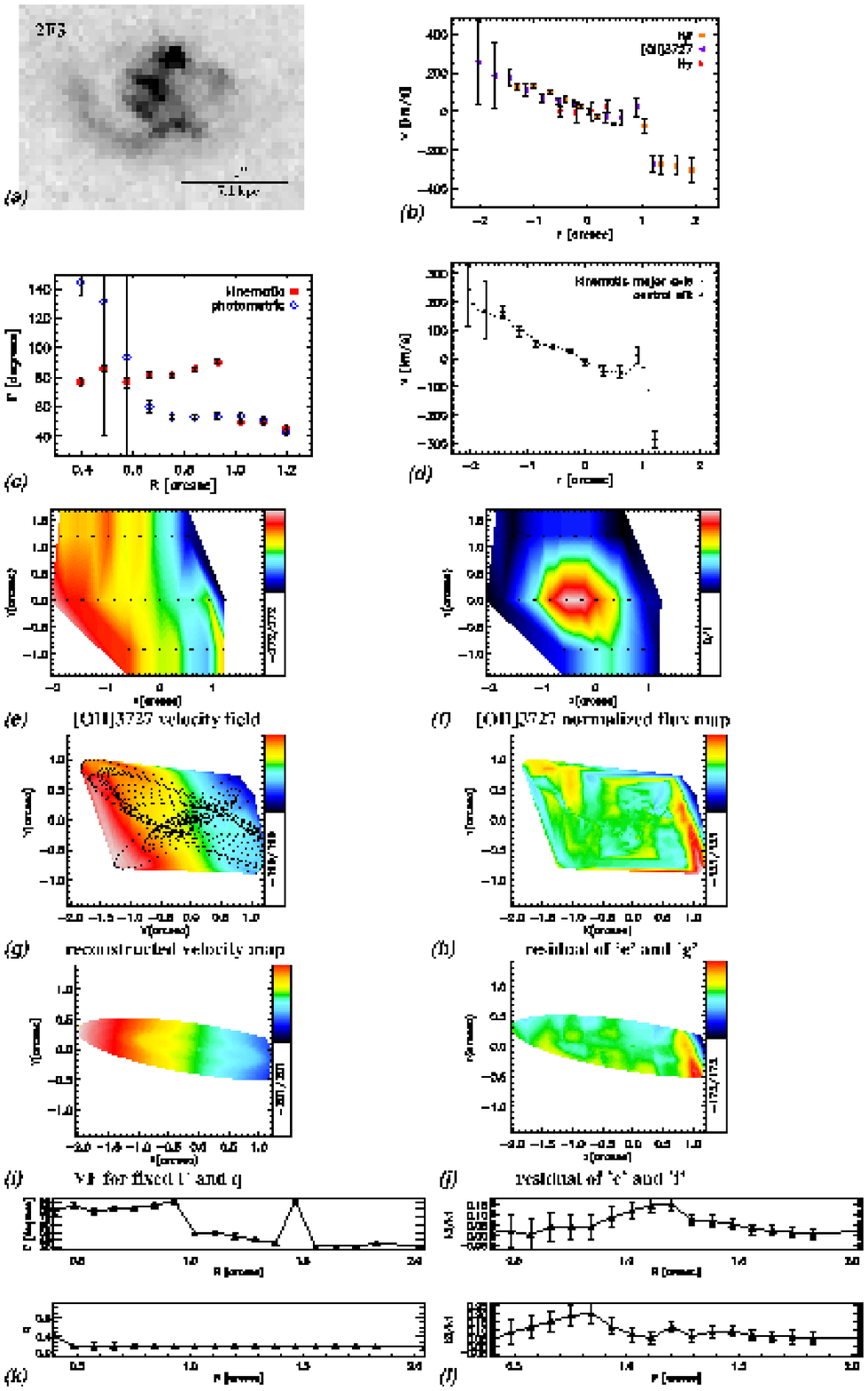}
 \caption{{\bf a)} HST-ACS image of the galaxy in the $V$~band. 
 {\bf b)}~Rotation curves of different emission lines extracted along the central slit.
{\bf c)}~Position angles of kinematic and photometric axes as a function of radius.
{\bf d)}~Rotation curves extracted along the central slit and the kinematic major axis.
{\bf e)}~[OII]3727~velocity field. 
{\bf f)}~Normalized [OII]3727~flux map. 
{\bf g)}~Velocity map reconstructed using 6~harmonic terms.
{\bf h)}~Residual of the velocity map and the reconstructed map. 
{\bf i)}~Simple rotation map constructed for position angle and ellipticity fixed to their global values.
{\bf j)}~Residual of the velocity map and the simple rotation map.
{\bf k)}~Position angle and flattening as a function of radius. {\bf l)}~$k_{3}/k_{1}$ and $k_{5}/k_{1}$ (from the analysis where position angle and ellipticity are fixed to their global values) as a function of radius.}
         \label{gal2F5}
   \end{figure*}

   \begin{figure*}
   \centering
   \includegraphics[angle=0,width=13.5cm,clip]{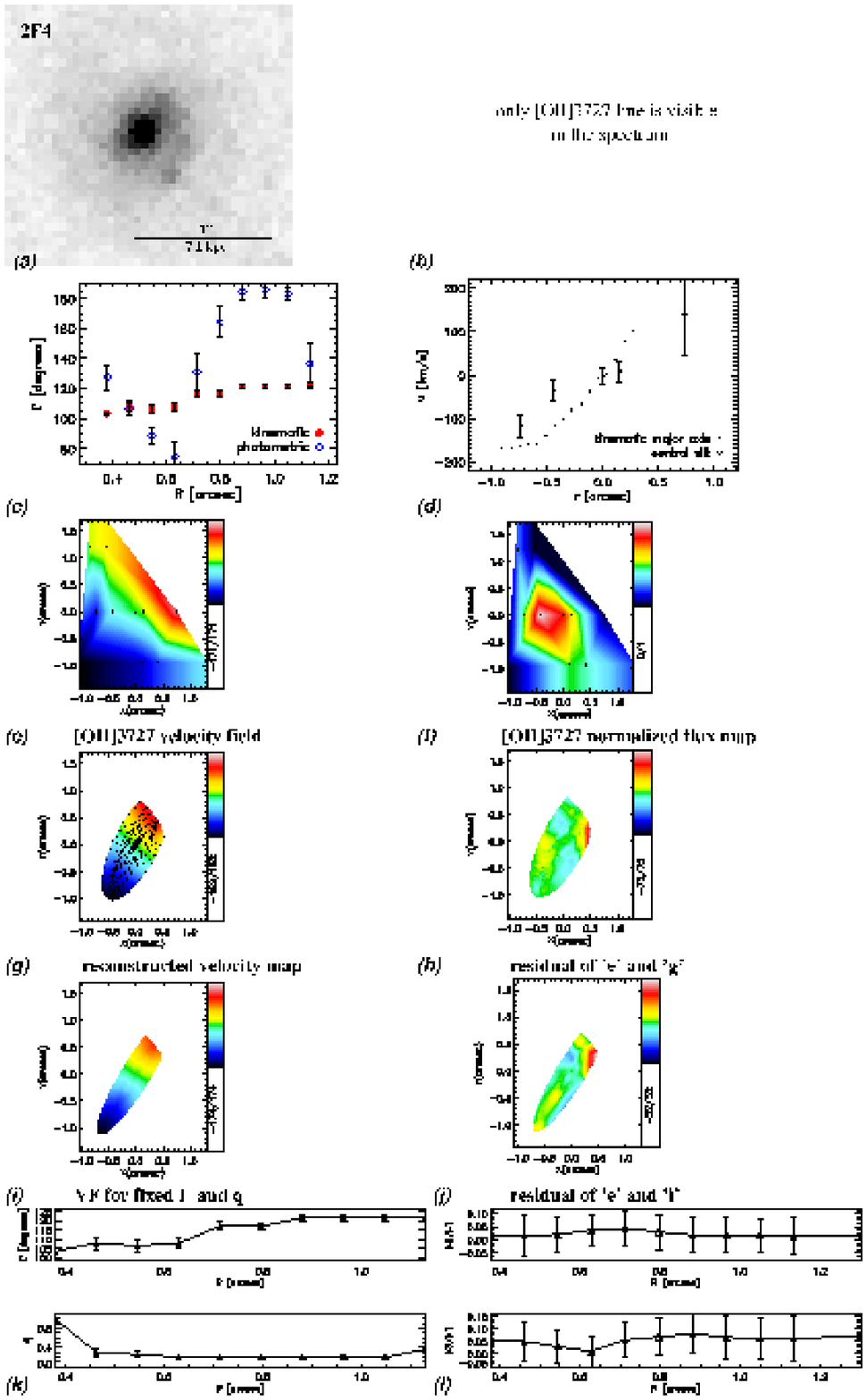}
 \caption{{\bf a)} HST-ACS image of the galaxy in the $V$~band. 
{\bf b)}~Rotation curves of different emission lines extracted along the central slit.
{\bf c)}~Position angles of kinematic and photometric axes as a function of radius.
{\bf d)}~Rotation curves extracted along the central slit and the kinematic major axis.
{\bf e)}~[OII]3727~velocity field. 
{\bf f)}~Normalized [OII]3727~flux map. 
{\bf g)}~Velocity map reconstructed using 6~harmonic terms.
{\bf h)}~Residual of the velocity map and the reconstructed map. 
{\bf i)}~Simple rotation map constructed for position angle and ellipticity fixed to their global values.
{\bf j)}~Residual of the velocity map and the simple rotation map.
{\bf k)}~Position angle and flattening as a function of radius. 
{\bf l)}~$k_{3}/k_{1}$ and $k_{5}/k_{1}$ (from the analysis where position angle and ellipticity are fixed to their global values) as a function of radius.}
         \label{gal2F6}
   \end{figure*}

   \begin{figure*}
   \centering
   \includegraphics[angle=0,width=13.3cm,clip]{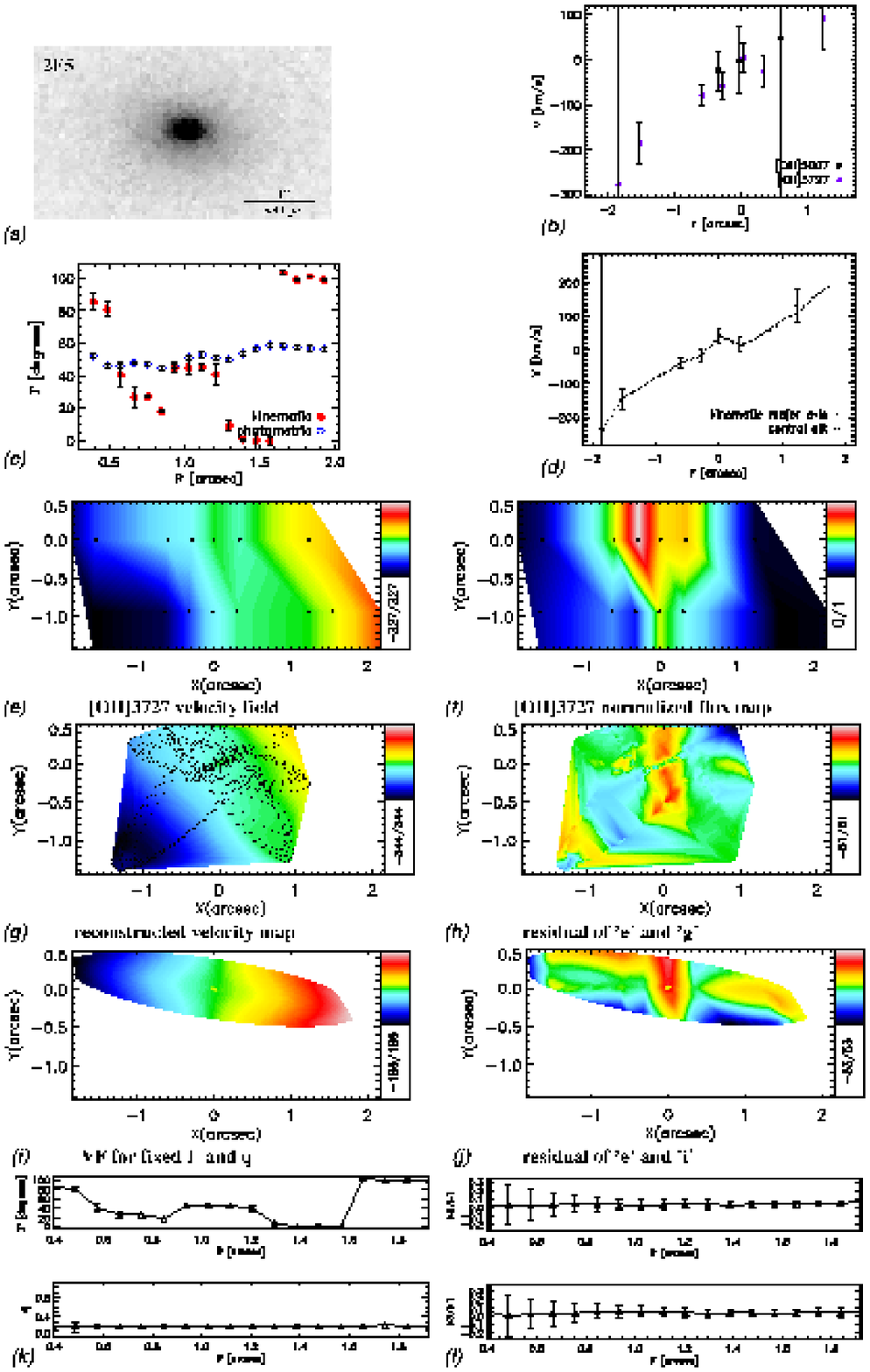}
 \caption{{\bf a)} HST-ACS image of the galaxy in the $V$~band. 
{\bf b)}~Rotation curves of different emission lines extracted along the central slit.
{\bf c)}~Position angles of kinematic and photometric axes as a function of radius.
{\bf d)}~Rotation curves extracted along the central slit and the kinematic major axis.
{\bf e)}~[OII]3727~velocity field. 
{\bf f)}~Normalized [OII]3727~flux map.
{\bf g)}~Velocity map reconstructed using 6~harmonic terms.
{\bf h)}~Residual of the velocity map and the reconstructed map. 
{\bf i)}~Simple rotation map constructed for position angle and ellipticity fixed to their global values.
{\bf j)}~Residual of the velocity map and the simple rotation map.
{\bf k)}~Position angle and flattening as a function of radius. {\bf l)}~$k_{3}/k_{1}$ and $k_{5}/k_{1}$ (from the analysis where position angle and ellipticity are fixed to their global values) as a function of radius.}
         \label{gal2F7}
   \end{figure*}

   \begin{figure*}
   \centering
   \includegraphics[angle=0,width=14.3cm,clip]{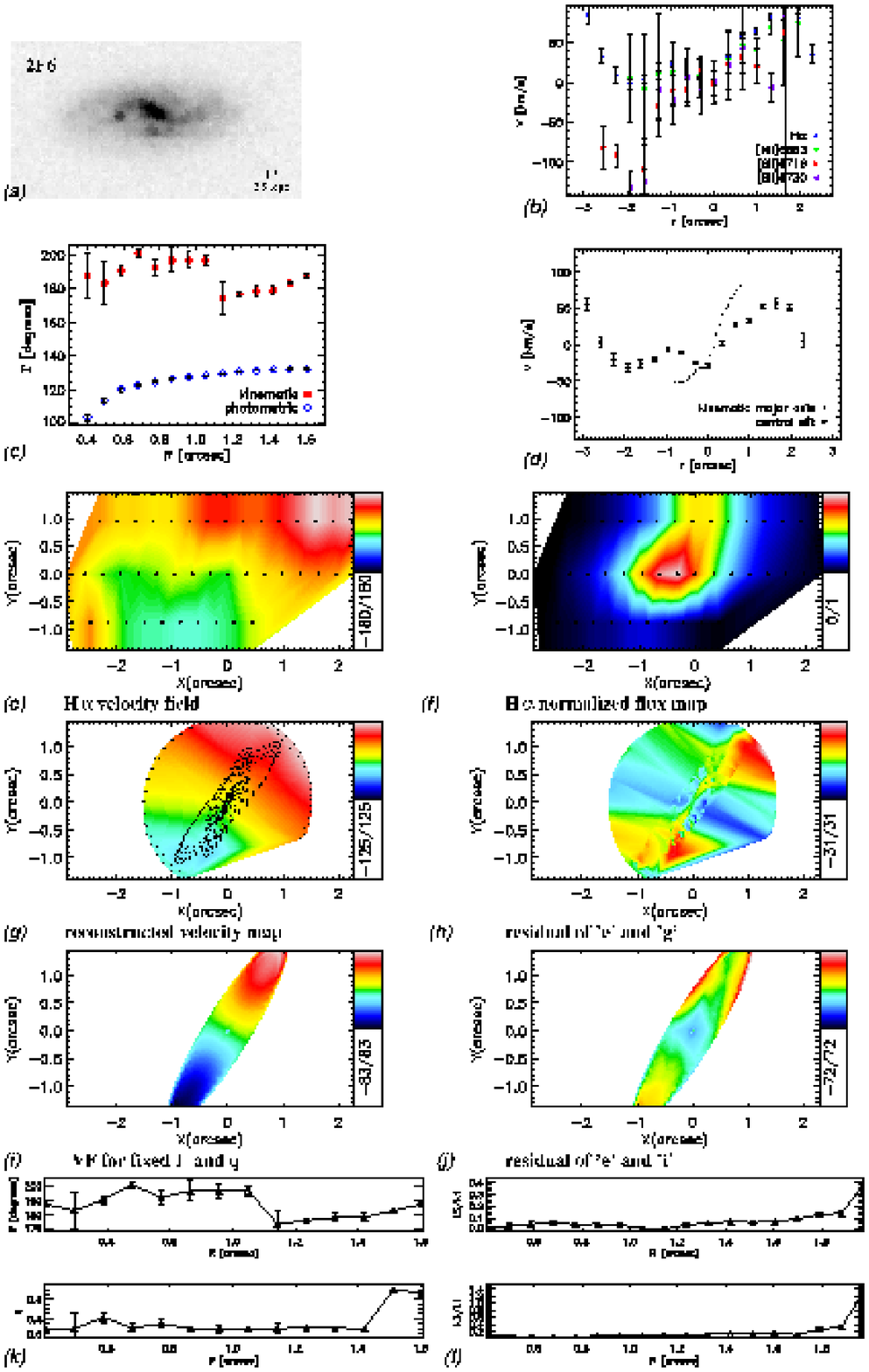}
 \caption{{\bf a)} HST-ACS image of the galaxy in the $V$~band. 
{\bf b)}~Rotation curves of different emission lines extracted along the central slit.
{\bf c)}~Position angles of kinematic and photometric axes as a function of radius.
{\bf d)}~Rotation curves extracted along the central slit and the kinematic major axis.
{\bf e)}~H$\alpha$~velocity field. 
{\bf f)}~Normalized H$\alpha$~flux map. 
{\bf g)}~Velocity map reconstructed using 6~harmonic terms.
{\bf h)}~Residual of the velocity map and the reconstructed map. 
{\bf i)}~Simple rotation map constructed for position angle and ellipticity fixed to their global values.
{\bf j)}~Residual of the velocity map and the simple rotation map.
{\bf k)}~Position angle and flattening as a function of radius. {\bf l)}~$k_{3}/k_{1}$ and $k_{5}/k_{1}$ (from the analysis where position angle and ellipticity are fixed to their global values) as a function of radius.}
         \label{gal2F9}
   \end{figure*}

   \begin{figure*}
   \centering
   \includegraphics[angle=0,width=13cm,clip]{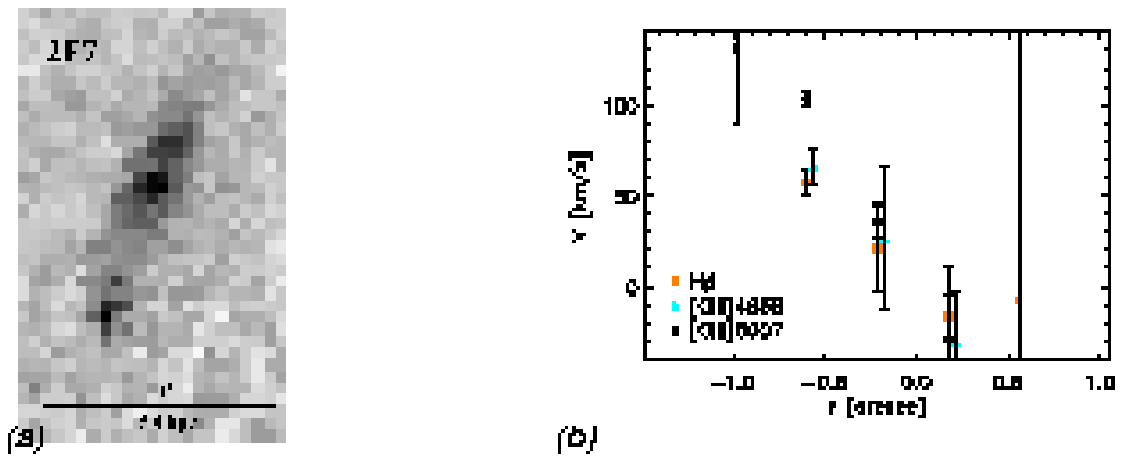}
 \caption{{\bf a)} HST-ACS image of the galaxy in the $V$~band. 
{\bf b)}~Rotation curves of different emission lines extracted along the central slit.}
         \label{gal2F11}
   \end{figure*}

   \begin{figure*}
   \centering
   \includegraphics[angle=0,width=13cm,clip]{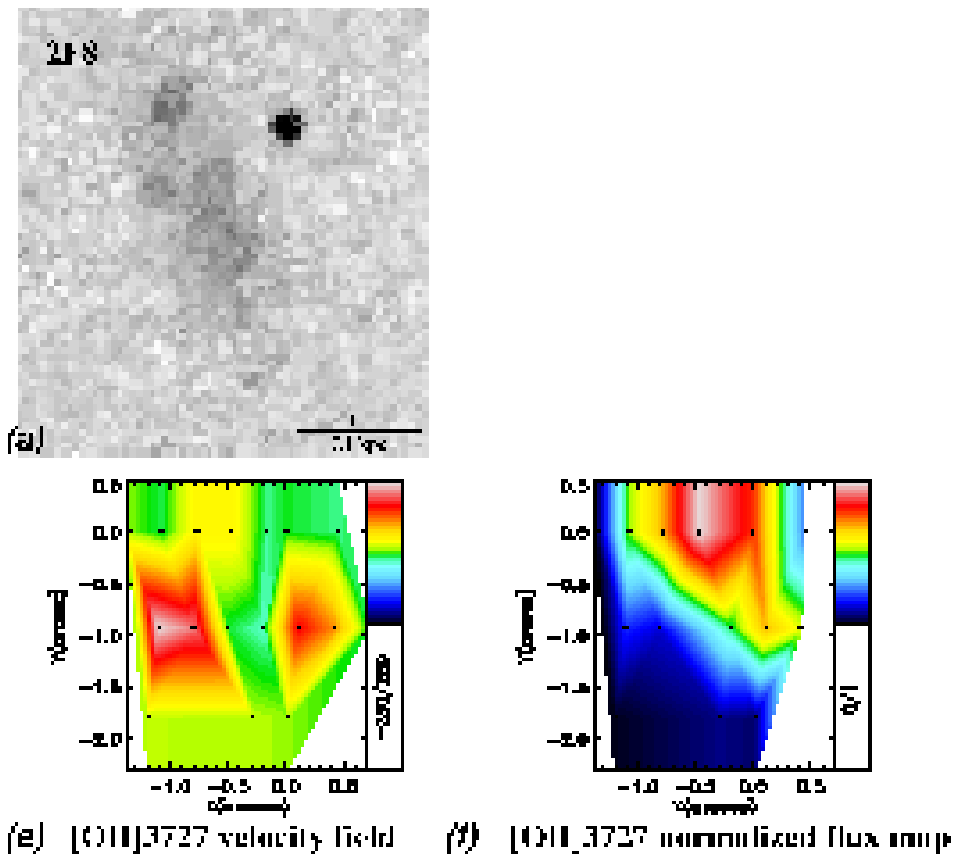}
 \caption{{\bf a)} HST-ACS image of the galaxy in the $V$~band. 
{\bf e)}~[OII]3727~velocity field. 
{\bf f)}~Normalized [OII]3727~flux map.}
         \label{gal2F15}
   \end{figure*}

   \begin{figure*}
   \centering
   \includegraphics[angle=0,width=14.7cm,clip]{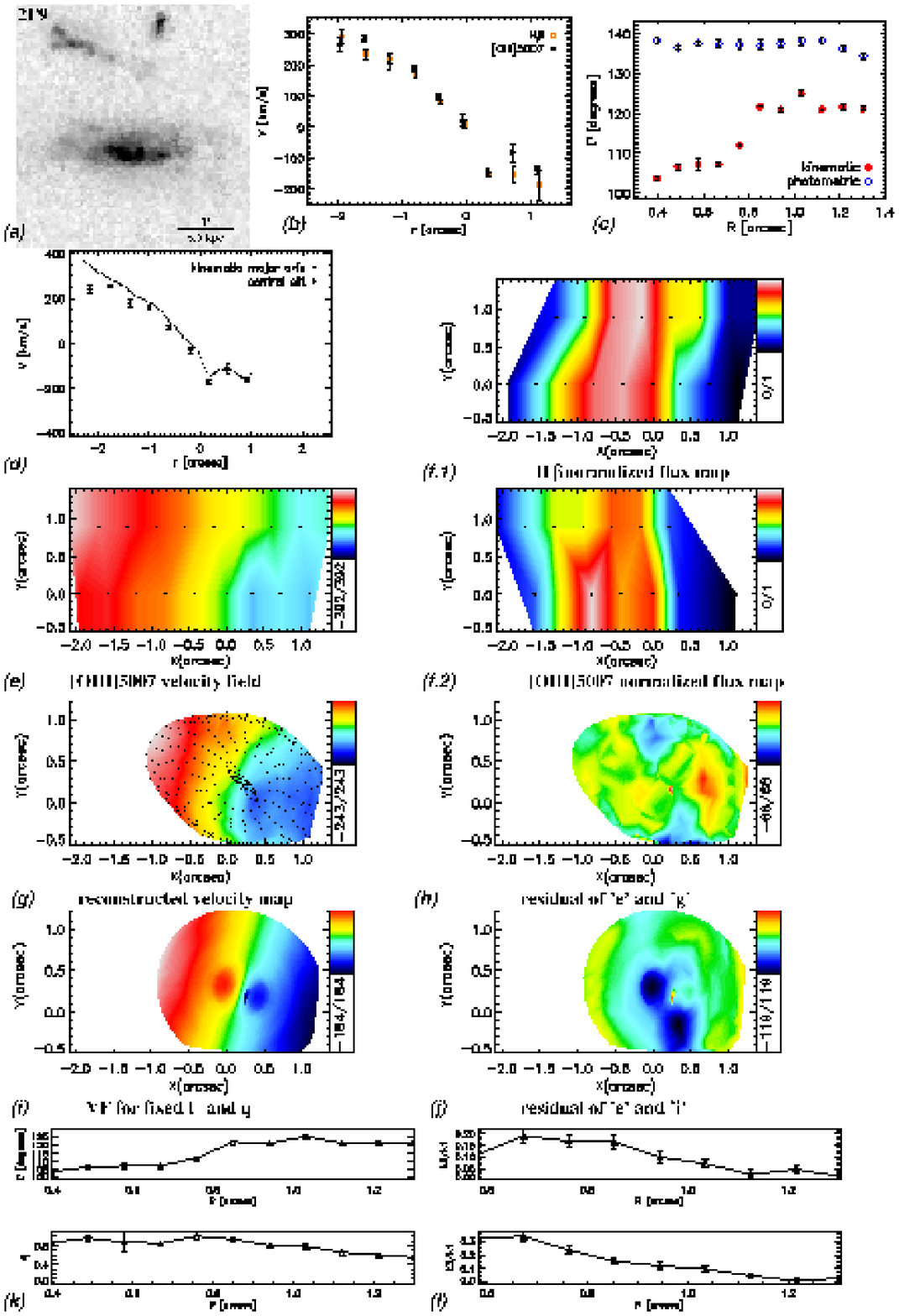}
 \caption{{\bf a)} HST-ACS image of the galaxy in the $V$~band. 
{\bf b)}~Rotation curves of different emission lines extracted along the central slit.
{\bf c)}~Position angles of kinematic and photometric axes as a function of radius.
{\bf d)}~Rotation curves extracted along the central slit and the kinematic major axis.
{\bf e)}~[OIII]5007~velocity field. 
{\bf f.1)}~Normalized H$\beta$~flux map.
{\bf f.2)}~Normalized [OIII]5007~flux map. 
{\bf g)}~Velocity map reconstructed using 6~harmonic terms.
{\bf h)}~Residual of the velocity map and the reconstructed map.
{\bf i)}~Simple rotation map constructed for position angle and ellipticity fixed to their global values.
{\bf j)}~Residual of the velocity map and the simple rotation map.
{\bf k)}~Position angle and flattening as a function of radius. {\bf l)}~$k_{3}/k_{1}$ and $k_{5}/k_{1}$ (from the analysis where position angle and ellipticity are fixed to their global values) as a function of radius.}
         \label{gal2F10}
   \end{figure*}

   \begin{figure*}
   \centering
   \includegraphics[angle=0,width=16.7cm,clip]{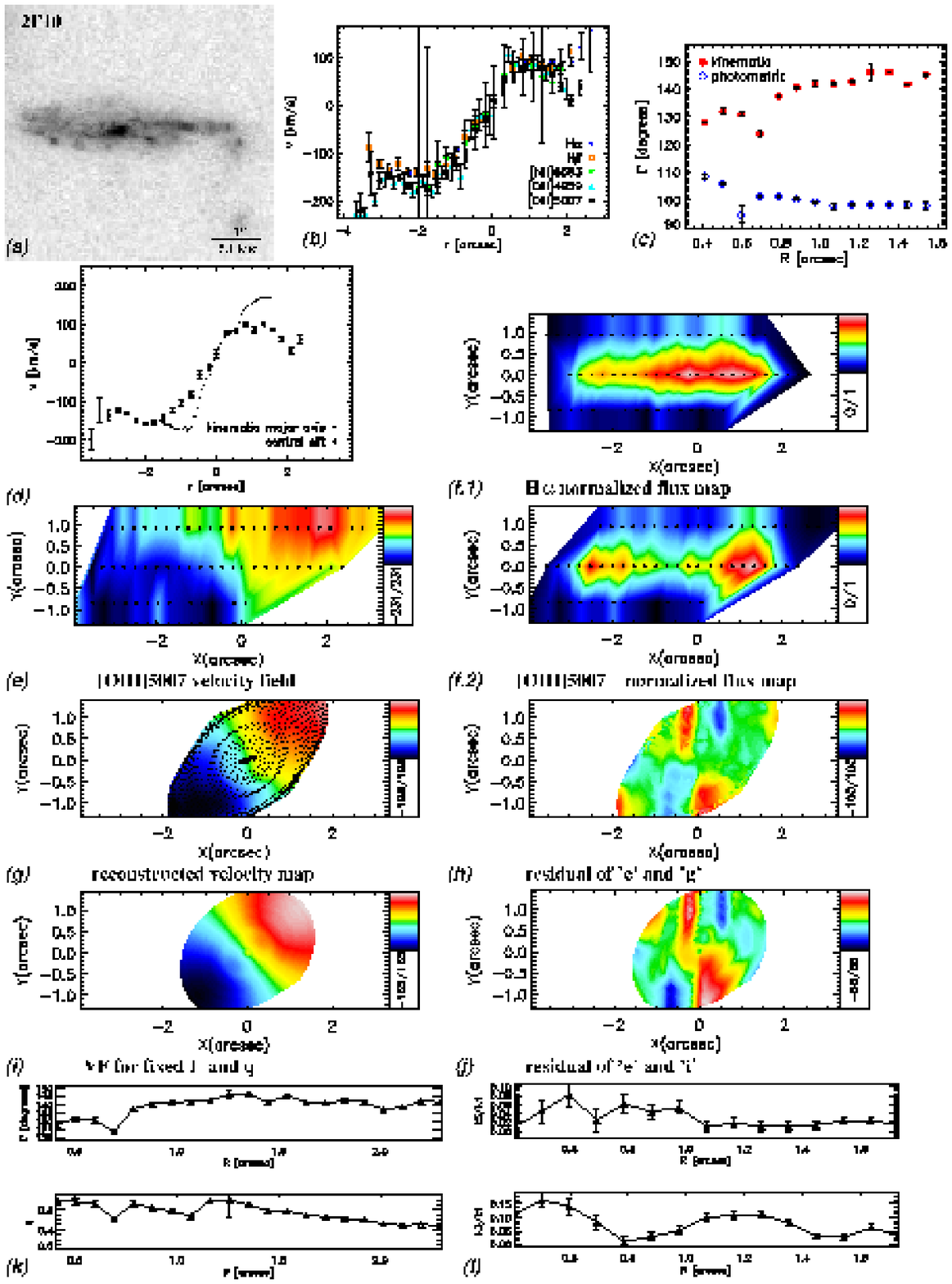}
 \caption{{\bf a)} HST-ACS image of the galaxy in the $V$~band. 
{\bf b)}~Rotation curves of different emission lines extracted along the central slit.
{\bf c)}~Position angles of kinematic and photometric axes as a function of radius.
{\bf d)}~Rotation curves extracted along the central slit and the kinematic major axis.
{\bf e)}~[OIII]5007~velocity field. 
{\bf f.1)}~Normalized H$\alpha$~flux map.
{\bf f.2)}~Normalized [OIII]5007~flux map.
{\bf g)}~Velocity map reconstructed using 6~harmonic terms.
{\bf h)}~Residual of the velocity map and the reconstructed map. 
{\bf i)}~Simple rotation map constructed for position angle and ellipticity fixed to their global values.
{\bf j)}~Residual of the velocity map and the simple rotation map.
{\bf k)}~Position angle and flattening as a function of radius. {\bf l)}~$k_{3}/k_{1}$ and $k_{5}/k_{1}$ (from the analysis where position angle and ellipticity are fixed to their global values) as a function of radius.}
         \label{gal2F13}
   \end{figure*}

   \begin{figure*}
   \centering
   \includegraphics[angle=0,width=16cm,clip]{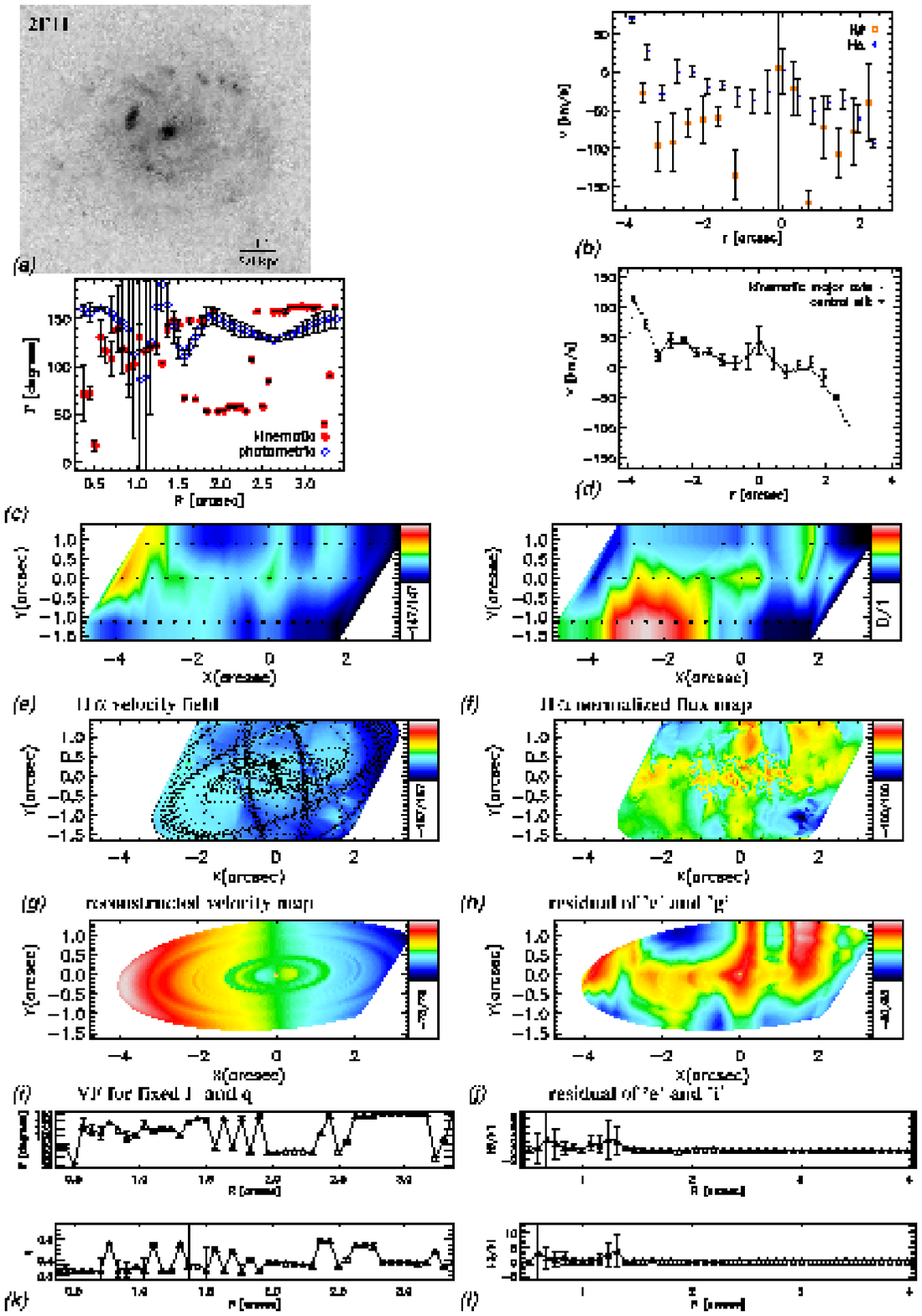}
 \caption{{\bf a)} HST-ACS image of the galaxy in the $V$~band. 
{\bf b)}~Rotation curves of different emission lines extracted along the central slit.
{\bf c)}~Position angles of kinematic and photometric axes as a function of radius.
{\bf d)}~Rotation curves extracted along the central slit and the kinematic major axis.
{\bf e)}~H$\alpha$~velocity field. 
{\bf f)}~Normalized H$\alpha$~flux map.
{\bf g)}~Velocity map reconstructed using 6~harmonic terms.
{\bf h)}~Residual of the velocity map and the reconstructed map. 
{\bf i)}~Simple rotation map constructed for position angle and ellipticity fixed to their global values.
{\bf j)}~Residual of the velocity map and the simple rotation map.
{\bf k)}~Position angle and flattening as a function of radius. {\bf l)}~$k_{3}/k_{1}$ and $k_{5}/k_{1}$ (from the analysis where position angle and ellipticity are fixed to their global values) as a function of radius.}
         \label{gal2F14}
   \end{figure*}

   \begin{figure*}
   \centering
   \includegraphics[angle=0,width=13.5cm,clip]{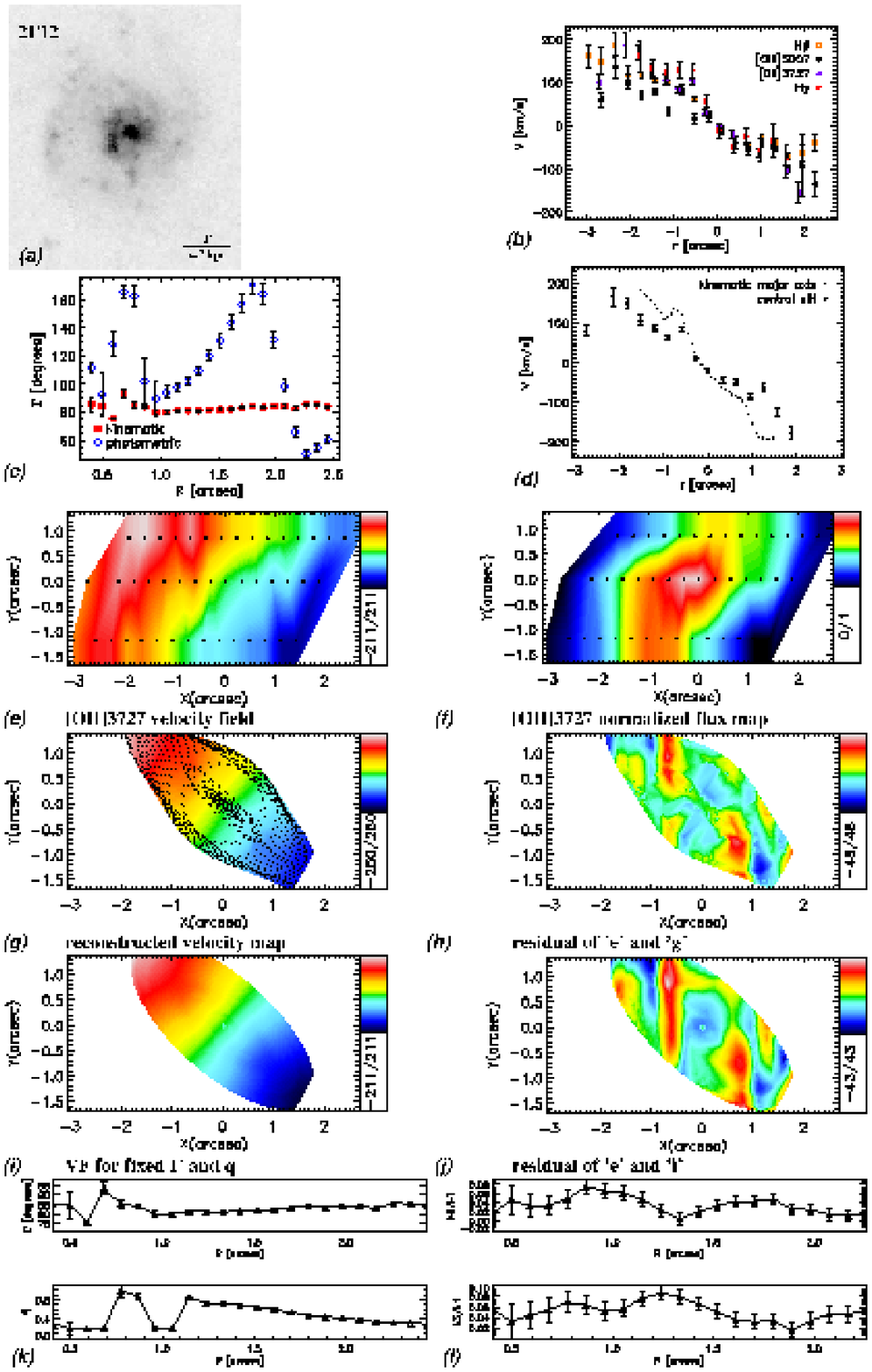}
 \caption{{\bf a)} HST-ACS image of the galaxy in the $V$~band. 
{\bf b)}~Rotation curves of different emission lines extracted along the central slit.
{\bf c)}~Position angles of kinematic and photometric axes as a function of radius.
{\bf d)}~Rotation curves extracted along the central slit and the kinematic major axis.
{\bf e)}~[OII]3727~velocity field. 
{\bf f)}~Normalized [OII]3727~flux map.
{\bf g)}~Velocity map reconstructed using 6~harmonic terms.
{\bf h)}~Residual of the velocity map and the reconstructed map. 
{\bf i)}~Simple rotation map constructed for position angle and ellipticity fixed to their global values.
{\bf j)}~Residual of the velocity map and the simple rotation map.
{\bf k)}~Position angle and flattening as a function of radius. 
{\bf l)}~$k_{3}/k_{1}$ and $k_{5}/k_{1}$ (from the analysis where position angle and ellipticity are fixed to their global values) as a function of radius.}
         \label{gal2F16}
   \end{figure*}

   \begin{figure*}
   \centering
   \includegraphics[angle=0,width=16cm,clip]{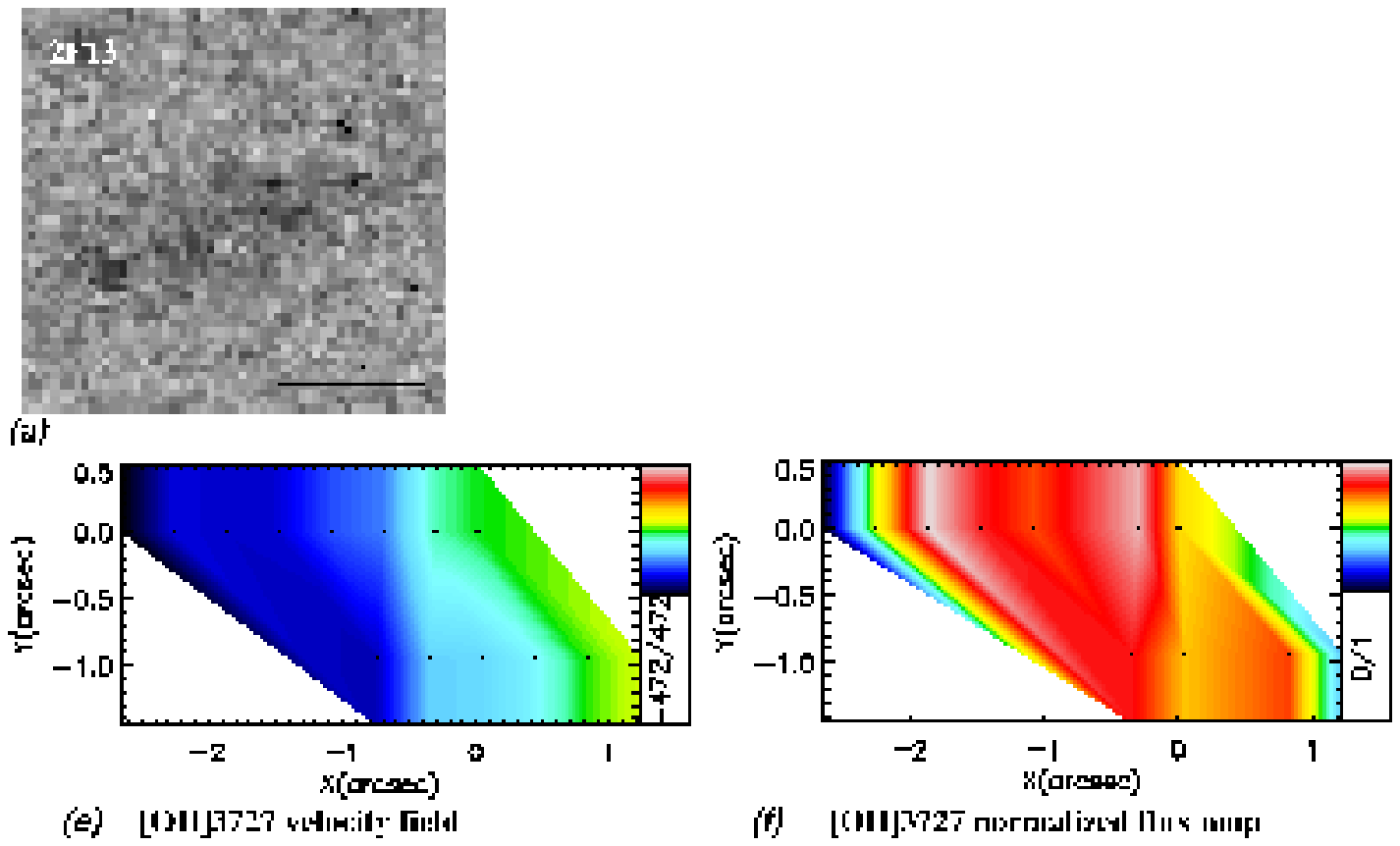}
 \caption{{\bf a)} HST-ACS image of the galaxy in the $V$~band. 
{\bf e)}~[OII]3727~velocity field. 
{\bf f)}~Normalized [OII]3727~flux map.}
         \label{gal2F17}
   \end{figure*}

   \begin{figure*}
   \centering
   \includegraphics[angle=0,width=16cm,clip]{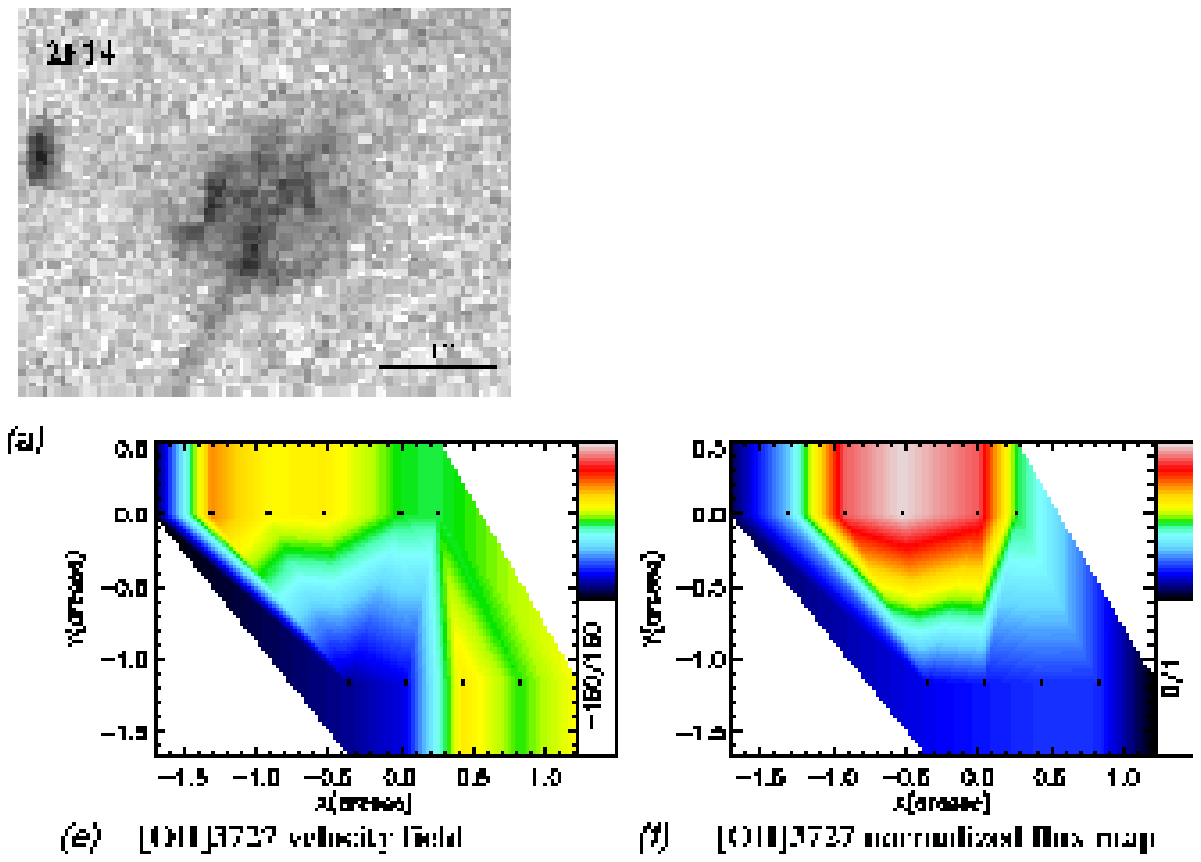}
 \caption{{\bf a)} HST-ACS image of the galaxy in the $V$~band. 
{\bf e)}~[OII]3727~velocity field. 
{\bf f)}~Normalized [OII]3727~flux map.}
         \label{gal2F18}
   \end{figure*}

   \begin{figure*}
   \centering
   \includegraphics[angle=0,width=15.7cm,clip]{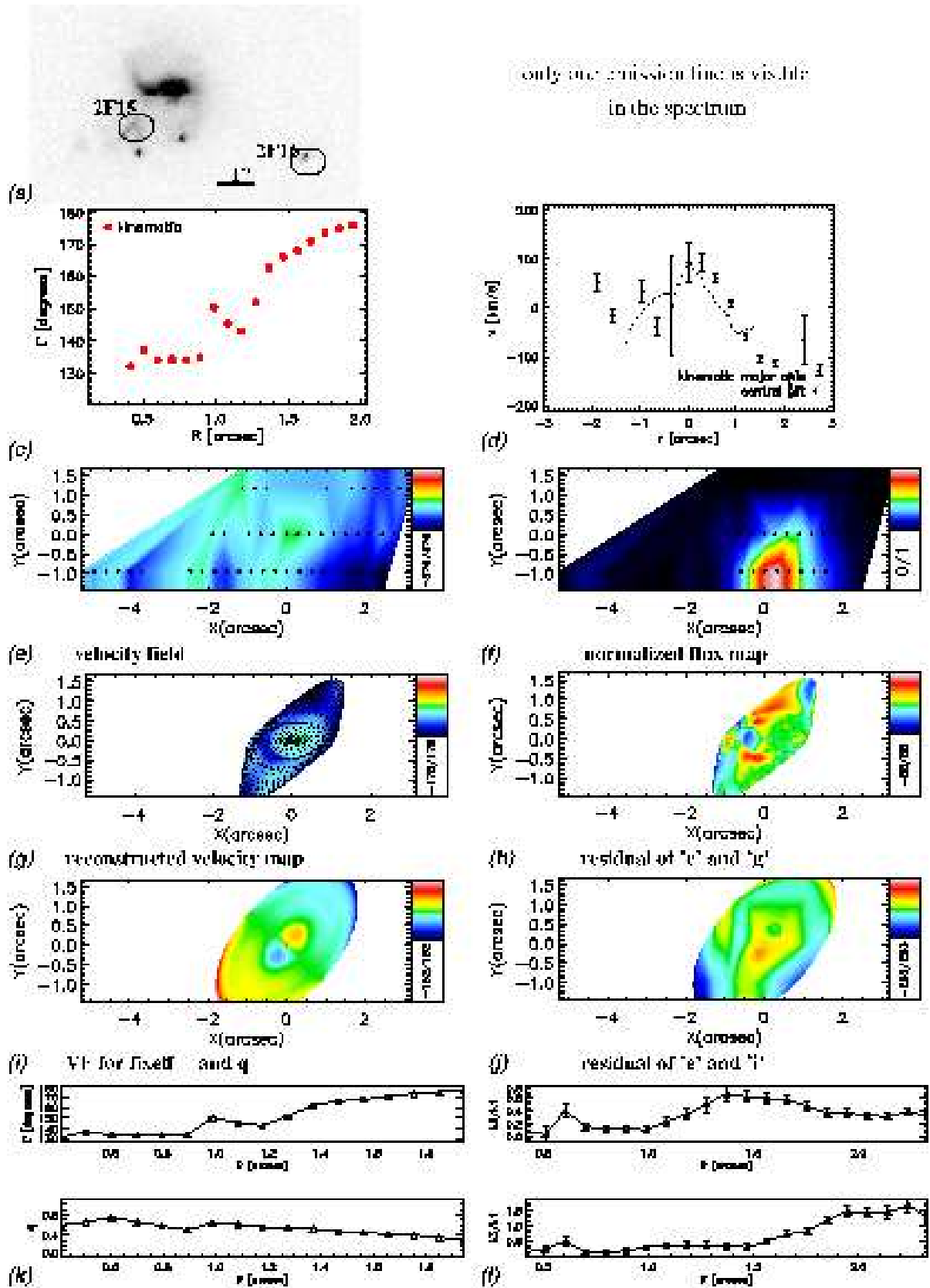}
 \caption{{\bf a)} HST-ACS image of the objects in the $V$~band. 
{\bf c)}~Position angle of the kinematic axis as a function of radius.
{\bf d)}~Rotation curves extracted along the central slit and the kinematic major axis.
{\bf e)}~Velocity field of the emission line which could not be identified.  The map belongs to 2F15 and 2F16 together.
{\bf f)}~Normalized flux map of the emission line which could not be identified.  The map belongs to 2F15 and 2F16 together.
{\bf g)}~Velocity map reconstructed using 6~harmonic terms.
{\bf h)}~Residual of the velocity map and the reconstructed map. 
{\bf i)}~Simple rotation map constructed for position angle and ellipticity fixed to their global values.
{\bf j)}~Residual of the velocity map and the simple rotation map.
{\bf k)}~Position angle and flattening as a function of radius. {\bf l)}~$k_{3}/k_{1}$ and $k_{5}/k_{1}$ (from the analysis where position angle and ellipticity are fixed to their global values) as a function of radius.}
         \label{gal2F20}
   \end{figure*}

   \begin{figure*}
   \centering
   \includegraphics[angle=0,width=16cm,clip]{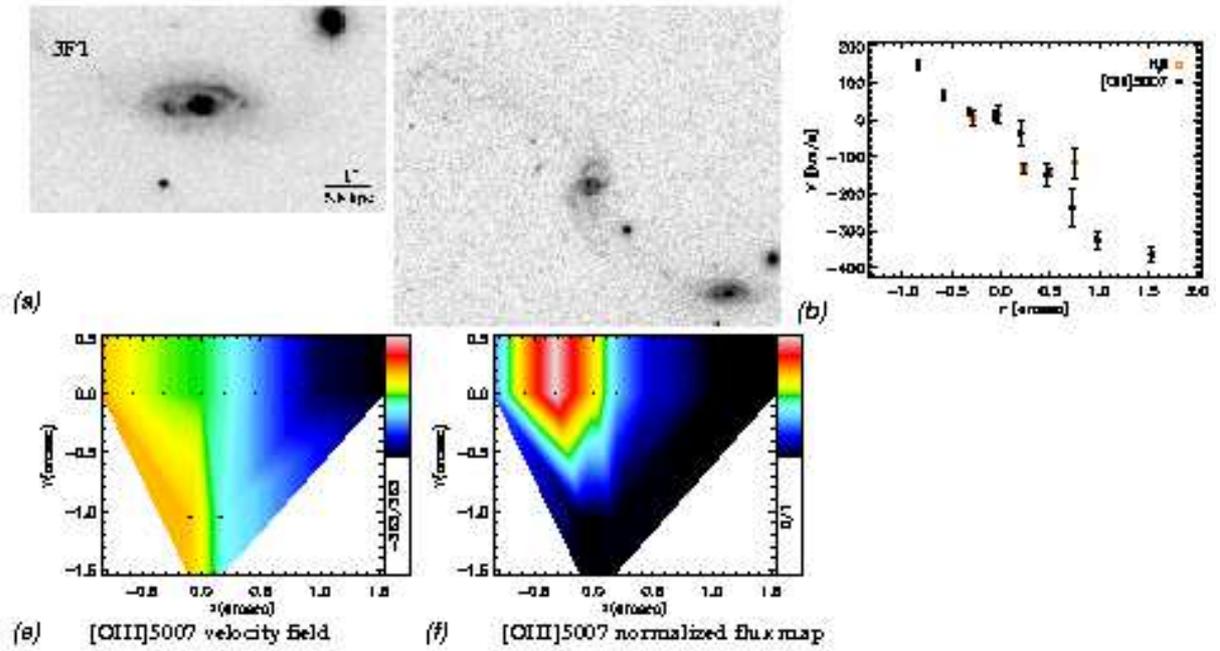}
 \caption{{\bf a.1)} HST-ACS image of the galaxy in the $V$~band. {\bf a.2)} Image showing the galaxy together with its companion. 
{\bf b)}~Rotation curves of different emission lines extracted along the central slit.
{\bf e)}~[OIII]5007~velocity field. 
{\bf f)}~Normalized [OIII]5007~flux map.}
         \label{gal3F1}
   \end{figure*}

   \begin{figure*}
   \centering
   \includegraphics[angle=0,width=13cm,clip]{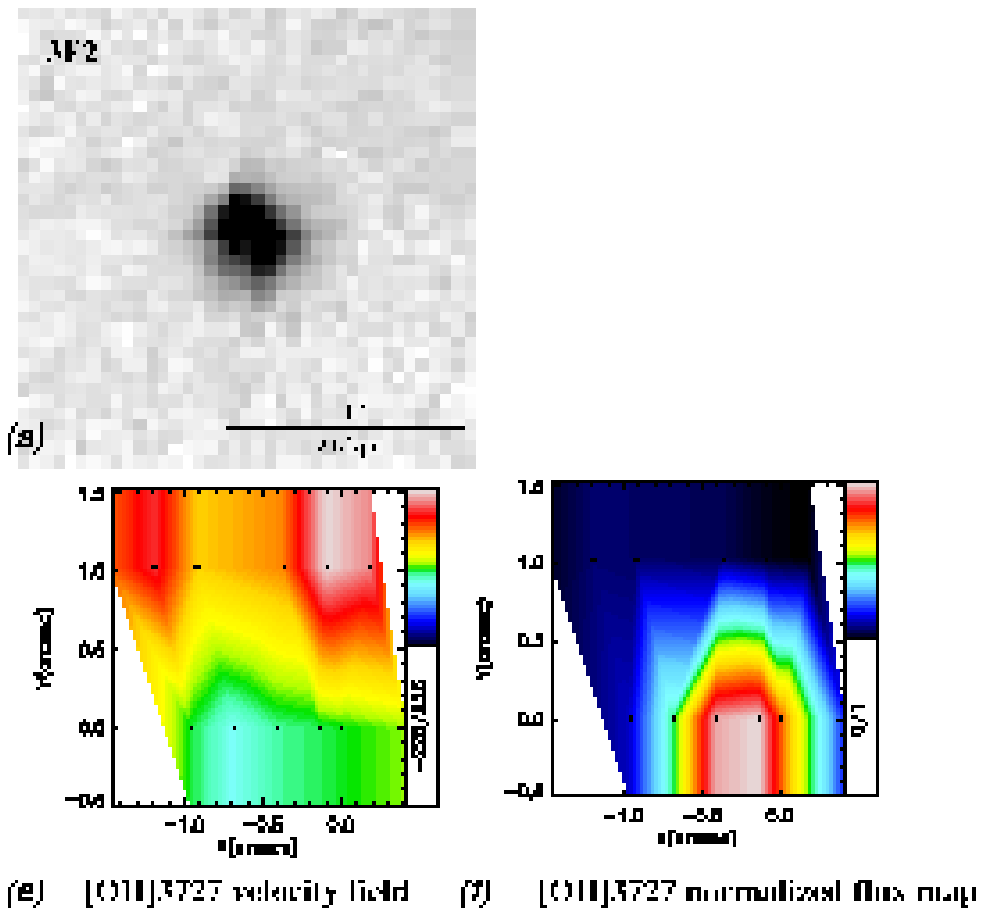}
 \caption{{\bf a)} HST-ACS image of the galaxy in the $V$~band. 
{\bf e)}~[OII]3727~velocity field. 
{\bf f)}~Normalized [OII]3727~flux map.}
         \label{gal3F4}
   \end{figure*}
\clearpage

   \begin{figure*}
   \centering
   \includegraphics[angle=0,width=14cm,clip]{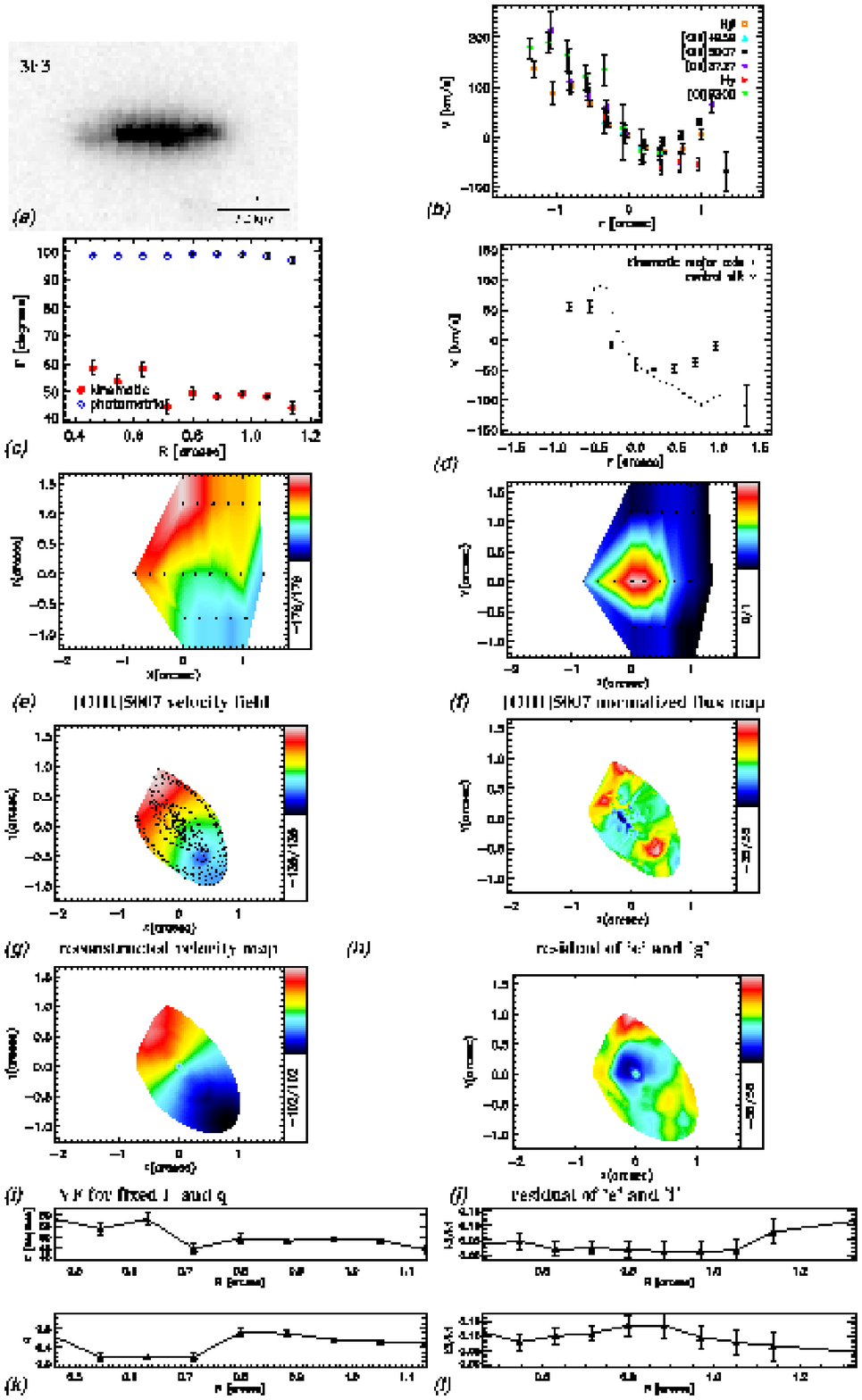}
 \caption{{\bf a)} HST-ACS image of the galaxy in the $V$~band. 
{\bf b)}~Rotation curves of different emission lines extracted along the central slit.
{\bf c)}~Position angles of kinematic and photometric axes as a function of radius.
{\bf d)}~Rotation curves extracted along the central slit and the kinematic major axis.
{\bf e)}~[OIII]5007~velocity field. 
{\bf f)}~Normalized [OIII]5007~flux map. 
{\bf g)}~Velocity map reconstructed using 6~harmonic terms.
{\bf h)}~Residual of the velocity map and the reconstructed map. 
{\bf i)}~Simple rotation map constructed for position angle and ellipticity fixed to their global values.
{\bf j)}~Residual of the velocity map and the simple rotation map.
{\bf k)}~Position angle and flattening as a function of radius. {\bf l)}~$k_{3}/k_{1}$ and $k_{5}/k_{1}$ (from the analysis where position angle and ellipticity are fixed to their global values) as a function of radius.}
         \label{gal3F2}
   \end{figure*}
\clearpage
   \begin{figure*}
   \centering
   \includegraphics[angle=0,width=15cm,clip]{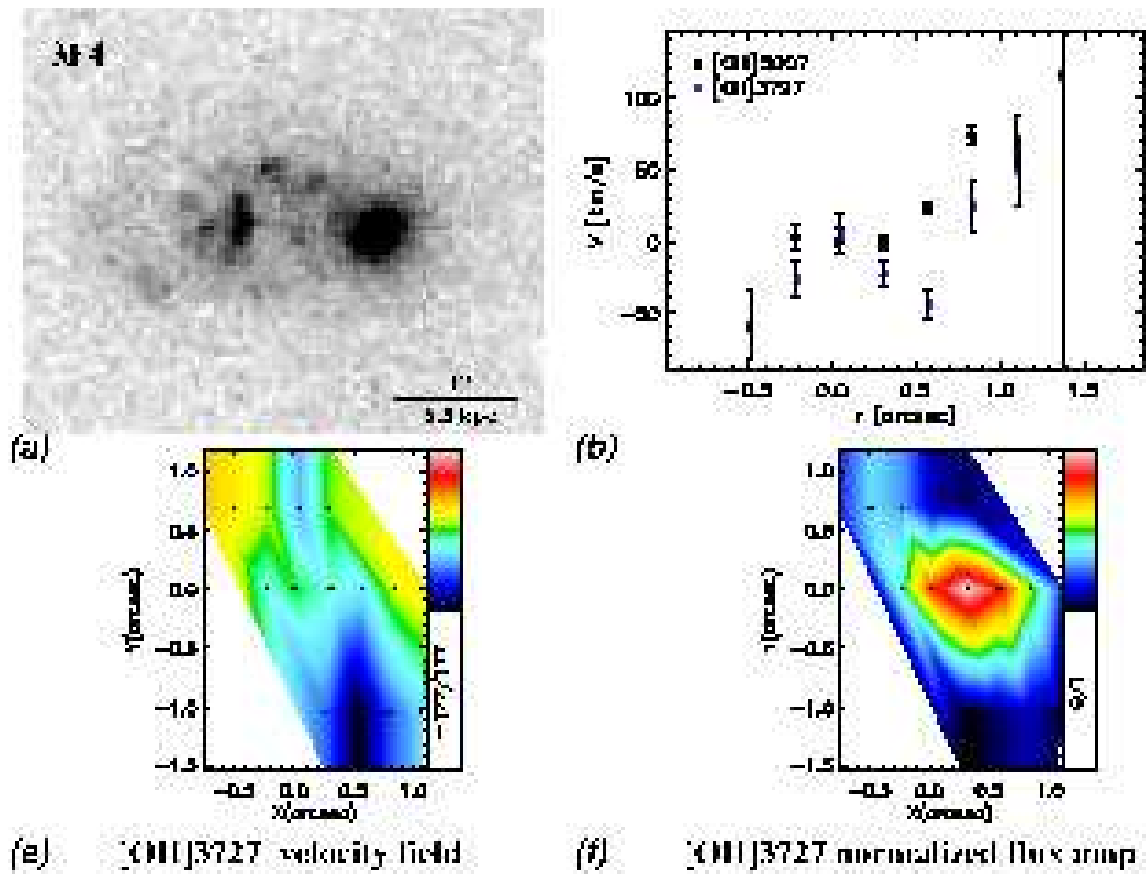}
 \caption{{\bf a)} HST-ACS image of the galaxy in the $V$~band. 
{\bf b)}~Rotation curves of different emission lines extracted along the central slit.
{\bf e)}~[OII]3727~velocity field. 
{\bf f)}~Normalized [OII]3727~flux map.}
         \label{gal3F6}
   \end{figure*}

   \begin{figure*}
   \centering
   \includegraphics[angle=0,width=15cm,clip]{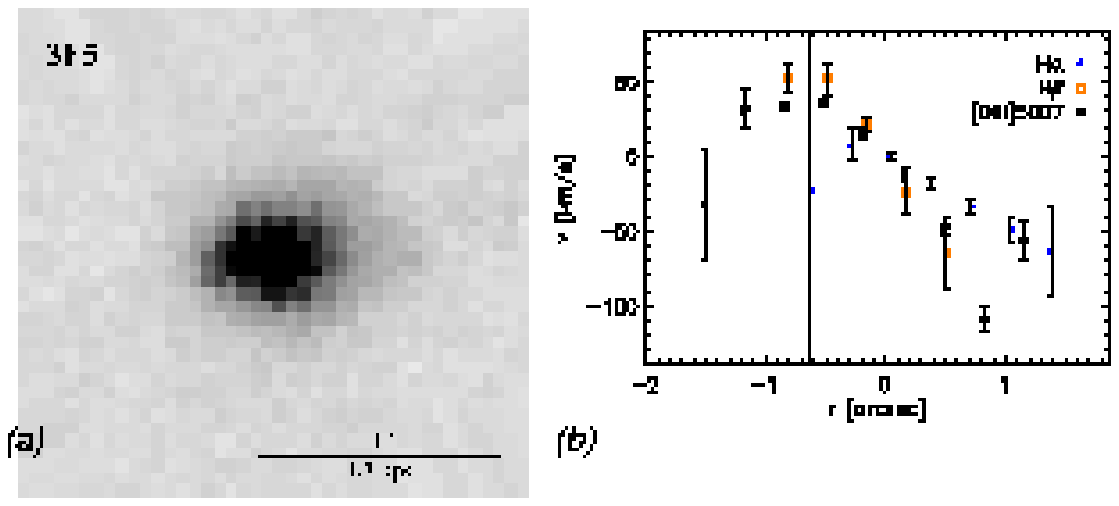}
 \caption{{\bf a)} HST-ACS image of the galaxy in the $V$~band. 
{\bf b)}~Rotation curves of different emission lines extracted along the central slit.}
         \label{gal3F12}
   \end{figure*}

   \begin{figure*}
   \centering
   \includegraphics[angle=0,width=14cm,clip]{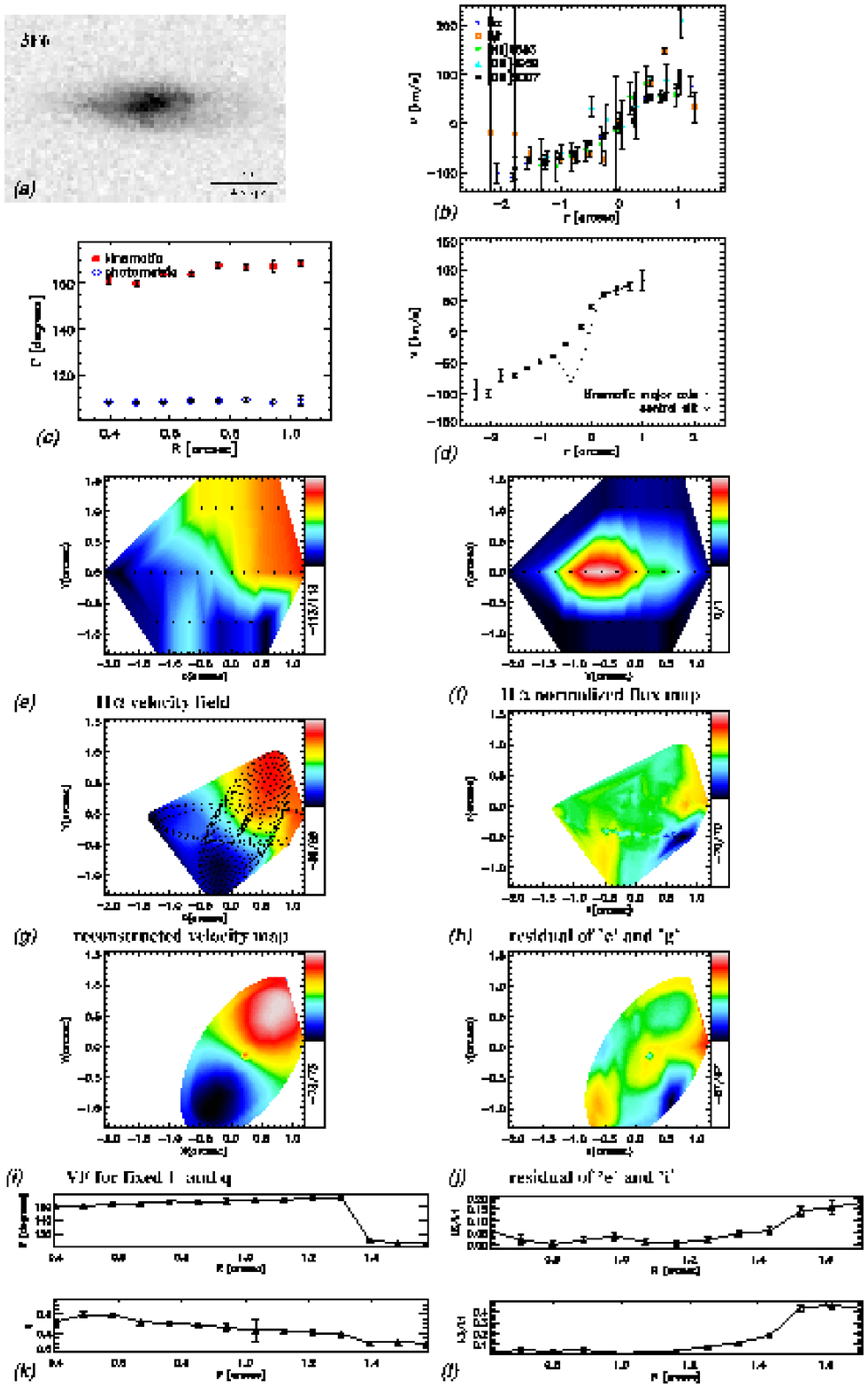}
 \caption{{\bf a)} HST-ACS image of the galaxy in the $V$~band. 
{\bf b)}~Rotation curves of different emission lines extracted along the central slit.
{\bf c)}~Position angles of kinematic and photometric axes as a function of radius.
{\bf d)}~Rotation curves extracted along the central slit and the kinematic major axis.
{\bf e)}~H$\alpha$~velocity field. 
{\bf f)}~Normalized H$\alpha$~flux map. 
{\bf g)}~Velocity map reconstructed using 6~harmonic terms.
{\bf h)}~Residual of the velocity map and the reconstructed map. 
{\bf i)}~Simple rotation map constructed for position angle and ellipticity fixed to their global values.
{\bf j)}~Residual of the velocity map and the simple rotation map.
{\bf k)}~Position angle and flattening as a function of radius. {\bf l)}~$k_{3}/k_{1}$ and $k_{5}/k_{1}$ (from the analysis where position angle and ellipticity are fixed to their global values) as a function of radius.}
         \label{gal3F5}
   \end{figure*}

   \begin{figure*}
   \centering
   \includegraphics[angle=0,width=14.2cm,clip]{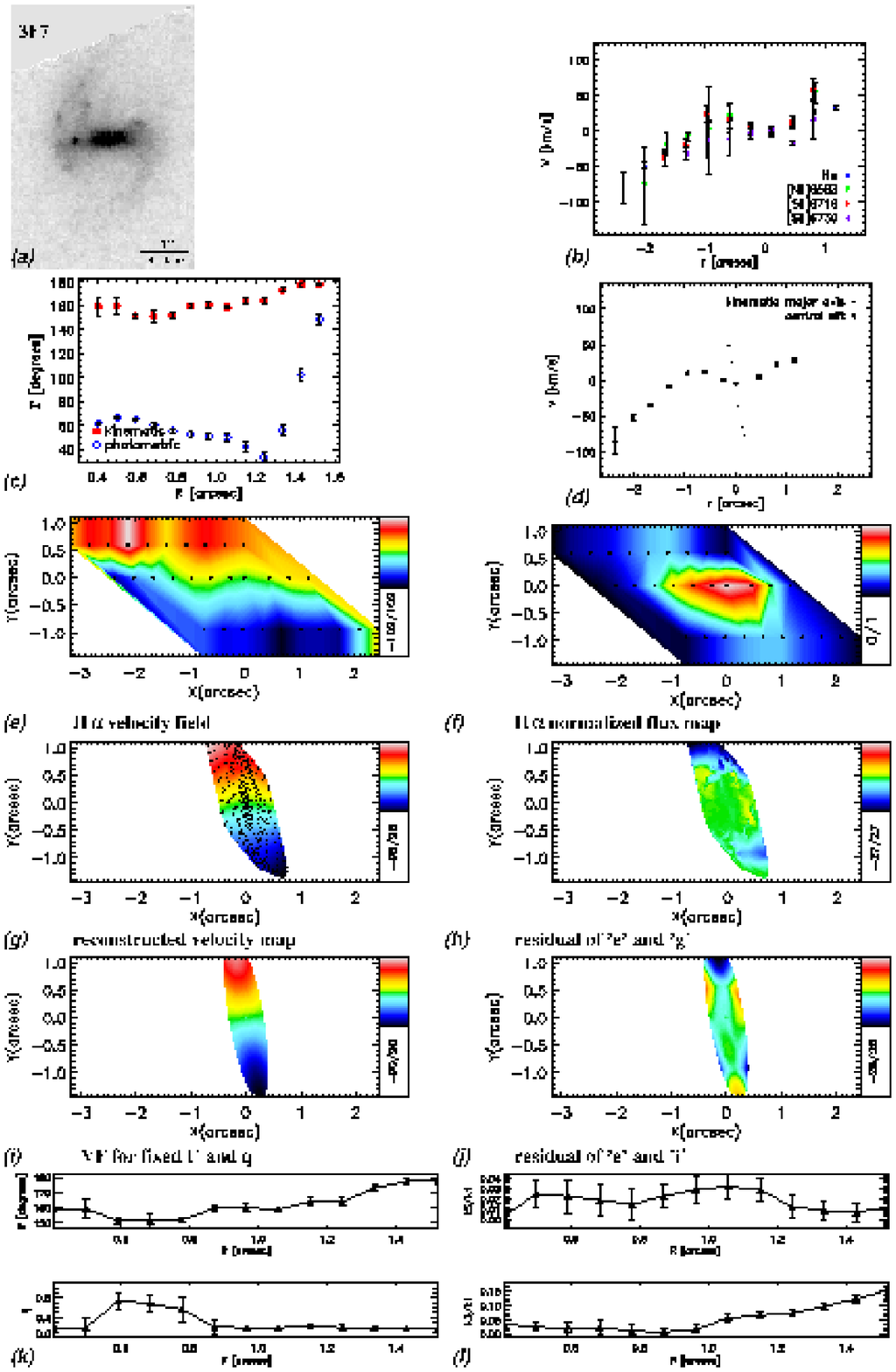}
 \caption{{\bf a)} HST-ACS image of the galaxy in the $V$~band. 
{\bf b)}~Rotation curves of different emission lines extracted along the central slit.
{\bf c)}~Position angles of kinematic and photometric axes as a function of radius.
{\bf d)}~Rotation curves extracted along the central slit and the kinematic major axis.
{\bf e)}~H$\alpha$~velocity field. 
{\bf f)}~Normalized H$\alpha$~flux map. 
{\bf g)}~Velocity map reconstructed using 6~harmonic terms.
{\bf h)}~Residual of the velocity map and the reconstructed map. 
{\bf i)}~Simple rotation map constructed for position angle and ellipticity fixed to their global values.
{\bf j)}~Residual of the velocity map and the simple rotation map.
{\bf k)}~Position angle and flattening as a function of radius. {\bf l)}~$k_{3}/k_{1}$ and $k_{5}/k_{1}$ (from the analysis where position angle and ellipticity are fixed to their global values) as a function of radius.}
         \label{gal3F7}
   \end{figure*}


   \begin{figure*}
   \centering
   \includegraphics[angle=0,width=16.3cm,clip]{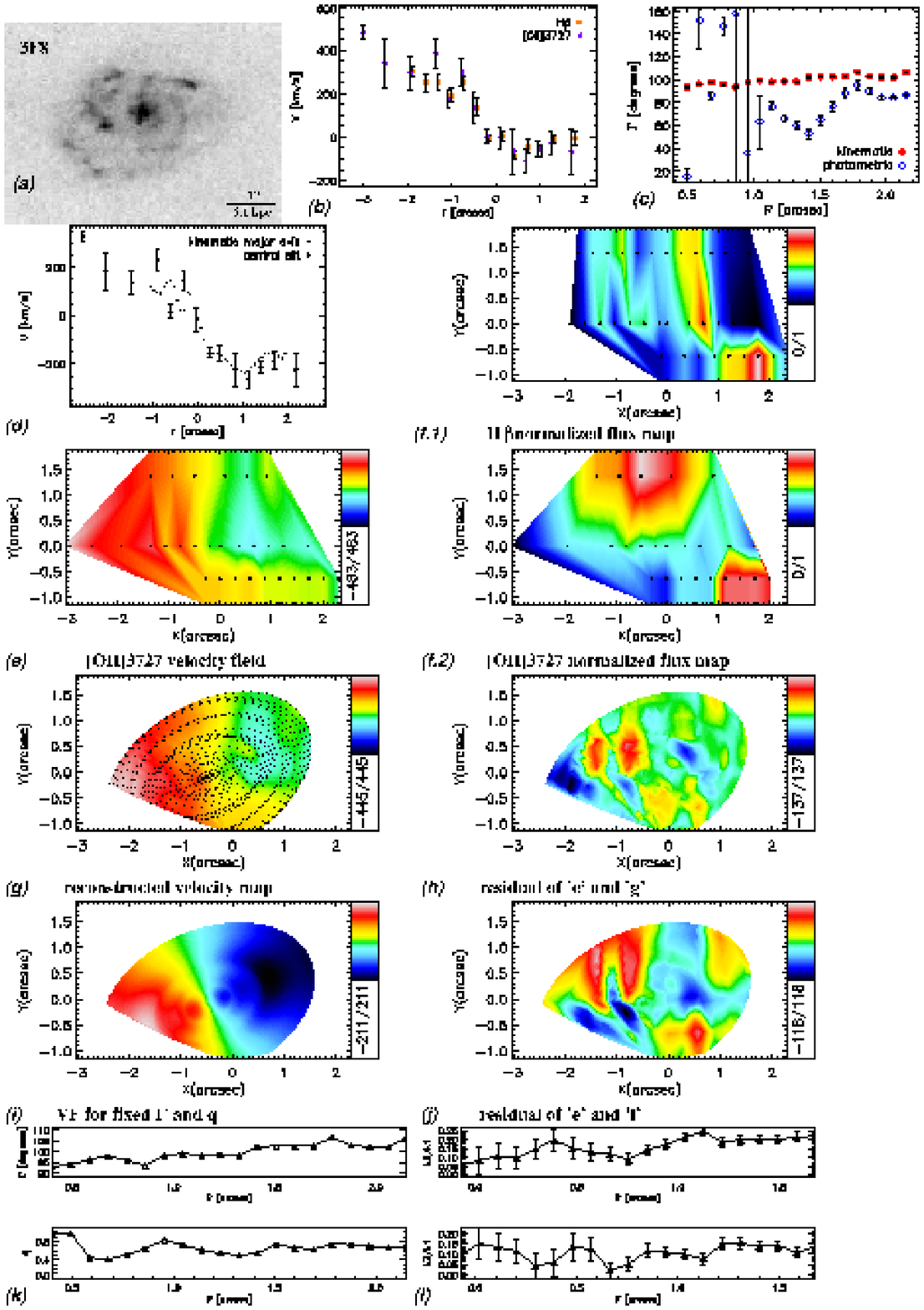}
 \caption{{\bf a)} HST-ACS image of the galaxy in the $V$~band. 
{\bf b)}~Rotation curves of different emission lines extracted along the central slit.
{\bf c)}~Position angles of kinematic and photometric axes as a function of radius.
{\bf d)}~Rotation curves extracted along the central slit and the kinematic major axis.
{\bf e)}~[OII]3727~velocity field. 
{\bf f.1)}~Normalized H$\beta$~flux map. 
{\bf f.2)}~Normalized [OII]3727~flux map. 
{\bf g)}~Velocity map reconstructed using 6~harmonic terms.
{\bf h)}~Residual of the velocity map and the reconstructed map. 
{\bf i)}~Simple rotation map constructed for position angle and ellipticity fixed to their global values.
{\bf j)}~Residual of the velocity map and the simple rotation map.
{\bf k)}~Position angle and flattening as a function of radius. {\bf l)}~$k_{3}/k_{1}$ and $k_{5}/k_{1}$ (from the analysis where position angle and ellipticity are fixed to their global values) as a function of radius.}
         \label{gal3F8}
   \end{figure*}

   \begin{figure*}
   \centering
   \includegraphics[angle=0,width=13.5cm,clip]{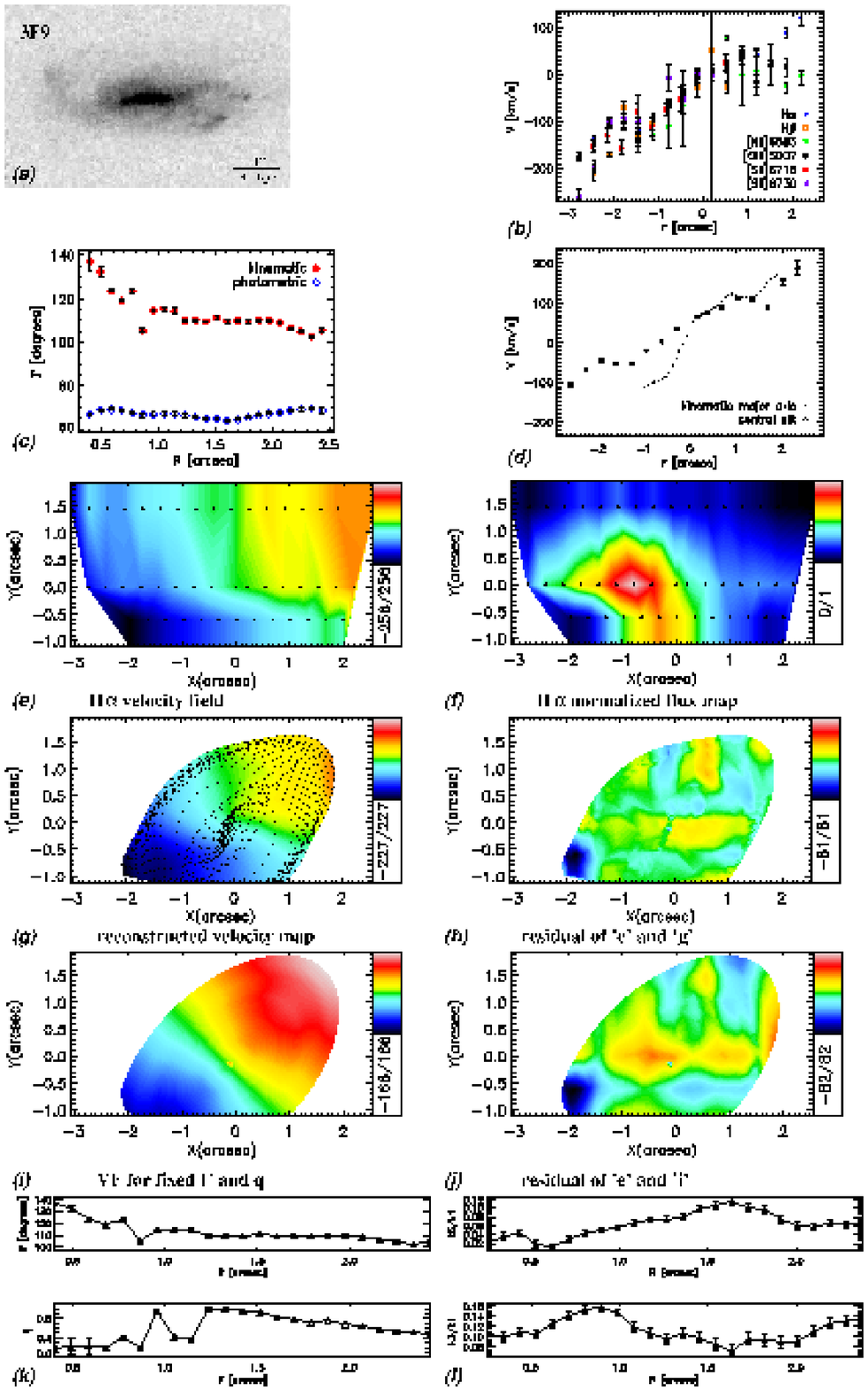}
 \caption{{\bf a)} HST-ACS image of the galaxy in the $V$~band. 
{\bf b)}~Rotation curves of different emission lines extracted along the central slit.
{\bf c)}~Position angles of kinematic and photometric axes as a function of radius.
{\bf d)}~Rotation curves extracted along the central slit and the kinematic major axis.
{\bf e)}~H$\alpha$~velocity field. 
{\bf f)}~Normalized H$\alpha$~flux map.
{\bf g)}~Velocity map reconstructed using 6~harmonic terms.
{\bf h)}~Residual of the velocity map and the reconstructed map. 
{\bf i)}~Simple rotation map constructed for position angle and ellipticity fixed to their global values.
{\bf j)}~Residual of the velocity map and the simple rotation map.
{\bf k)}~Position angle and flattening as a function of radius. {\bf l)}~$k_{3}/k_{1}$ and $k_{5}/k_{1}$ (from the analysis where position angle and ellipticity are fixed to their global values) as a function of radius.}
         \label{gal3F11}
   \end{figure*}

   \begin{figure*}
   \centering
   \includegraphics[angle=0,width=15cm,clip]{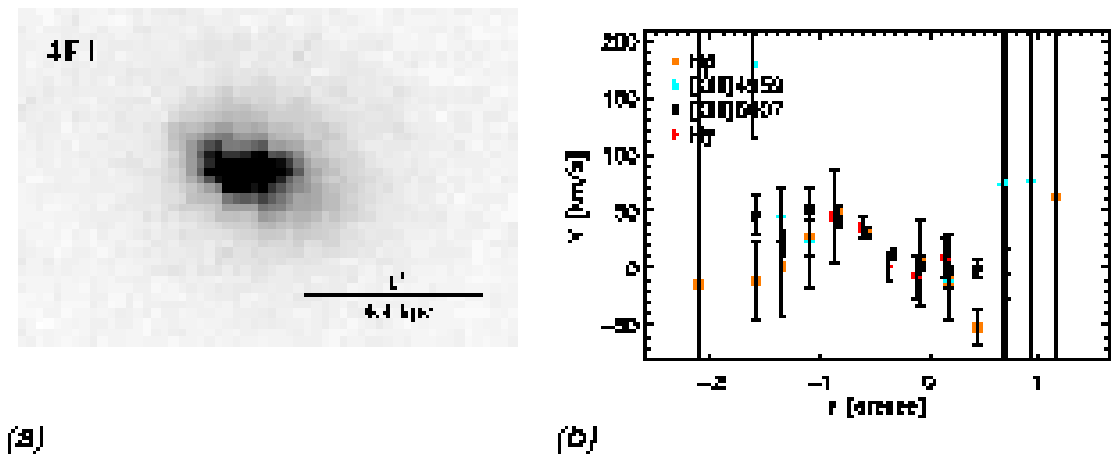}
 \caption{{\bf a)} HST-ACS image of the galaxy in the $V$~band. 
{\bf b)}~Rotation curves of different emission lines extracted along the central slit.}
         \label{gal4F1}
   \end{figure*}

   \begin{figure*}
   \centering
   \includegraphics[angle=0,width=15cm,clip]{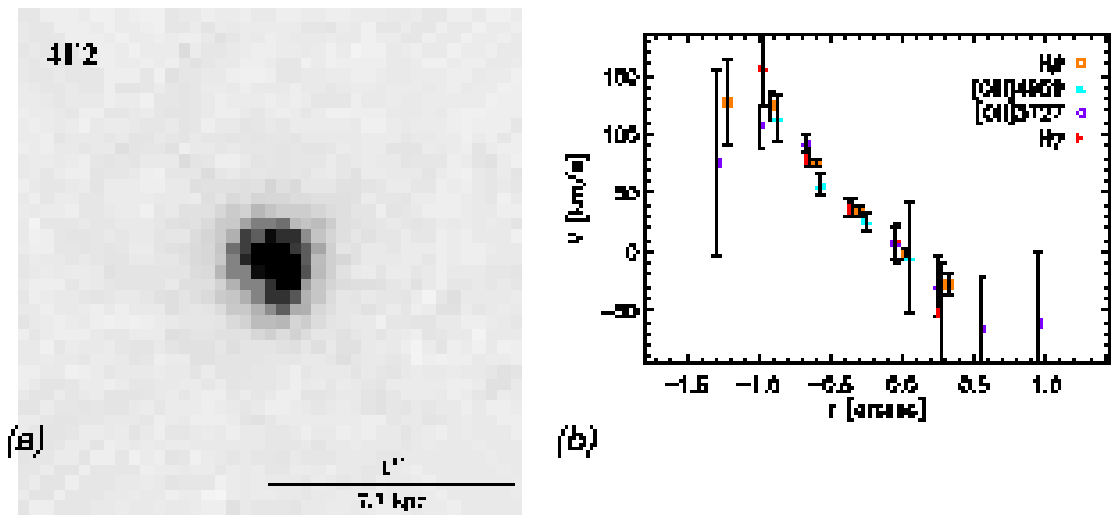}
 \caption{{\bf a)} HST-ACS image of the galaxy in the $V$~band. 
{\bf b)}~Rotation curves of different emission lines extracted along the central slit.}
         \label{gal4F3}
   \end{figure*}

   \begin{figure*}
   \centering
   \includegraphics[angle=0,width=13.5cm,clip]{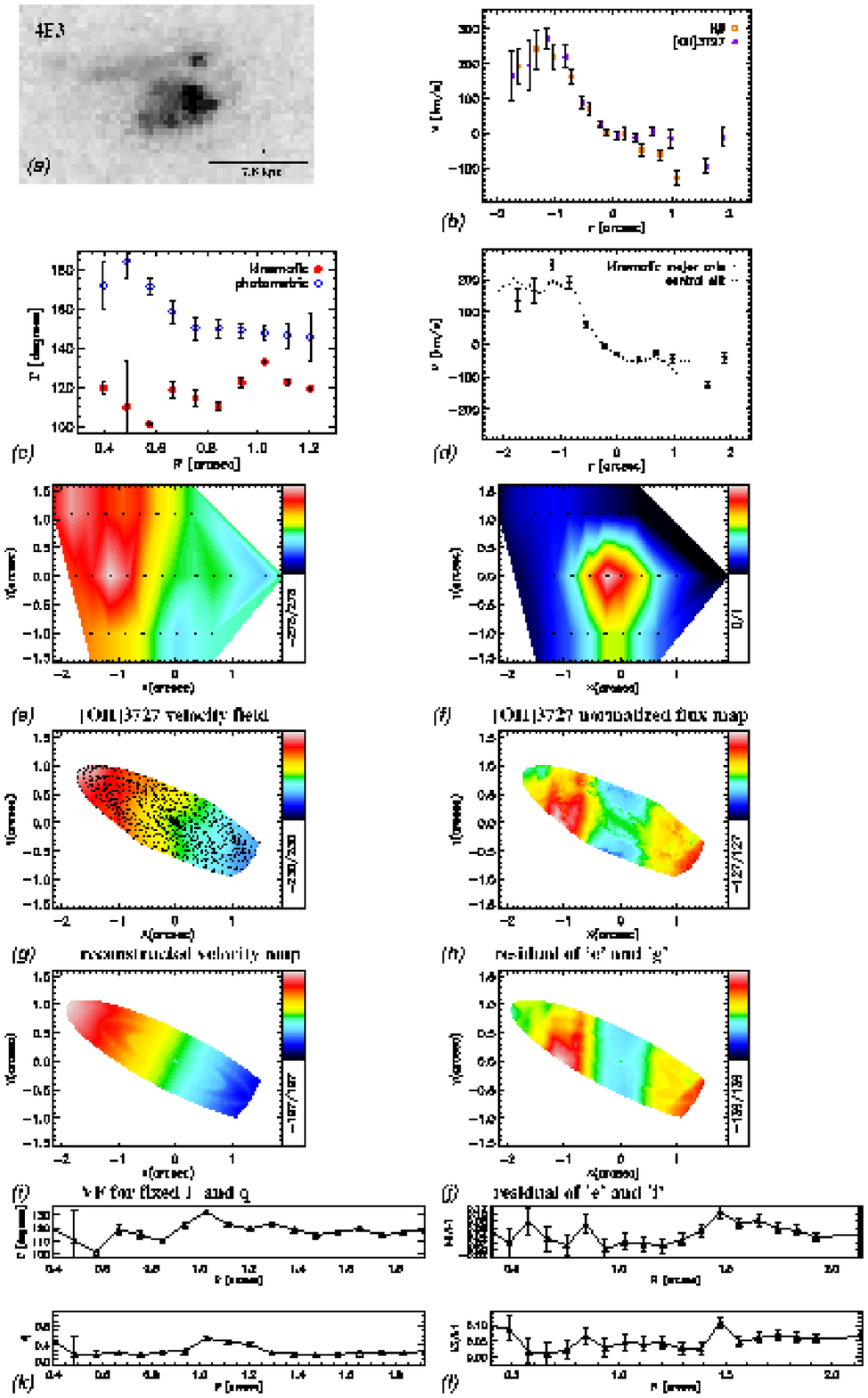}
 \caption{{\bf a)} HST-ACS image of the galaxy in the $V$~band. 
{\bf b)}~Rotation curves of different emission lines extracted along the central slit.
{\bf c)}~Position angles of kinematic and photometric axes as a function of radius.
{\bf d)}~Rotation curves extracted along the central slit and the kinematic major axis.
{\bf e)}~[OII]3727~velocity field. 
{\bf f)}~Normalized [OII]3727~flux map. 
{\bf g)}~Velocity map reconstructed using 6~harmonic terms.
{\bf h)}~Residual of the velocity map and the reconstructed map. 
{\bf i)}~Simple rotation map constructed for position angle and ellipticity fixed to their global values.
{\bf j)}~Residual of the velocity map and the simple rotation map.
{\bf k)}~Position angle and flattening as a function of radius. {\bf l)}~$k_{3}/k_{1}$ and $k_{5}/k_{1}$ (from the analysis where position angle and ellipticity are fixed to their global values) as a function of radius.}
         \label{gal4F2}
   \end{figure*}

   \begin{figure*}
   \centering
   \includegraphics[angle=0,width=13.5cm,clip]{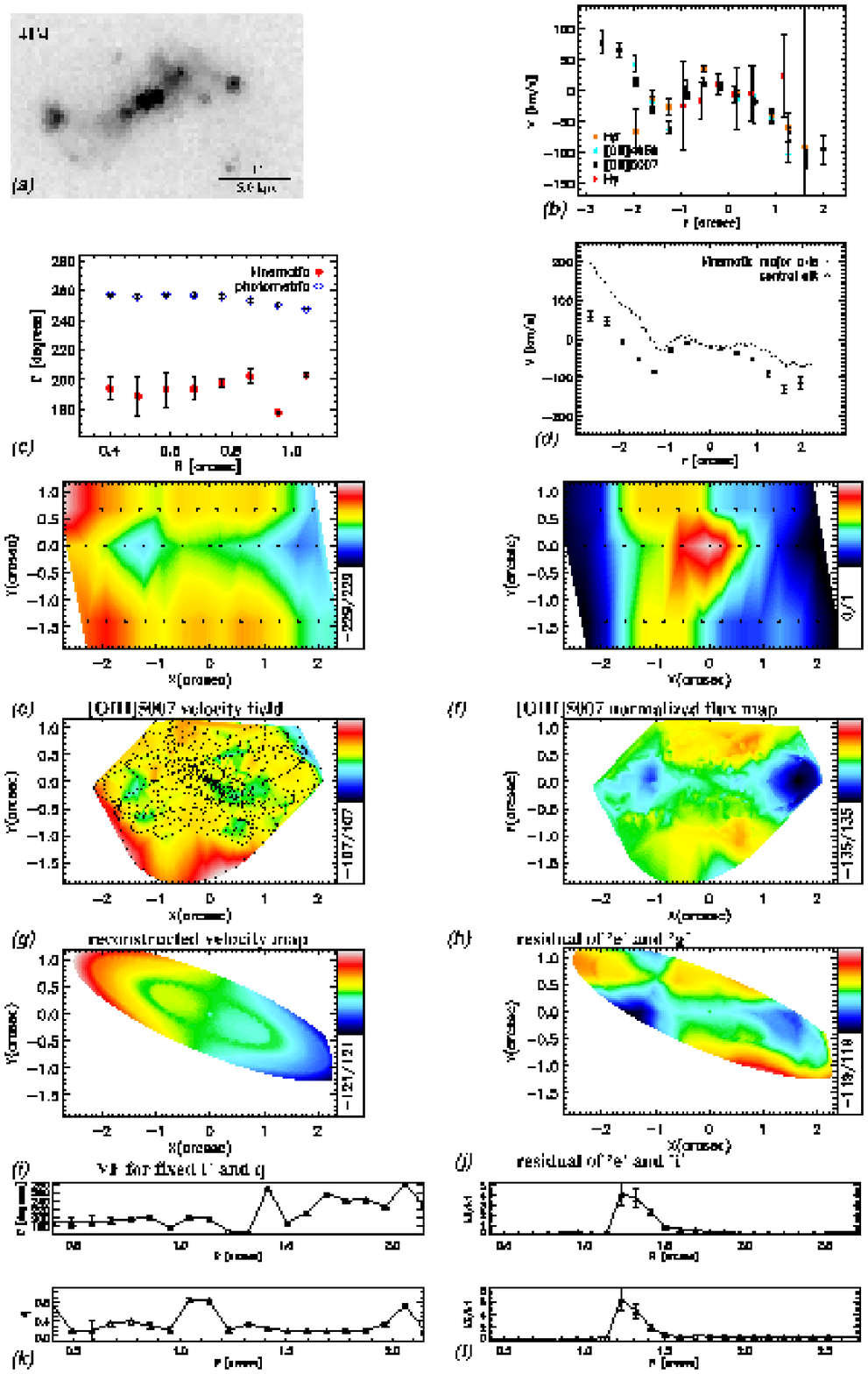}
 \caption{{\bf a)} HST-ACS image of the galaxy in the $V$~band. 
{\bf b)}~Rotation curves of different emission lines extracted along the central slit.
{\bf c)}~Position angles of kinematic and photometric axes as a function of radius.
{\bf d)}~Rotation curves extracted along the central slit and the kinematic major axis.
{\bf e)}~[OIII]5007~velocity field. 
{\bf f)}~Normalized [OIII]5007~flux map. 
{\bf g)}~Velocity map reconstructed using 6~harmonic terms.
{\bf h)}~Residual of the velocity map and the reconstructed map. 
{\bf i)}~Simple rotation map constructed for position angle and ellipticity fixed to their global values.
{\bf j)}~Residual of the velocity map and the simple rotation map.
{\bf k)}~Position angle and flattening as a function of radius. {\bf l)}~$k_{3}/k_{1}$ and $k_{5}/k_{1}$ (from the analysis where position angle and ellipticity are fixed to their global values) as a function of radius.}
         \label{gal4F4}
   \end{figure*}

   \begin{figure*}
   \centering
   \includegraphics[angle=0,width=14cm,clip]{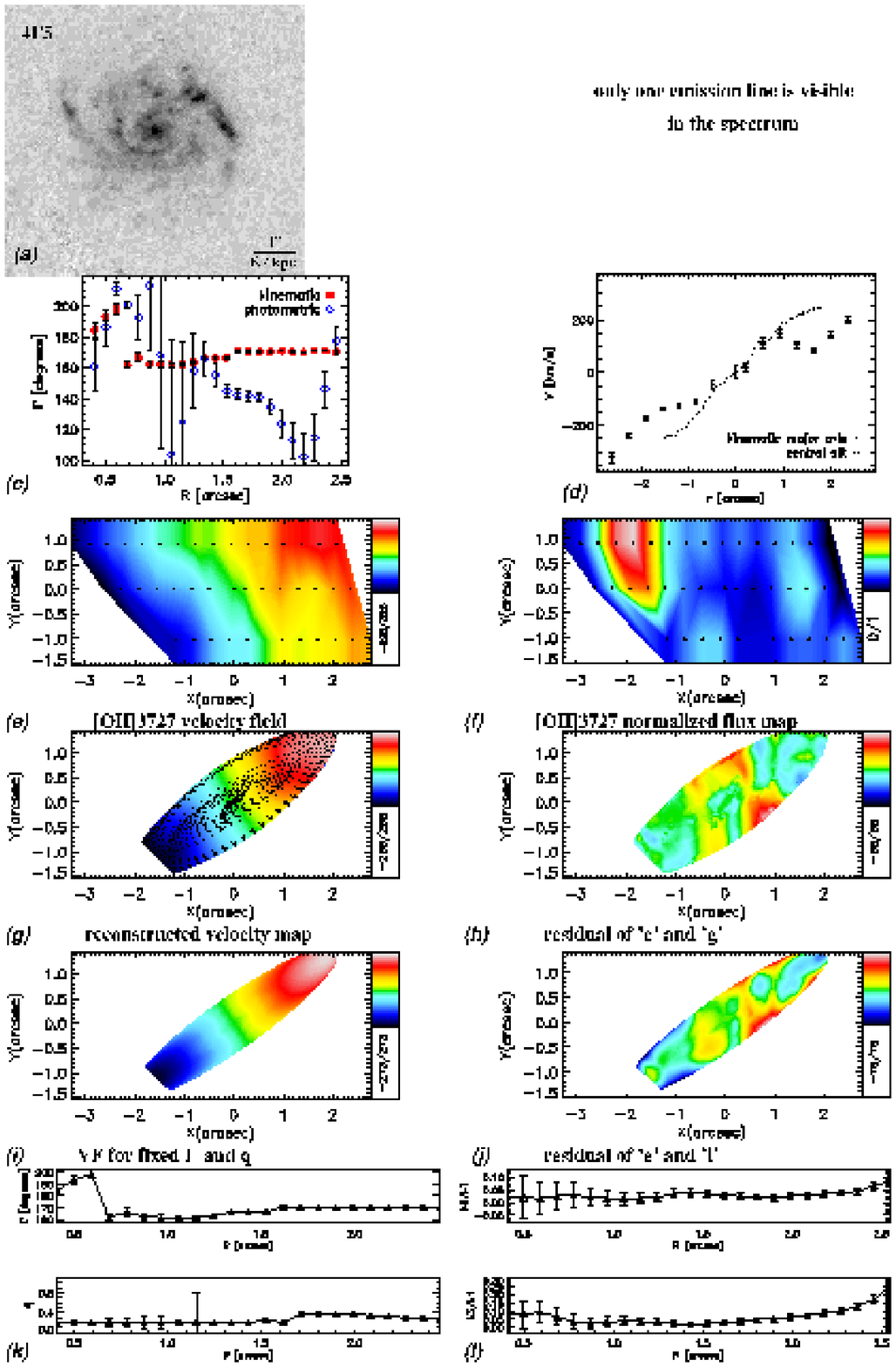}
 \caption{{\bf a)} HST-ACS image of the galaxy in the $V$~band. 
{\bf b)}~Rotation curves of different emission lines extracted along the central slit.
{\bf c)}~Position angles of kinematic and photometric axes as a function of radius.
{\bf d)}~Rotation curves extracted along the central slit and the kinematic major axis.
{\bf e)}~[OII]3727~velocity field.
{\bf f)}~Normalized [OII]3727~flux map.  
{\bf g)}~Velocity map reconstructed using 6~harmonic terms.
{\bf h)}~Residual of the velocity map and the reconstructed map. 
{\bf i)}~Simple rotation map constructed for position angle and ellipticity fixed to their global values.
{\bf j)}~Residual of the velocity map and the simple rotation map.
{\bf k)}~Position angle and flattening as a function of radius. {\bf l)}~$k_{3}/k_{1}$ and $k_{5}/k_{1}$ (from the analysis where position angle and ellipticity are fixed to their global values) as a function of radius.}
         \label{gal4F5}
   \end{figure*}

   \begin{figure*}
   \centering
   \includegraphics[angle=0,width=13cm,clip]{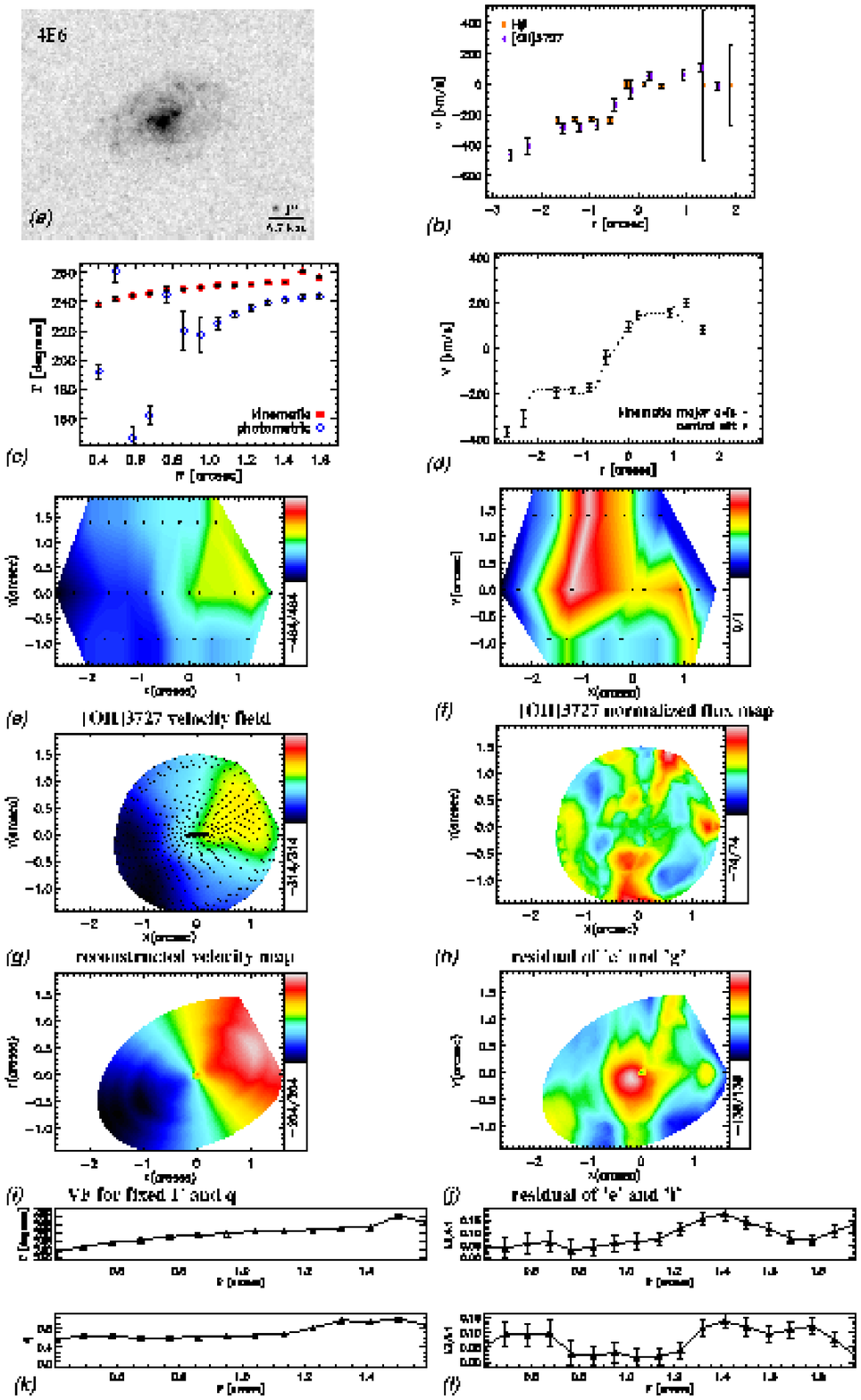}
 \caption{{\bf a)} HST-ACS image of the galaxy in the $V$~band. 
{\bf b)}~Rotation curves of different emission lines extracted along the central slit.
{\bf c)}~Position angles of kinematic and photometric axes as a function of radius.
{\bf d)}~Rotation curves extracted along the central slit and the kinematic major axis.
{\bf e)}~[OII]3727~velocity field. 
{\bf f)}~Normalized [OII]3727~flux map.
{\bf g)}~Velocity map reconstructed using 6~harmonic terms.
{\bf h)}~Residual of the velocity map and the reconstructed map. 
{\bf i)}~Simple rotation map constructed for position angle and ellipticity fixed to their global values.
{\bf j)}~Residual of the velocity map and the simple rotation map.
{\bf k)}~Position angle and flattening as a function of radius. {\bf l)}~$k_{3}/k_{1}$ and $k_{5}/k_{1}$ (from the analysis where position angle and ellipticity are fixed to their global values) as a function of radius.}
         \label{gal4F6}
   \end{figure*}

   \begin{figure*}
   \centering
   \includegraphics[angle=0,width=14cm,clip]{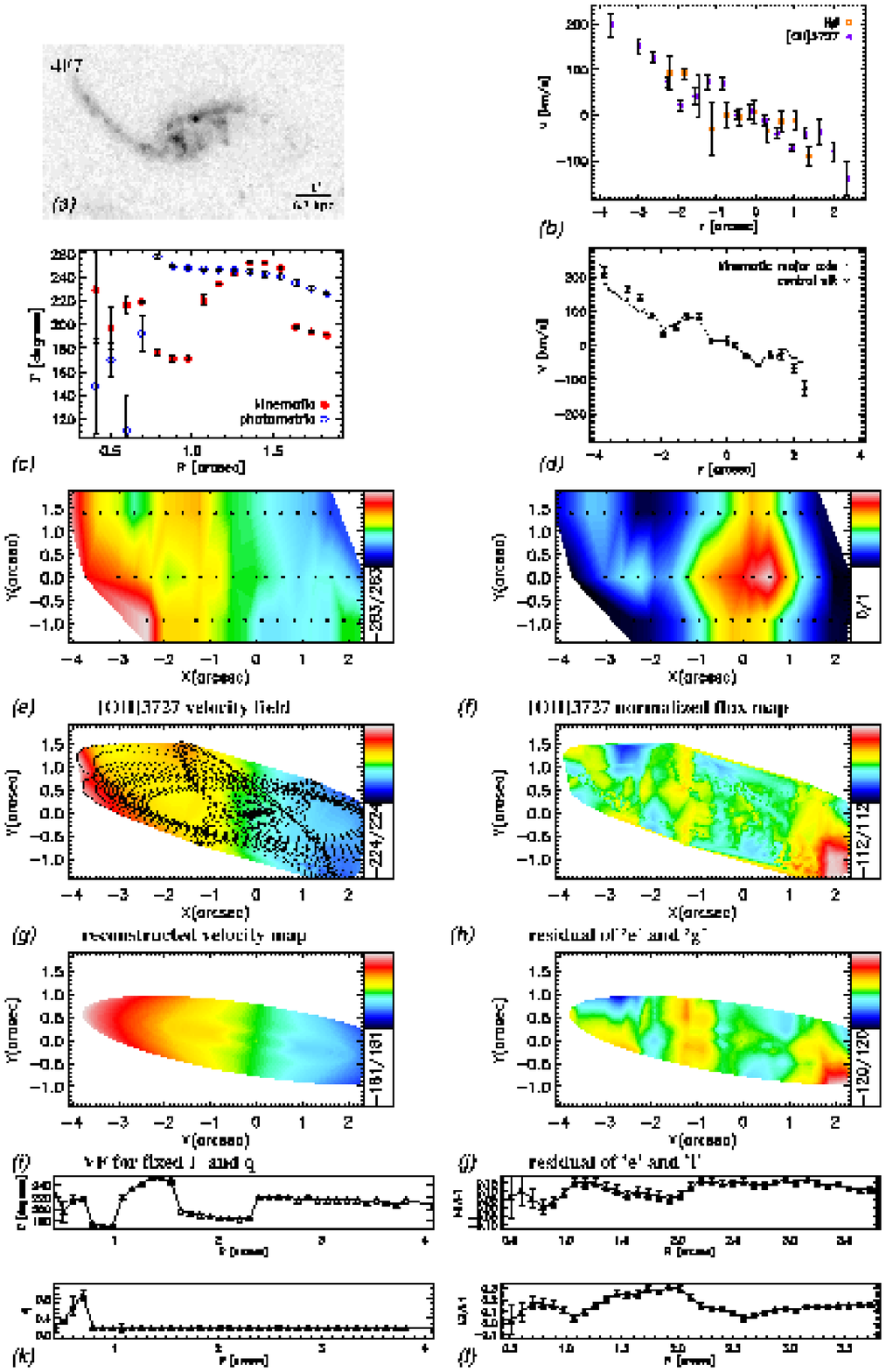}
 \caption{{\bf a)} HST-ACS image of the galaxy in the $V$~band. 
{\bf b)}~Rotation curves of different emission lines extracted along the central slit.
{\bf c)}~Position angles of kinematic and photometric axes as a function of radius.
{\bf d)}~Rotation curves extracted along the central slit and the kinematic major axis.
{\bf e)}~[OII]3727~velocity field. 
{\bf f)}~Normalized [OII]3727~flux map. 
{\bf g)}~Velocity map reconstructed using 6~harmonic terms.
{\bf h)}~Residual of the velocity map and the reconstructed map. 
{\bf i)}~Simple rotation map constructed for position angle and ellipticity fixed to their global values.
{\bf j)}~Residual of the velocity map and the simple rotation map.
{\bf k)}~Position angle and flattening as a function of radius. {\bf l)}~$k_{3}/k_{1}$ and $k_{5}/k_{1}$ (from the analysis where position angle and ellipticity are fixed to their global values) as a function of radius.}
         \label{gal4F7}
   \end{figure*}

\clearpage
   \begin{figure*}
   \centering
   \includegraphics[angle=0,width=17cm,clip]{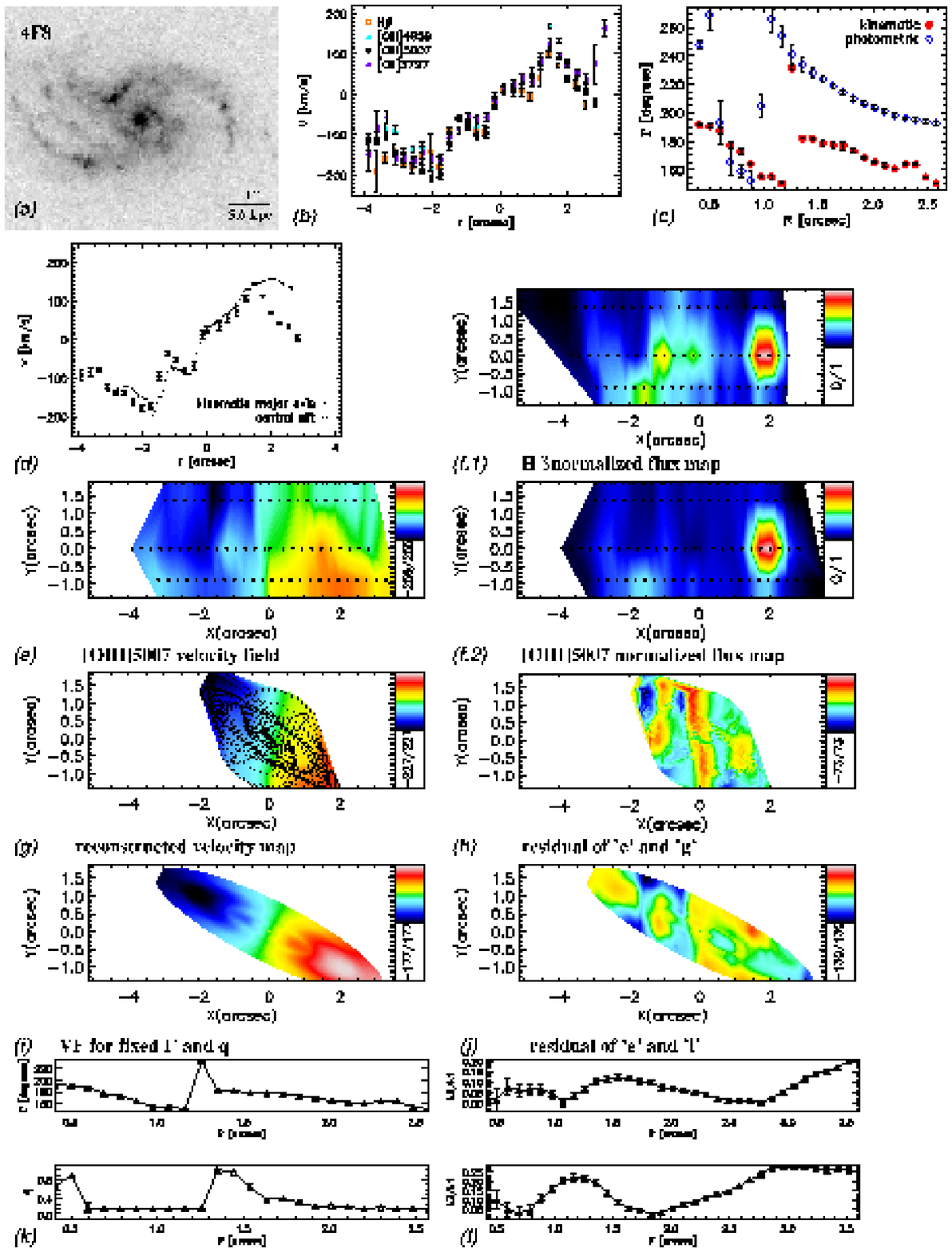}
 \caption{{\bf a)} HST-ACS image of the galaxy in the $V$~band. 
{\bf b)}~Rotation curves of different emission lines extracted along the central slit.
{\bf c)}~Position angles of kinematic and photometric axes as a function of radius.
{\bf d)}~Rotation curves extracted along the central slit and the kinematic major axis.
{\bf e)}~[OIII]5007~velocity field. 
{\bf f.1)}~Normalized H$\beta$~flux map. 
{\bf f.2)}~Normalized [OIII]5007~flux map. 
{\bf g)}~Velocity map reconstructed using 6~harmonic terms.
{\bf h)}~Residual of the velocity map and the reconstructed map. 
{\bf i)}~Simple rotation map constructed for position angle and ellipticity fixed to their global values.
{\bf j)}~Residual of the velocity map and the simple rotation map.
{\bf k)}~Position angle and flattening as a function of radius. {\bf l)}~$k_{3}/k_{1}$ and $k_{5}/k_{1}$ (from the analysis where position angle and ellipticity are fixed to their global values) as a function of radius.}
         \label{gal4F8}
   \end{figure*}

   \begin{figure*}
   \centering
   \includegraphics[angle=0,width=13.2cm,clip]{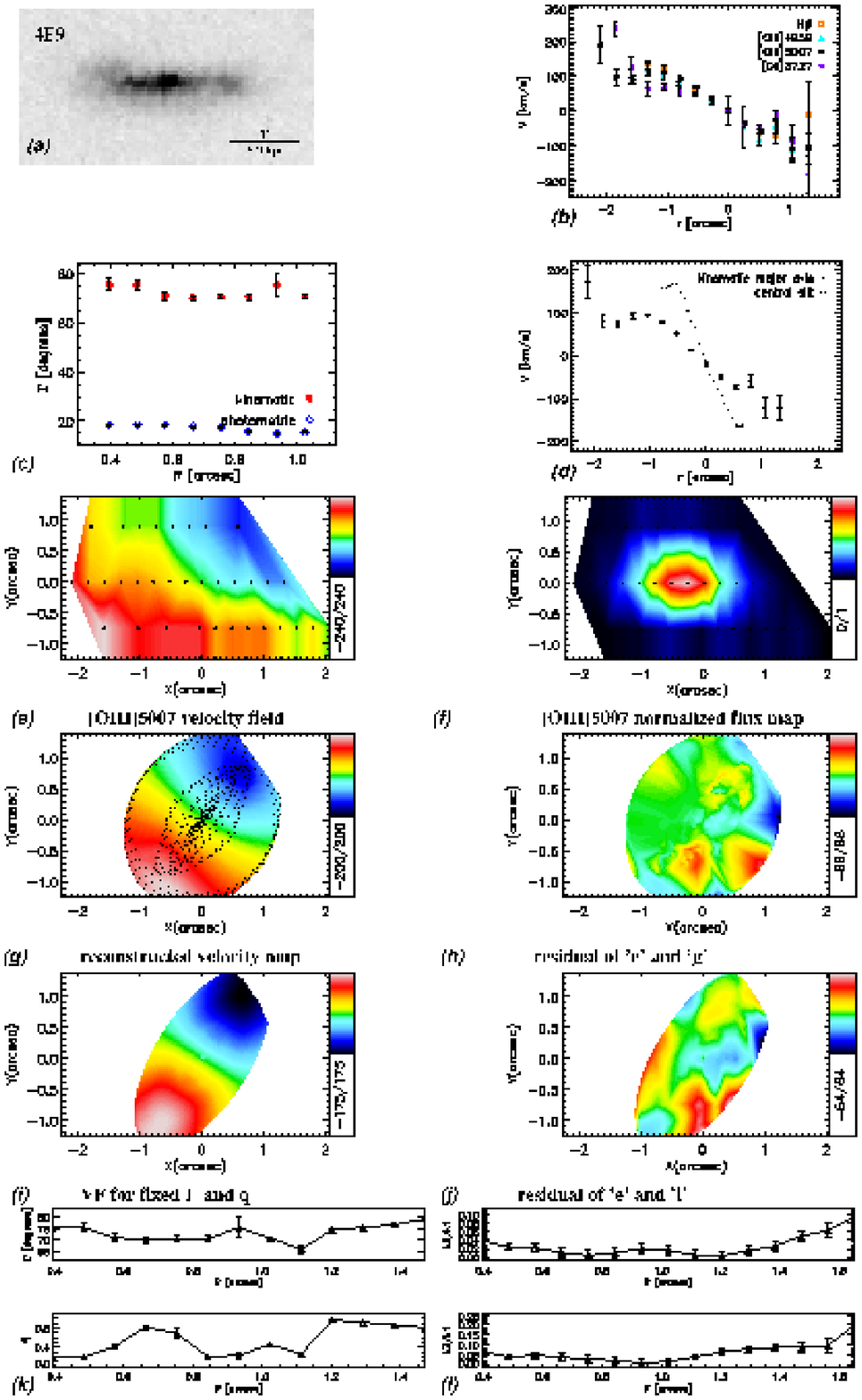}
 \caption{{\bf a)} HST-ACS image of the galaxy in the $V$~band. 
{\bf b)}~Rotation curves of different emission lines extracted along the central slit.
{\bf c)}~Position angles of kinematic and photometric axes as a function of radius.
{\bf d)}~Rotation curves extracted along the central slit and the kinematic major axis.
{\bf e)}~[OIII]5007~velocity field. 
{\bf f)}~Normalized [OIII]5007~flux map. 
{\bf g)}~Velocity map reconstructed using 6~harmonic terms.
{\bf h)}~Residual of the velocity map and the reconstructed map. 
{\bf i)}~Simple rotation map constructed for position angle and ellipticity fixed to their global values.
{\bf j)}~Residual of the velocity map and the simple rotation map.
{\bf k)}~Position angle and flattening as a function of radius. {\bf l)}~$k_{3}/k_{1}$ and $k_{5}/k_{1}$ (from the analysis where position angle and ellipticity are fixed to their global values) as a function of radius.}
         \label{gal4F9}
   \end{figure*}

   \begin{figure*}
   \centering
   \includegraphics[angle=0,width=16cm,clip]{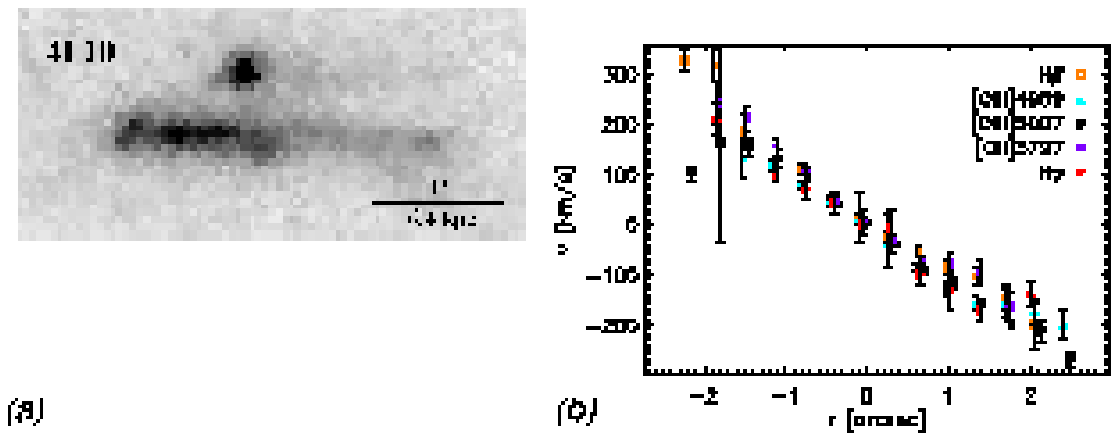}
 \caption{{\bf a)} HST-ACS image of the galaxy in the $V$~band. 
{\bf b)}~Rotation curves of different emission lines extracted along the central slit.}
         \label{gal4F10}
   \end{figure*}

   \begin{figure*}
   \centering
   \includegraphics[angle=0,width=16cm,clip]{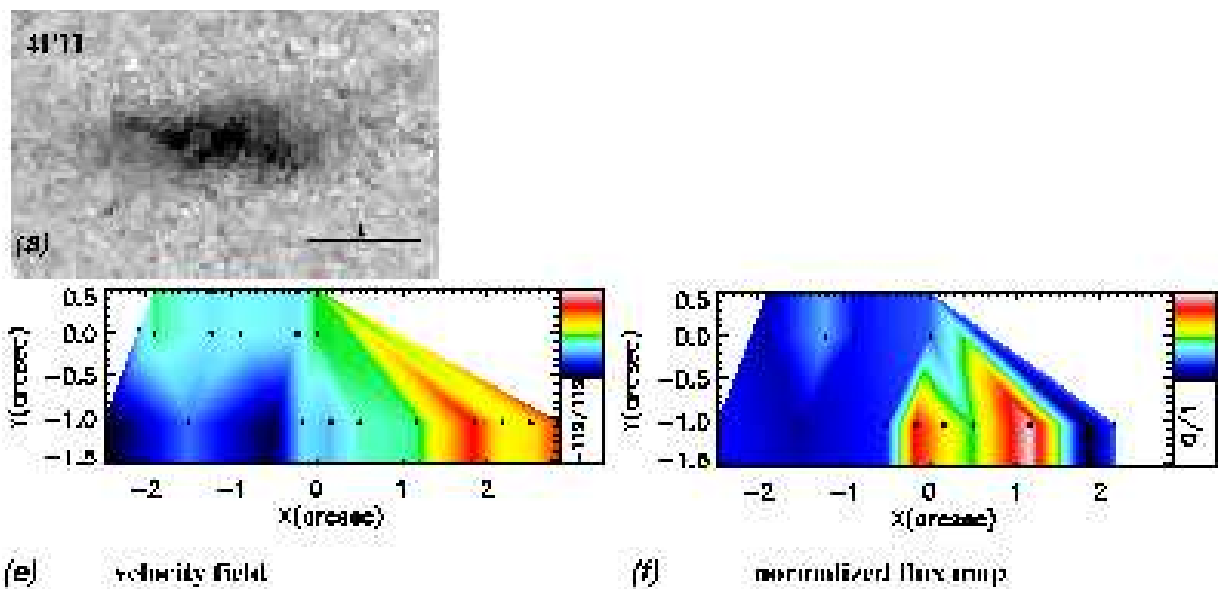}
 \caption{{\bf a)} HST-ACS image of the galaxy in the $V$~band. 
{\bf e)}~Velocity field constructed using the emission line which could not be identified.
{\bf f)}~Normalized flux map of the emission line which could not be identified.}
         \label{gal4F14}
   \end{figure*}

   \begin{figure*}
   \centering
   \includegraphics[angle=0,width=16.9cm,clip]{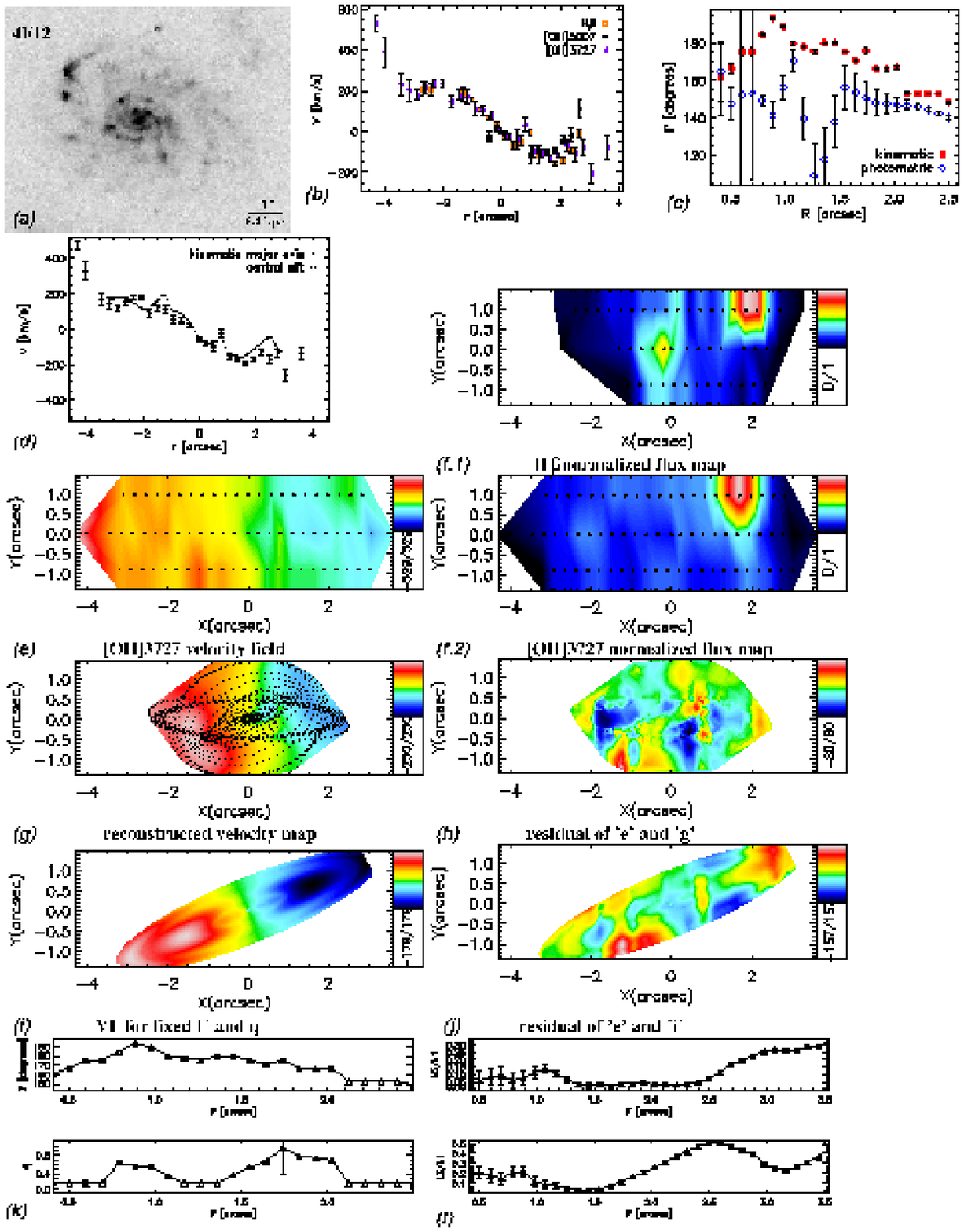}
 \caption{{\bf a)} HST-ACS image of the galaxy in the $V$~band. 
{\bf b)}~Rotation curves of different emission lines extracted along the central slit.
{\bf c)}~Position angles of kinematic and photometric axes as a function of radius.
{\bf d)}~Rotation curves extracted along the central slit and the kinematic major axis.
{\bf e)}~[OII]3727~velocity field. 
{\bf f.1)}~Normalized H$\beta$~flux map. 
{\bf f.2)}~Normalized [OII]3727~flux map. 
{\bf g)}~Velocity map reconstructed using 6~harmonic terms.
{\bf h)}~Residual of the velocity map and the reconstructed map. 
{\bf i)}~Simple rotation map constructed for position angle and ellipticity fixed to their global values.
{\bf j)}~Residual of the velocity map and the simple rotation map.
{\bf k)}~Position angle and flattening as a function of radius. {\bf l)}~$k_{3}/k_{1}$ and $k_{5}/k_{1}$ (from the analysis where position angle and ellipticity are fixed to their global values) as a function of radius.}
         \label{gal4F11}
   \end{figure*}

   \begin{figure*}
   \centering
   \includegraphics[angle=0,width=13.5cm,clip]{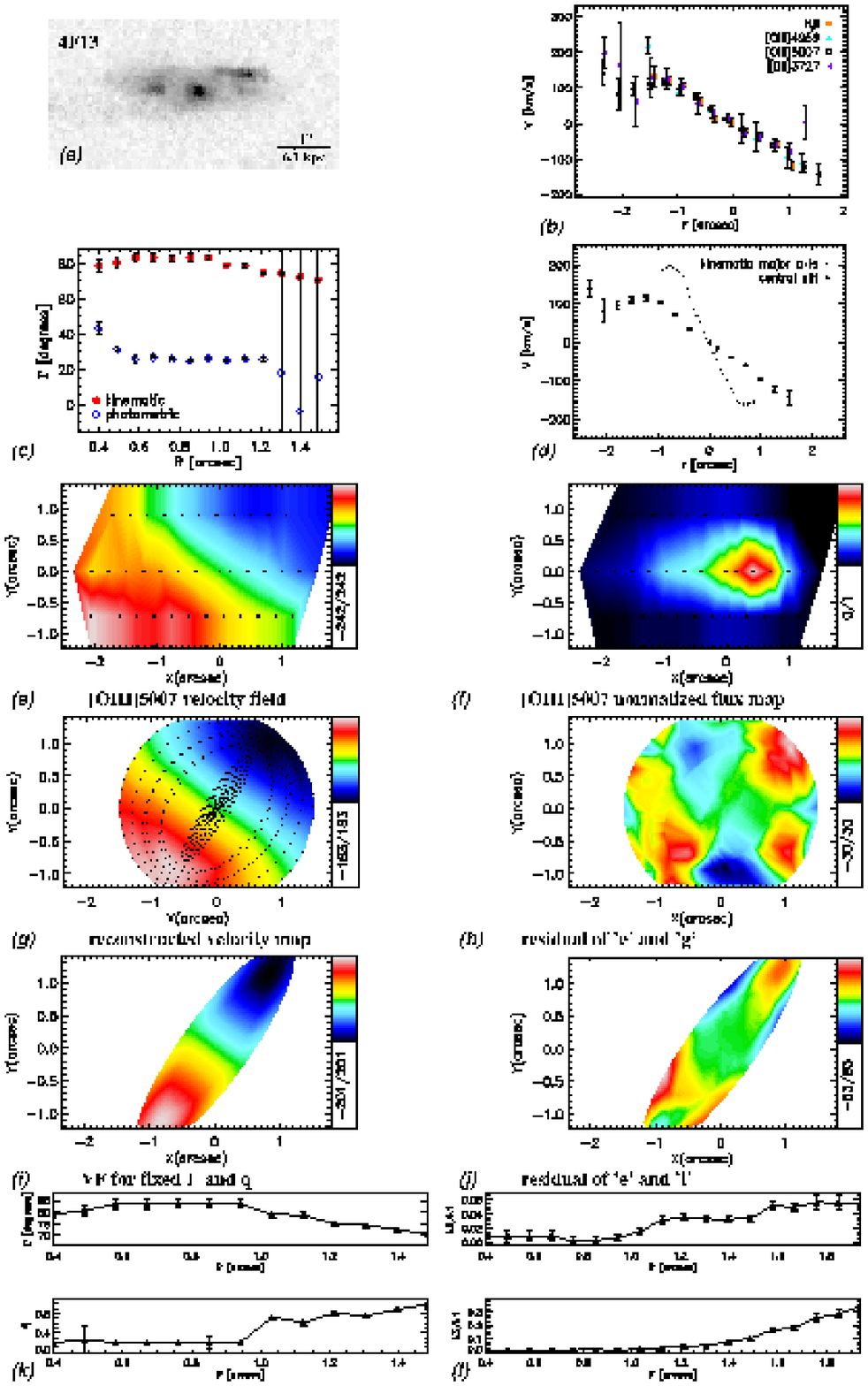}
 \caption{{\bf a)} HST-ACS image of the galaxy in the $V$~band. 
{\bf b)}~Rotation curves of different emission lines extracted along the central slit.
{\bf c)}~Position angles of kinematic and photometric axes as a function of radius.
{\bf d)}~Rotation curves extracted along the central slit and the kinematic major axis.
{\bf e)}~[OIII]5007~velocity field. 
{\bf f)}~Normalized [OIII]5007~flux map. 
{\bf g)}~Velocity map reconstructed using 6~harmonic terms.
{\bf h)}~Residual of the velocity map and the reconstructed map. 
{\bf i)}~Simple rotation map constructed for position angle and ellipticity fixed to their global values.
{\bf j)}~Residual of the velocity map and the simple rotation map.
{\bf k)}~Position angle and flattening as a function of radius. {\bf l)}~$k_{3}/k_{1}$ and $k_{5}/k_{1}$ (from the analysis where position angle and ellipticity are fixed to their global values) as a function of radius.}
         \label{gal4F13}
   \end{figure*}

     \begin{figure*}
   \centering
   \includegraphics[angle=0,width=16.7cm,clip]{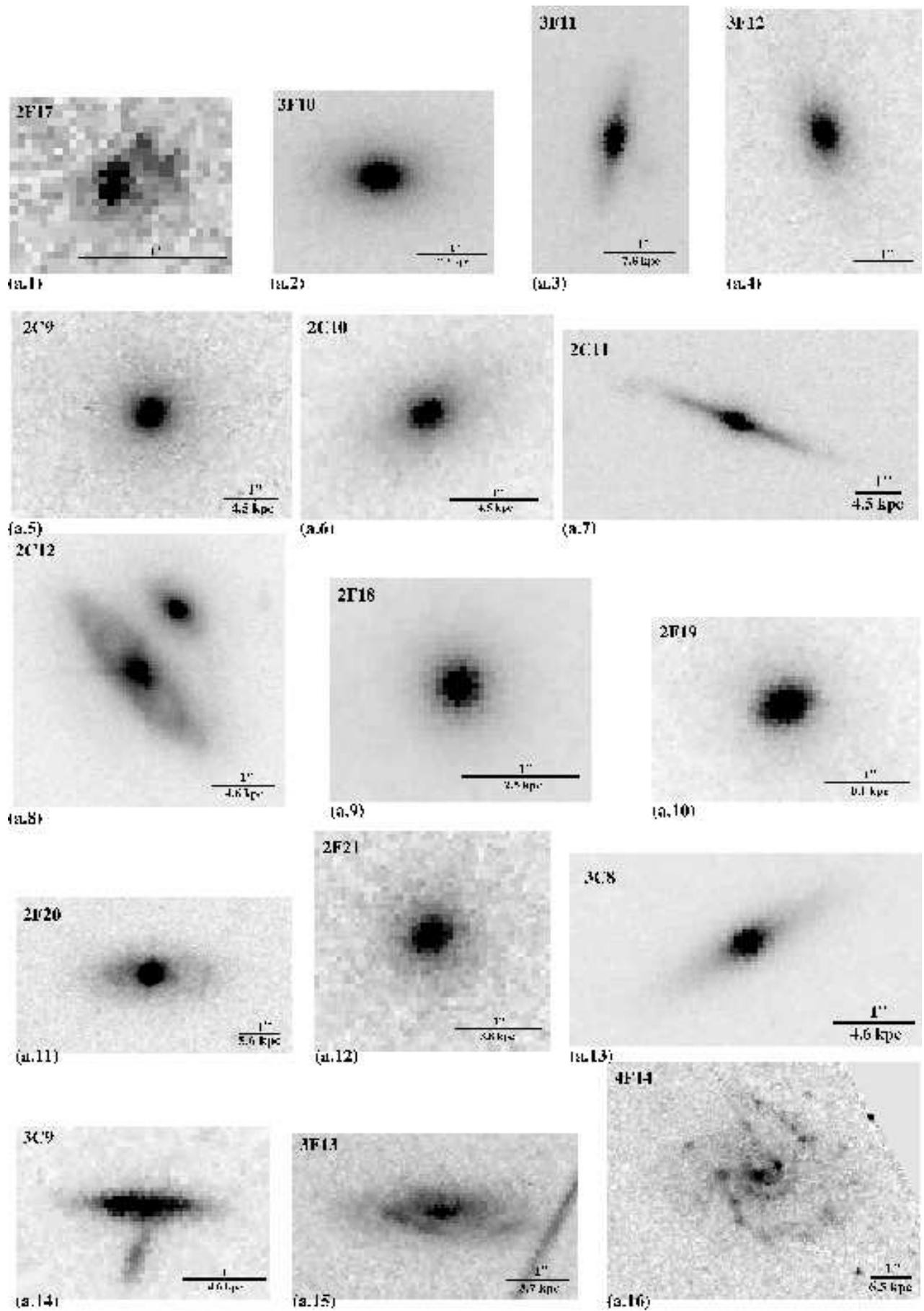}
 \caption{$V$~band HST-ACS image of galaxies that have very weak or no emission.}
         \label{no_em}
   \end{figure*} 

\clearpage

\section{Photometric tables \label{app_phot}}
\begin{table*}
\caption{Photometric parameters for the whole sample.}
\begin{tabular}{cccccccccccc}
\noalign{\medskip} \hline 
ID  &  $\alpha$ & $\delta$ & $M_B$ 	   &	   $B-V$ &     $V-R$    &	 $R-I$  &  $M_{stellar}[M_{\sun}]$  &	$A$  &    $C$  &     $G$  & $M_{20}$ \\
$(1)$ & $(2)$ &$(3)$ & $(4)$ &$(5)$ & $(6)$ &$(7)$ & $(8)$ &$(9)$ & $(10)$ & $(11)$ & $(12)$ \\
\hline
   1C1   &	 04:54:02.2	   &	    -02:57:10	   &	  	 -21.06 	 &	0.76	 &	     0.61	 &	  0.61     &	2.7e+10 &     0.21    &    0.41       & 	0.59  &     -2.06  \\	
   1C2   &	 04:54:17.2	   &	    -03:01:56	   &	  	 -20.91 	 &	0.68	 &	     0.59	 &	  0.55     &	1.7e+10 &     0.09    &    0.38       & 	0.51  &     -2.19  \\  
   1C3   &	 04:54:18.6	   &	    -03:01:03	   &	  	 -20.68 	 &	0.52	 &	     0.50	 &	  0.37     &	7.7e+09 &     0.13    &    0.41       & 	0.59  &     -2.04  \\  
   1C4   &	 04:54:17.6	   &	    -02:59:23	   &	  	 -20.37 	 &	0.42	 &	     0.44	 &	  0.21     &	3.2e+09 &     0.18    &    0.43       & 	0.64  &     -1.37  \\  
   1C5   &	 04:54:01.5	   &	    -02:59:24	   &	  	 -20.82 	 &	0.79	 &	     0.62	 &	  0.64     &	3.4e+10 &     0.19    &    0.57       & 	0.69  &     -2.26  \\  
   1C6   &	 04:54:01.3	   &	    -02:59:22	   &	  	 -20.59 	 &	0.46	 &	     0.49	 &	  0.42     &	8.3e+09 &     0.26    &    0.48       & 	0.61  &     -2.24  \\  
   1C7   &	 04:54:05.0	   &	    -02:59:40	   &	  	 -22.45 	 &	0.23	 &	     0.36	 &	  0.13     &	5.0e+09 &     0.38    &    0.31       & 	0.53  &     -1.23  \\  
   1C8   &	 04:54:04.4	   &	    -03:00:14	   &	  	 -21.68 	 &	0.48	 &	     0.49	 &	  0.40     &	1.4e+10 &     0.29    &    0.32       & 	0.48  &     -1.60  \\  
   1C9   &	 04:54:15.0	   &	    -03:00:22	   &	  	 -21.76 	 &	0.26	 &	     0.36	 &	  0.17     &	4.3e+09 &     0.41    &    0.38       & 	0.58  &     -1.48  \\  
  1C10   &	 04:54:17.7	   &	    -03:02:29	   &	  	 -20.05 	 &	0.18	 &	     0.36	 &	  0.07     &	1.1e+09 &     0.21    &    0.40       & 	0.55  &     -1.74  \\  
  1C11   &	 04:54:10.3	   &	    -03:00:46	   &	  	   --		 &	--	 &	     -- 	 &	  --	   &	-       &     0.20    &    0.40       & 	0.65  &     -2.24  \\ 
   1F1   &	 04:54:02.7	   &	    -02:58:41	   &	  	 -20.97 	 &	0.47	 &	     0.46	 &	  0.35     &	1.4e+10 &     0.24    &    0.25       & 	0.43  &     -1.27  \\  
   1F2   &	 04:54:03.8	   &	    -02:59:19	   &	  	 -21.96 	 &	0.43	 &	     0.46	 &	  0.32     &	1.4e+10 &     0.28    &    0.32       & 	0.54  &     -1.43  \\  
   1F3   &	 04:54:14.9	   &	    -02:58:17	   &	  	 -20.66 	 &	0.26	 &	     0.37	 &	  0.15     &	2.0e+09 &     0.35    &    0.32       & 	0.53  &     -1.24  \\  
   1F4   &	 04:54:07.8	   &	    -03:00:33	   &	  	 -18.78 	 &	0.28	 &	     0.29	 &	  0.30     &	3.8e+08 &     0.25    &    0.45       & 	0.58  &     -1.88  \\  
   1F5   &	 04:54:10.0	   &	    -03:00:28	   &	  	 -19.95 	 &	0.35	 &	     0.37	 &	  0.36     &	1.7e+09 &     0.22    &    0.35       & 	0.58  &     -1.97  \\  
   1F6   &	 04:54:18.9	   &	    -03:00:05	   &	  	 -18.56 	 &	0.39	 &	     0.41	 &	  0.28     &	2.8e+08 &     0.17    &    0.32       & 	0.47  &     -1.74  \\  
   1F7   &	 04:54:02.3	   &	    -02:58:10	   &	  	 -21.64 	 &	0.40	 &	     0.43	 &	  0.31     &	2.1e+10 &     0.56    &    0.23       & 	0.47  &     -1.32  \\  
   1F8   &	 04:54:10.4	   &	    -03:00:47	   &	  	   --		 &	--	 &	     -- 	 &	  --	   &	2.9e+10 &     0.26    &    0.28       & 	0.51  &     -1.41  \\ 
   1F9   &	 04:54:10.2	   &	    -03:01:57	   &	  	 -20.15 	 &	0.74	 &	     0.49	 &	  0.51     &	1.4e+10 &     0.15    &    0.61       & 	0.71  &     -2.42  \\  
  1F10   &	 04:54:00.5	   &	    -02:58:15	   &	  	 -19.67 	 &	0.33	 &	     0.39	 &	  0.22     &	1.4e+09 &     0.23    &    0.26       & 	0.40  &     -1.42  \\  
  1F11   &	 04:54:20.6	   &	    -03:00:16	   &	  	   --		 &	--	 &	     -- 	 &	  --	   &	-       &     0.21    &    0.36       & 	0.54  &     -1.93  \\ 
  2C1	 &	 10:10:29.0	   &	    -12:37:34	   &	  	 -19.90 	 &	0.62	 &	     0.47	 &	  0.45     &	5.3e+09 &     0.24    &    0.37       & 	0.58  &     -1.20  \\  
  2C2	 &	 10:10:20.5	   &	    -12:39:03	   &	  	 -20.77 	 &	0.64	 &	     0.52	 &	  0.49     &	1.3e+10 &     0.43    &    0.41       & 	0.60  &     -1.38  \\  
  2C3	 &	 10:10:46.8	   &	    -12:40:32	   &	  	 -21.22 	 &	0.60	 &	     0.47	 &	  0.45     &	1.4e+10 &     0.25    &    0.46       & 	0.60  &     -2.05  \\  
  2C4	 &	 10:10:25.9	   &	    -12:38:29	   &	  	    --  	 &	 --	 &	      --	 &	   --	   &	--      &     0.22    &    0.38       & 	0.51  &     -1.78  \\ 
  2C5	 &	 10:10:36.7	   &	    -12:40:04	   &	  	 -20.27 	 &	0.56	 &	     0.51	 &	  0.49     &	7.6e+09 &     0.50    &    0.33       & 	0.54  &     -1.41  \\  
  2C6	 &	 10:10:35.4	   &	    -12:39:59	   &	  	 -19.44 	 &	0.96	 &	     0.66	 &	  0.73     &	7.1e+09 &     0.16    &    0.26       & 	0.45  &     -1.18  \\  
  2C7   &	 10:10:22.4	   &	    -12:40:53	   &	  	 -19.86 	 &	0.35	 &	     0.43	 &	  0.31     &	3.0e+09 &     0.22    &    0.48       & 	0.64  &     -2.02  \\  
  2C8   &	 10:10:24.3	   &	    -12:37:44	   &	  	    --  	 &	 --	 &	      --	 &	   --	   &	--      &     0.28    &    0.49       & 	0.64  &     -1.81  \\ 
  2C9	 &	 10:10:28.6	   &	    -12:37:37	   &	  	 -19.99 	 &	0.79	 &	     0.49	 &	  0.53     &	8.3e+09 &     0.12    &    0.42       & 	0.51  &     -2.24  \\  
  2C10	 &	 10:10:45.9	   &	    -12:38:31	   &	  	 -19.28 	 &	0.75	 &	     0.49	 &	  0.51     &	5.9e+09 &     0.10    &    0.43       & 	0.54  &     -2.15  \\  
  2C11	 &	 10:10:27.8	   &	    -12:37:07	   &	  	 -19.93 	 &	0.88	 &	     0.51	 &	  0.60     &	1.5e+10 &     0.13    &    0.54       & 	0.66  &     -2.58  \\  
  2C12   &	 10:10:30.6	   &	    -12:37:23	   &	  	 -21.31 	 &	0.79	 &	     0.50	 &	  0.54     &	3.0e+10 &     0.15    &    0.41       & 	0.59  &     -1.96  \\  
  2F1	 &	 10:10:44.8	   &	    -12:38:31	   &	  	 -20.85 	 &	0.27	 &	     0.38	 &	  0.15     &	3.9e+09 &     0.60    &    0.32       & 	0.56  &     -0.97  \\  
  2F2	 &	 10:10:38.2	   &	    -12:38:14	   &	  	 -18.78 	 &	0.93	 &	     0.54	 &	  0.64     &	3.4e+09 &     0.09    &    0.41       & 	0.53  &     -2.04  \\  
  2F3	 &	 10:10:41.3	   &	    -12:39:21	   &	  	 -21.14 	 &	0.39	 &	     0.41	 &	  0.22     &	6.9e+09 &     0.33    &    0.30       & 	0.51  &     -1.38  \\  
  2F4	 &	 10:10:41.5	   &	    -12:39:19	   &	  	 -21.14 	 &	0.65	 &	     0.56	 &	  0.53     &	4.0e+10 &     0.18    &    0.41       & 	0.53  &     -2.14  \\  
  2F5	 &	 10:10:31.8	   &	    -12:41:03	   &	  	 -21.36 	 &	0.90	 &	     0.65	 &	  0.70     &	3.6e+10 &     0.10    &    0.41       & 	0.53  &     -2.26  \\  
  2F6	 &	 10:10:44.0	   &	    -12:36:35	   &	  	 -18.68 	 &	0.50	 &	     0.40	 &	  0.45     &	1.6e+09 &     0.15    &    0.34       & 	0.53  &     -1.71  \\  
  2F7   &	 10:10:34.2	   &	    -12:37:27	   &	  	 -16.55 	 &	0.53	 &	     0.55	 &	  0.54     &	5.1e+08 &     0.29    &    0.18       & 	0.39  &     -0.75  \\  
  2F8   &	 10:10:46.4	   &	    -12:40:36	   &	  	 -19.93 	 &	0.62	 &	     0.47	 &	  0.43     &	1.2e+10 &     0.17    &    0.13       & 	0.32  &     -0.77  \\  
  2F9   &	 10:10:33.4	   &	    -12:37:13	   &	  	 -20.40 	 &	0.72	 &	     0.55	 &	  0.56     &	1.0e+10 &     0.24    &    0.32       & 	0.47  &     -1.64  \\  
  2F10   &	 10:10:30.2	   &	    -12:37:26	   &	  	 -20.48 	 &	0.53	 &	     0.49	 &	  0.41     &	4.1e+09 &     0.26    &    0.23       & 	0.43  &     -1.18  \\  
  2F11   &	 10:10:33.8	   &	    -12:37:19	   &	  	 -21.60 	 &	0.65	 &	     0.50	 &	  0.49     &	6.9e+09 &     0.25    &    0.19       & 	0.35  &     -1.19  \\  
  2F12   &	 10:10:42.7	   &	    -12:40:21	   &	  	 -21.02 	 &	0.48	 &	     0.44	 &	  0.36     &	8.0e+09 &     0.28    &    0.39       & 	0.54  &     -1.92  \\  
  2F13   &	 10:10:36.4	   &	    -12:40:09	   &	  	    --  	 &	 --	 &	      --	 &	   --	   &	--      &     0.16    &    0.12       & 	0.35  &     -0.77  \\ 
  2F14   &	 10:10:36.9	   &	    -12:39:58	   &	  	    --  	 &	 --	 &	      --	 &	   --	   &	--      &     0.28    &    0.16       & 	0.34  &     -1.03  \\ 
  2F15   &	 10:10:20.5	   &	    -12:39:03	   &	  	    --  	 &	 --	 &	      --	 &	   --	   &	-- 	&         --	       &	--     &	     --        &	  --  \\ 
  2F16   &	 10:10:20.2	   &	    -12:39:01	   &	  	    --  	 &	 --	 &	      --	 &	   --	   &	-- 	&         --	       &	--     &	     --        &	  --  \\ 
  2F17   &	 10:10:32.2	   &	    -12:40:58	   &	  	    --  	 &	 --	 &	      --	 &	   --	   &	--      &     0.39    &    0.14       & 	0.42  &     -1.10  \\ 
  2F18	 &	 10:10:25.9	   &	    -12:38:28	   &	  	    --  	 &	 --	 &	      --	 &	   --	   &	2.7e+09 &     0.20    &    0.57       & 	0.67  &     -2.42  \\ 
  2F19	 &	 10:10:38.9	   &	    -12:38:22	   &	  	 -10.28 	 &	1.23	 &	     0.77	 &	  0.85     &	--      &     0.17    &    0.55       & 	0.62  &     -2.55  \\  
  2F20	 &	 10:10:40.4	   &	    -12:36:31	   &	  	 -20.89 	 &	0.84	 &	     0.62	 &	  0.65     &	3.2e+10 &     0.14    &    0.45       & 	0.57  &     -2.40  \\  
  2F21   &	 10:10:37.2	   &	    -12:39:57	   &	  	 -18.15 	 &	0.91	 &	     0.53	 &	  0.63     &	2.4e+09 &     0.13    &    0.38       & 	0.51  &     -2.05  \\  
  3C1	 &	 21:40:16.1	   &	    -23:37:10	   &	  	 -17.46 	 &	0.69	 &	     0.69	 &	  0.73     &	1.7e+09 &     0.08    &    0.33       & 	0.48  &     -1.64  \\  
  3C2	 &	 21:40:05.2	   &	    -23:40:44	   &	  	 -19.24 	 &	0.69	 &	     0.68	 &	  0.73     &	1.4e+10 &     0.33    &    0.33       & 	0.51  &     -1.45  \\  
  3C3	 &	 21:40:16.2	   &	    -23:37:13	   &	  	 -19.07 	 &	0.84	 &	     0.72	 &	  0.78     &	9.3e+09 &     0.12    &    0.36       & 	0.54  &     -1.94  \\  
  3C4	 &	 21:40:15.7	   &	    -23:38:53	   &	  	 -20.66 	 &	0.79	 &	     0.71	 &	  0.76     &	5.0e+10 &     0.39    &    0.46       & 	0.55  &     -2.29  \\  
  3C5	 &	 21:40:24.2	   &	    -23:36:51	   &	  	 -18.93 	 &	0.30	 &	     0.55	 &	  0.51     &	6.6e+09 &     0.24    &    0.26       & 	0.43  &     -1.23  \\  
\hline
\label{tabrun2}
\end{tabular}
\end{table*}
\begin{table*}
\begin{tabular}{cccccccccccc}
\noalign{\medskip} \hline 
ID  &  $\alpha$ & $\delta$ & $M_B$ 	   &	   $B-V$ &     $V-R$    &	 $R-I$  &  $M_{stellar}[M_{\sun}]$  &	$A$  &    $C$  &     $G$  & $M_{20}$ \\
$(1)$ & $(2)$ &$(3)$ & $(4)$ &$(5)$ & $(6)$ &$(7)$ & $(8)$ &$(9)$ & $(10)$ & $(11)$ & $(12)$ \\
\hline
  3C6	 &	 21:40:21.9	   &	    -23:41:46	   &	  	 -19.54 	 &	0.50	 &	     0.63	 &	  0.65     &	9.2e+09 &     0.41    &    0.45       & 	0.60  &     -1.99  \\  
  3C7	 &	 21:40:05.9	   &	    -23:36:45	   &	  	    --  	 &	 --	 &	      --	 &	   --	   &	--      &     0.18    &    0.31       & 	0.44  &     -1.88  \\ 
  3C8	 &	 21:40:28.1	   &	    -23:40:27	   &	  	    --  	 &	 --	 &	      --	 &	   --	   &	--      &     0.20    &    0.54       & 	0.65  &     -2.37  \\ 
  3C9	 &	 21:40:06.9	   &	    -23:37:01	   &	  	 -18.48 	 &	1.08	 &	     0.77	 &	  0.84     &	4.9e+09 &     0.33    &    0.36       & 	0.60  &     -1.44  \\  
  3F1	 &	 21:40:02.9	   &	    -23:37:49	   &	  	 -20.27 	 &	1.13	 &	     0.78	 &	  0.85     &	7.2e+10 &     0.22    &    0.40       & 	0.56  &     -2.16  \\  
  3F2	 &	 21:40:27.1	   &	    -23:38:07	   &	  	    --  	 &	 --	 &	      --	 &	   --	   &	--      &     0.26    &    0.28       & 	0.54  &     -1.51  \\ 
  3F3	 &	 21:40:24.9	   &	    -23:41:45	   &	  	 -17.39 	 &	0.54	 &	     0.64	 &	  0.67     &	3.0e+09 &     0.24    &    0.49       & 	0.67  &     -1.35  \\  
  3F4	 &	 21:40:17.9	   &	    -23:38:15	   &	  	 -20.20 	 &	0.75	 &	     0.70	 &	  0.75     &	3.3e+10 &     0.40    &    0.21       & 	0.46  &     -0.97  \\  
  3F5   &	 21:40:05.6	   &	    -23:36:40	   &	  	 -17.90 	 &	0.12	 &	     0.47	 &	  0.32     &	7.6e+08 &     0.26    &    0.44       & 	0.64  &     -1.98  \\  
  3F6	 &	 21:40:18.8	   &	    -23:38:21	   &	  	 -17.89 	 &	0.43	 &	     0.60	 &	  0.60     &	1.6e+09 &     0.13    &    0.31       & 	0.49  &     -1.76  \\  
  3F7	 &	 21:40:15.8	   &	    -23:36:30	   &	  	 -17.22 	 &	0.19	 &	     0.50	 &	  0.40     &	1.7e+09 &     0.29    &    0.38       & 	0.58  &     -1.92  \\  
  3F8	 &	 21:40:07.1	   &	    -23:40:25	   &	  	 -20.28 	 &	0.83	 &	     0.72	 &	  0.78     &	6.0e+10 &     0.32    &    0.24       & 	0.44  &     -1.16  \\  
  3F9   &	 21:40:22.8	   &	    -23:39:24	   &	  	 -16.79 	 &	0.28	 &	     0.55	 &	  0.50     &	1.7e+09 &     0.16    &    0.32       & 	0.47  &     -1.74  \\  
  3F10   &	 21:40:06.4	   &	    -23:36:54	   &	  	 -22.93 	 &	0.69	 &	     0.60	 &	  0.57     &	3.4e+10 &     0.11    &    0.54       & 	0.64  &     -2.66  \\  
  3F11	 &	 21:40:05.9	   &	    -23:36:45	   &	  	 -22.30 	 &	0.81	 &	     0.49	 &	  0.56     &	1.8e+11 &     0.22    &    0.62       & 	0.72  &     -2.49  \\  
  3F12   &	 21:40:19.2	   &	    -23:38:26	   &	  	    --  	 &	 --	 &	      --	 &	   --	   &	--      &     0.12    &    0.50       & 	0.60  &     -2.19  \\ 
  3F13	 &	 21:40:11.7	   &	    -23:38:10	   &	  	    --  	 &	 --	 &	      --	 &	   --	   &	--      &	  --	      &        --     & 	    --        & 	 --  \\ 
  4C1	 &	 04:13:13.8	   &	    -65:48:56	   &	  	 -20.04  	 &	0.46	 &	     0.46	 &	  0.29	   &	4.4e+09 &     0.20    &    0.26       & 	0.50  &     -1.53  \\  
  4C2	 &	 04:12:36.6	   &	    -65:49:52	   &	  	 -20.57 	 &	0.45	 &	     0.45	 &	  0.28     &	5.3e+09 &     0.29    &    0.41       & 	0.60  &     -1.68  \\  
  4C3	 &	 04:12:56.4	   &	    -65:51:13	   &	  	 -20.38 	 &	0.56	 &	     0.52	 &	  0.46     &	1.2e+10 &     0.21    &    0.41       & 	0.62  &     -1.39  \\  
  4F1	 &	 04:12:51.7	   &	    -65:50:43	   &	  	 -18.86 	 &	0.35	 &	     0.41	 &	  0.23     &	1.0e+09 &     0.29    &    0.48       & 	0.63  &     -1.54  \\  
  4F2	 &	 04:13:13.2	   &	    -65:49:24	   &	  	 -20.79 	 &	0.35	 &	     0.36	 &	  0.26     &	4.8e+09 &     0.28    &    0.61       & 	0.72  &     -1.84  \\  
  4F3	 &	 04:12:43.1	   &	    -65:52:02	   &	  	 -21.72 	 &	0.45	 &	     0.40	 &	  0.34     &	2.6e+10 &     0.48    &    0.33       & 	0.53  &     -1.15  \\  
  4F4	 &	 04:13:10.4	   &	    -65:52:20	   &	  	 -19.74 	 &	0.33	 &	     0.41	 &	  0.21     &	2.1e+09 &     0.38    &    0.33       & 	0.57  &     -1.13  \\  
  4F5	 &	 04:12:57.1	   &	    -65:52:49	   &	  	 -21.70 	 &	0.47	 &	     0.47	 &	  0.37     &	1.9e+10 &     0.25    &    0.19       & 	0.40  &     -0.95  \\  
  4F6	 &	 04:12:28.3	   &	    -65:50:30	   &	  	 -21.00 	 &	0.49	 &	     0.48	 &	  0.37     &	1.2e+10 &     0.23    &    0.30       & 	0.43  &     -1.88  \\  
  4F7	 &	 04:12:46.4	   &	    -65:51:28	   &	  	 -21.39 	 &	0.34	 &	     0.41	 &	  0.18     &	4.3e+09 &     0.44    &    0.19       & 	0.44  &     -1.20  \\  
  4F8	 &	 04:12:36.8	   &	    -65:53:07	   &	  	 -20.84 	 &	0.42	 &	     0.45	 &	  0.31     &	6.0e+09 &     0.33    &    0.24       & 	0.41  &     -1.08  \\  
  4F9	 &	 04:12:50.6	   &	    -65:47:45	   &	  	 -20.17 	 &	0.33	 &	     0.40	 &	  0.18     &	2.0e+09 &     0.32    &    0.32       & 	0.47  &     -1.59  \\  
  4F10   &	 04:12:36.6	   &	    -65:47:60	   &	  	 -20.03 	 &	0.28	 &	     0.38	 &	  0.16     &	1.4e+09 &     0.48    &    0.30       & 	0.48  &     -1.01  \\  
  4F11   &	 04:12:55.8	   &	    -65:49:38	   &	  	    --  	 &	 --	 &	      --	 &	   --	   &	--      &     0.12    &    0.23       & 	0.38  &     -1.49  \\ 
  4F12   &	 04:12:47.6	   &	    -65:53:35	   &	  	 -22.05 	 &	0.46	 &	     0.47	 &	  0.37     &	1.7e+10 &     0.35    &    0.26       & 	0.43  &     -0.99  \\  
  4F13   &	 04:13:07.1	   &	    -65:47:17	   &	  	 -20.00 	 &	0.31	 &	     0.40	 &	  0.16     &	1.5e+09 &     0.37    &    0.24       & 	0.46  &     -1.04  \\  
  4F14   &	 04:12:43.5	   &	    -65:51:42	   &	  	 -21.07 	 &	0.44	 &	     0.46	 &	  0.34     &	9.2e+09 &     0.38    &    0.22       & 	0.36  &     -1.27  \\

\hline
\end{tabular} \\
Column (1): object ID; Cols.~(2,~3): RA and Dec (J2000); Col.~(4): rest frame, galactic-extinction corrected Johnson-$B$ magnitudes ($k$-correction applied using the 
{\tt kcorrect} algorithm by \citet{BR07}); Cols.~(5$-$7): galactic extinction and k-corrected $B-V$, $V-R$ and $R-I$ colors measured from the FORS2/VLT images (The FORS2 filters B,
V and I are close approximations to the Bessel \citep{bes90} photometric system while R filter is a special filter for FORS2);
Col.~(8): Stellar mass (measured using the {\tt kcorrect} algorithm by \citet{BR07});
Col.~(9): asymmetry index;
Col.~(10): concentration index;
Col.~(11): Gini coefficient;
Col.~(12): $M_{20}$ index. \\
\end{table*}

\begin{table*}
\caption{Morphological parameters for the MS~1008, MS~2137 and Cl~0413 samples.}
\begin{tabular}{ccccccc}
\noalign{\medskip} \hline
ID & comp & $F606W$ & $R_{\rm e}/R_{\rm d} ({\rm kpc})$ & $n$ & $q$ & PA \\
$(1)$ & $(2)$ &$(3)$ & $(4)$ &$(5)$ & $(6)$ &$(7)$  \\
\hline 	   
	2C1	   &	 exp disk   &		 20.57    $\pm$    0.00 	  &	     4.14           $\pm$	   0.01  &				       --	&	    0.15	$\pm$	     0.00	&	61.22   $\pm$  	   0.05   	\\
	2C2	   &	 exp disk   &		 20.62    $\pm$    0.00 	  &	     3.17           $\pm$	   0.03  &				       --	&	    0.68	$\pm$	     0.00	&	61.64   $\pm$  	   0.67   	\\
	2C3	   &	 exp disk   &		 19.86    $\pm$    0.00 	  &	     5.15           $\pm$	   0.03  &				       --	&	    0.87	$\pm$	     0.00	&	-11.41   $\pm$     1.36   	\\
	2C4	   &	 exp disk   &		 22.66    $\pm$    0.01 	  &	     2.37           $\pm$	   0.08  &				       --	&	    0.46	$\pm$	     0.01	&	 -84.68   $\pm$  	   0.99   	\\
		   &	 S\'{e}rsic bulge   &	 23.15   $\pm$     0.02  &		     1.07         $\pm$ 	   0.02  &	    0.49	$\pm$	     0.01	&	    0.40       $\pm$	   0.00 	&      -79.40	     $\pm$	 0.44		\\
	2C5	   &	 exp disk   &		 20.55    $\pm$    0.01 	  &	     2.39           $\pm$	   0.02  &				       --	&	    0.83	$\pm$	     0.00	&	71.68   $\pm$  	   1.82   	\\
	2C6	   &	 exp disk   &		 21.13    $\pm$    0.00 	  &	     5.36           $\pm$	   0.03  &				       --	&	    0.16	$\pm$	     0.00	&	 -4.39   $\pm$  	   0.08   	\\
       2C7	   &	 exp disk   &		 21.29    $\pm$    0.02 	  &	     1.54           $\pm$	   0.01  &				       --	&	    0.73	$\pm$	     0.01	&	 -30.79   $\pm$  	   1.14   	\\
       	   	   &	 S\'{e}rsic bulge   &	 21.72   $\pm$     0.03  &		     2.13         $\pm$ 	   0.03  &	    1.89	$\pm$	     0.03	&	    0.23       $\pm$	   0.00 	&      -20.02	     $\pm$	 0.13		\\
       2C8	   &	 exp disk   &		 21.46    $\pm$    0.01 	  &	     1.42           $\pm$	   0.01  &				       --	&	    0.64	$\pm$	     0.00	&	 -48.14   $\pm$  	   1.20   	\\
       	   	   &	 S\'{e}rsic bulge   &	 22.34   $\pm$     0.03  &		     1.05         $\pm$ 	   0.02  &	    1.55	$\pm$	     0.05	&	    0.28       $\pm$	   0.01 	&	-75.94	     $\pm$	 0.46		\\
	2C9	   &	 exp disk   &		 21.43    $\pm$    0.06 	  &	     5.21           $\pm$	   0.12  &				       --	&	    0.73	$\pm$	     0.02	&	43.26   $\pm$  	   2.12   	\\
	2C10	   &	 exp disk   &		 21.50    $\pm$    0.04 	  &	     2.90           $\pm$	   0.03  &				       --	&	    0.55	$\pm$	     0.00	&	-13.84   $\pm$  	   0.41   	\\
		   &	 S\'{e}rsic bulge   &	 22.64   $\pm$     0.13  &		     0.94         $\pm$ 	   0.17  &	    3.44	$\pm$	     0.39	&	    0.69       $\pm$	   0.01 	&      -20.33	     $\pm$	 1.58		\\
	2C11	   &	 exp disk   &		 21.47    $\pm$    0.01 	  &	     3.92           $\pm$	   0.05  &				       --	&	    0.07	$\pm$	     0.00	&	23.44   $\pm$  	   0.05   	\\
		   &	 S\'{e}rsic bulge   &	 20.98   $\pm$     0.01  &		     2.10         $\pm$ 	   0.06  &	    4.63	$\pm$	     0.07	&	    0.49       $\pm$	   0.00 	&     21.09	     $\pm$	 0.38		\\
       2C12	   &	 exp disk   &		 19.92    $\pm$    0.00 	  &	     3.30           $\pm$	   0.01  &				       --	&	    0.38	$\pm$	     0.00	&	54.83   $\pm$  	   0.10   	\\
	2F1	   &	 exp disk   &		 22.06    $\pm$    0.01 	  &	     2.08           $\pm$	   0.01  &				       --	&	    0.34	$\pm$	     0.00	&	 -59.02   $\pm$  	   0.24   	\\
	2F2	   &	 exp disk   &		 21.26    $\pm$    0.01 	  &	     2.25           $\pm$	   0.02  &				       --	&	    0.55	$\pm$	     0.00	&	 -37.44   $\pm$  	   0.41   	\\
		   &	 S\'{e}rsic bulge   &	 22.15   $\pm$     0.03  &	             3.62         $\pm$ 	   0.00  &	    6.64	$\pm$	     0.23	&	    0.59       $\pm$	   0.01 	&      -37.83	     $\pm$	 1.75		\\
	2F3	   &	 exp disk   &		 21.80    $\pm$    0.01 	  &	     3.50           $\pm$	   0.06  &				       --	&	    0.66	$\pm$	     0.01	&	53.34   $\pm$  	   0.98   	\\
	2F4	   &	 exp disk   &		 22.31    $\pm$    0.04 	  &	     1.99           $\pm$	   0.03  &				       --	&	    0.92	$\pm$	     0.02	&	 7.72   $\pm$  	   8.31   	\\
		   &	 S\'{e}rsic bulge   &	 23.30   $\pm$     0.26  &		     1.80         $\pm$ 	   1.04  &	    8.00	$\pm$	     2.30	&	    0.55       $\pm$	   0.04 	&      -80.62	     $\pm$	 2.45		\\
	2F5	   &	 exp disk   &		 21.62    $\pm$    0.02 	  &	     5.85           $\pm$	   0.11  &				       --	&	    0.61	$\pm$	     0.01	&	55.29   $\pm$  	   1.34   	\\
		   &	 S\'{e}rsic bulge   &	 20.56   $\pm$     0.06  &	            21.20         $\pm$ 	   2.19  &	    8.00	$\pm$	     0.22	&	    0.65       $\pm$	   0.01 	&     49.96	     $\pm$	 0.74		\\
	2F6	   &	 exp disk   &		 20.18    $\pm$    0.00 	  &	     2.21           $\pm$	   0.01  &				       --	&	    0.40	$\pm$	     0.00	&	 -51.62   $\pm$  	   0.13   	\\
       2F7	   &	 exp disk   &		 24.39    $\pm$    0.03 	  &	     1.61           $\pm$	   0.06  &				       --	&	    0.21	$\pm$	     0.01	&	27.67   $\pm$  	   0.78   	\\
       2F8	   &	 exp disk   &		 22.72    $\pm$    0.03 	  &	     5.37           $\pm$	   0.23  &				       --	&	    0.41	$\pm$	     0.01	&	 -16.63   $\pm$  	   1.58   	\\
       2F9	   &	 exp disk   &		 21.62    $\pm$    0.01 	  &	     3.45           $\pm$	   0.05  &				       --	&	    0.33	$\pm$	     0.00	&	 -42.86   $\pm$  	   0.38   	\\
       2F10	   &	 exp disk   &		 20.87    $\pm$    0.00 	  &	     6.08           $\pm$	   0.04  &				       --	&	    0.19	$\pm$	     0.00	&	  -79.34   $\pm$  	   0.10   	\\
       2F11	   &	 exp disk   &		 20.13    $\pm$    0.01 	  &	     5.64           $\pm$	   0.05  &				       --	&	    0.94	$\pm$	     0.01	&	 -18.94   $\pm$  	   5.13   	\\
       2F12	   &	 exp disk   &		 19.84    $\pm$    0.00 	  &	     2.77           $\pm$	   0.01  &				       --	&	    0.94	$\pm$	     0.00	&	 -69.08   $\pm$  	   1.84   	\\
       2F13	   &	 exp disk   &		 23.79    $\pm$    0.05 	  &	     0\farcs76   $\pm$	    	0\farcs05  &				       --	&	    0.30	$\pm$	     0.02	&	63.86   $\pm$  	   1.62   	\\
       2F14	   &	 exp disk   &		 22.25    $\pm$    0.06 	  &	     0\farcs55   $\pm$ 		0\farcs03  &  				   --	    	&		0.74	    $\pm$	 0.00	    &	    85.43   $\pm$	       0.00   	\\
       2F17	   &	 exp disk   &		 24.33    $\pm$    0.02 	  &	     0\farcs16    $\pm$ 	0\farcs01  &  				   --	    	&		0.63	    $\pm$	 0.02	    &	    -88.64   $\pm$	       3.00   	\\
	2F18	   &	 exp disk   &		 22.45    $\pm$    0.04 	  &	     0.71           $\pm$	   0.02  &				       --	&	    0.85	$\pm$	     0.02	&	61.53   $\pm$  	   6.92   	\\
		   &	 S\'{e}rsic bulge   &	 20.62   $\pm$     0.01  &	             1.38         $\pm$ 	   0.05  &	    5.32	$\pm$	     0.12	&	    0.91       $\pm$	   0.01 	&     45.36	     $\pm$	 3.13		\\
	2F19	   &	 exp disk   &		 21.22    $\pm$    0.01 	  &	     0.07           $\pm$	   0.00  &				       --	&	    0.79	$\pm$	     0.01	&	76.85   $\pm$  	   1.27   	\\
		   &	 S\'{e}rsic bulge   &	 21.68   $\pm$     0.01  &		     0.02         $\pm$ 	   0.00  &	    1.77	$\pm$	     0.03	&	    0.74       $\pm$	   0.00 	&      -33.67	     $\pm$	 0.67		\\
	2F20	   &	 exp disk   &		 20.94    $\pm$    0.00 	  &	     3.80           $\pm$	   0.03  &				       --	&	    0.43	$\pm$	     0.00	&	 -84.05   $\pm$  	   0.22   	\\
		   &	 S\'{e}rsic bulge   &	 22.26   $\pm$     0.01  &		     0.67         $\pm$ 	   0.01  &	    1.23	$\pm$	     0.03	&	    0.85       $\pm$	   0.01 	&      -25.12	     $\pm$	 2.00		\\
       2F21	   &	 exp disk   &		 23.31    $\pm$    0.07 	  &	     1.63           $\pm$	   0.08  &				       --	&	    0.57	$\pm$	     0.03	&	 -9.64   $\pm$  	   2.74   	\\
       	   	   &	 S\'{e}rsic bulge   &	 21.78   $\pm$     0.03  &	             4.99         $\pm$ 	   0.61  &	    4.59	$\pm$	     0.32	&	    0.68       $\pm$	   0.01 	&     77.47	     $\pm$	 1.58		\\
	3C1	   &	 exp disk   &		 22.86    $\pm$    0.01 	  &	     1.19           $\pm$	   0.02  &				       --	&	    0.43	$\pm$	     0.00	&	  -85.27   $\pm$  	   0.56   	\\
	3C2	   &	 exp disk   &		 20.40    $\pm$    0.00 	  &	     3.81           $\pm$	   0.02  &				       --	&	    0.45	$\pm$	     0.00	&	63.61   $\pm$  	   0.16   	\\
	3C3	   &	 exp disk   &		 20.91    $\pm$    0.00 	  &	     2.74           $\pm$	   0.01  &				       --	&	    0.41	$\pm$	     0.00	&	 -30.27   $\pm$  	   0.13   	\\
		   &	 S\'{e}rsic bulge   &	 24.05   $\pm$     0.03  &		     0.65         $\pm$ 	   0.02  &	    0.99	$\pm$	     0.06	&	    0.66       $\pm$	   0.02 	&     19.67	     $\pm$	 3.16		\\
	3C4	   &	 exp disk   &		 19.76    $\pm$    0.02 	  &	     4.25           $\pm$	   0.03  &				       --	&	    0.51	$\pm$	     0.00	&	 81.07   $\pm$  	   0.21   	\\
		   &	 S\'{e}rsic bulge   &	 20.11   $\pm$     0.04  &	             2.33         $\pm$ 	   0.15  &	    4.73	$\pm$	     0.15	&	    0.35       $\pm$	   0.00 	&     74.28	     $\pm$	 0.12		\\
	3C5	   &	 exp disk   &		 20.70    $\pm$    0.00 	  &	     3.83           $\pm$	   0.02  &				       --	&	    0.65	$\pm$	     0.00	&	26.97   $\pm$  	   0.46   	\\
	3C6	   &	 exp disk   &		 20.83    $\pm$    0.00 	  &	     2.42           $\pm$	   0.01  &				       --	&	    0.60	$\pm$	     0.00	&	 -36.02   $\pm$  	   0.28   	\\

\hline
\label{tab4}
\end{tabular}
\end{table*}
\begin{table*}
\begin{tabular}{ccccccccccc}
\noalign{\medskip} \hline 
ID & comp & $F606W$ & $R_{\rm e}/R_{\rm d} ({\rm kpc})$ & $n$ & $q$ & PA \\
$(1)$ & $(2)$ &$(3)$ & $(4)$ &$(5)$ & $(6)$ &$(7)$  \\
\hline
	3C7	   &	 exp disk   &		 21.58    $\pm$    0.01 	  &	     3.59           $\pm$	   0.03  &				       --	&	    0.36	$\pm$	     0.00	&	  84.27   $\pm$  	   0.24   	\\
		   &	 S\'{e}rsic bulge   &	 25.08   $\pm$     0.05  &		     0.67         $\pm$ 	   0.02  &	    0.40	$\pm$	     0.08	&	    0.76       $\pm$	   0.03 	&     59.11	     $\pm$	 7.07		\\
	3C8	   &	 exp disk   &		 21.38    $\pm$    0.01 	  &	     2.30           $\pm$	   0.02  &				       --	&	    0.25	$\pm$	     0.00	&	-0.15   $\pm$  	   0.10   	\\
		   &	 S\'{e}rsic bulge   &	 21.91   $\pm$     0.01  &		     0.41         $\pm$ 	   0.01  &	    2.36	$\pm$	     0.06	&	    0.77       $\pm$	   0.01 	&     -0.01	     $\pm$	 1.27		\\
	3C9	   &	 exp disk   &		 22.04    $\pm$    0.01 	  &	     1.51           $\pm$	   0.01  &				       --	&	    0.21	$\pm$	     0.00	&	57.93   $\pm$  	   0.12   	\\
	3F1	   &	 exp disk   &		 20.76    $\pm$    0.09 	  &	     4.53           $\pm$	   0.35  &				       --	&	    0.47	$\pm$	     0.02	&	  88.06   $\pm$  	   2.63   	\\
	3F2	   &	 S\'{e}rsic         &	 23.86   $\pm$     0.01  &		     0.37         $\pm$ 	   0.01  &	    0.49	$\pm$	     0.03	&	    0.86       $\pm$	   0.01 	&     21.10	     $\pm$	 4.63		\\
	3F3	   &	 exp disk   &		 21.21    $\pm$    0.01 	  &	     1.52           $\pm$	   0.01  &				       --	&	    0.52	$\pm$	     0.00	&	  -85.25   $\pm$  	   0.40   	\\
	3F4	   &	 S\'{e}rsic galaxy   &	 22.07   $\pm$     0.04  &	             4.27         $\pm$ 	   0.23  &	    3.69	$\pm$	     0.11	&	    0.92       $\pm$	   0.01 	&	63.62	     $\pm$	 5.77		\\
       3F5	   &	 exp disk   &		 22.91    $\pm$    0.08 	  &	     0.62           $\pm$	   0.01  &				       --	&	    0.56	$\pm$	   0.01 	&	 -22.48   $\pm$  	   1.33   	\\
       	   	   &	 S\'{e}rsic bulge   &	 23.08   $\pm$     0.09  &		     0.88         $\pm$ 	   0.06  &	    3.09	$\pm$	     0.33	&	    0.74       $\pm$	   0.02 	&      -42.64	     $\pm$	 6.24		\\
	3F6	   &	 exp disk   &		 21.96    $\pm$    0.01 	  &	     2.16           $\pm$	   0.02  &				       --	&	    0.30	$\pm$	     0.00	&	 -69.12   $\pm$  	   0.20   	\\
	3F7	   &	 exp disk   &		 20.78    $\pm$    0.01 	  &	     3.34           $\pm$	   0.04  &				       --	&	    0.65	$\pm$	   0.01 	&	 -12.81   $\pm$  	   0.82   	\\
	3F8	   &	 exp disk   &		 20.47    $\pm$    0.00 	  &	     5.28           $\pm$	   0.03  &				       --	&	    0.64	$\pm$	   0.00 	&	 82.54   $\pm$  	   0.43   	\\
       3F9	   &	 exp disk   &		 20.85    $\pm$    0.01 	  &	     2.98           $\pm$	   0.01  &				       --	&	    0.34	$\pm$	   0.00 	&	64.12   $\pm$  	   0.15   	\\

       3F10	   &	 exp disk   &		 21.80    $\pm$    0.05 	  &	     7.51           $\pm$	   0.18  &				       --	&	    0.89	$\pm$	   0.02 	&	 -64.26   $\pm$  	   8.70   	\\
       	   	   &	 S\'{e}rsic bulge   &	 19.46   $\pm$     0.02  &	            17.68         $\pm$ 	   0.75  &	    8.00	$\pm$	     0.11	&	    0.60       $\pm$	   0.00 	&      -32.79	     $\pm$	 0.24		\\
	3F11	   &	 exp disk   &		 22.63    $\pm$    0.03 	  &	     2.74           $\pm$	   0.03  &				       --	&	    0.13	$\pm$	   0.00 	&	44.70   $\pm$  	   0.13   	\\
		   &	 S\'{e}rsic bulge   &	 21.14   $\pm$     0.01  &		     3.00         $\pm$ 	   0.09  &	    7.05	$\pm$	     0.18	&	    0.56       $\pm$	   0.01 	&     45.90	     $\pm$	 0.47		\\
       3F12	   &	 exp disk   &		 22.63    $\pm$    0.07 	  &	      0\farcs79     $\pm$	   0\farcs03   &			      --	&	    0.34	$\pm$	   0.01 	&	37.03   $\pm$  	   0.71   	\\
       	   	   &	 S\'{e}rsic bulge   &	 21.64   $\pm$     0.05  &	              0\farcs54         $\pm$ 	   0\farcs04  &	    4.02	$\pm$	     0.13	&	    0.63       $\pm$	   0.01 	&     35.39	     $\pm$	 0.70		\\
	4C1	   &	 exp disk   &		 22.16    $\pm$    0.01 	  &	     2.99           $\pm$	   0.03  &				       --	&	    0.27	$\pm$	   0.00 	&	 25.29   $\pm$  	   0.21   	\\
	4C2	   &	 exp disk   &		 21.88    $\pm$    0.00 	  &	     1.67           $\pm$	   0.01  &				       --	&	    0.59	$\pm$	   0.00 	&	-56.86   $\pm$  	   0.40   	\\
	4C3	   &	 exp disk   &		 22.40    $\pm$    0.01 	  &	     2.38           $\pm$	   0.06  &				       --	&	    0.86	$\pm$	   0.01 	&	 39.80   $\pm$  	   3.35   	\\
		   &	 S\'{e}rsic bulge   &	 22.63   $\pm$     0.03  &		     2.01         $\pm$ 	   0.01  &	    0.20	$\pm$	     0.01	&	    0.87       $\pm$	   0.01 	&     57.35	     $\pm$	 1.61		\\
	4F1	   &	 exp disk   &		 22.15    $\pm$    0.00 	  &	     1.14           $\pm$	   0.01  &				       --	&	    0.74	$\pm$	   0.01 	&	 -14.87   $\pm$  	   1.05   	\\
		   &	 S\'{e}rsic bulge   &	 23.03   $\pm$     0.01  &		     0.71         $\pm$ 	   0.00  &	    0.48	$\pm$	     0.02	&	    0.43       $\pm$	   0.00 	&      -6.59	     $\pm$	 0.49		\\
	4F2	   &	 exp disk   &		 24.75    $\pm$    0.11 	  &	     0.36           $\pm$	   0.01  &				       --	&	    0.11	$\pm$	   0.18 	&	 24.15   $\pm$  	   1.65   	\\
		   &	 S\'{e}rsic bulge   &	 23.27   $\pm$     0.01  &		     0.72         $\pm$ 	   0.01  &	    1.87	$\pm$	     0.08	&	    0.81       $\pm$	   0.01 	&     -36.89	     $\pm$	 3.40		\\
	4F3	   &	 S\'{e}rsic galaxy   &	 22.51    $\pm$    0.01 	  &	     4.13           $\pm$	   0.05  &		0.85	$\pm$	0.02       --	&	    0.79	$\pm$	   0.01 	&	 -32.88   $\pm$  	   1.64   	\\
	4F4	   &	 exp disk   &		 21.71    $\pm$    0.01 	  &	     3.66           $\pm$	   0.03  &				       --	&	    0.28	$\pm$	   0.00 	&	 59.17   $\pm$  	   0.23   	\\
	4F5	   &	 exp disk   &		 20.84    $\pm$    0.00 	  &	     6.46           $\pm$	   0.03  &				       --	&	    0.82	$\pm$	   0.01 	&	 -34.24   $\pm$  	   1.22   	\\
	4F6	   &	 exp disk   &		 21.71    $\pm$    0.04 	  &	     4.68           $\pm$	   0.07  &				       --	&	    0.60	$\pm$	   0.01 	&	 60.91   $\pm$  	   0.72   	\\
		   &	 S\'{e}rsic bulge   &	 23.60   $\pm$     0.29  &	             4.00         $\pm$ 	   1.18  &	    2.63	$\pm$	     0.34	&	    0.69       $\pm$	   0.03 	&     -66.50	     $\pm$	 4.63		\\
	4F7	   &	 exp disk   &		 21.57    $\pm$    0.01 	  &	     5.98           $\pm$	   0.05  &				       --	&	    0.40	$\pm$	   0.00 	&	 69.82   $\pm$  	   0.25   	\\
	4F8	   &	 exp disk   &		 20.59    $\pm$    0.00 	  &	     6.25           $\pm$	   0.04  &				       --	&	    0.67	$\pm$	   0.00 	&	  13.23   $\pm$  	   0.48   	\\
		   &	 S\'{e}rsic bulge   &	 24.26   $\pm$     0.27  &		     2.73         $\pm$ 	   1.66  &	    8.00	$\pm$	     2.54	&	    0.24       $\pm$	   0.03 	&      84.10	     $\pm$	 1.66		\\
	4F9	   &	 exp disk   &		 21.71    $\pm$    0.00 	  &	     3.21           $\pm$	   0.00  &				       --	&	    0.22	$\pm$	   0.00 	&	 16.39   $\pm$  	   0.00   	\\
       4F10	   &	 S\'{e}rsic galaxy  &	 22.64   $\pm$     0.01  &	             5.84         $\pm$ 	   0.06  &	    0.84	$\pm$	     0.02	&	    0.17       $\pm$	   0.00 	&      42.87	     $\pm$	 0.14		\\
       4F11	   &	 exp disk   &		 22.64    $\pm$    0.01 	  &	      0\farcs54     $\pm$	    0\farcs01  &			       --	&	    0.31	$\pm$	   0.00       &       -40.26   $\pm$		 0.44		\\
       4F12	   &	 exp disk   &		 20.39    $\pm$    0.02 	  &	     7.42           $\pm$	   0.04  &				       --	&	    0.71	$\pm$	   0.00 	&	-41.83   $\pm$  	   0.61   	\\
       4F13	   &	 exp disk   &		 21.84    $\pm$    0.01 	  &	     4.48           $\pm$	   0.04  &				       --	&	    0.30	$\pm$	   0.00       &        27.72   $\pm$		 0.22		\\
       4F14	   &	 exp disk   &		 21.14    $\pm$    0.01 	  &	     5.85           $\pm$	   0.05  &				       --	&	    0.78	$\pm$	   0.01       &       -32.58   $\pm$		 1.13		\\

\hline
\end{tabular}
\\Column (1): object ID;
Col.~(2): component;
Col.~(3): total magnitude;
Col.~(4): effective radius of the bulge/scale length of the disk;
Col.~(5): S\'{e}rsic index of the bulge profile;
Col.~(6): flattening;
Col.~(7): position angle of the disk measured from North through East.\\
The photometric zero point of the magnitude measurements is $Z_{\rm p}^{F606W}=26.398$.  The parameters of 2F14 are unreliable since there is a diffraction spike on
top of it.  The HST image of 3F7 is incomplete and therefore, its parameters are also unreliable.  The disk scale length of these galaxies is not used in the analysis. 
3F13 could not be analyzed because of the
diffraction spike on the image.
Col.~(4) is given in arcseconds in case the redshift of a galaxy is unknown.  
\end{table*}

\end{document}